%% file: aanda.tex
\documentclass[]{aa} 

\usepackage[]{graphicx}
\usepackage{subfig}
\usepackage{xcolor}
\usepackage[varg]{txfonts}
\usepackage{lscape}
\usepackage{natbib}
\usepackage{rotating}
\usepackage{multirow}
\usepackage{hhline}
\usepackage{makecell}

\usepackage[perpage,para,symbol]{footmisc}
\usepackage{tabularx}
\usepackage{booktabs}
\usepackage{caption}
\usepackage{threeparttable}
\usepackage{multirow}
\usepackage[abs]{overpic}
\usepackage{enumerate}
\usepackage[]{textpos}   
\usepackage{color}

\def\deg{\ifmmode^\circ\else$^\circ$\fi}
\def\alphaTF{\ifmmode{\alpha_{\mathrm{\,{\small TF}}}}\else{$\alpha_{\mathrm{\,{\small TF}}}$}\fi}

\def\Msun{\ifmmode{\mathrm M_\odot}\else{M$_\odot$}\fi}
\newcommand{\rbreak}{\ensuremath{R_\mathrm{break }}}
\newcommand{\mubreak}{\ensuremath{\mu_\mathrm{break}}}
\newcommand{\hi}{\ensuremath{h_{\mathrm{i}}}}
\newcommand{\mui}{\ensuremath{\mu_{0,\mathrm{i}}}}
\newcommand{\ho}{\ensuremath{h_{\mathrm{o}}}}
\newcommand{\muo}{\ensuremath{\mu_{0,\mathrm{o}}}}
\newcommand{\risoph}{\ensuremath{R_\mathrm{23}}}
\newcommand{\magarc}{mag arcsec$^{\mathrm{-2}}$}
\usepackage{txfonts}
\begin{document}

\title{Evolution of the anti-truncated stellar profiles of S0 galaxies\\ 
since $z=0.6$ in the SHARDS survey}
\subtitle{I - Sample and Methods}
\titlerunning{Anti-truncated stellar profiles on S0 galaxies at $0.2<z<0.6$}
\authorrunning{Borlaff et al.}

\author{Alejandro Borlaff\inst{1,2,3}, M.~Carmen Eliche-Moral\inst{1,2}, John E. Beckman\inst{1,3,4}, Bogdan C. Ciambur\inst{5},\\ Pablo G. P\'{e}rez-Gonz\'{a}lez\inst{2}, Guillermo Barro\inst{6}, Antonio Cava\inst{7}, Nicolas Cardiel\inst{2,8}} 

\institute{
Instituto de Astrof\'{i}sica de Canarias, C/ V\'{i}a L\'actea, E-38200 La Laguna, Tenerife, Spain
\\\email{asborlaff@iac.es}
\and
Departamento de Astrof\'{\i}sica y CC.~de la Atm\'osfera, Universidad Complutense de Madrid, E-28040 Madrid, Spain
\and
Facultad de F\'{i}sica, Universidad de La Laguna, Avda. Astrof\'{i}sico Fco. S\'{a}nchez s/n, 38200, La Laguna, Tenerife, Spain
\and
Consejo Superior de Investigaciones Cient\'{i}ficas, Spain
\and 
Centre for Astrophysics and Supercomputing, Swinburne University of Technology, Hawthorn, VIC 3122, Australia
\and
University of California, 501 Campbell Hall, Berkeley, CA 94720 Santa Cruz, USA
\and
Observatoire de Gen{\`e}ve, Universit{\'e} de Gen{\`e}ve, 51 Ch. des Maillettes, 1290 Versoix, Switzerland
\and
Instituto de Física de Cantabria (CSIC - Universidad de Cantabria), Avenida de los Castros s/n, 39005 Santander, Spain
}
  \abstract
   {The controversy about the origin of the structure of early-type S0--E/S0 galaxies may be due to the difficulty of comparing surface brightness profiles with different depths, photometric corrections and PSF effects (almost always ignored).}
   {We aim to quantify the properties of Type-III (anti-truncated) discs in a sample of S0 galaxies at $0.2<z<0.6$. In this paper, we present the sample selection and describe in detail the methods to robustly trace the structure in their outskirts and correct for PSF effects.}
   {We have selected and classified a sample of 150 quiescent galaxies at $0.2<z<0.6$ in the GOODS-N field. We perform a quantitative structural analysis of 44 S0--E/S0 galaxies. We have corrected their surface brightness profiles for PSF distortions and analysed the biases in the structural and photometric parameters when the PSF correction is not applied. Additionally, we have developed {\tt{Elbow}}, an automatic statistical method to determine whether a possible break is significant - or not - and its type and made it publicly available.}
   {We found 14 anti-truncated S0--E/S0 galaxies in the range $0.2<z<0.6$ ($\sim 30$\% of the final sample). This fraction is similar to the those reported in the local Universe. In our sample, $\sim25\%$ of the Type-III breaks observed in PSF-uncorrected profiles are artifacts, and their profiles turn into a Type I after PSF correction. PSF effects also soften Type-II profiles. We found that the profiles of Type-I S0 and E/S0 galaxies of our sample are compatible with the inner profiles of the Type-III, in contrast with the outer profiles.}
   {We have obtained the first robust and reliable sample of 14 anti-truncated S0--E/S0 galaxies beyond the local Universe, in the range $0.2<z<0.6$. PSF effects significantly affect the shape of the surface brightness profiles in galaxy discs even in the case of the narrow PSF of HST/ACS images, so future studies on the subject should make an effort to correct them.}

   \keywords{Methods: observational -- methods: statistical -- galaxies: fundamental parameters -- galaxies: elliptical and lenticular, cD  --  galaxies: structure}
   \maketitle
%
%

\section{Introduction}
\label{Sec:Intro}
The origin and evolution of lenticular (or S0) galaxies is still a matter of debate. Traditionally they were classified as the natural transition between the elliptical galaxies and more complex galaxy types; spirals, in the Hubble morphological sequence \citep{1959HDP....53..275D,1961hag..book.....S}. The cause of this is that S0 galaxies usually appear without any spiral structure or signs of recent star formation in their discs. 
\citet{1970ApJ...160..811F} pointed out that the discs of spiral and S0 galaxies usually show surface brightness profiles that are (to first order approximation) reasonably well-fitted by an exponential function out to a certain radius, due to the stellar density decline as a function of galactocentric radius \citep{1963BAAA....6...41S}. 

Nevertheless, as deeper observations allowed the study of the outskirts of galactic discs, it was found that many lenticular galaxies do not follow a purely exponential profile along their whole observable radius \citep{2009IAUS..254..173S,2012ApJS..198....2K,2012A&AT...27..313I}. \citet{2005ApJ...626L..81E} pointed out that a significant fraction of S0 galaxies present light excesses in the outskirts of the discs, which also show an exponential decline but with a shallower slope than their inner discs. \citet[][E08 hereafter]{2008AJ....135...20E} expanded the stellar disc classification of galaxies from \citet{1970ApJ...160..811F}, according to the profile structure. Type-I discs are well modelled with a single radial exponential profile. Type-II galaxies present a down-bending profile, i.e. a brightness deficit in the outer parts of the disc with respect to the extrapolated trend of the inner regions beyond a given break radius. Type-III discs becomes less steep outside the break radius than the extrapolation of the exponential trend of the inner parts \citep[anti-truncation, see][]{2006A&A...454..759P}. E08 and \citet[][G11 hereafter]{2011AJ....142..145G} showed that the fraction of anti-truncated discs was higher for S0 galaxies that for any other morphological type, increasing from $\sim 10-20\%$ in Sc--Sd galaxies, up to $\sim 20-50\%$ in S0--Sa galaxies \citep[see also][]{2012A&AT...27..313I,2015MNRAS.447.1506M}. In the present study we will focus on the Type-III profiles.

A high fraction of the anti-truncated profiles are due entirely to disc structure, although $\sim 15\%$ might be due to outer stellar haloes \citep{2012MNRAS.419..669M}. In some cases, transitions between the inner and outer profiles are associated with structural components of the galaxy such as rings or lenses \citep{2014MNRAS.441.1992L}. Some authors \citep[e.g.][]{2012ApJ...759...98C} pointed out that these transitions might be caused by combinations of thin+thick discs with different radial scale-lengths. 
The role of environmental density in the formation of these structures is still a matter of debate. Some authors do not find a significant correlation between the presence of Type-III profiles and the environment \citep{2012MNRAS.419..669M} while other authors do \citep{2014MNRAS.441.1992L}. A recent study \citep{2016arXiv160508845P} found that the outer scale-lengths of both Type-II and Type-III profiles are $\sim 10\%$ larger in the cluster environment compared to the field. These authors also suggest that the Type-I profiles would be an intermediate step in the transformation of Type-II into Type-III profiles. Nevertheless, Type-III profiles are ubiquitous regardless of their environment \citep{2012ApJ...744L..11E,2016arXiv160508845P}, and the observed suppression of Type-II profiles in the clusters argues against a common origin for both Type-II and Type-III profiles. On the contrary, this suggests that the two types of profiles are caused by completely different phenomena \citep{2012ApJ...744L..11E,2012ApJ...758...41R,2014MNRAS.441.1992L}. 

Radially varying profiles of star formation have been proposed as a cause for both Type-II and Type-III profiles \citep{2006ApJ...636..712E}. However, most of the proposed mechanisms to explain Type-III profiles are based on different modes of gravitational interactions, such as close encounters \citep{2006ApJ...650L..33P}, accretion of dark matter sub halos \citep{2009ApJ...700.1896K} or minor mergers \citep{2007ApJ...666..647Y}. In addition, \citet{2015MNRAS.448L..99H} also proposed radial mass redistribution as a cause for the different types of profiles due to different initial angular momentum of the host halo. \citet{2014A&A...570A.103B} tested whether major mergers can produce anti-truncated stellar profiles in S0 galaxies using hydrodynamical N-body simulations. They found that Type-III profiles of S0 remnants can be produced after a major merger event, and that those profiles obey similar scaling relations to those found in observed Type-III S0 galaxies. In a later paper, \citet{2015A&A...580A..33E} reported that these relations are similar to those found in Type-III spiral galaxies, which suggests that fundamental processes must be responsible for the formation of these structures in disc galaxies along the whole Hubble Sequence. However, no study has analysed the properties of anti-truncated S0 and E/S0 galaxies beyond the local universe to learn about their possible evolution, because cosmological dimming efficiently moves these structures towards even fainter (and prohibitive) surface brightness levels.  

The controversy that arises out of the comparison between different datasets may be due to the difficulty of comparing samples with different depths and surface brightness profile corrections. Measuring robust surface brightness profiles down to $\mu_{lim}\sim\, 27-28\,$mag arcsec$^\mathrm{-2}$ is a challenging task. \citet{2008MNRAS.388.1521D} showed that the detection of an apparent halo around the minor axis of an edge-on galaxy in the Hubble Ultra Deep Field can be largely explained by scattered light from the inner regions. This light spreading follows a point spread function (PSF, hereafter) whose shape depends on the telescope, instrument, filter and even time \citep{2014A&A...567A..97S,2015A&A...577A.106S}. It modifies the true light distribution of the object as a convolution, ``blurring'' the resulting image. The PSF effect scales with the intensity of the source, and it usually becomes smaller rapidly with increasing radii. Nevertheless, the contribution of the outer wings of the PSF profile can be significant in the outskirts of galaxies, where the light intensity is very faint. This would produce apparent light emission excesses where they otherwise would not be found. Thus estimating and correcting the PSF effect is a crucial step for any study of surface brightness profiles in the outskirts of galaxies, especially when looking for Type-III profiles.

In order to estimate the surface brightness profile of an object, the observer must know the shape of the PSF out to (at least) $1.5$ times the maximum size of the object in radius \citep{2014A&A...567A..97S}. To estimate precise measurements of the structural parameters of galaxies, it is not enough to simply scale the PSF profile to the central surface brightness of the profile \citep{2008MNRAS.388.1521D}. The observer must subtract the PSF contribution created by the whole 2D light distribution (or at least by a realistic model)  not only by the innermost region but also from the outer regions of the object. Direct image deconvolution methods such as the Richardson-Lucy \citep{2012A&A...539A.133P} are tempting, but inefficient (see \citet{2016arXiv161205122K} for a recent and less time-consuming method of direct deconvolution) and may lead to disruptions of the low surface brightness regions. In contrast to these, recent papers propose accurate methods for removing the PSF contribution by fitting 2D models of the galaxy taking into account the PSF \citep{2016ApJ...823..123T}. This method permits the derivation of reliable surface brightness profiles down to $\sim 31$ \magarc, according to these authors. In the present work we will follow a similar procedure, even though our profiles are limited to radii where the profiles are $3-4$ \magarc\ brighter, where the effects of the PSF are less significant. 

In order to shed light on the evolution of these structures, we have selected a sample of galaxies from the red sequence and then we have identified the S0 and E/S0 galaxies at $0.2 < z < 0.6$ by studying their star formation rate, morphology and analysing their surface brightness profiles. We have quantified the properties of Type-III discs in this sample and compared them to the Type-III profiles from the samples at $z\simeq0$. In this paper, we present the sample selection and describe in detail the statistical methods to trace the structure in their outskirts and correct for PSF effects. We will analyse the scaling relations at $0.2 < z < 0.6$ and compare them with their local counterparts in a forthcoming paper (Borlaff et al., in prep.).

The outline of this paper is as follows. The methodology is described in detail in Sect. 2. The results are presented and discussed in Sect. 3. The final conclusions can be found in Sect. 4. We tabulate the general properties of the initial red galaxy sample in Appendix A. Appendix B contains the profile classifications for the S0 -- E/S0 sample and their structural and photometric parameters. Appendix C describes the final sample of anti-truncated S0 and E/S0 galaxies individually and shows the RGB images used for the morphological classification as well as the decompositions performed to the profiles. We assume a concordant cosmology \citep[$\Omega_{\mathrm{M}} = 0.3,\Omega_{\mathrm{\Lambda}}=0.7, H_{0}= $70 km s$^{-1}$ Mpc$^{-1}$, see][]{2007ApJS..170..377S}. All magnitudes are in the AB system unless otherwise noted.

\section{Methods}
\label{Sec:Methods}

\begin{figure*}[]
\begin{center}
\includegraphics[width=0.96\textwidth]{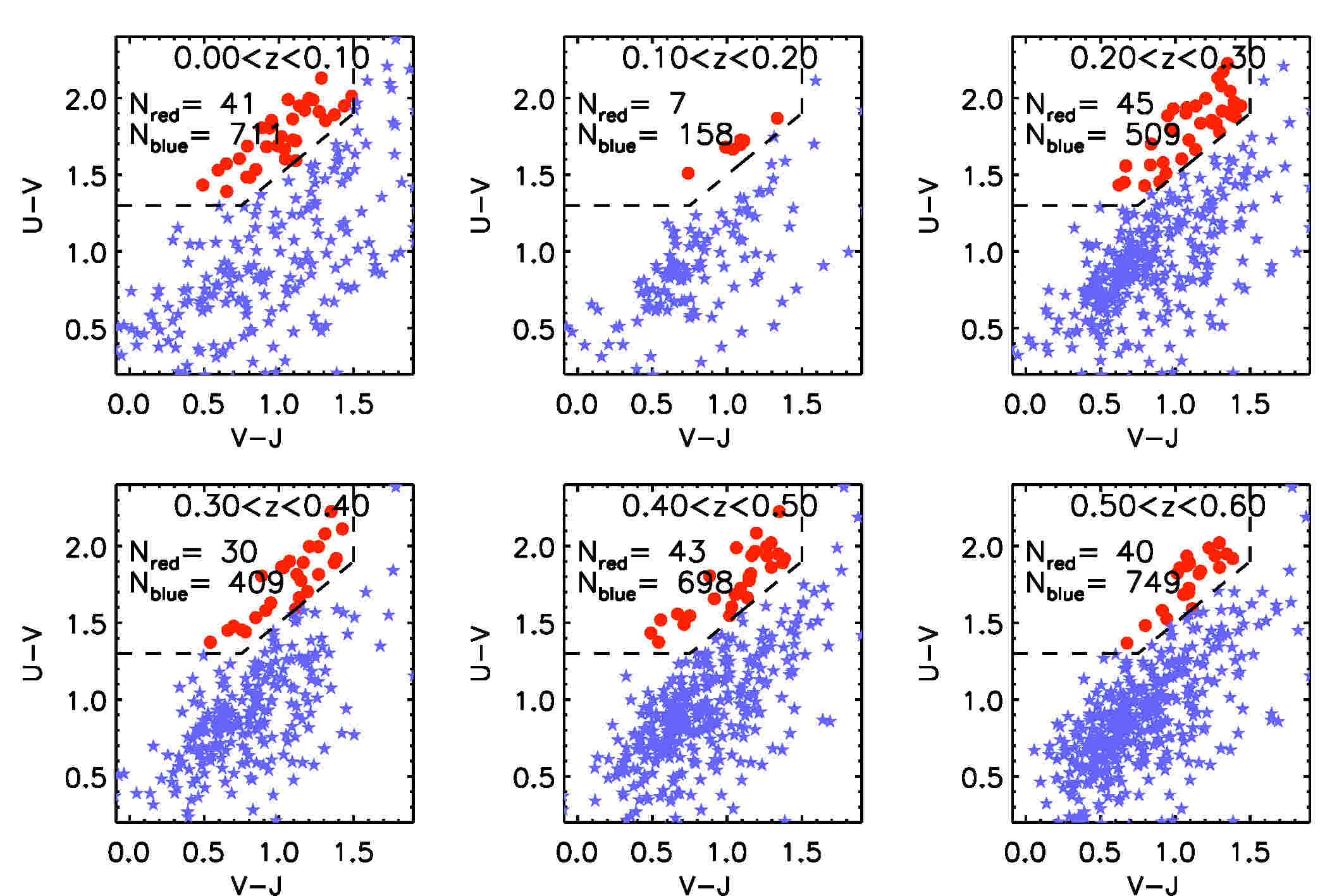}
\caption[$(U-V) vs. (V-J)$ colour-colour diagram sample selection]{Initial sample selection with the $(U-V)$ vs. $(V-J)$ colour-colour diagram by redshift intervals up to $z=0.6$ \emph{Dashed line}: \citet{2011ApJ...735...86W, 2012ApJ...754L..29W} boundaries for identifying the quiescent galaxy population and the star-forming galaxies (see eqs.\,\ref{Whitaker1}, \ref{Whitaker2} and \ref{Whitaker3}). \emph{Red dots}: Quiescent galaxies from the initial SHARDS sample. \emph{Blue stars}: Star-forming galaxies from the initial SHARDS sample. The number of objects selected in each class and interval of redshift are given on the panels. [\emph{A colour version of the figure is available in the online edition.}]} 
\label{fig:UVJ_all}
\end{center}
\end{figure*}

In order to study the properties of the surface brightness profiles of non-local S0 -- E/S0 galaxies in the GOODS-N field at $z<0.6$, first we had to create a sample of quiescent candidates, that must include S0 and E/S0 galaxies, by definition. The data used are presented in Sect. 2.1 and the selection of a sample of quiescent candidates with $z<0.6$ is described in Sect.\,\ref{Subsec:Red_sample}. Secondly, we performed a visual morphological classification based on the data available for the selected objects (2D structure, presence of AGN activity, star formation rate (SFR), total stellar mass and surface brightness profile), as described in Sects.\,\ref{Subsec:Red_sample} and \,\ref{Subsec:AGN}. Thirdly, we modelled each object and corrected any possible PSF contributions to the surface brightness profiles. After this, we estimated the necessary corrections for dust extinction, cosmological dimming and K-correction for each S0 and E/S0 object (Sects.\,\ref{Sec:Photometric_corrections}). This is explained in detail in Sects.\,\ref{Subsec:Modelling}--\ref{Subsec:Methods_profiles} and \ref{Subsec:reliability_PSF_correction}. Finally, we performed the identification, characterisation and analysis of the structure of the components in the surface brightness profiles as described in Sect.\,\ref{Subsec:Elbow}. The efficiency and reliability of the PSF correction performed to the galaxies in our sample is checked in Sect.\,\ref{Subsec:reliability_PSF_correction}.


\subsection{Data description}
\subsubsection{SHARDS and the Rainbow Database}
\label{Subsec:SHARDS}

The Survey for High-$z$ Absorption Red and Dead Sources \citep[SHARDS,][]{2013ApJ...762...46P} is an ESO/GTC Large Program carried out with the OSIRIS instrument on the 10.4\,m Gran Telescopio Canarias (GTC). This survey obtained data between 5000 \AA\ and 9500 \AA\ for galaxies in the GOODS-N field down to magnitudes $m$ < 26.5 AB mag, in 25 medium band filters (FWHM $\sim$ 170 \AA). The main goal of this survey was to study the properties of stellar populations of massive galaxies in quiescent evolution at $z$ = 1.0 - 2.3 through the pseudospectra resulting from the data of this photometric medium band filter system (the set is equivalent to low resolution integral field spectroscopy, R $\sim50$) and deep ancillary broad-band data. The wavelengths include key absorption indexes such as Mg(UV) ($\lambda \sim 2800$ \AA), essential for the study of high-$z$ early-type galaxies. This technique allowed the SHARDS team to select quiescent objects, calculate star formation histories (SFH, hereafter) and photometric redshifts with a precision better than $\Delta z /(1+z) \sim\ 0.007$ \citep{2016MNRAS.457.3743D}. See \citet{2013ApJ...762...46P} for a detailed description. 

SHARDS data are available through the Rainbow Database \citep{2011ApJS..193...13B,2011ApJS..193...30B} in compilation with multiple photometric and spectroscopic data for several cosmological fields, such as GOODS-N and GOODS-S, COSMOS, or the Extended Groth Strip. The authors used all the available photometry to build spectral energy distributions (SEDs) from X-ray to radio. Using the SED fitting procedure, they derived photometric redshifts and estimates of parameters such as the stellar mass, the UV- and IR-based SFRs, the stellar population age and rest-frame magnitudes in different filters \citep{2005ApJ...630...82P,2008ApJ...675..234P,2013ApJ...762...46P}.

In this study, we have selected a sample of GOODS-N galaxies included in the SHARDS survey, that: 1) have low star formation rates, compatible with being early-type galaxies, 2) have total stellar masses similar to those of the available local samples ($M \sim 10^{\mathrm{9}}-10^{\mathrm{11}} M_\odot$), and 3) have morphologies compatible with being S0 -- E/S0 objects. We restricted the selection only to the objects with reliable SHARDS data because: 1) we wanted to ensure that the S0 and E/S0 galaxies included in the sample are really quiescent and have stellar masses comparable with the local sample, and 2) we aimed to create a sample of anti-truncated S0 and E/S0 galaxies that can be used in a forthcoming study to investigate the properties of the stellar populations, metallicities and star formation rates of their inner and outer profiles. We have used SHARDS data release 2 in the database (iDR2beta).

\subsubsection{HST/ACS data in the GOODS-N field}
\label{Subsec:GOODS-N}
We have derived, corrected and analysed the surface brightness profiles using the deepest images available in our dataset tracing the rest-frame $V$ and $R$ bands, i.e., those in the F775W band of HST/ACS. The GOODS-N field is a large cosmological field with HST/ACS imaging centred on the Hubble Deep Field North \citep{2003mglh.conf..324D}. The spatial resolution of these images is 0.06 arcsec/px and the $1\sigma$ fluctuation level of the surface brightness on a area of 1 arcsec$^{\mathrm{2}}$ is $27.7$ mag. The average FWHM for the F775W filter mosaic is 0.11 arcsec \citep[][see Table 6]{2014ApJS..214...24S}. In order to derive the surface brightness profiles for the SHARDS objects, we relied on the ACS F775W mosaic from the ACS V2.0 data release \citep[HST Cycle 11, program IDs 9425 and 9583,][]{2004ApJ...600L..93G} available at the 3D-HST project webpage\footnote[4]{3D-HST - A Spectroscopic Galaxy Evolution Survey with the Hubble Space Telescope: http://3dhst.research.yale.edu/Home.html} \citep{2014ApJS..214...24S}. The zeropoints of the images are available in Table 6 of \citeauthor{2014ApJS..214...24S}, such that:
\begin{equation} \label{HSTmagnitude}
\mu_{\mathrm{F775W}} = -2.5 \cdot \log_{10}(\mathrm{counts}/\mathrm{area})+25.671
\end{equation}

The reduction process of the raw data was performed by using the CALACS software, available at the Space Telescope Science Institute site \citep{1999AAS...194.0807P}. This pipeline accounts for the bias subtraction, gain correction, and flat-fielding correction of each individual exposure. The corrected (flt.fits) individual images were then drizzled to the final mosaic through the Multidrizzle software package \citep{2003hstc.conf..337K}, and corrected for geometric distortion at the same time, removing cosmic-rays and performing sky-subtraction. For more details on the process of reduction and the data, we refer the reader to \citet{2004ApJ...600L..93G}.

We set the upper limit on redshift for the initial sample at $z=0.6$, in order to cover; with the F775W filter from HST/ACS, a similar wavelength range as the local universe observations we wanted to compare with (usually, in the $V$ or $R$ bands). This band traces similar wavelengths as the $R$ band from $z=0.1$ to $z\sim0.3$ and is equivalent to F606W (tracing approximately $V$) between $z=0.2$ and $z=0.6$. Thus, we selected the initial sample within $0.2<z<0.6$. In order to compare the surface brightness profiles between the local sample and ours, we also performed a photometric K-correction to our data to transform them into the rest-frame $R$ filter (see Sect.\,\ref{Sec:Photometric_corrections}).

\subsection{Red sample selection and morphological classification}
\label{Subsec:Red_sample}
In the local universe and beyond, galaxies are divided into two main classes: star-forming galaxies and quiescent galaxies \citep{2000ApJ...541...95V, 2006ApJ...647..853W, 2007MNRAS.377.1717K, 2009ApJ...691.1879W,2011ApJ...739...24B,2011ApJ...735...86W}. These two populations of galaxies show very different distributions in rest-frame colour-magnitude diagrams: quiescent galaxies usually have high luminosities and red colours while the star-forming galaxies present lower luminosities and bluer colours. This means that the two populations of galaxies appear as two well separated sequences in the $(U-V)$ vs. $(V-J)$ colour-colour diagram, so we can select a quiescent sample (that includes S0 and E/S0 galaxies, in which we are interested) by using the following boundary-relations \citep{2011ApJ...735...86W, 2012ApJ...754L..29W}: 

\begin{equation} \label{Whitaker1}
  U - V > 0.8 \times (V - J) + 0.7
\end{equation}
\begin{equation} \label{Whitaker2}
  U - V > 1.3 
\end{equation}
\begin{equation} \label{Whitaker3}
  V - J < 1.5  
\end{equation}

The galaxies selected within these boundaries present very red colours, either because they depleted their gas reservoir and cannot create new stars or because they are extremely extinguished by interstellar dust. We selected all sources from the GOODS-N field in the SHARDS catalogue (excluding those classified as stars) which present redshifts $0.2<z<0.6$, and we have used the previous boundaries in the colour-colour diagram to select those in the red sequence. We have excluded objects with $z<0.2$ in the catalogue because of the high uncertainties detected in the estimates of
the photometric redshifts at these distances \citep{2011ApJS..193...30B}. In addition to this, we used visual morphology classification as well as NIR and UV SFR to prevent reddened spiral and irregular galaxies from being included in the sample of quiescent galaxies (see Sect.\,\ref{Subsec:Results_SFR}). 
In Fig.\,\ref{fig:UVJ_all} we show an example of this first selection process. Rest-frame $U$, $V$ and $J$ magnitudes as well as the redshifts and stellar masses were taken from the Rainbow Database (see Sect.\,\ref{Subsec:SHARDS}). The panel shows in red dots those SHARDS objects within the red sample according to the criteria outlined previously. With blue stars, we represent the discarded objects. For the sake of completeness we show the diagrams corresponding to the interval $z=[0, 0.6]$ with $\Delta z=0.1$. We removed 8 objects with $0.2<z<0.6$: 7 of them were artifacts and spurious detections (SHARDS10007608, SHARDS10009632, SHARDS10012462, SHARDS10012556,  SHARDS20013873, SHARDS20001446 and SHARDS20009949) and 1 object was clearly affected by nearby star contamination (SHARDS20008526). Finally we identified 150 objects as "red" (mostly quiescent, as we will show in Sect.\,\ref{Subsec:Results_SFR}) in total between $0.2<z<0.6$. The sample presents median value of $z=0.41$. We note that the $\sim 75\%$ of the objects have $z>0.27$. The red objects in our sample and their main properties (coordinates, morphological type, photometric and spectroscopic redshifts, stellar mass, SFR, rest-frame absolute $K$ and $V$ magnitudes, extinction and K-correction for the F775W filter) are provided in Table \ref{tab:redsample} in Appendix \ref{Appendix:redsample}.

\begin{figure*}[]
 \begin{center}
\includegraphics[width=0.88\textwidth]{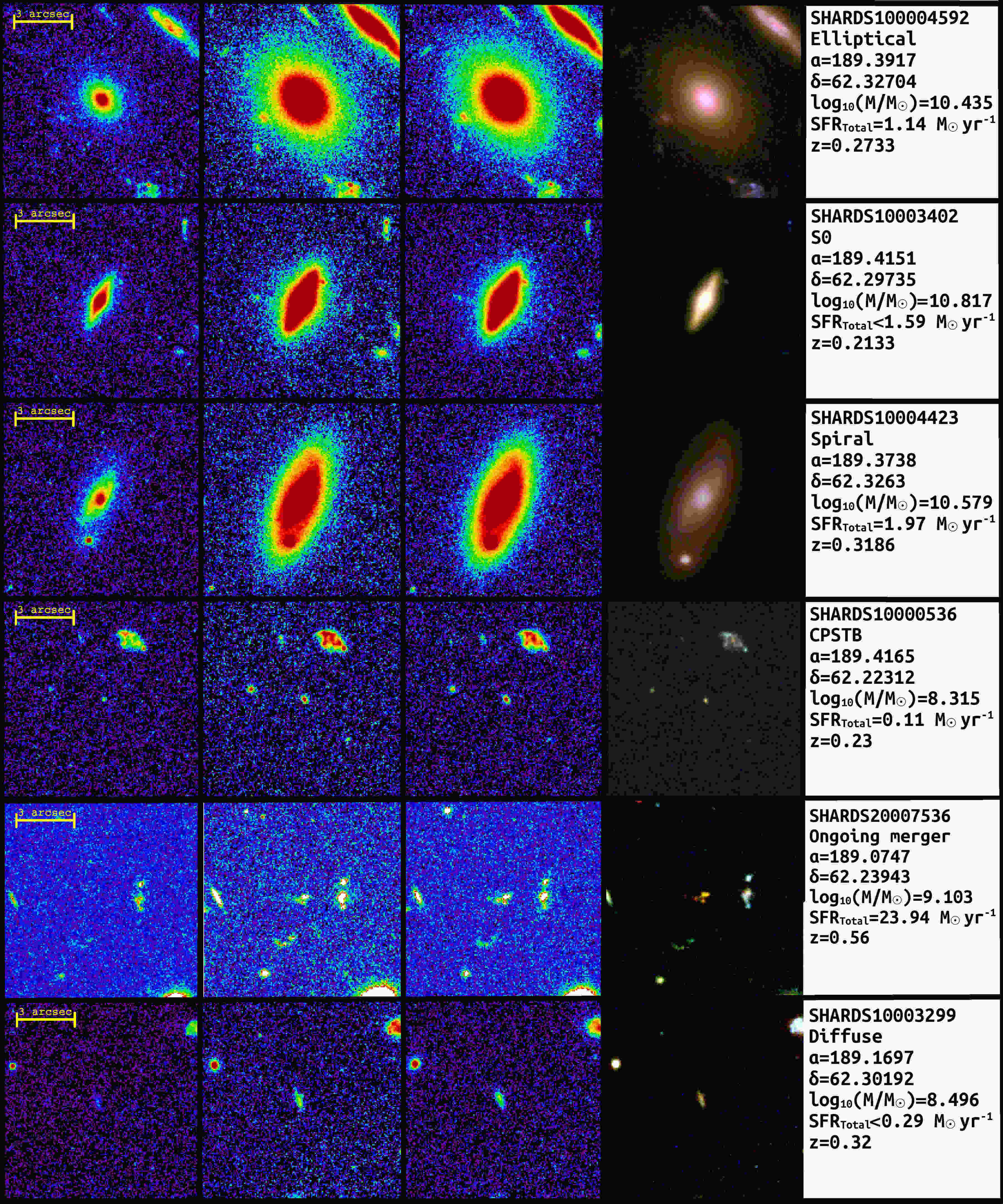}
\caption[]{F435W, F606W and F775W images of some objects from the initial selected red sample, used for the morphological classification. \emph{Columns, from left to right:} 1) F435W image, 2) F606W image, 3) F775W image, 4) RGB colour image, using the previous filters for the blue, green and red colours, respectively, 5) information panel with the identifier of the object in the SHARDS data release 2, position ($\alpha, \delta$), total stellar mass, SFR and $z$. [\emph{A colour version of the figure is available in the online edition.}]
}
\label{fig:morph_all}
\end{center}
\end{figure*}

We have performed a visual morphological classification of the objects based on the HST/ACS images in the F435W, F606W and F775W bands, and a false colour combined image of these three ones. In Fig.\,\ref{fig:morph_all} we present a summary of the morphological classification process by showing the available information of one object of each morphological class that we have considered, which we describe below. From left to right, the first three columns represent the images of each object in the three filters used for this analysis: F435W (blue), F606W (green) and F775W (red). We used the F435W filter in order to detect possible star formation regions that may not be detectable in the other redder filters. This allowed us to discard from the red sample the dust-reddened spirals that had not been rejected using the colour-colour criteria, or contamination sources which were very bright in the bluer bands but difficult to identify in the redder bands, such as small satellites, tidal tails or minor mergers in the outskirts of the objects. The RGB images shown in the fourth column were created by the combination of the previous three filters. These RGB images were used for the morphological classification, together with the individual filter images. In addition, preliminary surface brightness profiles (i.e, without PSF correction) as well as some characteristic parameters of each object (see the right column of Fig.\,\ref{fig:morph_all}) were used for the classification process. The images used during the classification presented a field-of-view (FoV, hereafter) of $7 \times\ R_{\mathrm{kr}}$ were $R_{\mathrm{kr}}$ is the Kron radius \citep{1980ApJS...43..305K} of the object (available from the Rainbow Database, see Sect.\,\ref{Subsec:SHARDS}). It was defined in this way in order to show a sufficient portion of the surrounding areas around the object to identify possible interactions, close companions or sources of diffuse light contamination. 

We classified the objects of the initially selected red sample morphologically according to the following criteria (see an example of each morphological type in Fig.\,\ref{fig:morph_all}):
\begin{itemize}
\item A. Morphological types assigned to well-resolved galaxies:\\
  \begin{enumerate}
  
  \item \textbf{Elliptical galaxies (E):} They do not present any signs of a disc component, but a smooth light distribution surrounding a bright compact core with very low ellipticity. The surface brightness profiles usually decrease with a steeper and non-constant slope when compared to an exponential profile. The surface brightness profiles are usually well-fitted with a single S\'{e}rsic profile with index $n \sim\ 3-5$. They do not show appreciable star formation signs, bars, or spiral arms.   

  \item \textbf{S0 -- E/S0 galaxies:} They show a noticeable disc component with a smooth light distribution, without any signs of spiral structure or star-forming regions. For those objects where the dominant component is the disc, we have assigned the S0 classification. On the contrary, the objects with a dominant central bulge or halo were classified as E/S0. Some S0 galaxies at low inclination may be mistaken for elliptical galaxies in the visual classification. To avoid contamination between the types, we refined our visual classification by using the surface brightness profiles and the multi-component {\tt{GALFIT3.0}} analysis, as detailed in Sect.\,\ref{Subsec:Modelling}. 
  
  \item \textbf{Spiral galaxies (Sp):} They present a disc component and a spiral pattern with clear star forming regions detected in the F435W band. 
  
  \item \textbf{Ongoing mergers (OM):} These objects present highly irregular and distorted light distributions in the images. The progenitors can still be identified separately. Close pairs without clear signs of tidal interaction were not included in this category and we have classified them as two independent galaxies. 
  
  \end{enumerate}
   \vspace{0.25cm}
   
\item B. Morphological types associated with low $S/N$ or low resolution data:\\
  \begin{enumerate}
  
  \item \textbf{Compact post-starbursts (CPSTB)} and \textbf{green peas (GP):} These are very compact objects. The spatial resolution is not enough to resolve their structure or different components. Some of these objects present recent merger signs. The CPSTBs present extremely red colours, possibly caused by extreme dust extinction. In contrast, the GPs usually present green colours in the false RGB images used to classify them. We adopted the ``green peas'' nomenclature following \citet{2009MNRAS.399.1191C}, but in order to confirm the objects within this class as real GPs according to their definition we would require specific observations, which are beyond the scope of this paper. 
  
  \item \textbf{Diffuse galaxies (DF):} We classify as diffuse galaxies those objects with very low $S/N$ ratio, usually with elongated morphologies. Faint spirals or interacting disc galaxies are possible candidates to appear as diffuse galaxies. They fall into the red sequence probably because of dust extinction. The surface brightness profiles are too faint and noisy due to the low apparent luminosities.  
  \end{enumerate}
\end{itemize}

As explained before, S0 galaxies can be confused with elliptical galaxies, especially those at low inclination. The discs of spiral and S0 galaxies usually show surface brightness profiles that follow an exponential law \citep{1970ApJ...160..811F} out to a certain radius. In contrast, elliptical galaxies show a steeper surface brightness decline, usually well fitted by a $n\sim 4$ Sérsic profile (i.e. a de Vaucouleurs profile). We exploited these structural characteristics to distinguish S0 from E galaxies even at low inclination. Thus, for the elliptical, S0 and E/S0 galaxies identified visually, we refined the morphological classification using the surface brightness profiles, as well as the 2D decomposition, in order to ensure that we are not including elliptical objects in the final S0 -- E/S0 sample (see Sect.\,\ref{Subsec:Modelling}). We found 50 S0 and E/S0 galaxies within our initial sample of 150 red galaxies (38 S0 and 12 E/S0). We could correct for PSF effects 44 out of these 50 S0 and E/S0 galaxies, as explained in Sect.\,\ref{Subsec:Modelling}. The remaining objects were removed from the final sample because they did not present stable bulge+disc 2D decompositions (SHARDS10001928, SHARDS10002901, SHARDS10005029,
SHARDS10008552, SHARDS20000858 and SHARDS20004273). The morphologies of our red galaxy sample and their global properties are described in detail in Sects. \,\ref{Subsec:Results_statistics} and \ref{Subsec:Results_SFR}. The final morphological classification for all objects in our red sample is provided in Table \ref{tab:redsample} of Appendix \ref{Appendix:redsample}. The RGB images of the objects finally classified as S0 and E/S0 that could be corrected for PSF effects are provided in Appendix \ref{Sec:TypeIIIcomments} (complete version available in colour in the online edition).

\subsection{GALFIT modelling and PSF correction}
\label{Subsec:Modelling}

\begin{figure}[]
\begin{center}
\includegraphics[width=0.5\textwidth,clip, trim=1.3cm 1cm 0.0cm 1.5cm]{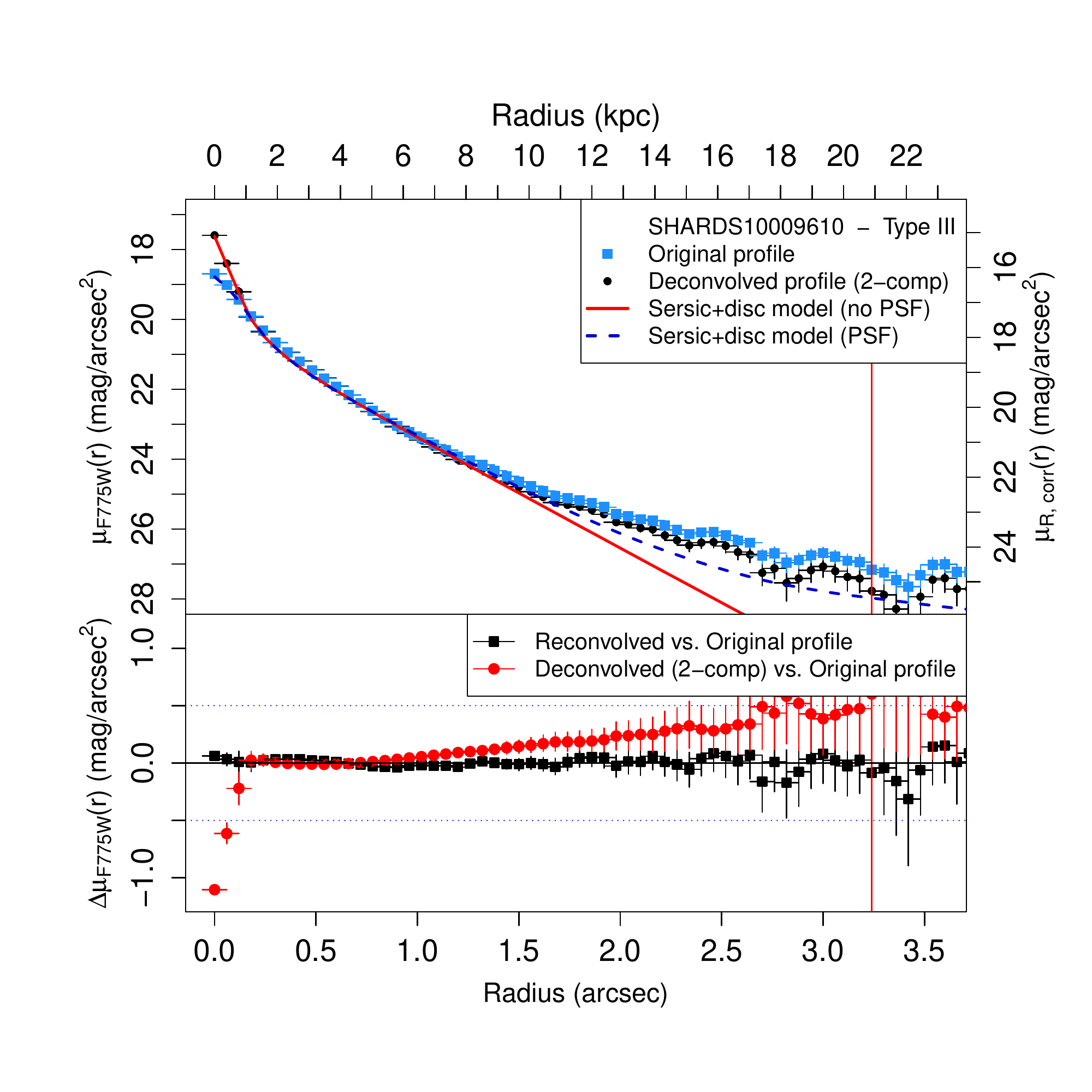}
\caption{\textbf{Top panel:} surface brightness profiles of the Type-III S0 galaxy SHARDS10009610. \emph{Blue squares:} surface brightness profile of the original image. \emph{Black dots:} the same for the PSF-corrected image (black). We represent the magnitude scale for the observed F775W band (left axis) and for the rest-frame $R$ band (right axis). The red solid and blue dashed lines correspond to the models fitted during the deconvolution and used for checking the visual morphological selection (see the legend). \textbf{Bottom panel:} Residuals between the original and corrected profiles. \emph{Red circles:} difference between the original and the PSF-corrected profiles. Notice that the PSF effects create a systematic upturn beyond $\sim 1.0$ arcsec. \emph{Black squares:} difference between the original profile and the reconvolved PSF-corrected image profile. The vertical red line represents the limiting radius. [\emph{A colour version of the figure is available in the online edition.}]} 
\label{fig:Profile_residuals}
\end{center}
\end{figure}

\begin{figure*}[]
 \begin{center}
\includegraphics[width=\textwidth]{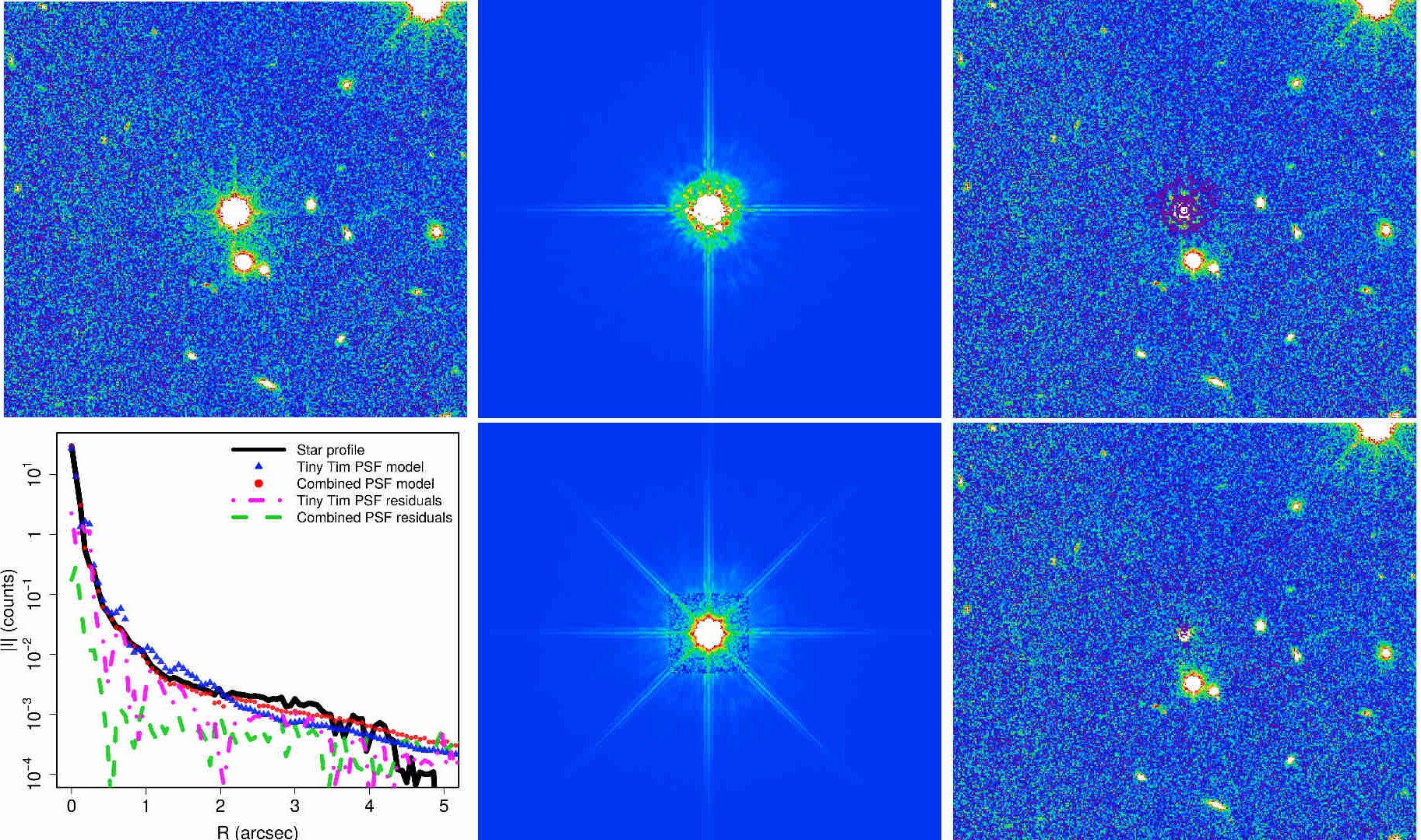}
\caption[]{Comparison of the subtraction of a star with {\tt{GALFIT3.0}}, using a 15 arcsec Tiny Tim PSF (upper row) and our combined PSF (lower row). \emph{Top left panel:} F775W image of a star. \emph{Bottom left panel:} Intensity profiles for the star, the PSF models and the residual images after the subtraction of the Tiny Tim PSF and after the subtraction using our combined PSF (see the legend for details). \emph{Middle column:} PSF scaled models for the Tiny Tim model (top) and the combined PSF (bottom). \emph{Right column:} Residuals of the subtraction for the Tiny Tim model (top) and the combined PSF (bottom). Note that the residuals for the wings in case of using our combined PSF are smaller at all radii than those resulting from using the Tiny Tim PSF, despite the common central underestimate. The FoV of all the images is $24\times24$ arcsec$^{2}$. All images are in the same colour scale. [\emph{A colour version of the figure is available in the online edition.}]
} 
\label{fig:PSF_test}
\end{center}
\end{figure*}

In order to estimate and correct the amount of dispersed light by the PSF in the HST/ACS images, especially in the outskirts of the galaxies, we cannot simply perform a numerical deconvolution of the image, because it would affect the light distribution in the low surface brightness region (see Sect.\,\ref{Sec:Intro}). We have followed a similar procedure to that presented in \citet{2013MNRAS.431.1121T}. 

First, we performed careful masking in all the images, in order to avoid contamination by foreground objects, non-symmetric components such as satellites, compact bright regions or close stars or their PSF spikes. We created these masks in two steps. We performed an automatic masking by using SExtractor \citep{1996A&AS..117..393B}, which is later manually checked. We manually extended the masks of the largest foreground objects with circular masks in order to prevent the scattered light from contaminating the surface brightness profiles of the galaxies. We estimated the limits of each manual masking by analysing the isophotal map on smoothed images. We gradually increased the smoothing kernel from 2 up to 8-10 pixels, checking on each step the possible features that appeared. By using the isophotal contours, we checked the possible distortions due to contamination sources or extreme changes in the position angle and ellipticity in each object. Finally we used the masked images (centred on each object, with a FoV of $100 \times 100$ arcsec) to measure individually the sky noise distribution, which defines for each object the limiting magnitude of the profile. See some examples of isophotal maps and masked regions in the figures of Appendix \ref{Sec:TypeIIIcomments}. The effects of PSF correction on the resulting profiles are analysed in Sect.\,\ref{Subsec:PSF_effect}.

We used {\tt{GALFIT3.0}} \citep{2002AJ....124..266P} to fit two sets of models for each object within our sample of 50 S0 and E/S0 galaxies: a single free Sérsic model and a free Sérsic model + exponential profile. In this step, we discarded from the S0 and E/S0 sample those objects which: 1) cannot be fitted with a stable model, due to small size or irregularities, 2) are well fitted with a single (high $n$) Sérsic profile, so the profile corresponds to an elliptical galaxy rather than an E/S0 or S0 (see Sect.\,\ref{Subsec:Red_sample}), or 3) do not drastically reduce the residuals of the fit by adding a second component to the first one. This revision was performed individually by three co-authors, checking the differences between the modelled and the original profiles. In this phase we distinguished between the elliptical galaxies from face-on S0 galaxies by analysing their surface brightness profiles. In order to do that, we have estimated for all the objects two different 2D decompositions: 1) using a single Sérsic profile and 2) with a Sérsic bulge + exponential disc profile. Note that each 2D decomposition implies a different PSF deconvolution. We analysed the residual profile of each object, and identified those which do not drastically improve with the addition of a disc component and present a smooth decrease of the slope with the galactocentric radius. Finally we confirmed that the 9 objects that visually presented elliptical morphologies really were ellipticals according to their surface brightness profiles. We show the surface brightness profiles, with the corresponding models and residual profiles, for the S0 and E/S0 galaxies with successful PSF correction in Appendix \ref{Sec:TypeIIIcomments}. 

We have estimated the errors of the parameters of each model by performing Monte Carlo simulations over the original images, varying each pixel value within the noise level and fitting sequentially. We note that we assume 1, 2 and 3 $\sigma$ limits as those values that enclose 68\%, 95\% and 99.7\% of the probability distribution of the sample. In the case of the noise level of our images the sample are the sky-dominated pixels. We define as sky-dominated pixels any pixel that was not identified as part of a source in the masks generated before to create the {\tt{GALFIT3.0}} models and the surface brightness profiles. We estimated the noise level as the upper $1 \sigma$ fluctuation of the sky background distribution. This value is equivalent to the standard deviation if we assume that the sky background follows a nearly Gaussian distribution with mean 0 and dispersion equal to the measured noise level. By providing {\tt{GALFIT3.0}} a PSF model for the image, the program tries to fit the best model that, after convolution with the PSF, matches the original image. This allows us to estimate the amount of diffuse light scattered into the outskirts of the image. We estimate the deconvolved image through the following image operation:

\begin{equation}
\label{Deconvolution_1}
Residuals  = Image \: raw - PSF \ast Model_{GALFIT}
\end{equation}
\begin{equation}
\label{Deconvolution_2}
Deconvolved \: final \: image  = Model_{GALFIT} + Residuals
\end{equation}

The 3DHST team provides a star-stacked PSF, but the model size is very small for our purposes ($\sim 4$ arcsec in diameter) while we need at least a $\sim 25$ arcsec diameter PSF to perform any kind of PSF subtraction \citep{2014A&A...567A..97S,2015A&A...577A.106S}. Otherwise, the reconvolved model (i.e., the fitted PSF-free model, again convolved by the PSF) would systematically underestimate the amount of dispersed light at radii larger than half the diameter of the PSF. This limit in size was calculated by multiple iterations in the surface brightness profile fitting, and roughly corresponds to $1.5$ times the limiting radius at $S/N = 3$ in the surface brightness profiles of the larger galaxies in our sample. We show an example of the deconvolution and modelling in Fig.\,\ref{fig:Profile_residuals}. In the top panel, we represent the surface brightness profiles of the original and PSF-corrected image, as well as the profiles of the models used for the deconvolution. Notice that the deconvolved profile is systematically biased towards brighter magnitudes in the centre and dimmer magnitudes in the outskirts. This effect is clear in the bottom panel, where we show the differences between the original and the PSF-corrected profiles (red dots). Notice that the original and reconvolved profile (black squares) are nearly identical along the whole profile, showing that the PSF correction and the fit are adequate, under the assumption that the PSF model is robust and realistic (we demonstrate this in Fig.\,\ref{fig:PSF_test}, see below). 

GOODS-N is an extremely star-clean field, so PSF modelling using star stacking may lead to noisy and biased distributions in the outskirts of the profiles due to the lack of bright stars in the field. Instead, we have created a model of the PSF using the Tiny Tim software \citep{2011SPIE.8127E..0JK} for the PSF outskirts, combined with a GOODS-N stacked star PSF for the inner regions of the PSF provided by 3DHST. Tiny Tim is a modelling tool for generating HST model PSFs for multiple cameras and filters. The synthetic PSFs generated by Tiny Tim have been previously used in similar studies \citep[See][]{2008MNRAS.388.1521D,2013MNRAS.431.1121T}. GOODS-N has a covering grid of 3$\times$5 individual ACS pointings, with some overlap to check the photometric and astrometry consistency between individual pointings. There is a rotation of the field of 45º between the odd-numbered epochs and the even-numbered epochs. In order to approximate the effect of the combination of rotated images in the final mosaic, we combined two Tiny Tim PSF model images of 29 arcsec in diameter, rotated 45º degrees \citep{2016ApJ...823..123T}. After that, the centre of the resulting PSF model was replaced by the natural star PSF model of 3DHST, which is reliable only out to 4 arcsec. We measure a FWHM of our combined PSF model is 0.09 arcsec. As mentioned in Sect.\,\ref{Subsec:GOODS-N}, in \citet{2014ApJS..214...24S} (see Table 6), the authors report an average FWHM of 0.11 arcsec for the F775W band mosaic of GOODS-N.

In Fig.\,\ref{fig:PSF_test} we show a quality test to show that our PSF model is robust and realistic. The upper panels show the test using a standard PSF created by Tiny Tim for the F775W filter, whereas the lower panels show the test using our combined PSF. We have fitted each PSF to a well-centred field star in the GOODS-N image, using {\tt{GALFIT3.0}}. The residuals of the two cases are shown in the right panels, and the residuals profile in the bottom left panel. We find that our combined PSF model significantly reduces the residuals of the PSF modelling of the star when compared to the original Tiny Tim model. The success of the subtraction is significant especially for the intermediate and external parts where the characteristic spikes and rings of the HST PSF are removed almost completely. In addition, the remaining residuals in the centre do not show circular symmetry, which would introduce contamination in our azimuthally averaged profiles. 

Additionally, in order to test on each object the validity of the PSF-deconvolved model obtained, we again convolve the PSF-deconvolved image by the PSF model (or reconvolved image), and compute the differences between the surface brightness profiles derived from the original and reconvolved images. We found that the differences between the original and the reconvolved profiles were less than $1\sigma$ in the surface brightness profiles up to the limiting radius in each case. Thus, we concluded that the applied PSF correction is robust and self-consistent. 

\subsection{Surface brightness profiles}
\label{Subsec:Methods_profiles}

\begin{figure*}[]
 \begin{center}
\includegraphics[width=0.49\textwidth]{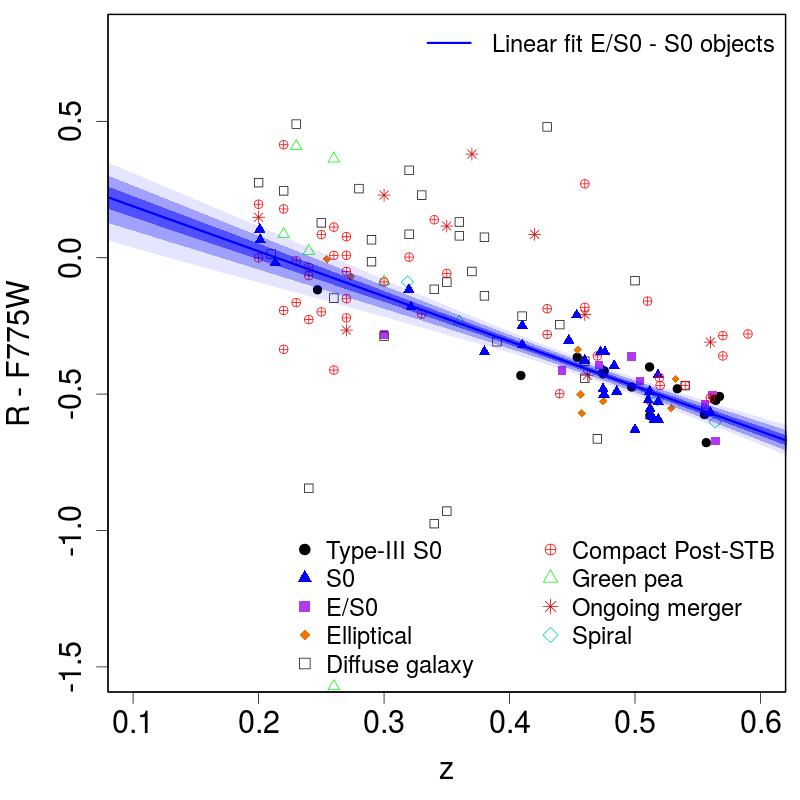}
\includegraphics[width=0.49\textwidth]{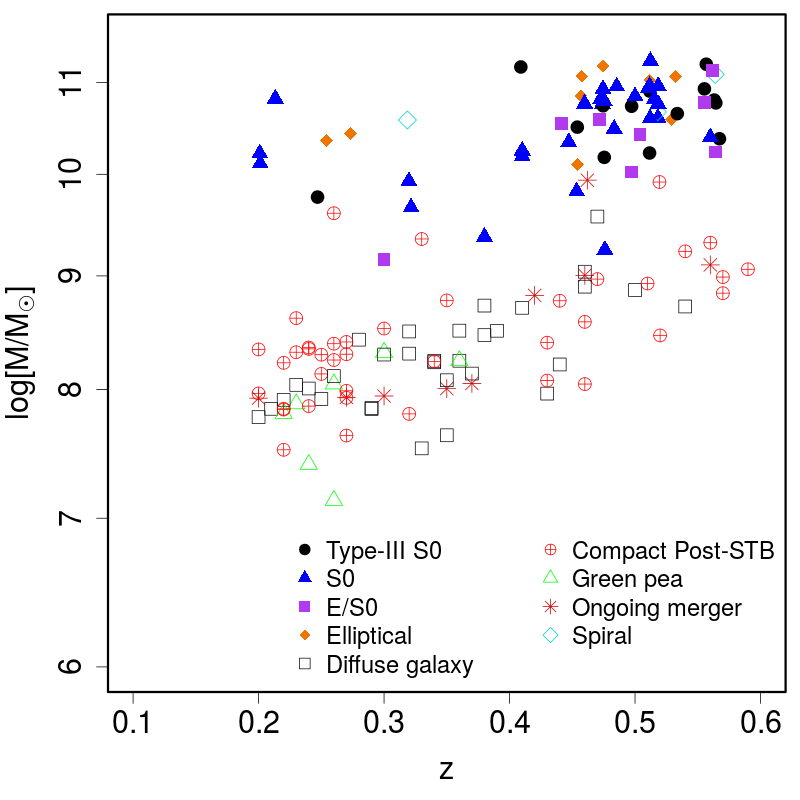}

\caption[]{\emph{Left panel:} Estimated K-correction for the F775W filters to the $R$ band as a function of $z$. It is important to highlight that the linear fit applies to the observed correlation of the S0 and E/S0 galaxies of the sample. The intensity levels of the linear fit represent the 1, 2 and 3$\sigma$ confidence regions of the fit. \emph{Right panel:} Stellar mass for the initial red sequence sample vs. $z$. Consult the legend in the figure for the morphological classification. [A colour version of the figure is available in the online edition.]} 
\label{fig:Kcorr_zcomp}
\end{center}
\end{figure*}

We derived the surface brightness profiles by two different methods, depending on the inclination of the object. For the low inclined objects we relied on the {\tt{ellipse}} task from IRAF. We fixed the position angle and ellipticity of the ellipses to the mean values obtained with SExtractor in the galaxy disc on the original image, with the limiting threshold set at $1\sigma$. By doing so, we avoid unstable solutions for the position angle and ellipticity, especially common in the outskirts, where the $S/N$ ratio is low (see E08) and where we will be centring our attention. As a side effect, the innermost parts of the profiles (first 5-6 pixels in radius) may be slightly biased towards lower values, but we discarded this region in the vast majority of the cases, so it does not affect our results. On the other hand, the nearly edge-on objects were analysed by using {\tt{ISOFIT}} \citep{2015ApJ...810..120C}. This task replaces the angular parameter that defines quasi-elliptical isophotes in polar coordinates for the eccentric anomaly. By doing this, {\tt{ISOFIT}} provides more accurate modelling of galaxies with non-elliptical shapes, such as disc galaxies viewed edge-on, and thus recovers more accurate surface brightness profiles. We have used free position angle and ellipticity in this case.  

In order to separate the edge-on objects from the rest of the sample, we have used the results from the 2D decomposition models performed previously for the deconvolution process. The Monte Carlo simulations that we performed on the {\tt{GALFIT3.0}} models provide the probability distribution for the axis ratio ($q=b/a$) of the disc component. We assumed an intrinsic axis ratio for the S0 galaxies of $q_{0} \sim 0.25$ \citep{2014MNRAS.444.3340W}, with the exception of those objects with $q<0.25$ where we used $q_{0}=0.1$, which is the lower limit of the $q_{0}$ distribution from \citet{2014MNRAS.444.3340W}. We note that we apply this ratio only as a median value in order to select objects with high inclinations, and it is not used to extract conclusions for single objects. The values of the inclination angle for each object are included in Table \ref{tab:fits_psforr} in Appendix \ref{Appendix:Fits_params}. The median dispersion for these values are $\pm3$ degrees. We identify five objects with $q_{0}<0.25$ as nearly edge-on galaxies (SHARDS10000845, SHARDS10001727, SHARDS20000827, SHARDS20003134 and SHARDS20011817). In addition to these, after visual classification, we added SHARDS10002351 and SHARDS10003402 to the edge-on sample. For them, the surface brightness profiles have been derived with {\tt{ISOFIT}} instead of {\tt{ellipse}}, as commented above.

We defined the outer limit of the profile on the major axis as the outermost elliptical bin with an integrated intensity with $S/N > 3$. The $1\sigma$ sky noise level was defined for each azimuthal bin as:
\begin{equation}
\label{SN_ratio}
\sigma_{\mathrm{sky, bin}}  = \frac{\sigma_{\mathrm{sky, pixel}}}{\sqrt{N}} 
\end{equation}

\noindent where $\sigma_{\mathrm{sky, pixel}}$ represents the standard deviation of the background, measured on the masked images and $N$ is the number of pixels included in each radial bin. The sky noise for each ellipse is then combined by addition in quadrature with the intrinsic dispersion along the azimuthal direction for each ellipse. The upper and lower limits of the magnitude values have been calculated by performing Monte Carlo simulations of the final noise distribution for each bin. The median limiting magnitude for our profiles is $\mu_{\mathrm{F775W,lim}} = 27.092^{+0.024}_{-0.032}$ \magarc\ (the uncertainties refer to the percentiles 84.1 and 15.6, equivalent to $\pm 1 \sigma$ in a Gaussian distribution).

\subsection{Photometric corrections}
\label{Sec:Photometric_corrections}

The rest-frame wavelength range for a filter varies as a function of $z$, so we need to apply a K-correction \citep{1956AJ.....61...97H,1968ApJ...154...21O} to the photometry of our profiles. We have performed the K-correction directly by estimating the variation of the colour between the apparent magnitude estimated from the fitted SED in the rest-frame filter we want to compare with ($R$ band) and the observed magnitude in the F775W filter. The rest frame (synthetic) $R$ and the observed F775W apparent magnitudes for each object of our sample are available at the Rainbow Database. Rest-frame magnitudes have been calculated by direct integration over the best-fit SED for each object \citep{2013ApJ...762...46P}. In the left panel of Fig.\,\ref{fig:Kcorr_zcomp} we show the K-correction derived for the S0 and E/S0 galaxies of our sample, as a function of $z$. It nearly follows a linear relation for S0 and E/S0 galaxies, which we have fitted by using 10,000 bootstrapping simulations of a least squares fit, resulting in: 
\begin{equation} \label{eq:RSteidel}
\mathrm{R} - \mathrm{F775W} = 0.340^{+0.033}_{-0.042} - 1.619^{+0.065}_{-0.072} \cdot\ z,
\end{equation}
with a Spearman correlation coefficient of $\rho = -0.838$, and a Pearson test probability $p<2.2\times10^{-16}$. We calculate the K-corrections for each object by using the previous linear relation, which applies a median correction as a function of $z$. The final K-corrections for each object are shown in Table \ref{tab:redsample} in Appendix \ref{Appendix:redsample} (the median uncertainty for $R -$ F775W is $\pm0.060$). Therefore, the final surface brightness profiles of our $0.2<z<0.6$ sample are in the rest-frame $R$ band, as the available data on local Type-III S0s by E08 and G11.

In order to check the goodness of the Rainbow photometric redshift for our subsample of S0 and E/S0 galaxies at $0.2<z<0.6$, we have estimated the difference between the spectroscopic and the photometric redshifts for those objects where the former is available (47 out of the selected 50 S0 and E/S0 galaxies have spectroscopic redshift). We found an excellent overall agreement between the spectroscopic $z$ and the photometric $z$ with a median dispersion in our subsample of $\Delta z/(1+z) = 0.0024$, meaning that the photometric $z$ estimated in Rainbow from SED fitting are very good for the galaxies of our sample. A notable exception is the object SHARDS10000845 ($z_{\mathrm{photo}}=0.36$, but $z_{\mathrm{spec}}=0.5123$). The spectroscopic redshift flag at the Rainbow Database is 4, which means a very reliable value for $z_{\mathrm{spec}}$. Nevertheless, this discrepancy does not affect the results, as the $z_{\mathrm{spec}}$ was used instead of the $z_{\mathrm{photo}}$ result in all the calculations when available. Only three S0 -- E/S0 galaxies did not present available $z_{\mathrm{spec}}$ values (SHARDS10005029, SHARDS10008552 and SHARDS20011817). However, the two first cases were removed from the final sample (see the reasons in Sect.\,\ref{Subsec:Red_sample}) and the last one was finally classified as an S0 galaxy with a Type-I profile. 

\begin{figure*}[!htbp]
\begin{center}

\includegraphics[width=0.99\textwidth]{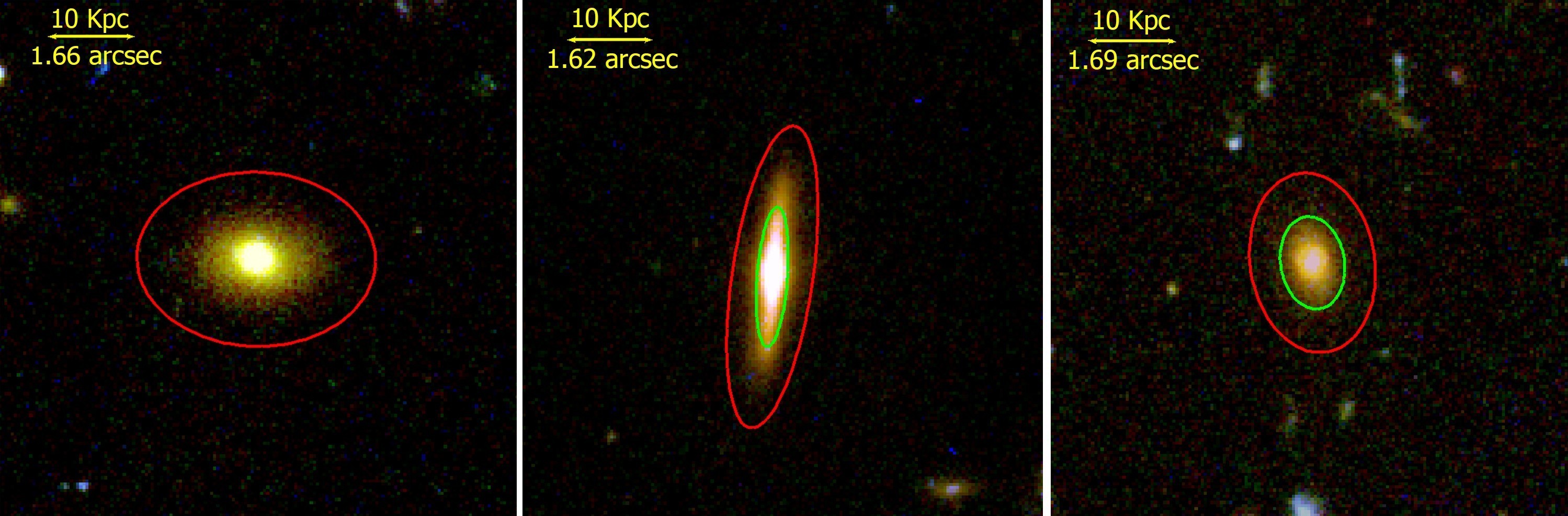}

\includegraphics[width=0.33\textwidth, clip, trim=0cm 0cm 1cm 1cm]{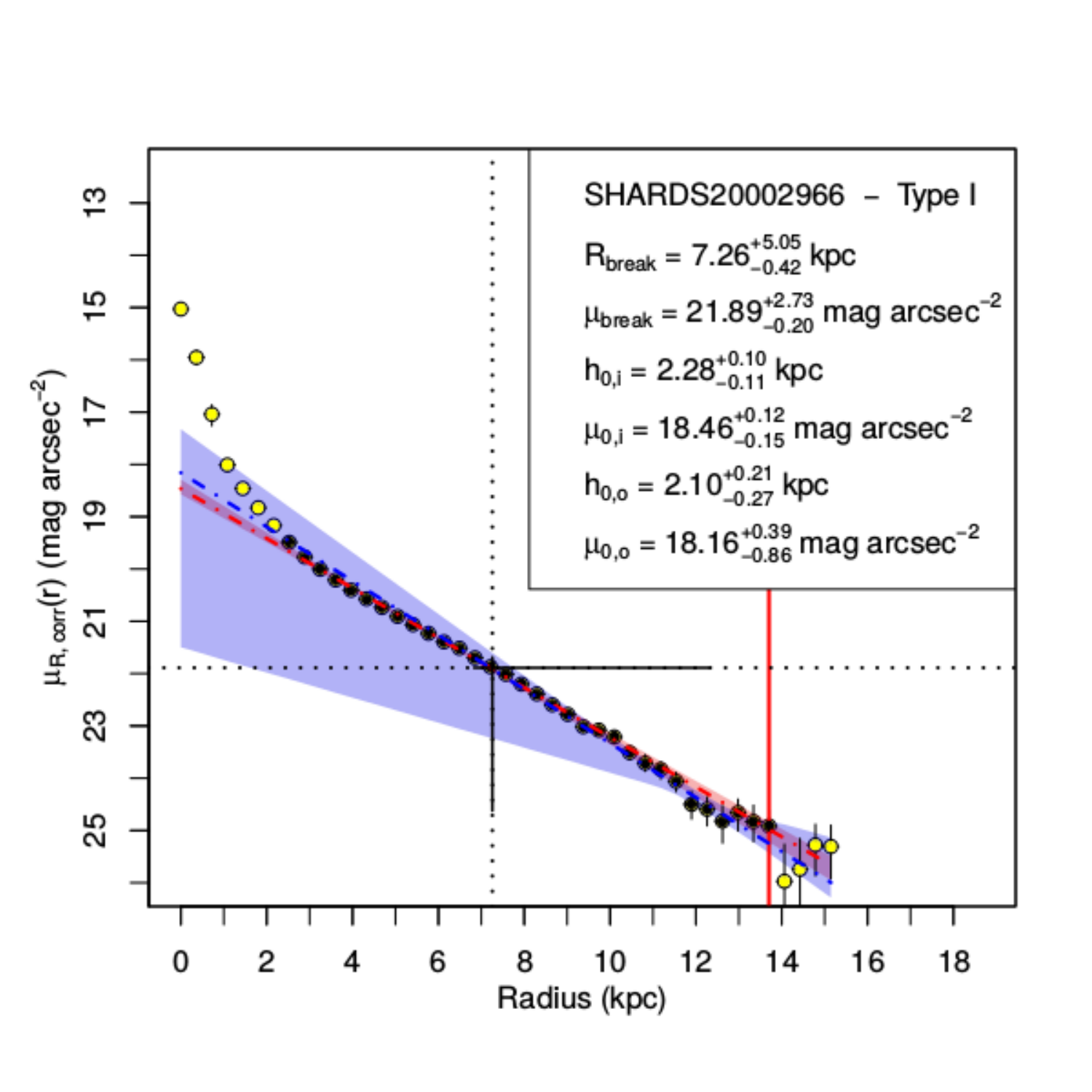}
\includegraphics[width=0.33\textwidth, clip, trim=0cm 0cm 1cm 1cm]{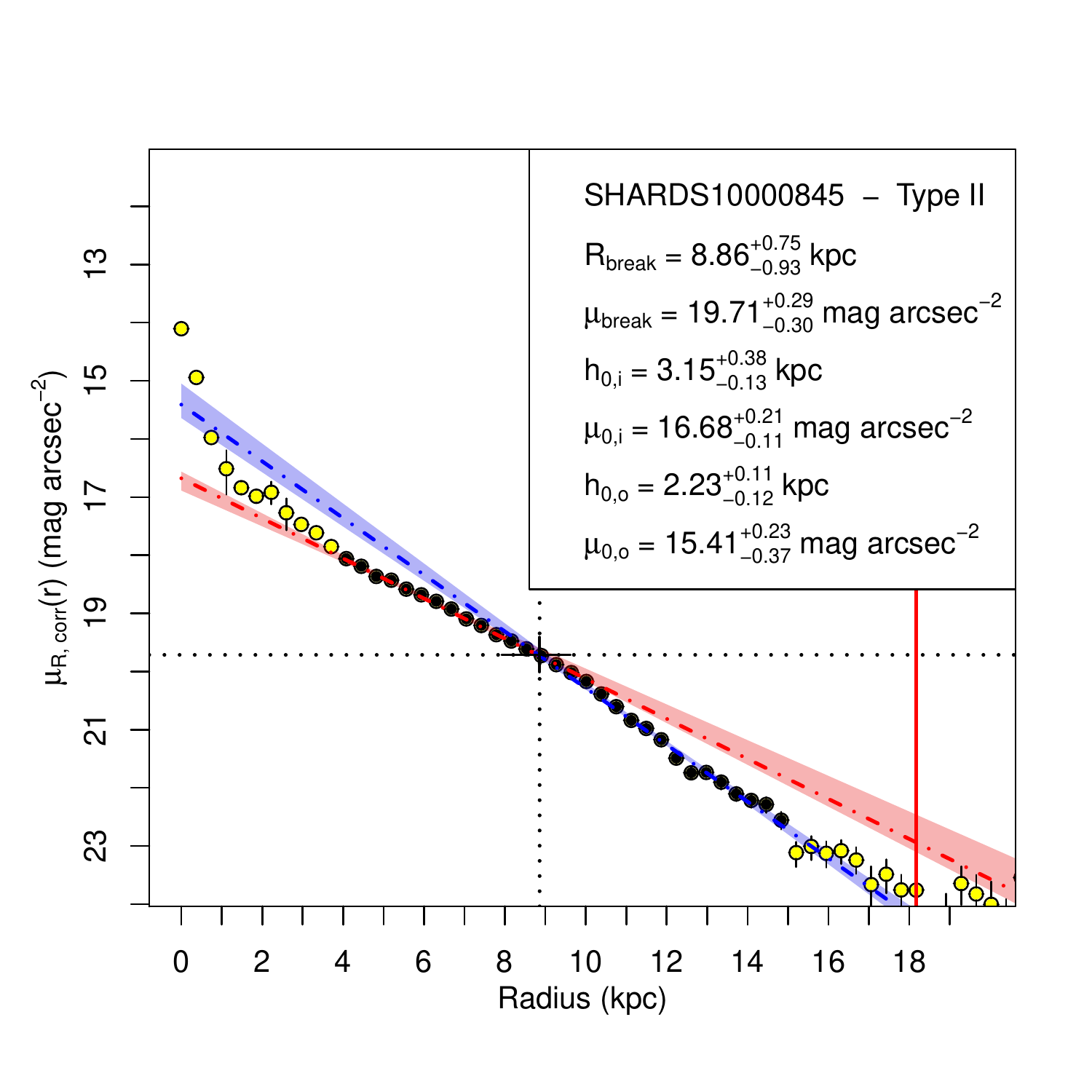}
\includegraphics[width=0.33\textwidth, clip, trim=0cm 0cm 1cm 1cm]{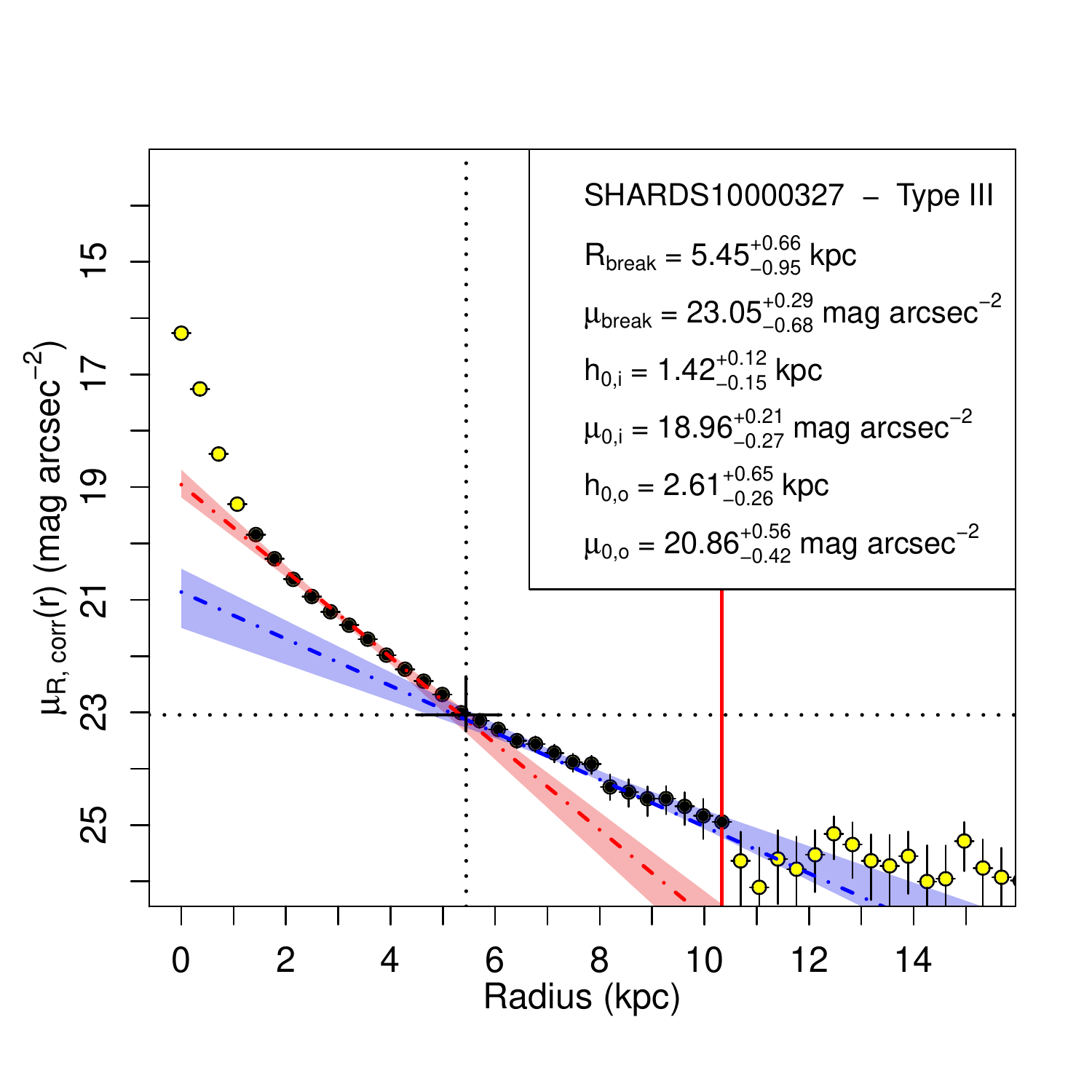}

\begin{overpic}[width=0.33\textwidth, clip, trim=0cm 0cm 1cm 1cm]
{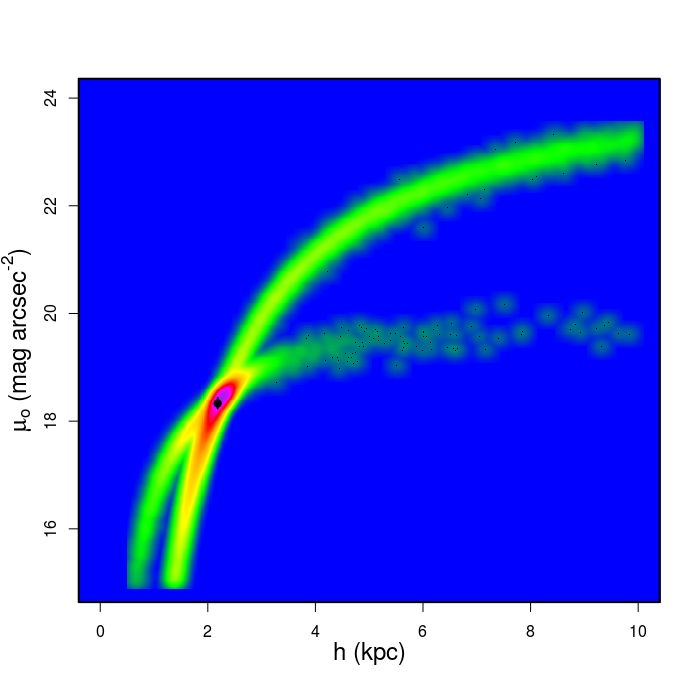}
\put(100,40){\color{yellow} \textbf{PDDs of h vs. $\mu_{0}$}}
\put(100,30){\color{yellow} \textbf{Type-I profile}}
\end{overpic}
\begin{overpic}[width=0.33\textwidth, clip, trim=0cm 0cm 1cm 1cm]{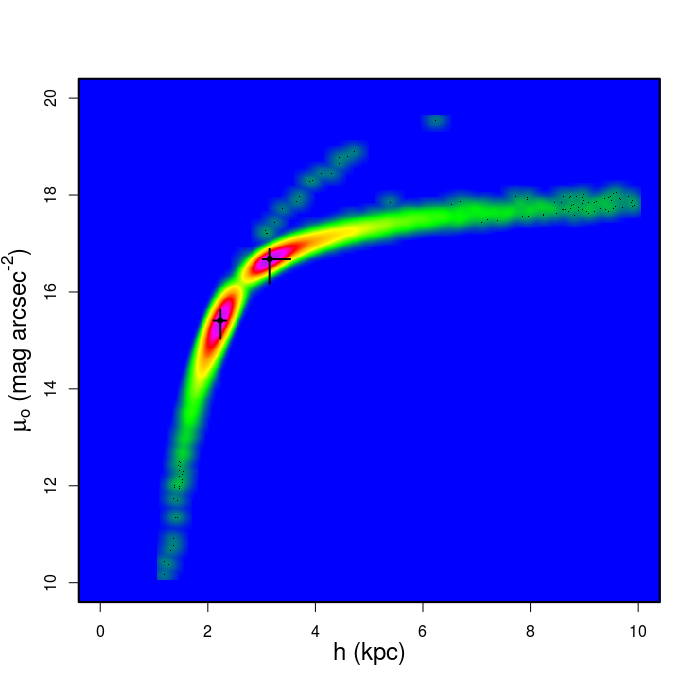}
\put(100,40){\color{yellow} \textbf{PDDs of h vs. $\mu_{0}$}}
\put(100,30){\color{yellow} \textbf{Type-II profile}}
\end{overpic}
\begin{overpic}[width=0.33\textwidth, clip, trim=0cm 0cm 1cm 1cm]{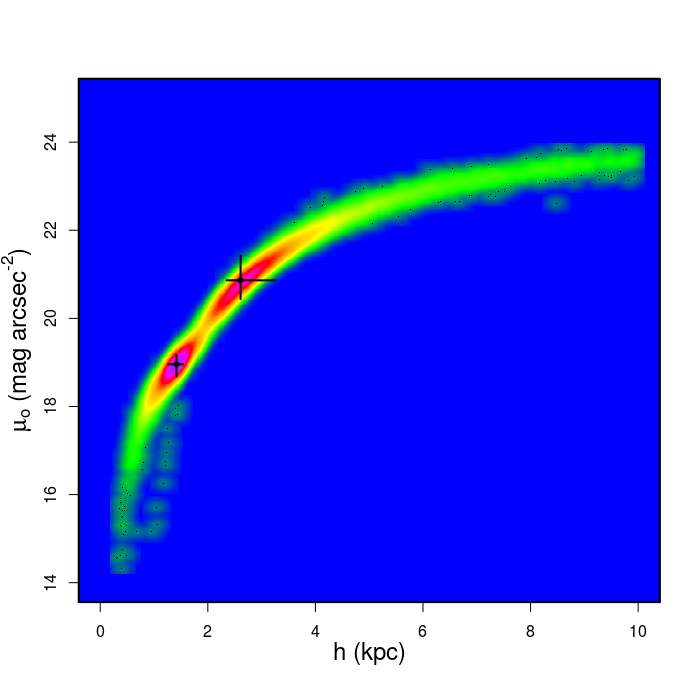}
\put(100,40){\color{yellow} \textbf{PDDs of h vs. $\mu_{0}$}}
\put(100,30){\color{yellow} \textbf{Type-III profile}}
\end{overpic}
\caption{Examples of the three kinds of disc profiles in our sample of S0 and E/S0 galaxies and their analysis. \emph{Columns, from left to right:} 1)  Type I: SHARDS20002966, 2) Type II: SHARDS10000845, 3) Type III: SHARDS10000327. \emph{Upper row:} RGB composed images of the objects using the HST/ACS filters F435W (blue), F606W (green) and F775W (red). The total FoV is $10\times 10$ arcsec$^{2}$. The red ellipse marks where the surface brightness profile presents a $S/N = 3$. The green ellipse indicates the location of the break if the profile is a Type II or III. The yellow bar represents 10 kpc in physical size. \emph{Middle row:} Surface brightness profiles in the rest-frame $R$ band corrected for the dust extinction, cosmological dimming and K-correction of the corresponding objects. Only the black filled points where included in the final break fit. The yellow points were removed from the fit to avoid bulge light contamination or are outside the limiting radius. The red and blue dashed-dotted lines represent the best fits to the inner and outer regions of the disc, and the coloured uncertainty regions around them represent the $1\sigma$ confidence interval of each fit, respectively. The red vertical line represents the limiting radius for considering data points in the fit ($S/N=3$). The dotted lines indicate the most probable break location in the profile (note the high uncertainty in the Type-I profile). \emph{Lower row:} 2D probability density distributions (PDDs) for the scale-length $h$ and central surface brightness $\mu_{0}$ decompositions. Note that in the case of SHARDS1000327 and SHARDS10000845 the PDDs show two peaks where the inner and outer profiles are located, while in SHARDS20002966 the solutions for the two profiles peak around an unique point. The rainbow-like colour scale represents probability density, with redder colours indicating higher values. [\emph{A colour version of the figure is available in the online edition.}]} 
\label{fig:Profile_types}
\end{center}
\end{figure*}

We corrected our profiles for Milky Way dust extinction too. To estimate the necessary corrections, we used the dust reddening maps from \citet{1998ApJ...500..525S}, available through the NASA/IPAC Infrared Science Archive\footnote[3]{http://irsa.ipac.caltech.edu/applications/DUST/}. We extracted the extinction factor $E(B-V)$ for each object at its position. We assumed a Landolt $V$ band extinction in magnitudes $A_{V} = 3.315 \times E(B-V)$, and the ratio $A_{\mathrm{F775W}}/A_{V} = 0.65$ \citep{1999PASP..111...63F,2005ApJ...619..931I}. We corrected for the calculated extinction correction for each object in the sample. The values obtained are available in Table \ref{tab:redsample} in Appendix \ref{Appendix:redsample}. 

Finally, we have corrected the profile for cosmological dimming, which entails a factor ${\Delta \mu = -10\cdot \log_{10} (1+z)}$. We use the best $z$ value for each object available in the Rainbow Database to perform this correction (see Table \ref{tab:redsample} in Appendix \ref{Appendix:redsample}). 

\subsection{Elbow: automated break analysis of disc surface brightness profiles}
\label{Subsec:Elbow}

\begin{figure*}[]
\begin{center}
\includegraphics[width=0.33\textwidth, clip, trim=0cm 0.7cm 0.5cm 1.5cm]{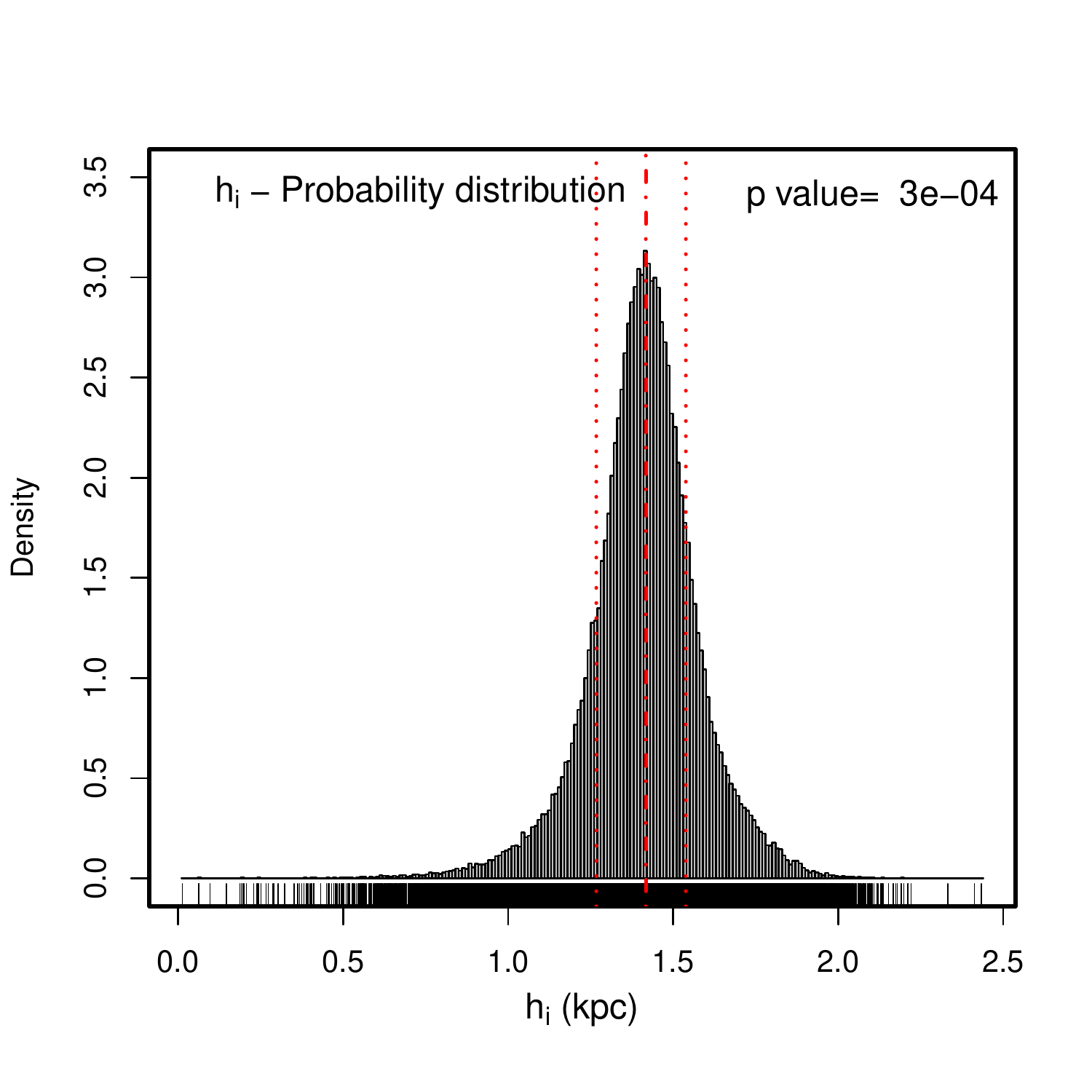}
\includegraphics[width=0.33\textwidth, clip, trim=0cm 0.7cm 0.5cm 1.5cm]{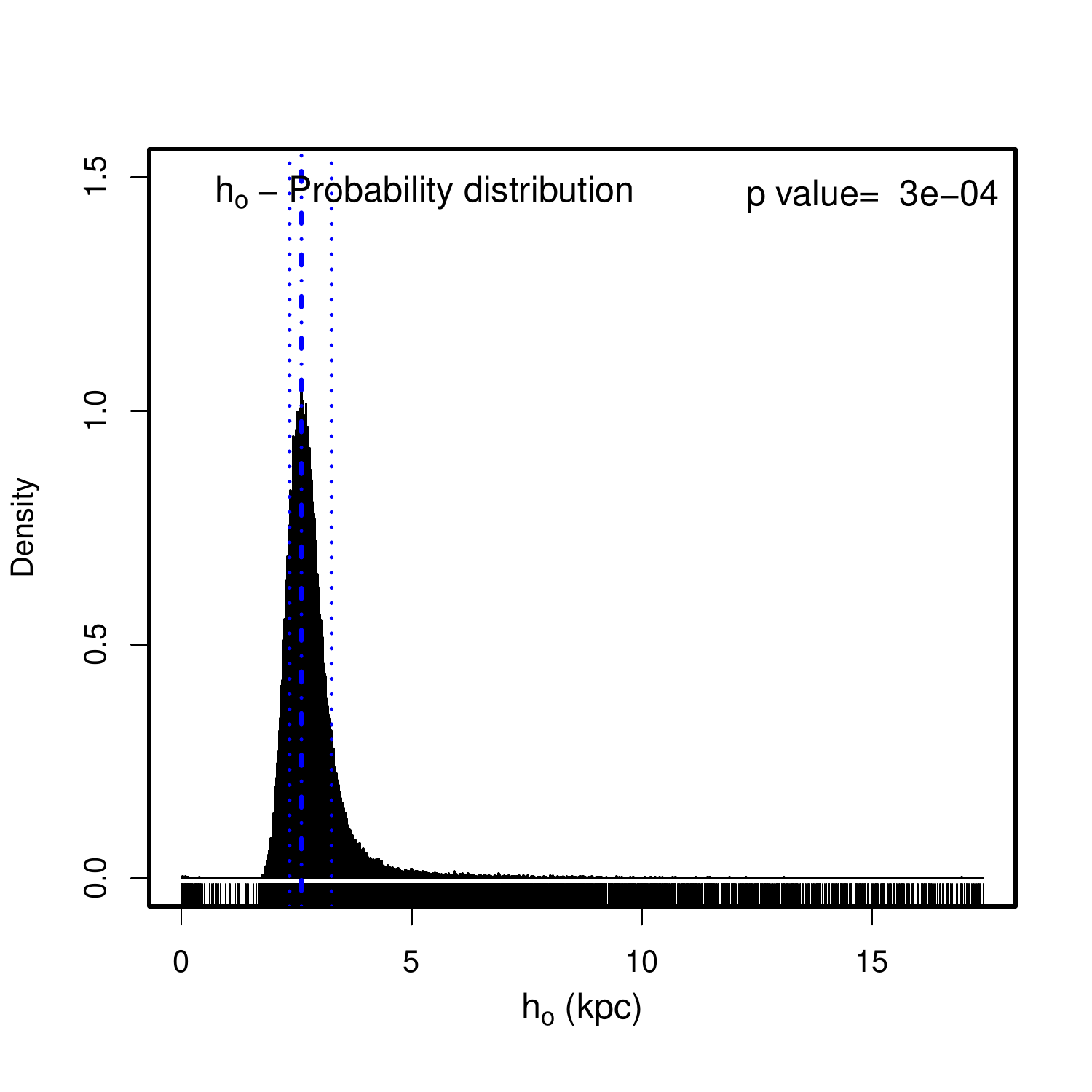}
\includegraphics[width=0.33\textwidth, clip, trim=0cm 0.7cm 0.5cm 1.5cm]{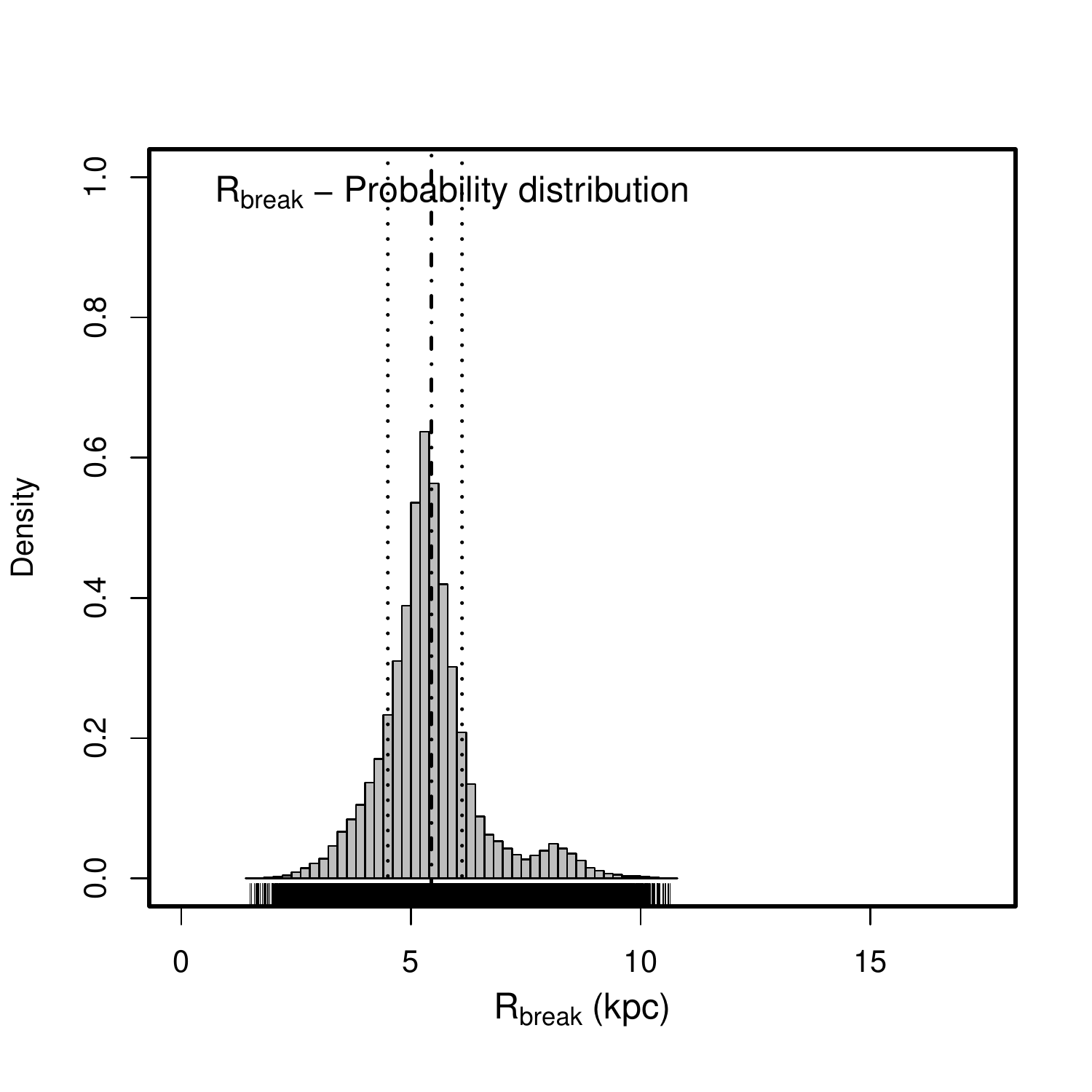}

\includegraphics[width=0.33\textwidth, clip, trim=0cm 0.7cm 0.5cm 1.5cm]{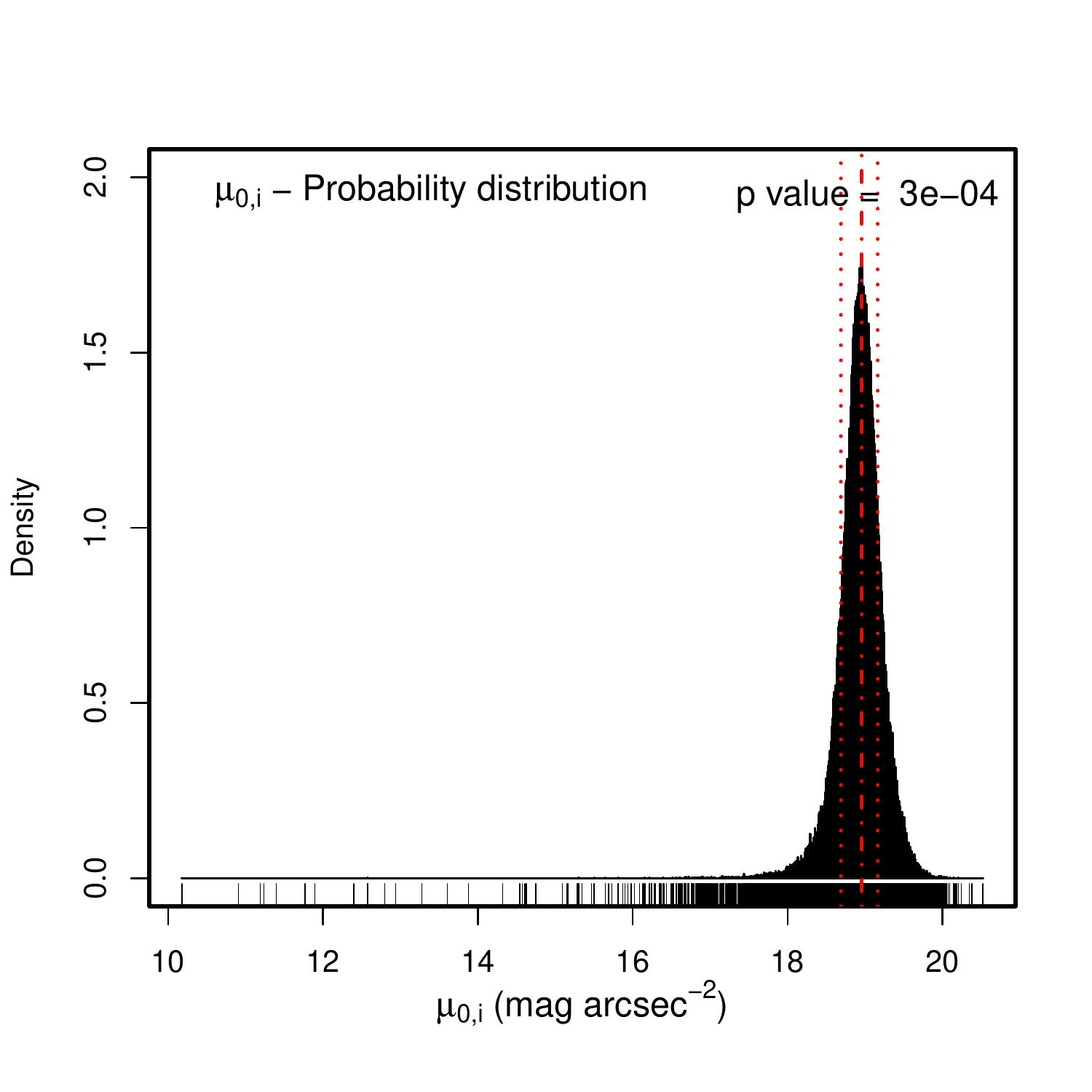}
\includegraphics[width=0.33\textwidth, clip, trim=0cm 0.7cm 0.5cm 1.5cm]{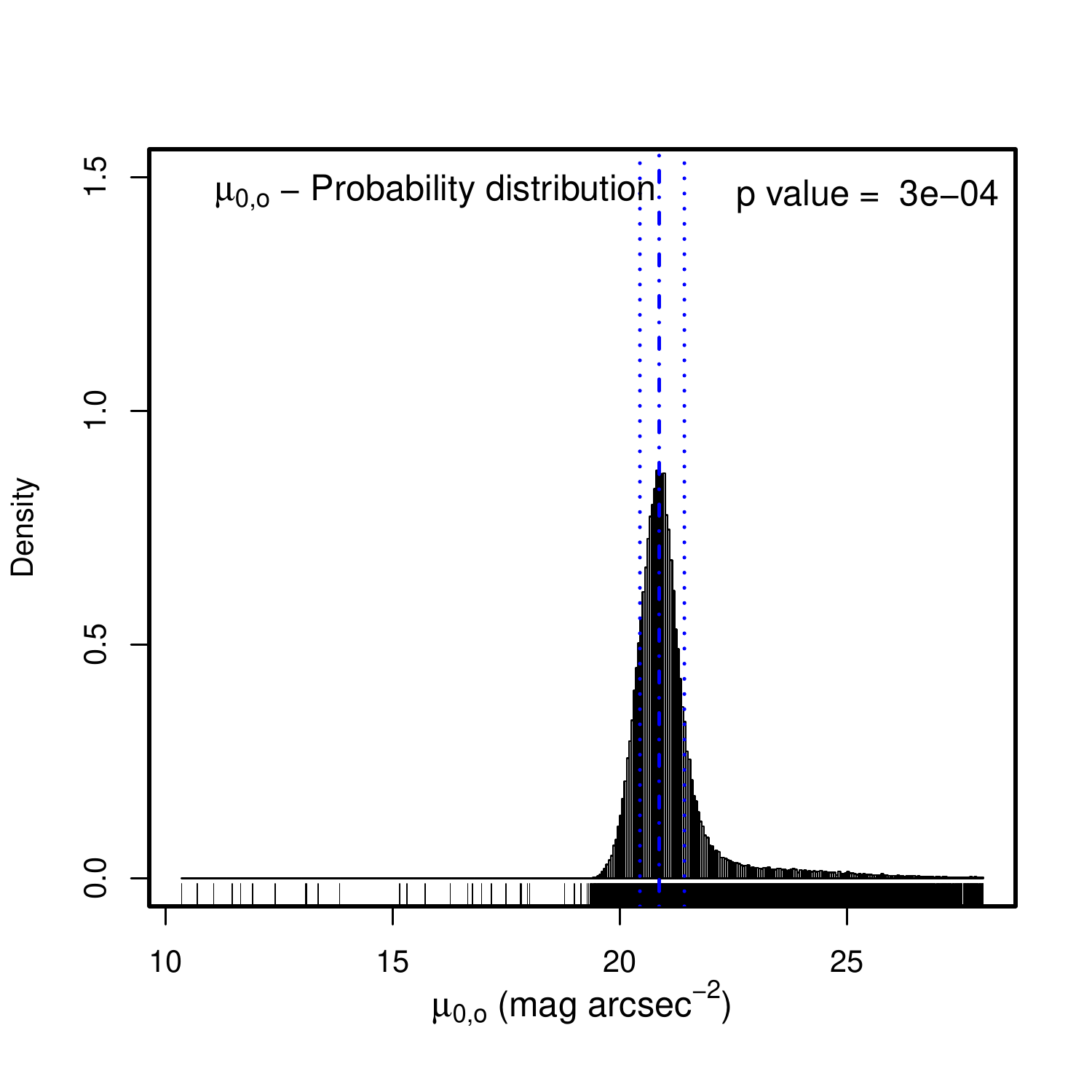}
\includegraphics[width=0.33\textwidth, clip, trim=0cm 0.7cm 0.5cm 1.5cm]{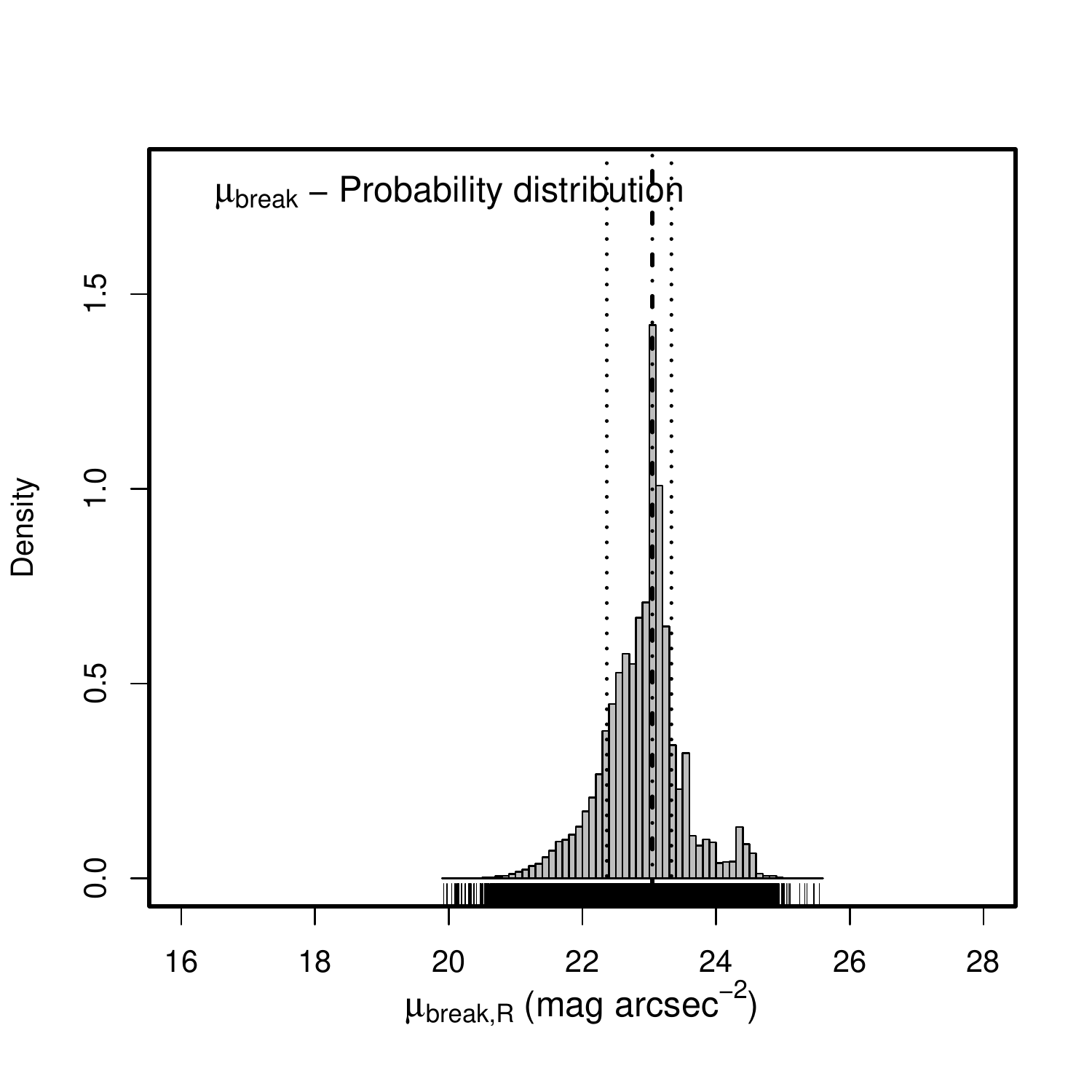}

\caption{Probability density distributions of the surface brightness profile fit for the Type-III S0 galaxy SHARDS10000327. \emph{Upper row, from left to right:} scale-length for the inner disc profile \hi\ (kpc), scale-length for the outer profile \ho\ (kpc), break radius \rbreak\ (kpc). \emph{Lower row, from left to right:} central surface brightness profile of the inner profile \mui\ (\magarc), central surface brightness profile of the outer profile \muo\ (\magarc), surface brightness at the break radius (\mubreak). The dashed-dotted lines represent the best solutions for each parameter, and the dotted lines mask the limits of the 68\% ($1 \sigma$) confidence region. [\emph{A colour version of the figure is available in the online edition.}] }    
\label{fig:histograms_elbow}
\end{center}
\end{figure*}

After the surface brightness profile calculation, we proceeded to classify the profiles into the three types described in Sect.\,\ref{Sec:Intro}, paying special attention to those objects showing a Type-III profile. We have performed this classification using two procedures: 1) simple visual classification and 2) an automated break analysis via an algorithm written by ourselves in $R$\footnote[2]{R: A language and environment for statistical computing. http://www.R-project.org.} \citep{CITER} that we called {\tt{Elbow}}. {\tt{Elbow}} accepts as input the surface brightness profile and the errors in both magnitude and position of each bin, as well as minimum and maximum radial limits for the analysis. The tasks of the program are: 1) to estimate the best double exponential fit to a surface brightness profile and 2) to calculate the probability that the slopes of the inner and outer fitted profiles are statistically different. The algorithm works by carrying out Monte Carlo + Bootstrapping simulations on the input profile, using the uncertainties provided by the user for each data point in both radius and magnitude. This is a two-steps process. First, for each simulation, {\tt{Elbow}} generates a new profile by random re-sampling with replacement (bootstrapping). This implies that the new profile contains the same number of points than the original one, but some of them are not included in each simulation. Secondly, we replace each data point with the values from a Gaussian distribution with mean value equal to the original data point and $\sigma$ equal to the original error (Monte Carlo). This is done for both variables ($r$ and $\mu$). We take advantage of the combination of the bootstrapping and Monte Carlo methods in order to take into account the different uncertainties associated to each point and obtain an accurate probability distribution of the parameters at the same time. One of the main benefits of the re-sampling methods is to avoid any assumptions of normality on the sample and hence obtain a more accurate distribution for certain statistics. In addition, the shape of the resulting probability distributions for the parameters give us information about the presence of outliers and irregularities. This method of fitting breaks in linear trends via bootstrapping was already tested in \citet{1998A&AS..127..597C} to detect changes in the slope of the line-strength gradients of Mg$_{2}$ and the $\lambda4000\AA$ break as a function of the galactocentric radius in central cluster galaxies, and more recently by \citet{2016A&A...585A..47M} to detect breaks in surface brightness profiles. 

We have avoided the inner parts of the profile - which are dominated by the bulge emission - by using the bulge model fitted with the {\tt{GALFIT3.0}} and removing from the fit the section of the profile where the bulge dominates the emission over the exponential profile. For several objects this minimum limit was not enough to properly mask the central part of the profile (because of additional inner components), and it had to be set to a more restrictive and larger radius by visual inspection of the profile and the model decompositions. Similarly, we have cut the profile of several objects before reaching the limiting radius, due to the presence of irregularities or possible contamination sources that we wanted to avoid in the final profile fit.

For each simulation, {\tt{Elbow}} minimises the residual distances of the input observational profile to a broken profile modelled as: 

\begin{equation} \label{eq:broken_exp}
\mu(r) = 
\begin{cases}
\mui + \cfrac{2.5}{\ln(10)}\cdot\cfrac{r}{\hi} & \text{if } r < \rbreak, \vspace{0.25cm}\\
\muo + \cfrac{2.5}{\ln(10)}\cdot\cfrac{r}{\ho} & \text{if } r > \rbreak,
\end{cases}
\end{equation}

\begin{equation} \label{eq:rbreak}
\rbreak = \hi\cdot\ho\cdot\cfrac{\ln{(10)}}{2.5}\cdot\cfrac{\mui-\muo}{\hi-\ho}
\end{equation}

\noindent hence we obtain the probability distribution of the characteristic parameters for both inner and outer profiles, as well as the \rbreak\ probability distribution. We estimated the maximum likelihood solution for \hi, \mui, \ho\ and \muo\ (or best solutions) as the mode of the corresponding distribution. For each simulation, we calculate a value for \rbreak\ following Eq.\,\ref{eq:rbreak}. The best solution of \rbreak\ was defined as the value corresponding to the maximum likelihood solution for \hi, \mui, \ho\ and \muo. Thus, \rbreak\ is a dependent parameter of the inner and outer fits to the profile. In contrast to that, the corresponding \mubreak\ values were calculated from the interpolation of the input profile at the break position $R=\rbreak$.

{\tt{Elbow}} estimates the probability density distributions (PDDs, hereafter) as the distribution of each parameter resulting from the simulations ($\sim100,000$ in our case). We have used these PDDs to estimate the central values of each distribution (maximum probability values, which is the mode of the distribution, as commented above) and the uncertainty intervals (for $1, 2$ and $3\sigma$). We have also used these PDDs to estimate the probability that the inner and outer profiles are significantly different. This is:
\begin{equation} \label{eq:elbow_ho}
H_{0,1}: \hi = \ho,
\end{equation}
\begin{equation} \label{eq:elbow_muo}
H_{0,2}: \mui = \muo.
\end{equation}

Our null hypothesis was that the two profiles are not different, so that the inner slope and the outer slope would be the same ($H_{0,1}$) or that the central magnitude of the inner profile is compatible with that of the outer profile ($H_{0,2}$). Thus the probability for this test consists of two estimators: 1) the fraction of simulations that gave an $\ho > \hi$ in the case of a Type-III profile (and $\ho < \hi$ in the case of a Type-II profile) and 2) the fraction of simulations that gave $\muo > \mui$ in the case of a Type-III profile (and $\muo < \mui$ in the case of a Type-II profile). These two $p$-values are extremely consistent and in most cases they present equivalent or equal results (see Table \ref{tab:fits_psforr}). Notice that this test does not only take into account the presence of a noticeable excess of light in the outskirts of the galaxy, but it also would give negative results if this light distribution is not well fitted by a double exponential disc function. This was the case in several objects with irregularities in the outer regions of their profiles.

After that, we ran the simulations and identified those objects with significant breaks in their final profiles. Finally we visually checked the profiles and the fits, and classified all the objects in the three different classes (Type I, II or III). We assume a $p$-value of $10^{-2}$ as a limit value to accept a profile as a Type III, which corresponds to a $99\%$ confidence level that the PDDs of the characteristic parameters of the profile are not compatible with a Type-I profile. The same was applied for the Type-II profiles. 

In Fig.\,\ref{fig:Profile_types} we show three examples of the final break analysis performed to the deconvolved profiles. The top panels show the composed RGB images of each object with the limiting radius represented as a red ellipse. The intermediate panels of Fig.\,\ref{fig:Profile_types} represent the corresponding surface brightness profile in the $R$ band corrected for dust extinction, cosmological dimming and K-correction of each object. We represent using black filled points those radial bins that were included in the final break fit. Notice that we systematically avoided the innermost regions due the dominant bulge emission. The red and blue dashed-dotted lines represent the best fits to the inner and outer regions. The dotted vertical and horizontal black lines represent the best \rbreak\ and \mubreak\ respectively, with their $1\sigma$ confidence regions marked with black solid lines. The best results obtained for the inner and outer fitted profiles along with their respective confidence intervals are indicated in the intermediate panels. 

For the profile of the object in the first column of panels of Fig.\,\ref{fig:Profile_types} (SHARDS20002966) the confidence regions of the inner (red) and outer (blue) profiles completely overlap (see the left bottom panel). The $p$-value associated with this profile is $p=0.224$, so the null hypothesis is not discarded and therefore, we do not find any difference between the inner and outer profiles, meaning that this galaxy has a Type-I profile. On the contrary, SHARDS10000845 (second column of panels) presents a clear Type-II profile. The associated $p$-value calculated with {\tt{Elbow}} for this profile is $p=3.4 \times 10^{-4}$,  thus we can discard the null hypothesis and the break is statistically significant. A clear Type-III case is SHARDS10000327  (see the third column of panels in Fig.\,\ref{fig:Profile_types}). The associated $p$-value for this Type-III break is $p=3.0 \times 10^{-4}$, thus the break is statistically significant. The lower row of panels of Fig.\,\ref{fig:Profile_types} represents the 2D probability distribution (PDD) of the scale-length $h$ and central surface brightness $\mu_{0}$ for the three cases. The colour scale is linear and represents
density of solutions of both inner (\mui, \hi) and outer profiles (\muo, \ho) in an arbitrary scale. These 2D histograms represent where the fitted solutions concentrate for each profile, in order to find the best central surface brightness and scale-length. In the case of SHARDS10000327 and SHARDS10000845 (which present a clear break) the solutions converge on a bimodal distribution with two maxima, marked with black crosses. On the contrary, the PDD of SHARDS20002966 only presents one clear maximum, meaning that it is a Type-I profile. 

In Fig.\,\ref{fig:histograms_elbow} we show the PDDs for each parameter of the surface brightness profile of SHARDS10000327 (\hi, \ho\ and \rbreak\ in the upper row, and \mui, \muo\ and \mubreak\ in the lower row). We represent with dashed-dotted lines the maximum likelihood solution for \hi, \mui, \ho\ and \muo. The PDD for \rbreak\ shows the solution derived from these best-fit parameters following Eq.\,\ref{eq:rbreak}. Finally, we show the corresponding PDD for the \mubreak\ measured over the surface brightness profile with the best fit solution marked also with a vertical dashed-dotted line. As detailed before, SHARDS10000327 shows a Type-III surface brightness profile. This causes that the PDD for \hi\ is centred at a lower value that the PDD for \ho. An analogous result is found for \mui\ and \muo. As a consequence of this, the histograms for \rbreak\ and \mubreak\ show single-peaked and Gaussian-like PDDs, meaning that the break is well-defined and statistically significant.

The best fitting results for the 44 S0 and E/S0 galaxies at $z<0.6$ within our sample, along with the confidence intervals and $p$-values, are available in Table \ref{tab:fits_psforr} in Appendix \ref{Appendix:Fits_params}. 

\subsection{Efficiency and reliability of the PSF correction}
\label{Subsec:reliability_PSF_correction}
\begin{figure*}[]
\begin{center}
\includegraphics[width=0.346\textwidth, clip, trim=0cm 2.5cm 1.5cm 3.3cm]{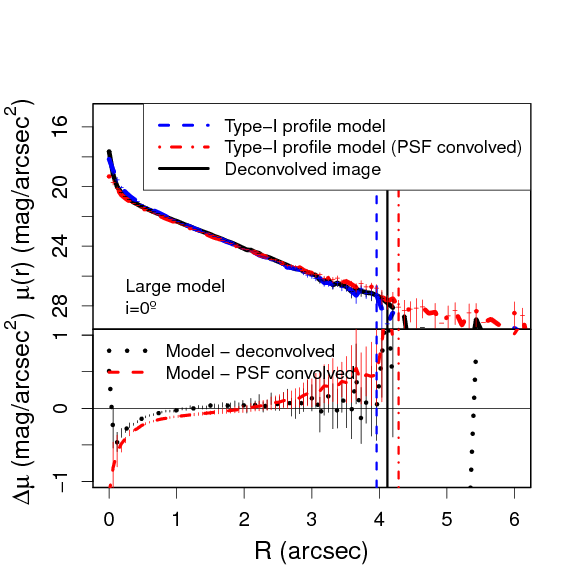}
\includegraphics[width=0.3\textwidth, clip,  trim=2.5cm 2.5cm 1.5cm 3.3cm]{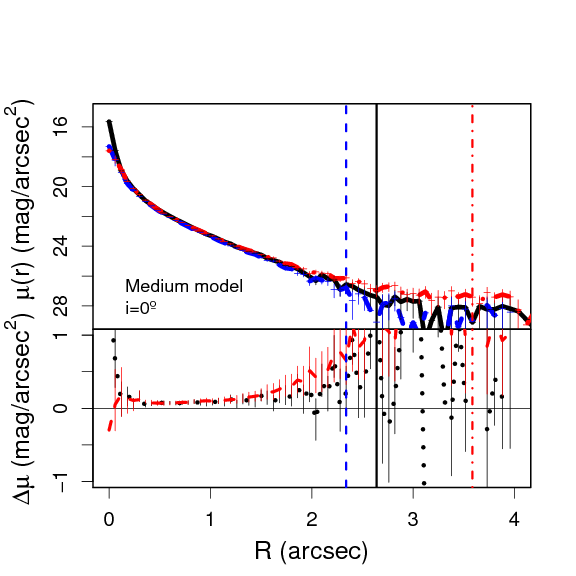}
\includegraphics[width=0.3\textwidth, clip,  trim=2.5cm 2.5cm 1.5cm 3.3cm]{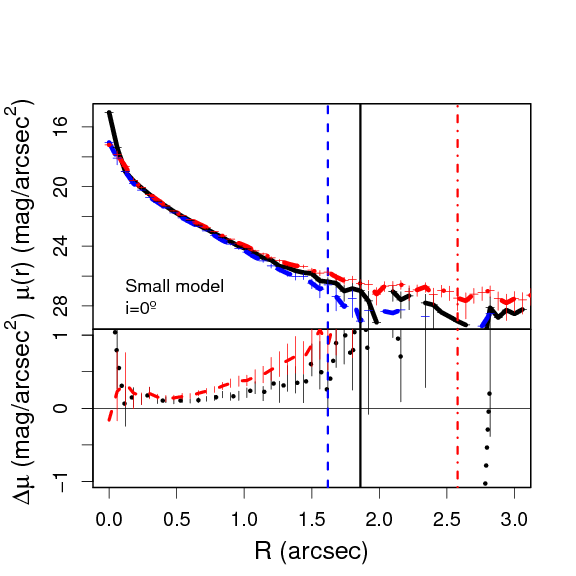}

\includegraphics[width=0.346\textwidth, clip, trim=0cm 2.5cm 1.5cm 3.3cm]{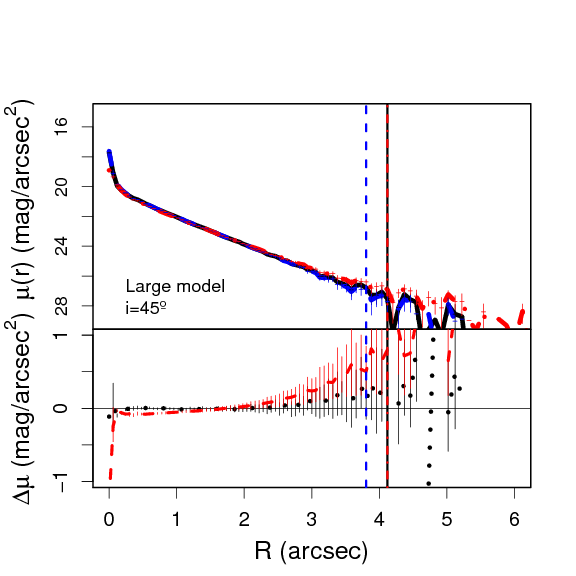}
\includegraphics[width=0.3\textwidth, clip,  trim=2.5cm 2.5cm 1.5cm 3.3cm]{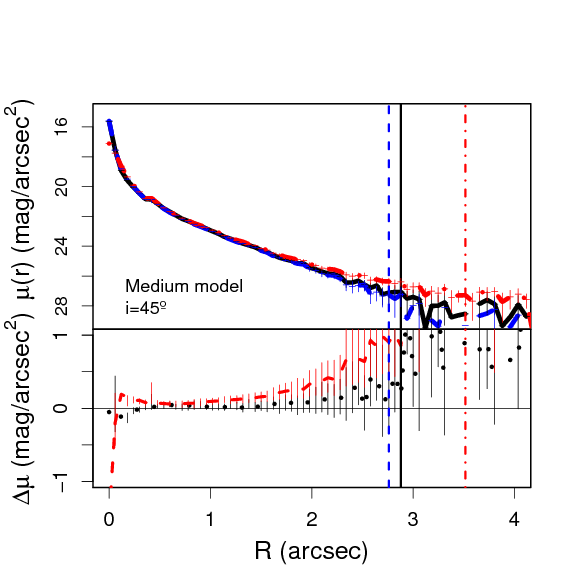}
\includegraphics[width=0.3\textwidth, clip,  trim=2.5cm 2.5cm 1.5cm 3.3cm]{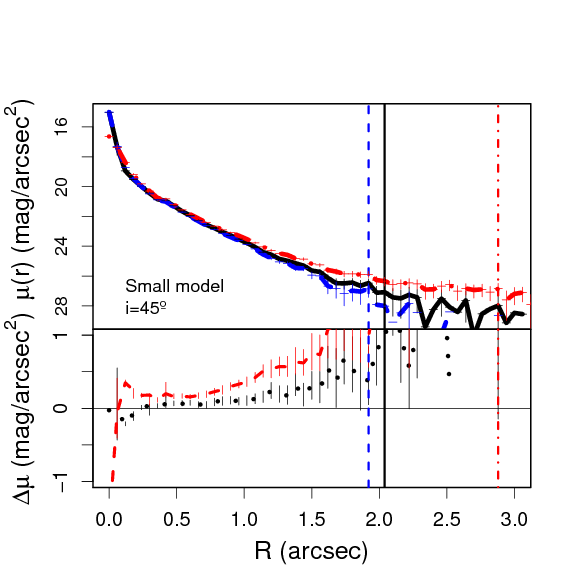}

\includegraphics[width=0.346\textwidth, clip, trim=0cm 0cm 1.5cm 3.3cm]{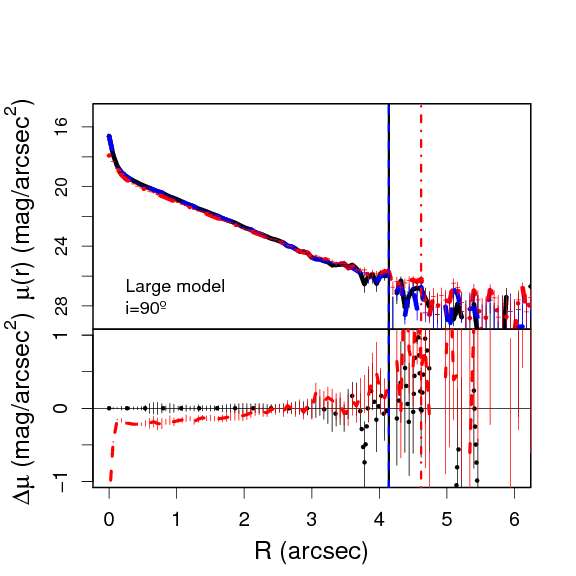}
\includegraphics[width=0.3\textwidth, clip,  trim=2.5cm 0cm 1.5cm 3.3cm]{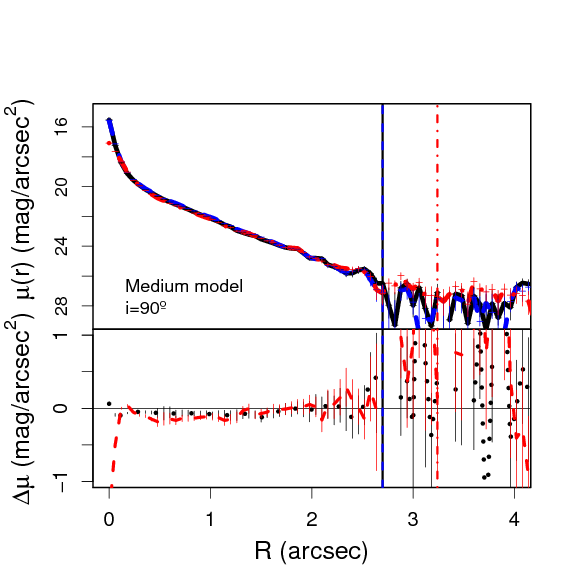}
\includegraphics[width=0.3\textwidth, clip,  trim=2.5cm 0cm 1.5cm 3.3cm]{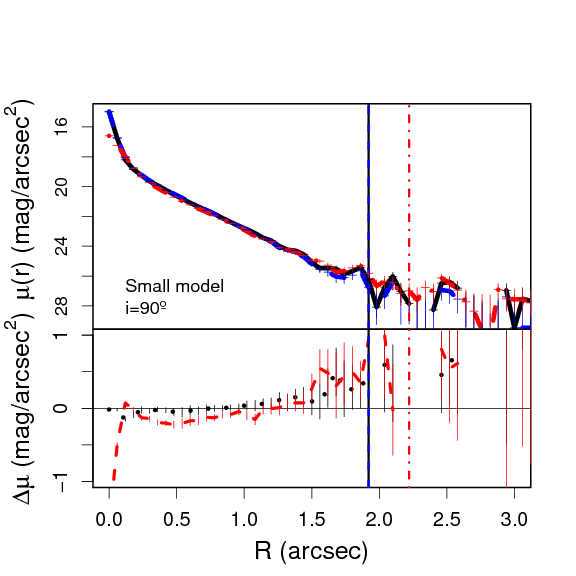}

\caption{Surface brightness profiles of the synthetic images of galaxies: \emph{Top panels:} $i=0º$. \emph{Middle panels:} $i=45º$. \emph{Bottom panels:} $i=90º$. \emph{Left column:} large sized model. \emph{Middle column:} medium sized model. \emph{Right column:} small sized model. \emph{Blue dashed profile:} non-PSF affected surface brightness profile (real profile). \emph{Red dashed-dotted profile:} PSF-affected surface brightness profile (analogous to our original uncorrected data). \emph{Black solid profile:} PSF-corrected surface brightness profile. The vertical solid lines represent the limiting radius of each profile (\emph{Solid black line:} PSF-corrected profile, \emph{Dashed blue line:} non-PSF affected model profile, \emph{Dashed-dotted red line:} PSF-convolved model profile). \emph{Lower panels:} \emph{Black dotted line} Differences between the non-PSF affected and the corrected surface brightness profile. \emph{Dashed line} Differences between the non-PSF affected and the PSF affected surface brightness profile. [\emph{A colour version of the figure is available in the online edition.}]}     
\label{fig:PSF_effects}
\end{center}
\end{figure*}

We additionally tested the reliability and efficiency of the PSF correction process by including synthetic images of galaxies with intrinsic exponential profiles into the GOODS-N image convolved with our PSF. After that, we analysed the type of profile which is recovered following the same procedure that we have used with real ones. To do this, we generated 9 synthetic galaxy models with bulge + pure Type-I exponential discs. We chose three different size ranges: small, medium and large, with $R_{\mathrm{eff,bulge}}= 0.06, 0.12$ and $0.18$ arcsec and $h=0.24,$ $0.42$ and $0.6$ arcsec respectively. We used three different inclinations (0º, 45º and 90º) in order to test the variation of the PSF effects and the dependence of the result of the deconvolution with disc inclination. The models were generated with {\tt{GALFIT3.0}}.

The synthetic images were generated by convolving the models created with {\tt{GALFIT3.0}} with our PSF model (see Sect.\,\ref{Subsec:Modelling}). After this, we included the resulting image in a region free of detections in the GOODS-N mosaic. By doing so, we took into account: 1) the possible presence of background sources not resolved or with low $S/N$ ratio that were not detected in the masking process, and 2) the noise distribution, which had to be similar to the one in the real images. We did not find any differences with the results described next when a pure Gaussian noise distribution was added to the images, which indicates that the sky level and the background sources were not affecting our profiles. After the synthetic galaxies are generated, we used the same procedure as in the case of the real images (see Sects.\,\ref{Subsec:Methods_profiles}--\,\ref{Subsec:Elbow}). 

In Fig.\,\ref{fig:PSF_effects} we show the surface brightness profiles of our synthetic Type-I models convolved (analogous to our original data) and not (i.e., the real profile) by the PSF of our data (blue and red, respectively) and the surface brightness profiles of the deconvolved images (black), i.e., the profiles recovered after using the same extraction and analysis procedures as in our real data. The vertical solid lines represent the radius where $S/N = 3$. The surface brightness profiles of the PSF-affected models (red) present clear excesses in the outskirts compared to the real (non-affected by the PSF) profiles (blue), which systematically increases the radial region where the object has a $S/N>3$. The main result is that, after the PSF correction (black), we successfully remove or reduce most of the scattered light, reconstructing the real shape of the profiles in the centre (blue) at the same time (see the differences between the real, original Type-I profiles and those recovered by our procedure in the corresponding subpanels, as a function of radius). 

We also use {\tt{Elbow}} to analyse whether the PSF wings can create apparent Type-III profiles in the synthetic Type-I galaxies that we have simulated. This test was also useful as a benchmark to test {\tt{Elbow}} on a synthetic model. The results of the analysis are summarised in Table \ref{tab:psf_test}. In the table we show the $p$-values associated with the likelihood of each surface brightness profile to be well-represented by a single exponential profile. We find that:
\begin{itemize}
    \item The amount of light scattered by the PSF to the outskirts of the synthetic objects is generally small, but it is detectable as a break by {\tt{Elbow}} in PSF-uncorrected profiles (especially in the face-on and 45º inclination cases of the small objects).
    
    \item The PSF deconvolution procedure efficiently removes the scattered light in the outskirts and recovers the inner profile at the same time.
\end{itemize}

Consequently, the deconvolved images are completely free of misclassified Type-III profiles in all ranges of sizes and orientations due to PSF effects. We can therefore conclude that the procedure that we have followed for extracting and correcting the surface brightness profiles for PSF effects is reliable and efficient for recovering the real profiles.   

\begin{table}

{\normalsize
\begin{center}
\begin{tabular}{ccccc}
\toprule
Orientation  & PSF & Large & Medium & Small\\
\midrule
\multirow{2}{*}{Face-on} & Uncorrected & 0.105 & 0.010 & $6\times 10^{-4}$\\\vspace{0.1cm}
& Deconvolved & 0.199 & 0.499 & 0.120 \\
\midrule
\multirow{2}{*}{45º} & Uncorrected & 0.119 & 0.013 & $2\times 10^{-4}$\\\vspace{0.1cm}
& Deconvolved & 0.489 & 0.185 & 0.079 \\
\midrule
\multirow{2}{*}{Edge-on} & Uncorrected & 0.009 & 0.131 & 0.085\\
& Deconvolved & 0.365 & 0.230 & 0.496\\

\bottomrule
\end{tabular}
\caption{Likelihood ($p$-values) for the surface brightness profiles of the synthetic models, so that the PSF uncorrected images and the deconvolved models present a Type-I (pure exponential) profile. The small, medium and large models present $R_{\mathrm{eff,bulge}}= 0.18,$ $0.12$ and $0.06$ arcsec and $h=0.6,$ $0.42$ and $0.24$ arcsec respectively. The models were deconvolved following the procedure described in Sect.\,\ref{Subsec:Modelling}, and analysed with {\tt{Elbow.}}}\label{tab:psf_test} 
\end{center}
}
\end{table}

\subsection{AGN in the initial sample}
\label{Subsec:AGN}
We detailed in the previous sections that the deconvolution procedure relies on the fact that the light distribution of the objects is sufficiently well approximated by a certain analytical function (in our case, different combinations of Sérsic profiles). For galaxies hosting a powerful active galactic nucleus (AGN), this may not be the case. The light distribution of the central regions of AGN host galaxies may be much brighter than those with no AGN emission. As a consequence of this, we may be systematically underestimating the PSF contribution in the outskirts.

\citet{2003AJ....126..539A} have identified X-ray emitters in the Chandra Deep Field North, which covers the entire GOODS-N field. They detected 503 sources in 7 X-ray bands between 0.5-8 keV. \citet{2003ApJ...584L..61B} found optical and infrared counterparts for the objects identified by \citet{2003AJ....126..539A}. In \citet{2004AJ....128.2048B}, the authors classified the 504 objects in the categories AGN, star-forming galaxies and Galactic stars by using their optical spectral classifications, radio morphologies, variability, X-ray-to-optical flux ratios, X-ray spectra and their intrinsic luminosities. In order to take into account the presence of AGN in our sample, we have cross-correlated our initial red sample with the AGN catalogues in GOODS-N. We used the optical coordinates from \citet{2003ApJ...584L..61B} to find the matching sources within 1 arcsec of separation. 

We found 9 matching sources, 4 of them classified as star-forming galaxy candidates (SHARDS10001344, SHARDS20000827, SHARDS20002935 and SHARDS20003134) and 5 of them classified as AGN (SHARDS10000827, SHARDS20002147, SHARDS20003119, SHARDS20003377 and SHARDS20004440). All of them are S0 or E/S0s in our classification, with the exception of SHARDS20003119, which was classified as a red spiral by us.  
The maximum distance between the matching sources in the Rainbow catalogue and the optical coordinates was less than 0.3 arcsec and we did not find any other possible counterparts until we increased the matching radius to 5 arcsec, well above the precision of the astrometry. 

The emission of a source smaller than the pixel size of the detector would create a 2D light distribution equal to the PSF, by definition. In order to account for the possible optical emission of the AGN in these objects, we have added a PSF component to their {\tt{GALFIT3.0}} model. Therefore, we fitted the AGN, the bulge component and the disc as the sum of a PSF, a free Sérsic and an exponential ($n=1$) Sérsic profile. The model fitting was performed inside out, starting from a PSF + free Sérsic profile for the inner regions and then adding the disc component once a stable solution was reached for the inner regions. In some cases, we needed to fix several parameters alternatively, such as total fluxes or the centres of the components. Finally, we inspected the residuals and the solution, paying attention to the innermost regions. Due to unavoidable degeneracies between the PSF and the bulge total flux as a result of the limited spatial resolution, we have flagged the AGN objects in our sample in Table \ref{tab:fits_psforr} in Appendix \ref{Appendix:Fits_params}, although we have kept them in the final sample of S0--E/S0 galaxies. 

Furthermore, the classification of 4 S0 galaxies by \citet{2004AJ....128.2048B} as star-forming galaxies appears to be in contradiction with the results from the SFR calculated using the data in the Rainbow Database (see Sect.\,\ref{Subsec:Results_SFR}). We checked ancillary data for each one of these objects for signs of star-formation. 

SHARDS20000827 was included in the sample from \citet{2005ApJ...633..174T} (Table 1, ID 1267). These authors do not report any [OII] or [H$\delta$] emission lines in a $S/N=32.12$ spectrum and the galaxy was also classified as S0. This object was also studied in \citet{2007MNRAS.377..203G} (Table 1, ID 81). They do not find any IR emission excess above the stellar expectation. They pointed out that X-ray emission of this object may be due to hot gas and low mass X-ray binaries.

SHARDS10001344 was also classified as an IR-faint object in \citeauthor{2007MNRAS.377..203G}, (Table 1, ID 354). These authors propose a post-starburst classification (possibly due to a past merger event) as a cause for the observed X-ray emission, although the observed IR luminosity is compatible with those of early-type galaxies.

The observed wavelength of the $H\alpha$ emission line for SHARDS20002935 is $\lambda_{H\alpha} \sim 9677$ \AA\ ($z=0.4745$), which is in the observable spectral window of the WFC3 grism in the NIR (G141). The Rainbow Database contains a G141 1D spectrum, which we analysed looking for any signs of emission lines. We found an $H\alpha$ emission line candidate at $\lambda = 9692$ \AA, with a total flux of $f_{H\alpha}=1.61^{+1.61}_{-0.82}\times 10^{-16}$ erg cm$^{-2}$ s$^{-1}$. This emission entails a total $H{\alpha}$ luminosity in of $L_{H{\alpha}}=1.34^{+1.36}_{-0.69}\times 10^{40}$ erg s$^{-1}$. Assuming the expression given by \citet{1994ApJ...435...22K}: 
\begin{equation}
\mathrm{SFR_ {H\alpha}(M_\sun\,yr^{-1})} = 7.9\times 10^{-42}\,L_{H_\alpha} \mathrm{(erg\,s^{-1})},
\end{equation}
we estimate a SFR$_{H\alpha} = 1.06^{+1.07}_{-0.54}$ M$_{\sun}$ yr$^{-1}$. This value is very similar to the total SFR obtained by the combination of the NIR and 2800 \AA\ emission and thus compatible with a quiescent galaxy (SFR$=3.05$ M$_{\sun}$ yr$^{-1}$ and $\log_{10}M/M_{\odot}$ = 10.92, see Sect.\,\ref{Subsec:Results_SFR}).

The Team Keck Treasury Redshift Survey \citep[TKRS,][]{2004AJ....127.3121W} has an available optical spectrum between 5000 -  10000 \AA\ of SHARDS20003134 (ID 4554). Simple visual inspection of the data was sufficient to discard any signs of emission lines associated with star formation.

Finally, \citet{2007ApJ...667..826P} uses SED fitting procedures to estimate photometric redshift and separate galaxy types. According to their results the SEDs of SHARDS10001344, SHARDS20000827, SHARDS20002935 and SHARDS20003134 are compatible with those of early-type galaxies. 

In conclusion, we have identified the objects and corrected the images and the profiles of the sources with possible AGN activity accounting for the central AGN light emission. In addition to this, the ancillary data studied supports our morphological classification, and confirms that the SFR levels of the objects included in our sample are compatible with those of early-type galaxies. Therefore, these 4 AGN with S0 morphology were finally kept in the final S0 -- E/S0 sample.


\section{Results and discussion}
\label{Sec:Results}

\subsection{Statistics of the red galaxy sample by morphology}
\label{Subsec:Results_statistics}

After the revision performed to the visual morphological classification by checking the images and the PSF-corrected surface brightness profiles, we find that, from the original sample of 150 red galaxies at $0.2<z<0.6$, 38 were finally classified as S0 ($25.3\%$) and 12 as E/S0 ($8.0\%$) - this is a total of $33.3\%$ S0 and E/S0 objects in the red galaxy sample. Additionally, 40 sources were classified as compact post-starburst objects ($26.6\%$), 32 objects have been classified as diffuse objects ($21.3\%$), 9 objects present clear signs of being in interaction with others ($6.0\%$), 7 objects are classified as "green peas" ($4.7\%$), 9 objects have been confirmed as E galaxies ($6.0\%$), and we have also identified 3 spiral galaxies. 

The results of the morphological classification of the red galaxy sample selected at $0.2<z<0.6$ within the SHARDS catalogue in the GOODS-N field and the following physical parameters are presented in the Appendix A: position ($\alpha$, $\delta$), morphological type, photometric redshift ($z_{\mathrm{phot}}$), spectroscopic redshift when available ($z_{\mathrm{spec}}$), stellar mass, SFR, rest-frame absolute magnitude in the Johnson $V$ ($M_{V}$) and $K_{\mathrm{s}}$ bands ($M_{K}$), the extinction correction ($A_{\mathrm{F775W}}$) and the median ($R$ - F775W) colours used for K-correction. In conclusion, from 150 objects from the red sample we have removed 100 objects and we have obtained a final sample of 38 S0s and 12 E/S0s (50 objects in total) to search for Type-III disc profiles within them.  

\subsection{Star formation rates and masses as a function of the morphology}
\label{Subsec:Results_SFR}

\begin{figure*}[]
 \begin{center}
\includegraphics[width=\textwidth]{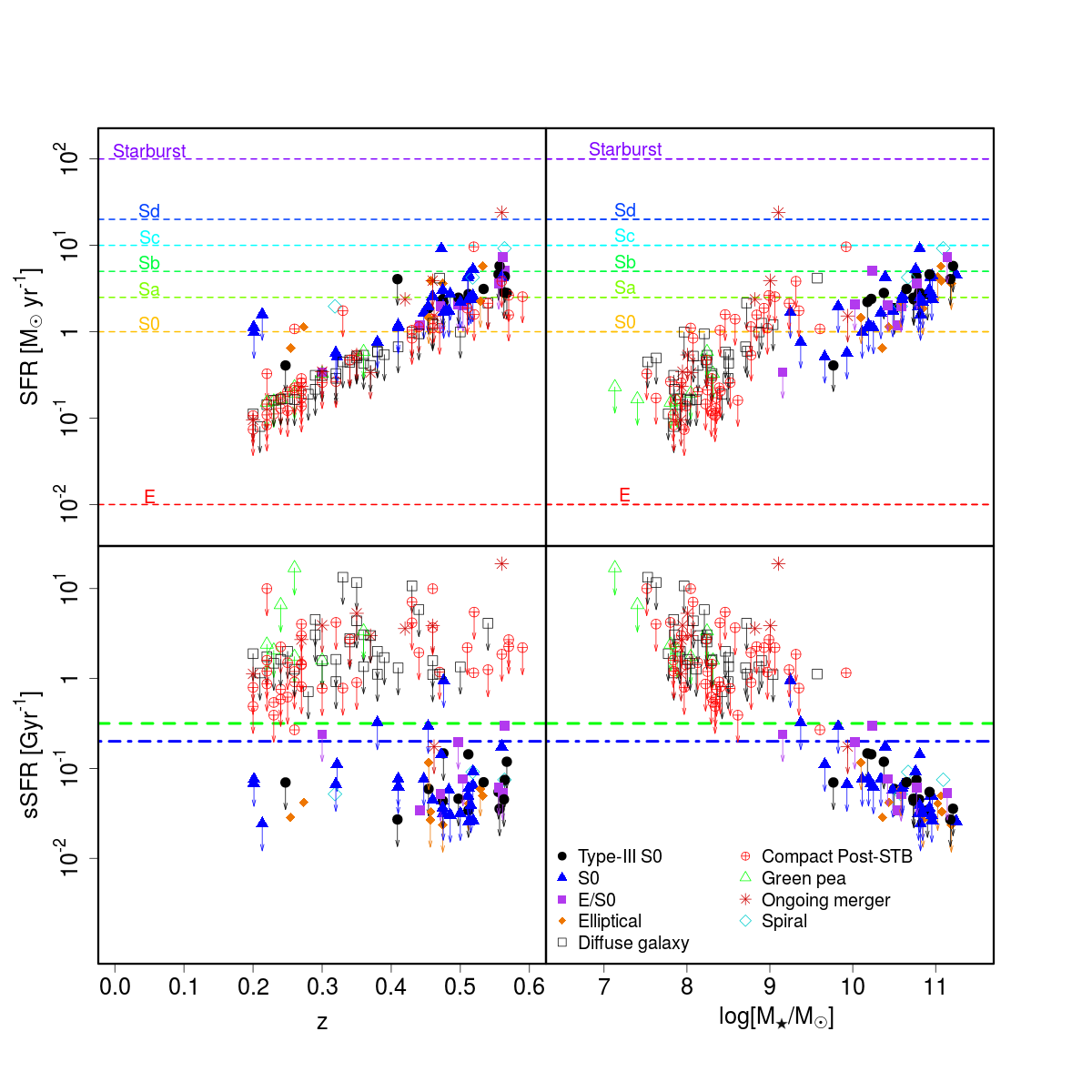}

\caption{SFR (top row) and sSFR (bottom row) for the objects of the initial red sequence sample, as a function of $z$ (left column) and the stellar mass (right column), according to their morphological types. Arrows indicate upper limits for the SFRs. The horizontal dashed lines in the top panels represent the typical values for the SFR for different morphological types in the local universe (see references in the text). Note that these values change with $z$ (see the text). The blue dash-dotted line and the green dashed line in the bottom panels correspond to two typical reference upper limits for the sSFR in early-type galaxies (see the text). Consult the legend in the figure for the morphological type. [\emph{A colour version of the figure is available in the online edition.}]} 
\label{fig:SFR}
\end{center}
\end{figure*}

\begin{figure*}[]
\begin{center}
\includegraphics[width=0.49\textwidth]{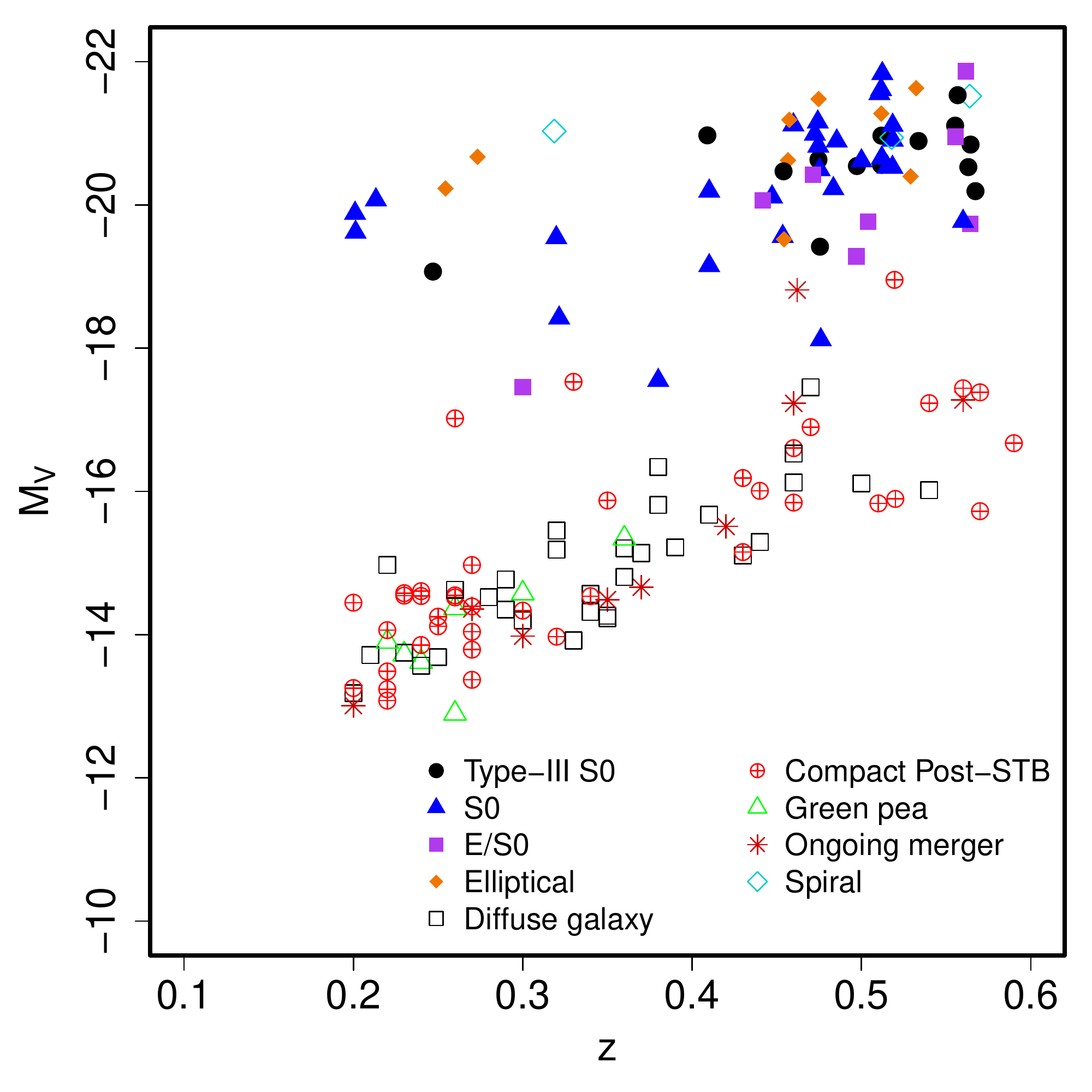}
\includegraphics[width=0.49\textwidth]{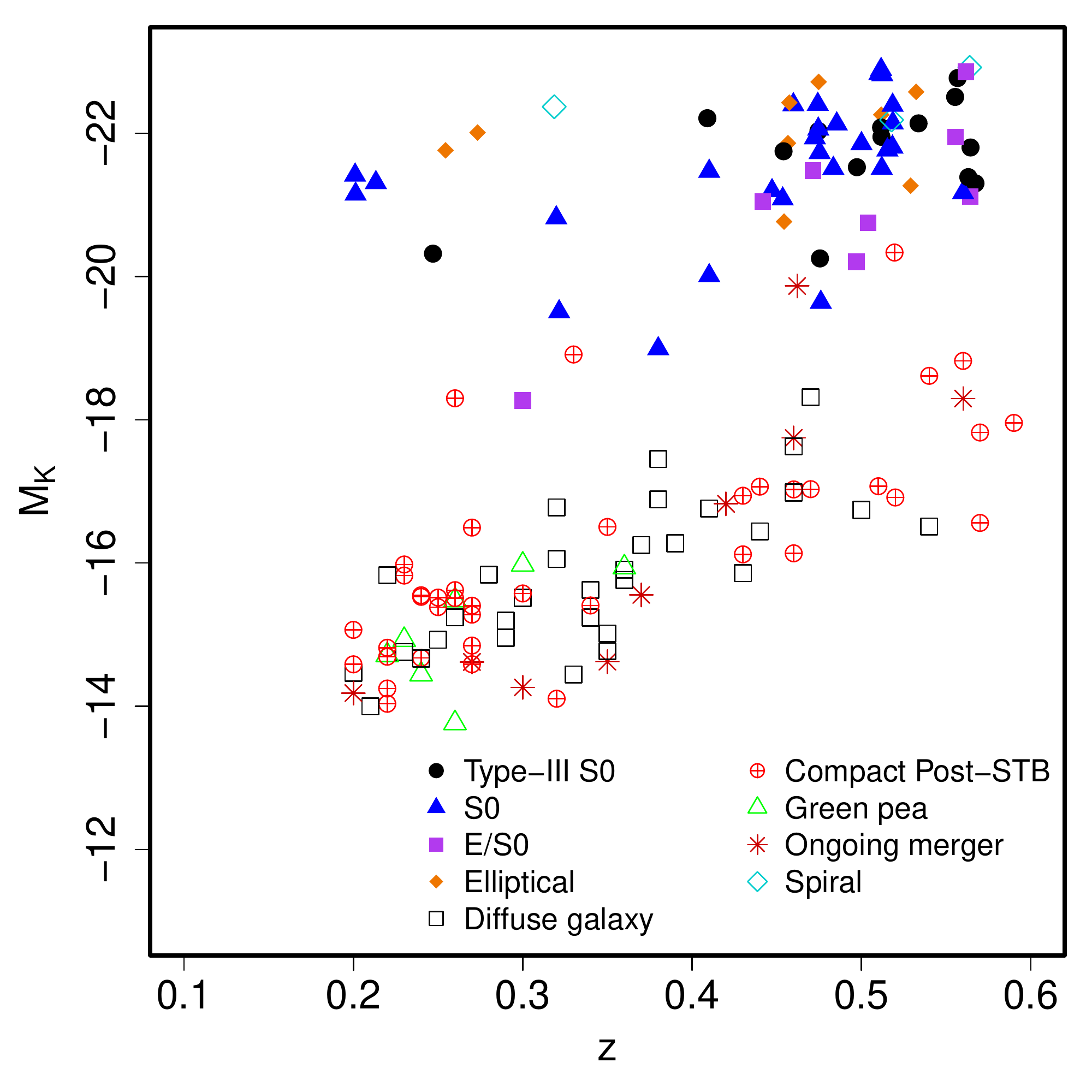}

\caption{Synthetic rest-frame absolute magnitude distribution of the objects in the red sample versus $z$, as a function of their morphological types. \emph{Left panel:} Johnson $V$ band rest-frame absolute magnitude. \emph{Right panel:} $K_{\mathrm{s}}$ band rest-frame absolute magnitude. [\emph{A colour version of the figure is available in the online edition.}]
} 
\label{fig:MvMk}
\end{center}
\end{figure*}

The right panel of Fig.\,\ref{fig:Kcorr_zcomp} represents the stellar masses of the objects of the red sequence sample vs. their redshift. We have used different symbols according to the assigned morphology, to compare the properties of our selected subsample of 50 S0 and E/S0 with those of all galaxies in the initial red sequence sample. The objects classified within the well-resolved high-$S/N$ morphological types (see Sect.\,\ref{Subsec:Red_sample}) present stellar masses two orders of magnitude higher than the low-$S/N$ counterparts. The S0 and E/S0 galaxies have stellar masses between $10^\mathrm{9}$ and $10^{\mathrm{11}}$ M$_{\odot}$. 

One of the fundamental characteristics of the S0 galaxies is their low SFR in the discs, by definition. We have selected our initial sample by using a colour-colour diagram that isolates the objects in the red sequence, which are good candidates for having old stellar populations and no recent star formation episodes (see Sect.\,\ref{Subsec:Red_sample}). In order to confirm their quiescence, we have used the available data in the Rainbow Database to ensure that the SFRs of the selected S0 and E/S0 objects are low, coherently with their morphological type. In Fig.\,\ref{fig:SFR} we represent the SFR (top panels) and the specific SFR, defined as sSFR=SFR/$M_{\star}$ (bottom panels) for the objects of the initial red sequence sample, as a function of $z$ (left column) and the stellar mass ($M_{\star}$, right column). The horizontal dashed lines represent the typical values for the SFR of different morphological types in the local universe \citep{1989ApJ...344..747T, 1989ApJ...347L..55Y,1991ARA&A..29..581Y,1994AJ....108.1186C,1998ARA&A..36..189K,2000AJ....120.1946S,2010ApJ...708..841W,2010ApJ...725L..62W,2014ApJ...783..135A}. The blue dotted-dashed line in the bottom panels corresponds to a conservative upper limit for the sSFR in quiescent galaxies \citep[sSFR $= 0.2$ Gyr$^\mathrm{-1}$, see][]{2016MNRAS.457.3743D}. The dashed green line represents a higher value for this limit of quiescence \citep[sSFR $= 0.32$ Gyr$^{-1}$, see][]{2013ApJ...765..104B}. The procedure to estimate the total SFR from the UV emission at 2800 \AA\ (caused by young stars) and the total IR emission between 8 and 1000 $\mu m$ (caused by dust re-emission) is fully detailed in \citet[][and references therein]{2013ApJ...765..104B}. 

The data available in the Rainbow database do not provide direct SFR values for most of our objects. In fact, 129 out of 150 only have upper limits for the SFR (represented by down-pointing arrows, see the top panels in Fig.\,\ref{fig:SFR}). We note that the observed correlation of the SFR upper limits is a systematic effect due to the detection limits in NIR and UV. Thus, for the objects at increasing $z$, we obtain higher upper limits. Nevertheless, this does not affect our aim, because we only needed to test whether the objects present low levels of star formation or not and the detection limits are low enough to do so. In the upper panels of Fig.\,\ref{fig:SFR}, it is shown that the galaxies classified as S0 and E/S0 present SFR levels close to the typical values of local S0-Sa galaxies, as expected for the quiescent nature of S0s and E/S0s, corroborating the morphological classification that we have performed.  Multiple studies indicate that galaxies were forming stars more actively in the past, regardless of their environment \citep{1996ApJ...460L...1L, 2005ApJ...619L..47S}. The typical SFR of S0 and E/S0 galaxies increases from the local universe to $z=0.8$ and it reaches values of $4$ M$_{\odot}$ yr$^{-1}$ at $z=0.6$ \citep{2001ApJ...562L..23M}. This is in good agreement with the SFR distribution of our S0 and E/S0 sample ($\langle$SFR$\rangle $ $=2.4^{+1.0}_{-1.9}$ M$_{\odot}$\,yr$^{-1}$).

We also show that the sSFR of those objects classified as S0 and E/S0 is lower than -or close to- the typical values usually considered as the upper limit to identify quiescent objects in the bottom panels of Fig.\,\ref{fig:SFR}. The only object that presents an upper limit for the sSFR higher than the reference values is SHARDS20002889 (SFR $<1.68$ M$_{\odot}$\,yr$^{-1}$, sSFR $<0.95$ Gyr$^{-1}$, $\log_{10}M/M_{\odot}$ = 9.25). Careful visual inspection of the images reveals a close bright galaxy in the F435W band, which has a clumpy structure and could be a face-on spiral galaxy. Traces of star formation in the outskirts of our object by this spiral galaxy may be biasing the sSFR derived from Rainbow data towards higher values. We note that the objects classified as diffuse, green peas, and compact post-starburst galaxies show higher sSFR than the S0 and E/S0s, clearly above the limits of quiescence. Those objects present similar values for the total SFR as our S0s and E/S0s but much lower stellar masses, thus resulting in higher sSFR values.

In Fig.\,\ref{fig:MvMk} we show the distribution of rest-frame absolute magnitudes in the Johnson $V$ band and $K_{\mathrm{s}}$ band, as a function of their redshift and the morphological type. We find that the objects classified as S0 or E/S0 present absolute $V$ band magnitudes between $-22<M_{\mathrm{V}}<-18$. Quiescent galaxies present colours with typical values of $(B-V)\sim1.0$ \citep{1996ApJ...467...38K,2001ApJ...550..212B}, so the objects classified as S0s and E/S0s in our sample, would present absolute magnitudes in the Johnson $B$ band of $-17>M_{\mathrm{B}}>-21$. The available data on local Type-III S0 galaxies present a similar range of absolute magnitudes in the $B$ band \citep{2008AJ....135...20E,2011AJ....142..145G}, so the S0 and E/S0 galaxies in our sample at $0.2<z<0.6$ are analogous to those in the available local samples, enabling the direct comparison with them. 

The right panel of Fig.\,\ref{fig:MvMk} shows the same diagram as in the left panel, but for the rest-frame $K_{\mathrm{s}}$ absolute magnitude. We find that the S0 and E/S0 galaxies of our sample present values between $-19>M_{\mathrm{K}}>-23$. By using the stellar masses in the Rainbow database we find a mass-to-light ratio in the $K$ band of $M/L_{\mathrm{K}} = 0.68^{+0.43}_{-0.16}\, M_{\odot}/L_{K,\odot}$. Assuming $(B-V) \sim 1.0$, \citet{2003ApJS..149..289B} predict a mass-to-light ratio in the $K_{\mathrm{s}}$ band of $M/L_{\mathrm{K}} \sim 0.85\,  M_{\odot}/L_{K,\odot}$ for quiescent galaxies, which is compatible with the results found for the objects in our sample.

We conclude that the SFR and sSFR of the objects that we have identified visually as S0s and E/S0s are characteristic of quiescent objects at the corresponding redshifts of each galaxy, therefore supporting our morphological classification.

\subsection{Types of disc profiles in the S0 sample}
\label{Subsec:Profiles}

\begin{figure}[]
 \begin{center}
\includegraphics[width=0.45\textwidth,clip, trim=0.0cm 0.2cm 1cm 2cm]{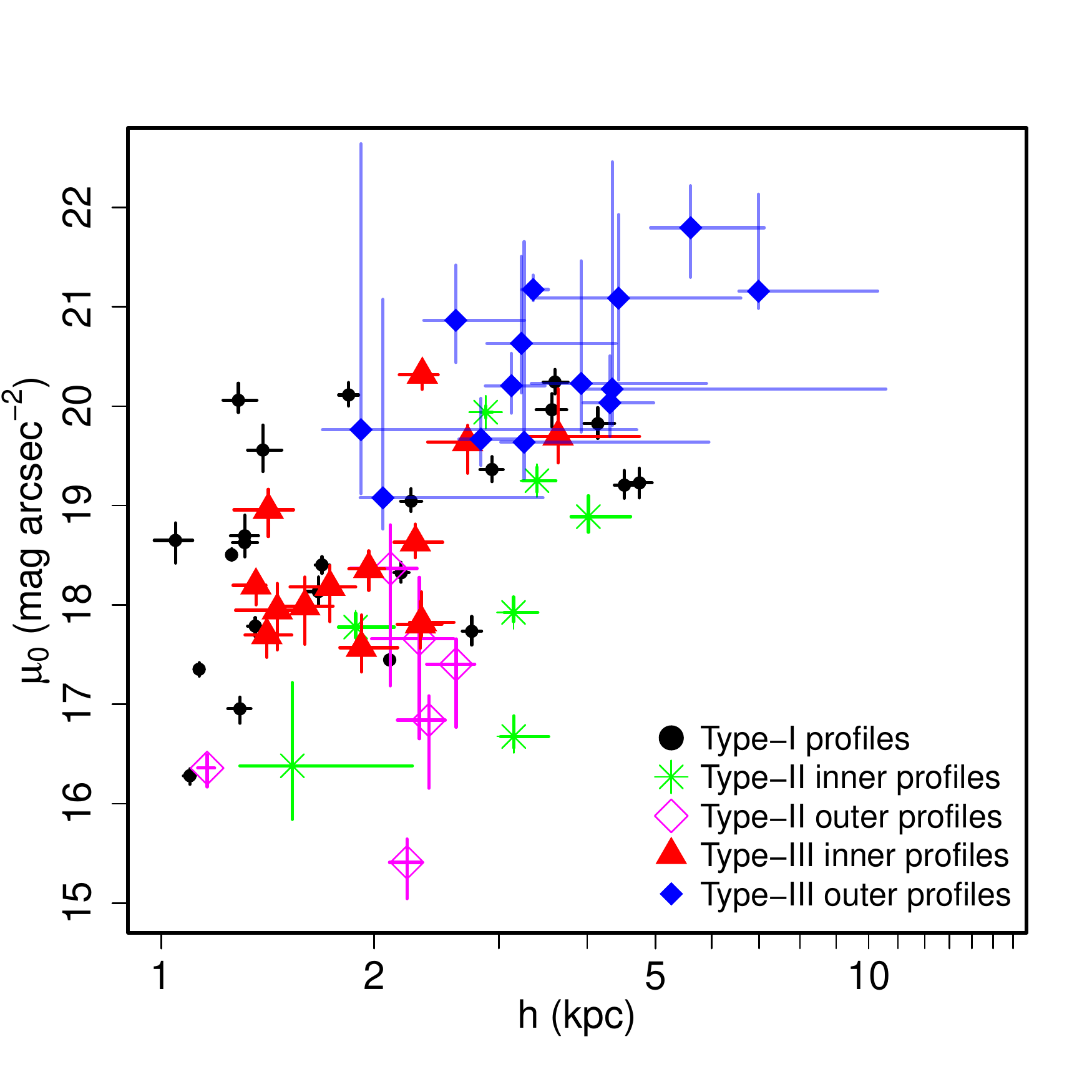}

\caption{Central surface brightness $\mu_{0}$ vs. scale-length $h$ diagram showing the distribution of the Type-I, Type-II and Type-III profile parameters. See the legend for details} 
\label{fig:mu_h_comp}
\end{center}
\end{figure}

\begin{figure*}[]
 \begin{center}
\includegraphics[width=\textwidth]{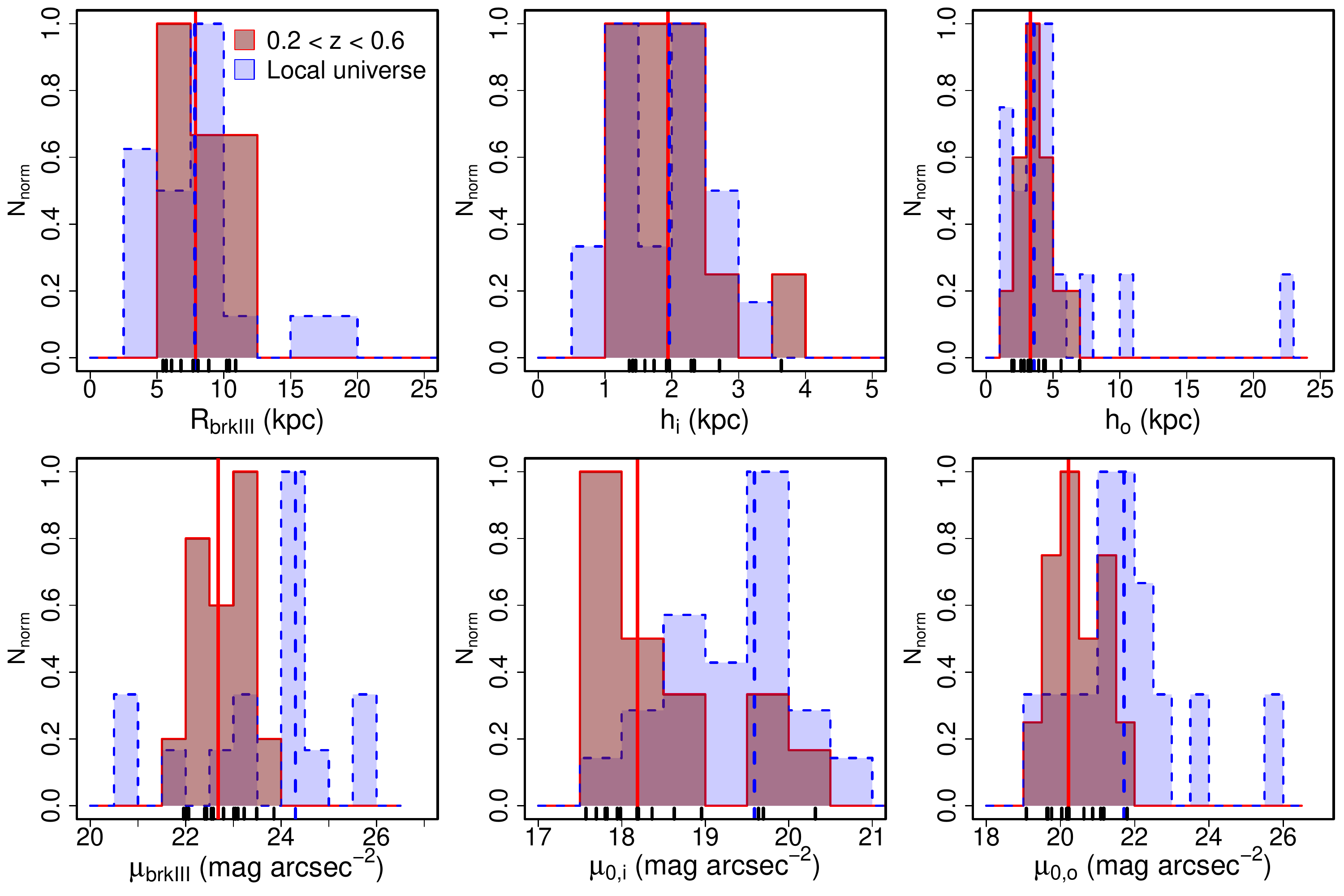}

\caption{Normalised distributions of the structural and photometric parameters of the Type-III breaks of the our sample of S0 and E/S0 at $0.2<z<0.6$, compared to those from the local universe (E08 and G11). \emph{Upper row, from left to right:} break radius \rbreak\ (kpc), scale-length of the inner disc profile \hi\ (kpc), scale-length of the outer profile \ho\ (kpc). \emph{Lower row, from left to right:} surface brightness at the break radius \mubreak\ (\magarc), central surface brightness of the inner disc profile \mui\ (\magarc), central surface brightness of the outer profile \muo\ (\magarc). The red solid histogram marks our sample of Type-III S0 and E/S0 galaxies at $0.2<z<0.6$. The blue dashed histogram represents the  distributions found for local universe sample from data by E08 and G11. [\emph{A colour version of the figure is available in the online edition.}]}  
\label{fig:param_histogram}
\end{center}
\end{figure*}

We created {\tt{GALFIT3.0}} models for 44 objects from the sample of 50 S0 and E/S0s ($\sim88\%$). We discarded the 6 remaining objects because they presented an apparent size too small and/or a {\tt{GALFIT3.0}} solution too unstable for a bulge+disc model (SHARDS10001928, SHARDS10002901, SHARDS10005029, SHARDS10008552, SHARDS20000858, SHARDS20004273). The presence of significant breaks in the S0 -- E/S0 sample was confirmed quantitatively by {\tt{Elbow}}. Using automatic fitting classification, we classified the galaxies into 3 types: Type I (23 objects, 52\% of the sample), Type II (7 objects, 16\% of the sample) and Type III (14 objects, 32\% of the sample). 

Due to the limiting depth of the observations, we find three different cases where a Type-III profile may be detectable by the automated break analysis routine {\tt{Elbow}}:
\begin{enumerate}
    \item The outer profile is clearly resolved, usually associated with bright values of the \mubreak\ and low $p$-values of {\tt{Elbow}} (e.g, SHARDS10000327).
    \item The outer profile is not clearly resolved, because it presents low values of the $\ho/\hi$ ratio or dimmer values of the \mubreak, that usually result in smaller detectable radial zones to fit (e.g, SHARDS10003647).
    \item The anti-truncation is observed on a bright region, but the outer profile presents irregularities or is not well represented by an exponential disc (e.g, SHARDS10001344). 
\end{enumerate}

In order to estimate the effects of the PSF on the profiles, we also analysed the original profiles without the PSF correction. We found that 6 anti-truncations detected in PSF-uncorrected images were removed from the profiles after the PSF correction (SHARDS10001269, SHARDS20002147, SHARDS20003377, SHARDS20003678, SHARDS20004440, and the apparent hybrid Type II+III profile, SHARDS10001847, which resulted to be a simple Type-II profile). Two Type-I profiles were classified as Type-II after the PSF subtraction (SHARDS10004777 and SHARDS20002995). None of the Type-II profiles was classified as Type I or Type III after PSF correction. Due to the shape of the PSF - that disperses light from the brighter and (almost always) inner parts of a source - this was an expected result, given that it is extremely unlikely for any object to present more light in the outskirts after PSF deconvolution.

Finally, we report that 14 out of a final sample of 44 S0 and E/S0 galaxies present anti-truncated surface stellar profiles once corrected for PSF effects. As explained above, this final sample of 44 objects only includes those S0 and E/S0 galaxies that were successfully corrected from PSF effects through 2D modelling and were not contaminated by FoV objects. Note that the $p$-values for the break test in SHARDS20003134 present very different values ($p_{\mathrm{h}}=2.0\times10^{-2}$ and $p_{\mathrm{\mu_{0}}}=7.6\times10^{-4}$). The profile clearly shows a Type-II break, although with an irregular inner profile, which could be the cause of the lack of agreement in this case. 13 out of the 14 the Type-III objects have $z$ values equal or higher than 0.4, with the exception of SHARDS20000593, which has a spectroscopic redshift of $z=0.247$. The lack of objects at the lower redshift range is expected since the $82.6\%$ of the S0 and E/S0 sample have redshifts higher than $z=0.4$. This means that we find a fraction of $\sim30\%$ of Type-III profiles in S0 and E/S0 galaxies at $0.4<z<0.6$. This fraction of Type-III profiles from the total S0 and E/S0 sample is slightly lower than the observed ratio of Type-III profiles in local S0 and E/S0 galaxies in E08 and G11 ($\sim 50\%$), but compatible to the $\sim 20-30\%$ detected in \citet{2014MNRAS.441.1992L,2016A&A...596A..25L}.

The final profile classification is available in Table \ref{tab:fits_psforr}. Comments on individual objects, RGB images and the piecewise fits to the PSF-corrected disc surface brightness profiles of the 44 S0 and E/S0 galaxies in our PSF-corrected sample are available in Appendix \ref{Sec:TypeIIIcomments}. 

Finally, we compare the surface brightness profiles of the identified Type-I galaxies with the surface brightness profiles of the Type-II and Type-III galaxies from our sample. We fit single exponential profiles to the Type-I profiles and estimate their values of scale length $h$ and central surface brightness $\mu_{0}$. We found a median value for the scale length of the Type-I galaxies of $h=1.69^{+0.42}_{-0.29}$ kpc, and the central surface brightness of $\mu_{0}=18.65^{+0.39}_{-0.14}$ \magarc.
In contrast, the median values for the scale-lengths of Type-II profiles are $\hi=3.15^{+0.00}_{-0.27}$ kpc and $\ho=2.23^{+0.09}_{-0.12}$ kpc, and $\mui=17.92^{+0.96}_{-0.15}$ \magarc, $\muo=16.84^{+0.56}_{-0.48}$ \magarc\ for the central surface brightness. In the case of Type-III profiles, these values are $\hi=1.94^{+0.21}_{-0.34}$ kpc and $\ho=3.31^{+0.13}_{-0.61}$ kpc, and $\mui=18.19^{+0.20}_{-0.28}$ \magarc, $\muo=20.21^{+0.10}_{-0.43}$ \magarc\ for the central surface brightness (see Table \ref{table:median_values}). As before, the median values were estimated by 50.000 1$\sigma$ bootstrapping + Monte Carlo simulations. In Fig.\,\ref{fig:mu_h_comp} we show the distribution of the central surface brightness $\mu_{0}$ vs. scale-length $h$  for each profile type. We have represented the inner and outer profile parameters by separate in the case of Type-II and Type-III galaxies. We found an overall agreement between the distributions of the inner parts from the Type-III profiles and the same parameters from Type-I profiles, which is not observed in the outer Type-III profiles or in the Type-II profile parameters. We conclude that the profiles of the Type-I S0 and E/S0 galaxies in our sample at $0.2<z<0.6$ have values for the characteristic parameters more similar to the inner discs of the Type-III galaxies than those of the outer profiles of Type-III galaxies (see Table \ref{table:median_values}).  

\begin{table}
{\small 
\begin{center}
\begin{tabular}{lllll}
\toprule
&\multicolumn{1}{c}{Parameter}& &$z\simeq0$&$0.2<z<0.6$\\
\midrule
a) & \multirow{1}{*}{<\rbreak>} & [kpc] & $7.8^{+1.1}_{-1.1}$  &  $7.88^{+1.10}_{-0.98}$\\[1.2ex]
b) & \multirow{1}{*}{<\hi>} & [kpc] & $1.96^{+0.49}_{-0.13}$ & $1.94^{+0.21}_{-0.34}$\\[1.2ex]
c) &\multirow{1}{*}{<\ho>} & [kpc] & $3.58^{+0.21}_{-0.83}$ & $3.31^{+0.13}_{-0.61}$\\[1.2ex]
\hline

\rule{0pt}{2.7ex}d)&\multirow{1}{*}{<\mubreak>} & [mag arcsec $^{-2}$] & $24.30^{+0.50}_{-0.10}$ & $22.68^{+0.17}_{-0.32}$\\[1.2ex]
e)&\multirow{1}{*}{<\mui>} & [mag arcsec $^{-2}$] & $19.59^{+0.52}_{-0.25}$ & $18.19^{+0.20}_{-0.28}$\\[1.2ex]
f)&\multirow{1}{*}{<\muo>} & [mag arcsec $^{-2}$] & $21.60^{+0.27}_{-0.30}$ & $20.21^{+0.10}_{-0.43}$\\[0.5ex]
\bottomrule
\end{tabular}
\caption{Median values of the structural and photometric parameters fitted to Type-III profiles of real S0s. \emph{Left column:} Values for the samples on the local Universe (E08, G11). \emph{Right column:} Values for the sample at $0.2<z<0.6$. \emph{Rows from top to bottom:} a) Break radius \rbreak\ (kpc), b) inner profile scale length \hi\ (kpc), c) outer profile scale length \ho\ (kpc), d) surface brightness at the break radius \mubreak\ (\magarc), e) central surface brightness of the inner profile \mui\ (\magarc), f) central surface brightness of the outer profile \muo\ (mag arcsec $^{-2}$).}\label{table:median_values}
\end{center}
}
\end{table}

\subsection{Comparison of anti-truncated discs of S0 galaxies at $0.2<z<0.6$ with local analogs}
\label{Subsec:Local_comp}

In Fig.\,\ref{fig:param_histogram} we show the distributions of the characteristic parameters of the Type-III surface brightness profiles of our sample of S0 and E/S0 galaxies at $0.2<z<0.6$, compared to those of local Type-III S0 galaxies by E08 and G11. This local sample consists 20 antitruncated S0 -- E/S0 galaxies. Thus the sample size is low but comparable to our own ($N=14$). The median values of each distribution are shown with solid vertical lines. The top panels represent the distribution of the structural parameters (\rbreak, \hi\ and \ho), and the lower panels represent the distribution of the photometric parameters (\mubreak, \mui\ and \muo). The median values of the distributions for each parameter, for both samples, are summarised in Table \ref{table:median_values}, and were estimated by 50.000 $1\sigma$ bootstrapping + Monte Carlo simulations. 

The distributions of the structural parameters \rbreak, \hi, and \ho\ in the top panels of Fig.\,\ref{fig:param_histogram} are very similar for the local and our $0.2<z<0.6$ sample, showing compatible median values (see also Table \ref{table:median_values}). We use the Anderson-Darling criterion \citep{10.2307/2288805} in order to test whether the parameters from both samples arose from a common unspecified distribution function (null hypothesis). The results indicate that there are no noticeable differences between the distributions of the structural parameters (the $p$-values are $p=0.463$ for \rbreak, $p=0.243$ for \hi\ and $p=0.542$ for \ho). We found that our sample at $0.2<z<0.6$ shows lower maximum values of \ho\ (i.e, the largest outer profile scale length in the local sample has a value $\ho=22.26$ kpc, while the maximum of the our higher redshift sample is $\ho \sim 7$ kpc). Similarly, the outermost break of the local sample presents a $\rbreak=18.6$ kpc compared with the maximum value $\rbreak=10.89^{+3.44}_{-0.44}$ kpc of the sample at $0.2<z<0.6$. This could be for two reasons: 1) the local sample contains deeper data than ours at $0.2<z<0.6$, and therefore, we could be losing the outermost parts of the corresponding analogs at $z\sim0.5$, or 2) the profiles of the local sample with such large values of \rbreak\ and \ho\ could be affected by diffuse PSF tails that were misidentified with real Type-III profiles, because these authors did not correct for PSF effects. Nevertheless, we note that the Anderson-Darling test did not find any noticeable differences between the general distributions of the structural parameters, thus the impact of these effects would require to be tested with larger samples. Additionally, we note that the Type-III S0 with the largest \rbreak\ from the sample at $0.2<z<0.6$ is also the only edge-on S0 galaxy with a Type-III profile.

The lower panels of Fig.\,\ref{fig:param_histogram} show the distributions of \mubreak, \mui\ and \muo. Contrary to the structural parameters, the photometric parameters of the two samples do not give either compatible distributions or median values. We apply again the Anderson-Darling criterion to test whether the parameters from both samples arose from a common parent sample. The $p$-values are $p=6.9\times10^{-3}$ for \mubreak, $p=4.4\times10^{-3}$ for \mui\ and $p=3.1\times10^{-3}$ for \muo. The results indicate that there are noticeable differences between the distributions of the photometric parameters of the samples of Type-III S0-E/S0 galaxies at $0.2<z<0.6$ when comparing to the local sample. The \mubreak, \mui\ and \muo\ values of the sample at $0.2<z<0.6$ are brighter than the local sample by $\sim 1.5$ \magarc\ ($\Delta\mubreak=-1.61^{+0.33}_{-0.53}$ \magarc, $\Delta\mui=-1.40^{+0.39}_{-0.56}$ \magarc, $\Delta\muo=-1.49^{+0.46}_{-0.46}$ \magarc). Deeper surface brightness profiles and a possible PSF contribution could possibly explain this result. But, in any case, we expect a general brightening of the surface brightness profiles of galaxies with increasing $z$, due to the stellar population evolution, which could also explain this difference \citep[see][Tapia et al. submitted and references therein]{2003MNRAS.344.1000B}.

\begin{figure*}[]
 \begin{center}
\includegraphics[width=\textwidth]{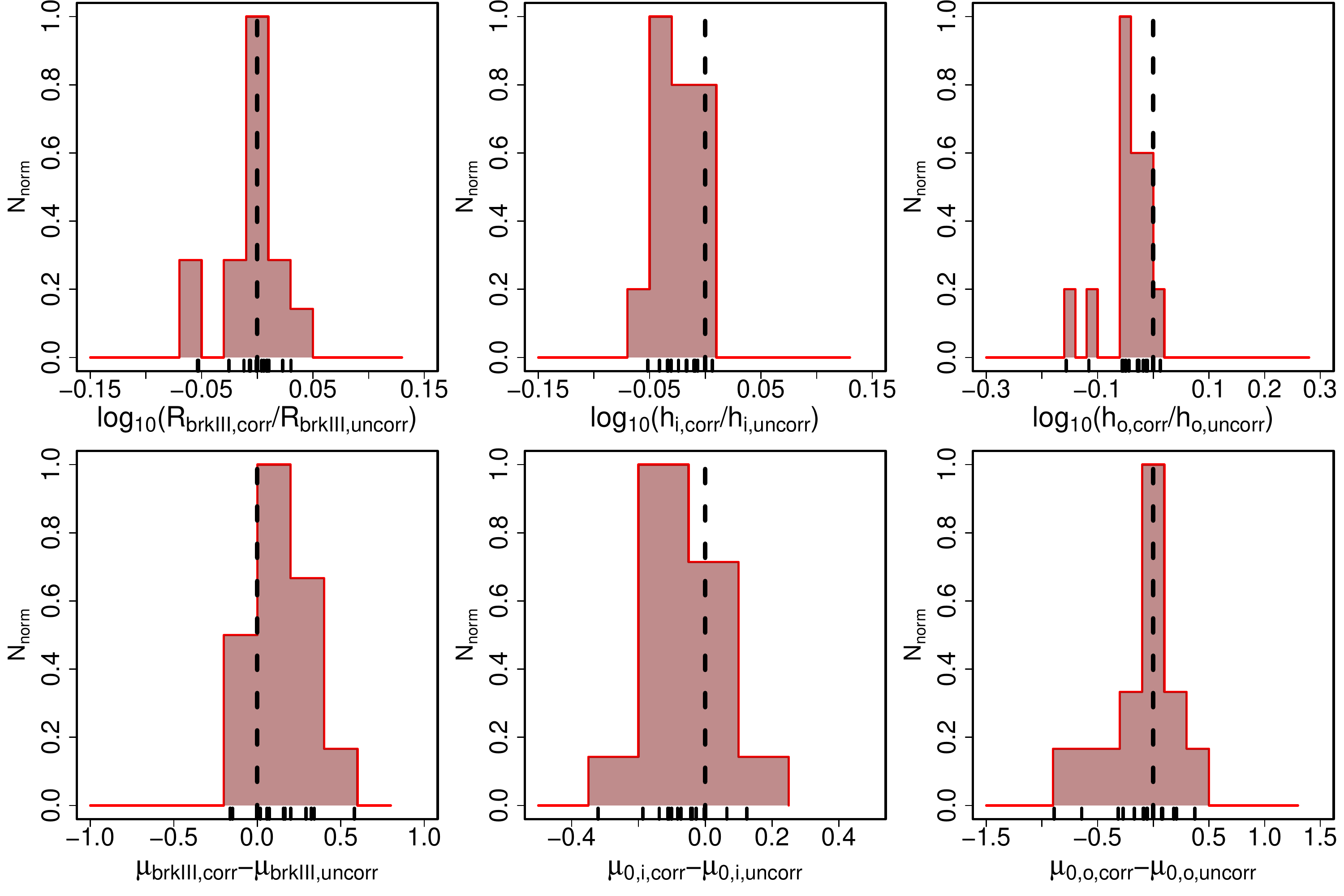}

\caption{Comparison between the PSF corrected and non-corrected values of the structural and photometric parameters of the Type-III S0 and E/S0 at $0.2<z<0.6$ sample. \emph{Upper row, from left to right:} decimal logarithm of the ratio between the PSF corrected structural parameters and the same parameters measured on the original profile: break radius \rbreak\ (kpc), scale-length of the inner profile \hi\ (kpc), scale-length of the outer profile \ho\ (kpc). \emph{Lower row, from left to right:} differences between the PSF corrected structural parameters and the same measured on the original profile: surface brightness at the break radius \mubreak\ (\magarc), central surface brightness of the inner profile \mui\ (\magarc), central surface brightness of the outer profile \muo\ (\magarc). The black dashed vertical line represents the ratio or difference where the PSF correction would not have any effect.} 
\label{fig:param_psf_histogram}
\end{center}
\end{figure*}


\subsection{Effects of the PSF on the profiles}
\label{Subsec:PSF_effect}

In order to test the effects of the scattered light on the surface brightness distribution, we have analysed the profiles of the original images without the PSF correction and compared them with the PSF-corrected profiles. In Fig.\,\ref{fig:param_psf_histogram} we compare the results of the structural and photometric parameters of the Type-III profiles in our S0 and E/S0 sample before and after being corrected by PSF contribution. The histograms in the upper row show the decimal logarithm of the ratio of the \rbreak, \hi\ and \ho\ values measured on the corrected profiles over those parameters measured on the uncorrected profiles. The dashed lines represent the value that would result if the PSF did not have any effect on the observed structure. We found that \ho\ can have significantly higher values in the uncorrected profiles than in the PSF-corrected ones ($\Delta \ho = {\ho}_{,corr} - {\ho}_{,uncorr} = -0.37^{+0.33}_{-0.37}$ kpc). Thus, the dispersed light can increase systematically the outer profile scale length. This also happens with the distribution of \hi\ ($\Delta \hi = -0.105^{+0.082}_{-0.083}$ kpc) although the results are less significant. In contrast, the distribution of ratios of \rbreak\ does not show any clear bias ($\Delta \rbreak = -0.087^{+0.54}_{-0.55}$ kpc).

In the lower panels of Fig.\,\ref{fig:param_psf_histogram}, we represent the differences between the values of \mubreak, \mui\ and \muo\ measured on the profiles corrected for PSF effects and the original (uncorrected) ones. Again, the dashed lines represent the position in case that the dispersed light would not have any effect on our profiles. We do not find any systematic effect in their distributions  ($\Delta \mubreak = 0.16^{+0.23}_{-0.23}$ \magarc, $\Delta \mui = -0.07^{+0.11}_{-0.11}$ \magarc, $\Delta \muo =-0.10^{+0.25}_{-0.26}$ \magarc). In conclusion, the inner and outer disc scale-lengths are the parameters that are most affected by the PSF scattered light.


\section{Conclusions}

We present the first sample of S0 and E/S0 galaxies with anti-truncated discs obtained beyond the local Universe (at $0.2<z<0.6$), on the basis of PSF-corrected surface brightness profiles. We have described in detail the selection procedure and analysis performed to characterise them on a sample of 150 red galaxies at $0.2<z<0.6$ on the GOODS-N field, by using both HST/ACS and SHARDS data from the Rainbow database. We have selected a sample of quiescent disc galaxies with visual S0 and E/S0 morphologies, and studied their SFRs, sSFRs, stellar masses and their surface brightness profiles. Additionally, we have corrected for PSF-scattered light in their F775W images, to obtain surface brightness profiles unbiased by the light dispersion produced by the PSF of our data. 

In order to estimate the break parameters accurately and perform a quantitative morphological classification of the surface brightness profiles, we have developed {\tt{Elbow},} a program to detect, classify and fit breaks in surface brightness profiles, which we make publicly available. We identified S0 and E/S0 objects at $0.2<z<0.6$, with exponential (Type I), truncated (Type II) and anti-truncated (Type III) profiles and fitted their different disc components with exponential functions. We compared the structural and photometric parameters from the sample of Type-III galaxies in the range $0.2<z<0.6$ with the local samples from previous studies (E08 and G11). Our main results are the following: 
\begin{enumerate}
    \item We report the first sample of Type-III profiles in S0 and E/S0 galaxies at $0.2<z<0.6$, corrected for PSF effects. We found that 14 out of 44 S0-E/S0 galaxies at $0.2<z<0.6$ after PSF corrections have antitruncated profiles ($\sim 30\%$). This fraction is similar to the those reported in the local Universe by \citet{
	2014MNRAS.441.1992L, 2016A&A...596A..25L}.
    
    \item As a result of correcting the profiles for scattered light by the PSF, we found that $\sim25-30\%$ of the apparent Type-III S0 discs detected in uncorrected profiles are false positives and can be explained by a combination of pure exponential profiles and scattered light from the inner regions. In two cases, we found significant down-bending breaks (Type-II profile) after the PSF deconvolution. 
    
    \item The structural parameters (\rbreak, \hi\ and \ho) of the Type-III profiles at $0.2<z<0.6$ present similar distributions to their local counterparts. In contrast, the photometric parameters (\mubreak, \mui\ and \muo) have values $\sim1.5$ \magarc\ brighter than the local sample from E08 and G11. 
    
    \item The PSF-dispersed light tends to increase the scale-lengths of the inner and outer disc profiles (\hi, \ho) in our sample of Type-III S0 and E/S0 galaxies, but it does not significantly affect either the central surface brightness values of the inner and outer discs (\mui, \muo) or the break location, \rbreak. 
    
    \item The profiles of Type-I of our S0 and E/S0 sample have characteristic values compatible with the inner profiles (\mui\ and \hi) of Type-III S0 and E/S0 galaxies, but they are not consistent with the equivalent parameters (\muo\ and \ho) of the outer profiles.

\end{enumerate}

The wings of the PSF tend to create an apparent excess of emission in the outskirts of the galaxies. This tends to increase both \hi\ and \ho\ in Type-III profiles. For any study that does not take into account the effects of the PSF on the profiles, this will yield a systematic bias in the results. This applies not only to Type-III profiles, but also to Type-II profiles, as commented above. We also show that the PSF-corrected images recover some Type-II breaks, which were not detectable in the original images. Therefore; in PSF-uncorrected data, an unknown fraction of the real Type-II profiles may be softened to the point of being misclassified as Type-I by the observers.

In this work we have presented {\tt{Elbow}}\footnote[1]{{\tt{Elbow}} is publicly available at GitHub (https://github.com/Borlaff/Elbow)}, a statistically robust and automated method to fit, classify the surface brightness profiles and calculate the likelihood associated with a certain break to exist (Types II and III) or not (Type I). This tool is free and we make it available for the scientific community. The non-linear nature of the surface brightness magnitudes demands a careful treatment of the uncertainties and error propagation, especially when measuring quantities in the low $S/N$ regime. In order to properly account for the uncertainties associated with the surface brightness profiles, {\tt{Elbow}} calculates the errors using non-parametric statistics, such as the bootstrap resampling and the Monte Carlo method. Future studies may analyse what percentage of the reported surface brightness profiles have been previously misidentified due to the lack of an appropriate statistical method. Moreover, the results from the present and other recent studies \citep[such as][and references in Sect.\,\ref{Sec:Intro}]{2014A&A...567A..97S,2016ApJ...823..123T} demonstrate that the PSF corrections are clearly necessary in this sort of studies, especially when working with low surface-brightness Type-III profiles.

According to the results presented here, the structure of the Type-III profiles of S0 galaxies at $0.2<z<0.6$ appears to be similar to the local ones. Nevertheless, the \mui, \muo\ and \mubreak\ have brighter values than the local sample by $-1.5$ \magarc. This could be due to one or several of the following reasons:
\begin{enumerate}
\item The difference may be due to the evolution of the stellar populations. This is expected to produce general brightening of the surface brightness profiles of galaxies with increasing $z$, sufficient to explain the difference between the samples \citep[see][Tapia et al. submitted and references therein]{2003MNRAS.344.1000B}. 

\item Due to the observational depth limit in the surface brightness, we could be biased to the innermost (and thus brighter) Type-III breaks. Nevertheless, we think this effect is negligible in our sample, because the \rbreak, \hi\ and \ho\ distributions of our $0.2<z<0.6$ S0 and E/S0 galaxies are highly compatible with those of the local sample, and only differ for the high values. 

\item The local sample studies traditionally have not corrected for PSF effects. The wings created by the PSF tend to create Type-III profiles with dimmer values of \muo\ and larger values of \ho. Still, although the scattered light may introduce systematic biases in the results - especially in the \ho\ values - it is highly unlikely that most Type-III profiles in the current local samples have been dramatically modified (or produced) by the PSF effects alone.
\end{enumerate}

Recent papers have thrown doubt on the real nature of some Type-III profiles, attributing them to central light scattered radially outwards by the PSF of the telescope optics. Here we have carried out a successful detailed procedure to correct this effect from a large sample of objects and get unbiased samples of discs with real breaks in their surface brightness profiles. As a result, we present the first sample of proven Type-III profiles of S0 and E/S0 objects beyond the local Universe, in the range $0.2<z<0.6$, and that will be analysed in detail in a forthcoming paper (Borlaff et al., in preparation). 


\begin{acknowledgements}
We thank the referee very much for comments that have significantly improved the quality of this work. Supported by the Ministerio de Econom\'{\i}a y Competitividad del Gobierno de España (MINECO) under project AYA2012-31277 and project P3/86 of the Instituto de Astrofisica de Canarias. PGP-G acknowledges support from MINECO grants AYA2015-70815-ERC and AYA2015-63650-P. This work has made use of the Rainbow Cosmological Surveys Database, which is operated by the Universidad Complutense de Madrid (UCM) partnered with the University of California Observatories at Santa Cruz (UCO/Lick,UCSC). We are deeply grateful to the SHARDS team, since this work would not have been possible without their efforts. Based on observations made with the Gran Telescopio Canarias (GTC) installed at the Spanish Observatorio del Roque de los Muchachos of the Instituto de Astrof\'{\i}sica de Canarias, in the island of La Palma. We thank all the GTC Staff for their support and enthusiasm with the SHARDS project. This work is based on observations taken by the 3D-HST Treasury Program (GO 12177 and 12328) with the NASA/ESA HST, which is operated by the Association of Universities for Research in Astronomy, Inc., under NASA contract NAS5-26555. We acknowledge Dr. Pilar Esquej for describing us the usual procedure followed to identify AGN hosts in our S0 sample. The authors thank David Velilla for his kind technical support, and Ignacio Trujillo for his advice with the PSF correction procedures. 
\end{acknowledgements}

\input{appendix.tex}

\bibliographystyle{aa}
\bibliography{borlaff_14.bib}{}

\end{document}

%% file: appendix.tex
\normalsize

\begin{appendix}
\onecolumn
\begin{landscape}

\section{Global properties of the initial sample of galaxies on the red sequence at $0.2<z<0.6$ in the GOODS-N field}
\label{Appendix:redsample}

\begin{center}
\begin{longtable}{llllllllllllc}
\caption{Properties of the initial red sample and morphological classification.}
\label{tab:redsample}
\\ 
\hline
\\\vspace{-0.75cm}\\
\multirow{2}{*}{Number} & \multirow{2}{*}{SHARDS ID}& $\alpha$ & $\delta$ &\multirow{2}{*}{Morph type} &  \multirow{2}{*}{$z_{phot}$} & \multirow{2}{*}{$z_{spec}$} & \multirow{2}{*}{$\log_{10}M/M_{\odot}$} & SFR & $M_{V}$ & $M_{K}$ & A$_\mathrm{F775W}$ & $R$ - F775W  \\
 & & ($^{\circ}$) & ($^{\circ}$) & & & & & $M_{\odot}/$yr & [mag] & [mag] &  [$10^{-2}$ mag] & [mag] \\
(1) & (2) & (3) & (4) & (5) & (6) & (7) & (8) & (9) & (10) & (11) & (12) & (13) \vspace{0.1cm}\\\hline
\\\vspace{-0.25cm}\\
\endfirsthead

\\ \hline
\\\vspace{-1cm}\\
\\\multirow{2}{*}{Number} & \multirow{2}{*}{SHARDS ID} & $\alpha$ & $\delta$& \multirow{2}{*}{Morph type} & \multirow{2}{*}{$z_{phot}$} & \multirow{2}{*}{$z_{spec}$} & \multirow{2}{*}{$\log_{10}M/M_{\odot}$} & SFR & $M_{V}$ & $M_{K}$ & A$_\mathrm{F775W}$ &  $R$ - F775W \\
 & & ($^{\circ}$) & ($^{\circ}$) & & & & & $M_{\odot}/$yr & [mag] & [mag] & [$10^{-2}$ mag] & [mag] \\
(1) & (2) & (3) & (4) & (5) & (6) & (7) & (8) & (9) & (10) & (11) & (12) & (13) \vspace{0.1cm}\\\hline
\\\vspace{-0.75cm}\\
\endhead

1&SHARDS10000327&189.3624&62.2135&E/S0&0.46&0.4755&$10.177\pm0.073$&-2.210&-19.42&-20.25&$-2.54^{+0.11}_{-0.17}$&-0.433\\
2&SHARDS10000478&189.3746&62.2169&E&0.52&0.5117&$11.02\pm0.10$&-4.308&-21.27&-22.25&$-2.52^{+0.13}_{-0.15}$&--\\
3&SHARDS10000488&189.3237&62.2208&DF&0.38&--&$8.73\pm0.13$&-0.595&-16.34&-17.45&$-2.564^{+0.086}_{-0.151}$&--\\
4&SHARDS10000536&189.4165&62.2231&CPSTB&0.23&--&$8.31\pm0.18$&0.112&-14.58&-15.98&$-2.52^{+0.11}_{-0.13}$&--\\
5&SHARDS10000737&189.3396&62.2263&E&0.47&0.4747&$11.188\pm0.081$&-3.615&-21.48&-22.72&$-2.54^{+0.11}_{-0.13}$&--\\
6&SHARDS10000762&189.3593&62.2297&S0&0.47&0.4742&$10.755\pm0.092$&2.484&-21.16&-22.40&$-2.54^{+0.11}_{-0.13}$&-0.430\\
7&SHARDS10000827&189.4341&62.2329&S0&0.51&0.5116&$10.223\pm0.060$&2.398&-20.56&-22.08&$-2.50^{+0.13}_{-0.13}$&-0.491\\
8&SHARDS10000840&189.4244&62.2330&E/S0&0.52&0.5118&$10.901\pm0.054$&-2.725&-20.97&-21.95&$-2.52^{+0.11}_{-0.13}$&-0.493\\
9&SHARDS10000845&189.4441&62.2332&S0&0.36&0.5123&$11.247\pm0.062$&4.570&-21.83&-22.82&$-2.50^{+0.13}_{-0.13}$&-0.493\\
10&SHARDS10000849&189.3590&62.2342&S0&0.47&0.4746&$10.736\pm0.084$&-2.371&-20.63&-22.03&$-2.54^{+0.11}_{-0.13}$&-0.432\\
11&SHARDS10001013&189.2899&62.2399&S0&0.47&0.4753&$10.793\pm0.063$&-1.976&-20.49&-21.73&$-2.59^{+0.11}_{-0.13}$&-0.433\\
12&SHARDS10001052&189.2839&62.2396&Sp&0.57&0.5639&$11.091\pm0.092$&9.220&-21.52&-22.92&$-2.59^{+0.11}_{-0.13}$&--\\
13&SHARDS10001058&189.4457&62.2420&CPSTB&0.47&--&$8.970\pm0.100$&-1.100&-16.90&-17.03&$-2.50^{+0.13}_{-0.13}$&--\\
14&SHARDS10001235&189.4768&62.2475&OM&0.38&0.4620&$9.939\pm0.080$&-1.521&-18.81&-19.87&$-2.50^{+0.13}_{-0.13}$&--\\
15&SHARDS10001269&189.4690&62.2470&S0&0.51&0.5116&$10.958\pm0.074$&-4.443&-21.61&-22.89&$-2.50^{+0.13}_{-0.13}$&-0.492\\
16&SHARDS10001314&189.2334&62.2486&S0&0.31&0.3215&$9.662\pm0.057$&-0.510&-18.42&-19.50&$-2.586^{+0.108}_{-0.086}$&-0.183\\
17&SHARDS10001344&189.2860&62.2504&S0&0.57&0.5673&$10.375\pm0.086$&-2.813&-20.19&-21.30&$-2.586^{+0.108}_{-0.086}$&-0.584\\
18&SHARDS10001350&189.4721&62.2483&S0&0.50&0.5106&$10.935\pm0.071$&-4.234&-21.55&-22.83&$-2.50^{+0.13}_{-0.13}$&-0.492\\
19&SHARDS10001629&189.4774&62.2575&E&0.45&0.4566&$10.846\pm0.058$&-1.869&-20.62&-21.86&$-2.50^{+0.13}_{-0.13}$&--\\
20&SHARDS10001648&189.4680&62.2582&S0&0.45&0.4540&$10.501\pm0.080$&-1.874&-20.47&-21.75&$-2.50^{+0.15}_{-0.13}$&-0.397\\
21&SHARDS10001727&189.2911&62.2567&S0&0.22&0.2013&$10.113\pm0.046$&-0.984&-19.62&-21.15&$-2.607^{+0.086}_{-0.086}$&0.014\\
22&SHARDS10001847&189.4818&62.2618&S0&0.46&0.4598&$10.761\pm0.096$&2.576&-21.12&-22.40&$-2.50^{+0.13}_{-0.13}$&-0.408\\
23&SHARDS10001928&189.4598&62.2647&S0&0.44&0.4535&$9.824\pm0.055$&-1.959&-19.48&-21.01&$-2.52^{+0.13}_{-0.15}$&-0.398\\
24&SHARDS10001971&189.4888&62.2633&E&0.46&0.4574&$11.072\pm0.088$&3.896&-21.19&-22.43&$-2.50^{+0.13}_{-0.13}$&--\\
25&SHARDS10002351&189.5065&62.2742&S0&0.48&0.4834&$10.483\pm0.047$&-1.745&-20.23&-21.51&$-2.48^{+0.15}_{-0.11}$&-0.446\\
26&SHARDS10002730&189.3097&62.2823&S0&0.44&0.4472&$10.339\pm0.062$&-1.660&-20.11&-21.20&$-2.63^{+0.11}_{-0.11}$&-0.386\\
27&SHARDS10002769&189.4826&62.2824&E/S0&0.44&0.4417&$10.539\pm0.074$&1.187&-20.06&-21.05&$-2.52^{+0.24}_{-0.15}$&-0.378\\
28&SHARDS10002901&189.3599&62.2869&E/S0&0.57&0.5642&$10.23\pm0.14$&5.088&-19.73&-21.11&$-2.61^{+0.15}_{-0.15}$&-0.579\\
29&SHARDS10002942&189.3608&62.2871&S0&0.57&0.5644&$10.77\pm0.12$&-4.363&-20.84&-21.80&$-2.61^{+0.15}_{-0.15}$&-0.578\\
30&SHARDS10003216&189.2486&62.3041&E/S0&0.49&0.4970&$10.024\pm0.059$&-2.081&-19.28&-20.20&$-2.650^{+0.108}_{-0.086}$&-0.468\\
31&SHARDS10003299&189.1697&62.3019&DF&0.32&--&$8.50\pm0.20$&-0.289&-15.19&-16.06&$-2.63^{+0.13}_{-0.13}$&--\vspace{0.5cm}\\
\hline\\ 
\pagebreak\\ 
32&SHARDS10003312&189.2352&62.3012&E/S0&0.49&0.4973&$10.731\pm0.084$&-2.475&-20.54&-21.53&$-2.650^{+0.108}_{-0.086}$&-0.468\\
33&SHARDS10003402&189.4151&62.2974&S0&0.20&0.2133&$10.817\pm0.062$&-1.592&-20.07&-21.31&$-2.59^{+0.17}_{-0.17}$&-0.007\\
34&SHARDS10003647&189.1583&62.2914&S0&0.56&0.5553&$10.926\pm0.068$&-4.616&-21.11&-22.51&$-2.61^{+0.15}_{-0.15}$&-0.563\\
35&SHARDS10004066&189.3262&62.3433&E&0.25&0.2544&$10.357\pm0.052$&0.649&-20.23&-21.76&$-2.69^{+0.17}_{-0.15}$&--\\
36&SHARDS10004234&189.3083&62.3435&E&0.53&0.5323&$11.064\pm0.061$&5.750&-21.63&-22.58&$-2.71^{+0.15}_{-0.15}$&--\\
37&SHARDS10004423&189.3738&62.3263&Sp&0.32&0.3186&$10.58\pm0.41$&1.968&-21.03&-22.37&$-2.65^{+0.15}_{-0.19}$&--\\
38&SHARDS10004592&189.3917&62.3270&E&0.27&0.2733&$10.435\pm0.070$&1.140&-20.67&-22.01&$-2.65^{+0.22}_{-0.19}$&--\\
39&SHARDS10004719&189.3806&62.3472&GP&0.30&--&$8.31\pm0.30$&-0.328&-14.58&-15.98&$-2.69^{+0.17}_{-0.22}$&--\\
40&SHARDS10004777&189.2822&62.3143&S0&0.24&0.2010&$10.219\pm0.041$&1.133&-19.88&-21.41&$-2.67^{+0.11}_{-0.11}$&0.014\\
41&SHARDS10005029&189.2326&62.2485&E/S0&0.30&--&$9.155\pm0.069$&-0.341&-17.45&-18.27&$-2.586^{+0.108}_{-0.086}$&-0.147\\
42&SHARDS10005391&189.4568&62.2852&CPSTB&0.23&--&$8.614\pm0.084$&-0.161&-14.55&-15.82&$-2.54^{+0.22}_{-0.17}$&--\\
43&SHARDS10005533&189.3459&62.2152&CPSTB&0.46&--&$8.58\pm0.18$&-1.399&-16.60&-17.03&$-2.54^{+0.11}_{-0.17}$&--\\
44&SHARDS10005598&189.3552&62.2282&DF&0.38&--&$8.46\pm0.18$&-0.584&-15.81&-16.89&$-2.54^{+0.11}_{-0.13}$&--\\
45&SHARDS10006521&189.3817&62.2372&DF&0.46&--&$9.04\pm0.22$&-1.208&-16.12&-16.98&$-2.54^{+0.11}_{-0.13}$&--\\
46&SHARDS10006744&189.4309&62.2823&CPSTB&0.22&--&$7.51\pm0.24$&-0.327&-13.08&-14.03&$-2.56^{+0.19}_{-0.19}$&--\\
47&SHARDS10006819&189.4873&62.2901&DF&0.25&--&$7.92\pm0.21$&-0.167&-13.69&-14.93&$-2.52^{+0.24}_{-0.15}$&--\\
48&SHARDS10006931&189.4210&62.3015&OM&0.37&--&$8.05\pm0.23$&-0.336&-14.66&-15.55&$-2.59^{+0.17}_{-0.17}$&--\\
49&SHARDS10007413&189.3688&62.3567&CPSTB&0.24&--&$8.34\pm0.21$&-0.129&-14.61&-15.53&$-2.71^{+0.17}_{-0.17}$&--\\
50&SHARDS10007672&189.4374&62.2383&DF&0.20&--&$7.77\pm0.36$&-0.112&-13.18&-14.47&$-2.50^{+0.13}_{-0.13}$&--\\
51&SHARDS10007737&189.1887&62.2536&GP&0.26&--&$7.13\pm0.30$&-0.229&-12.90&-13.76&$-2.59^{+0.15}_{-0.11}$&--\\
52&SHARDS10007846&189.2103&62.2692&CPSTB&0.20&--&$7.97\pm0.15$&-0.074&-13.25&-14.59&$-2.61^{+0.13}_{-0.11}$&--\\
53&SHARDS10007883&189.5146&62.2755&GP&0.26&--&$8.05\pm0.12$&-0.194&-14.37&-15.48&$-2.48^{+0.15}_{-0.11}$&--\\
54&SHARDS10007902&189.2121&62.2778&CPSTB&0.26&--&$8.39\pm0.33$&-0.224&-14.52&-15.62&$-2.61^{+0.15}_{-0.11}$&--\\
55&SHARDS10008552&189.4493&62.2348&S0&0.38&--&$9.372\pm0.099$&-0.761&-17.55&-18.99&$-2.50^{+0.13}_{-0.13}$&-0.279\\
56&SHARDS10008997&189.3576&62.2331&DF&0.41&--&$8.71\pm0.13$&-0.669&-15.67&-16.76&$-2.54^{+0.11}_{-0.13}$&--\\
57&SHARDS10009032&189.4216&62.2412&CPSTB&0.25&--&$8.13\pm0.26$&-0.202&-14.12&-15.52&$-2.52^{+0.13}_{-0.13}$&--\\
58&SHARDS10009131&189.4763&62.2589&DF&0.39&--&$8.50\pm0.12$&-0.543&-15.22&-16.28&$-2.50^{+0.13}_{-0.13}$&--\\
59&SHARDS10009411&189.3849&62.2949&DF&0.30&--&$8.29\pm0.26$&-0.311&-14.20&-15.51&$-2.61^{+0.15}_{-0.19}$&--\\
60&SHARDS10009577&189.3581&62.3522&DF&0.28&--&$8.42\pm0.22$&-0.189&-14.52&-15.84&$-2.71^{+0.17}_{-0.17}$&--\vspace{0.5cm}\\
\hline\\ 
\pagebreak\\ 
61&SHARDS10009610&189.1628&62.3059&S0&0.55&0.5568&$11.209\pm0.069$&-5.768&-21.53&-22.77&$-2.65^{+0.11}_{-0.15}$&-0.565\\
62&SHARDS10009819&189.2794&62.2468&CPSTB&0.27&--&$8.40\pm0.26$&-0.206&-14.04&-15.28&$-2.586^{+0.108}_{-0.086}$&--\\
63&SHARDS10009927&189.1876&62.2756&DF&0.21&--&$7.84\pm0.18$&-0.080&-13.71&-14.00&$-2.61^{+0.13}_{-0.11}$&--\\
64&SHARDS10010084&189.2336&62.3079&DF&0.34&--&$8.24\pm0.20$&-0.443&-14.57&-15.62&$-2.672^{+0.086}_{-0.108}$&--\\
65&SHARDS10010282&189.3884&62.2943&DF&0.36&--&$8.50\pm0.19$&-0.431&-14.81&-15.91&$-2.59^{+0.17}_{-0.17}$&--\\
66&SHARDS10010726&189.3233&62.2166&DF&0.54&--&$8.72\pm0.10$&-2.152&-16.02&-16.51&$-2.54^{+0.11}_{-0.15}$&--\\
67&SHARDS10010775&189.4538&62.2429&DF&0.34&--&$8.23\pm0.11$&-0.471&-14.32&-15.24&$-2.50^{+0.13}_{-0.13}$&--\\
68&SHARDS10011094&189.3024&62.3716&CPSTB&0.57&--&$8.99\pm0.18$&-2.633&-17.38&-17.82&$-2.76^{+0.13}_{-0.13}$&--\\
69&SHARDS10011171&189.3521&62.2071&CPSTB&0.32&--&$7.80\pm0.17$&-0.265&-13.97&-14.11&$-2.52^{+0.11}_{-0.24}$&--\\
70&SHARDS10011229&189.2272&62.2494&DF&0.43&--&$7.97\pm0.23$&-0.991&-15.11&-15.86&$-2.586^{+0.108}_{-0.086}$&--\\
71&SHARDS10011544&189.3378&62.2820&GP&0.24&--&$7.40\pm0.20$&-0.166&-13.62&-14.44&$-2.61^{+0.13}_{-0.15}$&--\\
72&SHARDS10012012&189.4065&62.2628&DF&0.35&--&$7.63\pm0.28$&-0.496&-14.26&-15.02&$-2.56^{+0.13}_{-0.15}$&--\\
73&SHARDS20000593&189.1656&62.1385&S0&0.24&0.2470&$9.766\pm0.038$&-0.407&-19.07&-20.32&$-2.43^{+0.15}_{-0.13}$&-0.060\\
74&SHARDS20000668&189.0924&62.1440&CPSTB&0.33&--&$9.351\pm0.087$&-1.754&-17.53&-18.91&$-2.46^{+0.11}_{-0.11}$&--\\
75&SHARDS20000827&189.0335&62.1475&S0&0.40&0.4090&$11.177\pm0.079$&-4.065&-20.97&-22.21&$-2.435^{+0.108}_{-0.086}$&-0.324\\
76&SHARDS20000858&189.2628&62.1502&S0&0.55&0.5600&$10.391\pm0.056$&4.298&-19.77&-21.17&$-2.46^{+0.15}_{-0.19}$&-0.571\\
77&SHARDS20001051&189.1201&62.1564&S0&0.52&0.5183&$10.602\pm0.061$&-2.607&-20.52&-21.81&$-2.46^{+0.15}_{-0.11}$&-0.502\\
78&SHARDS20001223&189.0208&62.1619&GP&0.23&--&$7.88\pm0.16$&-0.157&-13.72&-14.93&$-2.456^{+0.086}_{-0.086}$&--\\
79&SHARDS20001534&189.2412&62.1702&S0&0.40&0.4101&$10.249\pm0.062$&-1.112&-19.15&-20.01&$-2.50^{+0.13}_{-0.13}$&-0.327\\
80&SHARDS20002147&189.1471&62.1861&S0&0.40&0.4101&$10.190\pm0.057$&-1.183&-20.19&-21.47&$-2.50^{+0.13}_{-0.11}$&-0.327\\
81&SHARDS20002235&189.2818&62.1867&E/S0&0.46&0.4714&$10.583\pm0.057$&-1.989&-20.42&-21.48&$-2.52^{+0.11}_{-0.15}$&-0.426\\
82&SHARDS20002542&189.1283&62.1996&DF&0.32&--&$8.30\pm0.11$&-0.328&-15.45&-16.78&$-2.500^{+0.129}_{-0.065}$&--\\
83&SHARDS20002550&189.0022&62.1988&E/S0&0.55&0.5616&$11.139\pm0.090$&-7.355&-21.86&-22.86&$-2.478^{+0.086}_{-0.043}$&-0.573\\
84&SHARDS20002599&189.0414&62.2324&DF&0.22&--&$7.914\pm0.073$&-0.143&-14.97&-15.83&$-2.500^{+0.129}_{-0.065}$&--\\
85&SHARDS20002889&189.2019&62.2213&S0&0.47&0.4760&$9.25\pm0.12$&-1.681&-18.12&-19.64&$-2.564^{+0.086}_{-0.086}$&-0.434\\
86&SHARDS20002935&189.2003&62.2192&S0&0.47&0.4745&$10.924\pm0.068$&3.046&-20.82&-22.06&$-2.564^{+0.086}_{-0.086}$&-0.431\\
87&SHARDS20002966&189.1671&62.2182&S0&0.49&0.4854&$10.953\pm0.066$&-2.720&-20.89&-22.13&$-2.543^{+0.108}_{-0.086}$&-0.449\\
88&SHARDS20002995&189.1044&62.2168&S0&0.52&0.5185&$10.958\pm0.072$&2.404&-20.90&-22.14&$-2.521^{+0.108}_{-0.086}$&-0.503\\
89&SHARDS20003119&189.2360&62.2126&Sp&0.52&0.5178&$10.67\pm0.10$&4.251&-20.94&-22.19&$-2.564^{+0.086}_{-0.151}$&--\\
90&SHARDS20003134&189.0950&62.2166&S0&0.47&0.4725&$10.807\pm0.071$&9.135&-20.99&-21.94&$-2.521^{+0.108}_{-0.086}$&-0.427\\
91&SHARDS20003210&189.2523&62.2096&S0&0.56&0.5631&$10.800\pm0.063$&-2.855&-20.53&-21.39&$-2.564^{+0.086}_{-0.151}$&-0.576\vspace{0.5cm}\\
\hline\\ 
\pagebreak\\ 
92&SHARDS20003217&189.1158&62.2113&S0&0.52&0.5153&$10.808\pm0.066$&-2.485&-20.53&-21.77&$-2.521^{+0.108}_{-0.086}$&-0.499\\
93&SHARDS20003377&189.0637&62.2060&S0&0.32&0.3198&$9.930\pm0.061$&-0.568&-19.54&-20.82&$-2.500^{+0.129}_{-0.065}$&-0.180\\
94&SHARDS20003678&189.1938&62.1976&E/S0&0.50&0.5039&$10.421\pm0.061$&-2.014&-19.77&-20.75&$-2.52^{+0.11}_{-0.11}$&-0.480\\
95&SHARDS20003814&189.1394&62.2842&E&0.46&0.4543&$10.100\pm0.091$&-1.459&-19.52&-20.77&$-2.61^{+0.13}_{-0.15}$&--\\
96&SHARDS20003909&189.1583&62.2709&E&0.53&0.5290&$10.581\pm0.081$&-2.256&-20.40&-21.27&$-2.59^{+0.15}_{-0.13}$&--\\
97&SHARDS20003969&189.1071&62.2617&CPSTB&0.20&--&$8.339\pm0.080$&-0.107&-14.45&-15.07&$-2.56^{+0.15}_{-0.13}$&--\\
98&SHARDS20004069&189.0768&62.2552&OM&0.27&--&$7.93\pm0.22$&-0.233&-14.36&-14.62&$-2.54^{+0.11}_{-0.11}$&--\\
99&SHARDS20004113&189.1728&62.2557&CPSTB&0.27&--&$7.93\pm0.30$&-0.256&-13.79&-14.85&$-2.59^{+0.15}_{-0.13}$&--\\
100&SHARDS20004273&189.1426&62.2425&S0&0.52&0.5184&$10.758\pm0.086$&5.286&-21.11&-22.39&$-2.56^{+0.11}_{-0.11}$&-0.502\\
101&SHARDS20004275&189.1439&62.2424&CPSTB&0.26&--&$9.60\pm0.14$&1.078&-17.02&-18.30&$-2.56^{+0.11}_{-0.11}$&--\\
102&SHARDS20004328&189.1245&62.2393&CPSTB&0.52&0.5195&$9.920\pm0.049$&9.615&-18.95&-20.34&$-2.54^{+0.13}_{-0.11}$&--\\
103&SHARDS20004359&189.2638&62.2383&S0&0.51&0.5120&$10.597\pm0.091$&-2.385&-20.66&-21.51&$-2.59^{+0.11}_{-0.11}$&-0.493\\
104&SHARDS20004420&189.1728&62.2341&E/S0&0.55&0.5556&$10.774\pm0.069$&-3.640&-20.95&-21.94&$-2.56^{+0.11}_{-0.11}$&-0.565\\
105&SHARDS20004440&189.0736&62.2290&E/S0&0.53&0.5337&$10.649\pm0.085$&3.126&-20.89&-22.14&$-2.521^{+0.108}_{-0.086}$&-0.527\\
106&SHARDS20004586&189.1416&62.1314&OM&0.46&--&$9.01\pm0.27$&-3.901&-17.23&-17.75&$-2.43^{+0.13}_{-0.17}$&--\\
107&SHARDS20004857&189.0668&62.1637&CPSTB&0.44&--&$8.77\pm0.20$&-1.145&-16.01&-17.07&$-2.456^{+0.108}_{-0.065}$&--\\
108&SHARDS20005328&189.2706&62.1944&OM&0.30&--&$7.95\pm0.16$&-0.345&-13.98&-14.26&$-2.52^{+0.11}_{-0.15}$&--\\
109&SHARDS20005718&189.1152&62.2396&DF&0.29&--&$7.848\pm0.098$&-0.216&-14.77&-15.20&$-2.54^{+0.11}_{-0.11}$&--\\
110&SHARDS20006178&189.1600&62.1373&CPSTB&0.26&--&$8.25\pm0.21$&-0.211&-14.55&-15.51&$-2.43^{+0.15}_{-0.13}$&--\\
111&SHARDS20006262&189.1689&62.1475&DF&0.23&--&$8.04\pm0.25$&-0.161&-13.75&-14.76&$-2.46^{+0.15}_{-0.15}$&--\\
112&SHARDS20006283&189.2127&62.1492&CPSTB&0.35&--&$8.774\pm0.094$&-0.537&-15.87&-16.50&$-2.46^{+0.15}_{-0.15}$&--\\
113&SHARDS20006429&189.0748&62.1624&CPSTB&0.27&--&$7.99\pm0.22$&-0.138&-14.97&-16.49&$-2.456^{+0.108}_{-0.086}$&--\\
114&SHARDS20006437&189.0413&62.1627&CPSTB&0.43&--&$8.40\pm0.11$&-1.042&-16.19&-16.94&$-2.456^{+0.086}_{-0.086}$&--\\
115&SHARDS20006480&188.9992&62.1661&CPSTB&0.24&--&$8.36\pm0.25$&-0.172&-14.54&-15.55&$-2.456^{+0.086}_{-0.086}$&--\\
116&SHARDS20006824&188.9935&62.1913&CPSTB&0.27&--&$7.63\pm0.12$&-0.171&-13.37&-14.59&$-2.478^{+0.065}_{-0.043}$&--\\
117&SHARDS20006985&189.1612&62.2029&CPSTB&0.24&--&$7.86\pm0.16$&-0.164&-13.86&-14.67&$-2.521^{+0.108}_{-0.086}$&--\\
118&SHARDS20007009&189.1015&62.2041&DF&0.36&--&$8.24\pm0.14$&-0.535&-15.20&-15.76&$-2.500^{+0.129}_{-0.065}$&--\\
119&SHARDS20007308&189.1162&62.2174&DF&0.37&--&$8.13\pm0.13$&-0.408&-15.14&-16.25&$-2.521^{+0.129}_{-0.086}$&--\\
120&SHARDS20007309&189.2030&62.2174&GP&0.22&--&$7.80\pm0.19$&-0.152&-13.90&-14.71&$-2.564^{+0.086}_{-0.108}$&--\\
121&SHARDS20007395&189.1077&62.2634&DF&0.47&--&$9.57\pm0.13$&4.176&-17.45&-18.32&$-2.56^{+0.15}_{-0.13}$&--\\
122&SHARDS20007536&189.0747&62.2394&OM&0.56&--&$9.10\pm0.17$&23.937&-17.28&-18.30&$-2.521^{+0.129}_{-0.086}$&--\vspace{0.5cm}\\
\hline\\ 
\pagebreak\\
123&SHARDS20007647&189.1417&62.1105&CPSTB&0.25&--&$8.29\pm0.14$&-0.122&-14.25&-15.38&$-2.39^{+0.17}_{-0.15}$&--\\
124&SHARDS20008604&189.2621&62.2185&CPSTB&0.52&--&$8.46\pm0.11$&-1.583&-15.90&-16.92&$-2.564^{+0.086}_{-0.129}$&--\\
125&SHARDS20008789&189.2519&62.2283&OM&0.35&--&$8.01\pm0.17$&-0.539&-14.49&-14.62&$-2.59^{+0.11}_{-0.13}$&--\\
126&SHARDS20008811&189.2151&62.2298&CPSTB&0.22&--&$8.22\pm0.11$&-0.145&-13.49&-14.82&$-2.586^{+0.086}_{-0.108}$&--\\
127&SHARDS20009039&189.1674&62.2450&CPSTB&0.30&--&$8.52\pm0.28$&-0.258&-14.34&-15.57&$-2.56^{+0.13}_{-0.11}$&--\\
128&SHARDS20009057&189.1712&62.2444&GP&0.36&--&$8.24\pm0.17$&-0.586&-15.34&-15.94&$-2.56^{+0.13}_{-0.11}$&--\\
129&SHARDS20009101&189.1076&62.2419&DF&0.46&--&$8.90\pm0.11$&-1.254&-16.53&-17.63&$-2.54^{+0.11}_{-0.11}$&--\\
130&SHARDS20009404&189.1490&62.2593&CPSTB&0.59&--&$9.06\pm0.16$&-2.542&-16.67&-17.96&$-2.59^{+0.15}_{-0.13}$&--\\
131&SHARDS20009676&189.0200&62.1704&CPSTB&0.54&--&$9.23\pm0.13$&-2.136&-17.23&-18.61&$-2.456^{+0.086}_{-0.065}$&--\\
132&SHARDS20009942&189.1314&62.1832&DF&0.44&--&$8.21\pm0.23$&-0.947&-15.29&-16.44&$-2.50^{+0.13}_{-0.11}$&--\\
133&SHARDS20010138&189.1646&62.2442&DF&0.29&--&$7.84\pm0.13$&-0.313&-14.35&-14.96&$-2.56^{+0.13}_{-0.11}$&--\\
134&SHARDS20010154&189.0798&62.2310&CPSTB&0.56&--&$9.315\pm0.087$&-3.841&-17.44&-18.82&$-2.521^{+0.108}_{-0.086}$&--\\
135&SHARDS20010466&189.0968&62.2099&CPSTB&0.46&--&$8.04\pm0.25$&1.105&-15.84&-16.14&$-2.500^{+0.129}_{-0.065}$&--\\
136&SHARDS20010555&189.2050&62.2566&DF&0.26&--&$8.11\pm0.21$&-0.164&-14.63&-15.24&$-2.61^{+0.13}_{-0.13}$&--\\
137&SHARDS20010777&189.2606&62.2381&CPSTB&0.22&--&$7.84\pm0.21$&-0.083&-14.06&-14.69&$-2.59^{+0.11}_{-0.11}$&--\\
138&SHARDS20010793&189.2710&62.2300&DF&0.33&--&$7.53\pm0.24$&-0.448&-13.92&-14.44&$-2.59^{+0.11}_{-0.13}$&--\\
139&SHARDS20010969&189.1317&62.1697&CPSTB&0.43&--&$8.07\pm0.24$&-0.841&-15.15&-16.12&$-2.48^{+0.15}_{-0.11}$&--\\
140&SHARDS20011001&188.9709&62.1863&DF&0.50&--&$8.87\pm0.34$&-0.995&-16.11&-16.74&$-2.478^{+0.086}_{-0.065}$&--\\
141&SHARDS20011033&188.9454&62.1989&CPSTB&0.27&--&$8.30\pm0.21$&-0.287&-14.39&-15.40&$-2.478^{+0.108}_{-0.043}$&--\\
142&SHARDS20011279&189.1110&62.2383&CPSTB&0.51&--&$8.93\pm0.14$&-1.873&-15.83&-17.07&$-2.54^{+0.11}_{-0.11}$&--\\
143&SHARDS20011817&189.0578&62.1248&S0&0.50&--&$10.843\pm0.054$&-2.207&-20.62&-21.86&$-2.41^{+0.11}_{-0.13}$&-0.472\\
144&SHARDS20011862&189.0222&62.1519&OM&0.42&--&$8.82\pm0.22$&2.370&-15.51&-16.83&$-2.456^{+0.086}_{-0.108}$&--\\
145&SHARDS20012018&189.2125&62.2069&DF&0.35&--&$8.08\pm0.11$&-0.522&-14.23&-14.77&$-2.54^{+0.11}_{-0.13}$&--\\
146&SHARDS20012798&189.0583&62.2140&CPSTB&0.34&--&$8.24\pm0.20$&-0.469&-14.54&-15.41&$-2.500^{+0.129}_{-0.065}$&--\\
147&SHARDS20013012&189.0228&62.2195&CPSTB&0.57&--&$8.84\pm0.17$&-1.565&-15.72&-16.56&$-2.500^{+0.108}_{-0.065}$&--\\
148&SHARDS20013475&189.1848&62.2321&CPSTB&0.22&--&$7.83\pm0.15$&-0.109&-13.24&-14.25&$-2.56^{+0.11}_{-0.11}$&--\\
149&SHARDS20013783&189.1933&62.2540&DF&0.24&--&$8.01\pm0.22$&-0.166&-13.57&-14.67&$-2.59^{+0.15}_{-0.11}$&--\\
150&SHARDS20014088&189.1593&62.2771&OM&0.20&--&$7.93\pm0.21$&-0.095&-13.01&-14.18&$-2.61^{+0.13}_{-0.15}$&--\vspace{0.5cm}\\

\hline\vspace{-0.5cm}
\end{longtable}
\end{center}
\centering
\begin{minipage}[t]{25cm}
\emph{Columns}: (1) ID sample. (2) ID SHARDS (release DR2 beta). (3) Right ascension (degrees). (4) Declination (degrees). (5) Morphological classification: E (elliptical), E/S0, S0, Sp (spiral), CPSTB (compact post-starburst), OM (ongoing merger), GP (green pea), DF (diffuse galaxy). (6) Photometric redshift. (7) Spectroscopic redshift \emph{(when available)}. (8) Decimal logarithm of the stellar mass. (9) Total SFR \citep[from the emission in 24 $\mu m$ and 2800\AA\ provided in the Rainbow database, following the procedure described in][]{2013ApJ...765..104B}. Negative values correspond to upper limits following the notation on the \emph{Rainbow} Database (see Sect.\,\ref{Sec:Results}). (10) Absolute synthetic rest-frame magnitude in Johnson $V$ band from the Rainbow database (Sect.\,\ref{Sec:Photometric_corrections}). (11) Absolute rest-frame synthetic magnitude in the $K_{\mathrm{s}}$ band from the Rainbow database (Sect.\,\ref{Sec:Photometric_corrections}). (12) Milky Way extinction in the F775W band (magnitudes). (13) Estimated K-correction for the F775W filters to the Steidel $R$ band for the redshift of each galaxy, considering the fit in the Eq.\ref{eq:RSteidel} (see Sect.\,\ref{Sec:Photometric_corrections}).  
\end{minipage}
\end{landscape}

\clearpage
\newpage
\begin{landscape}

\section{Profile parameters of the sample of S0 and E/S0 galaxies at $0.2<z<0.6$}
\label{Appendix:Fits_params}
{\footnotesize\renewcommand{\arraystretch}{1.25}%

\begin{center}
\begin{longtable}{lccccccccccccc}


\caption{Profile classification and photometric parameters of S0 and E/S0 galaxies at $0.2<z<0.6$ in the GOODS-N field}
\label{tab:fits_psforr}
\\\hline
\vspace{-0.3cm}\\ 
\multirow{2}{*}{\#} & \multirow{2}{*}{ID}& \multirow{2}{*}{\thead{Profile \\ type}} & $i$ & \mubreak & \rbreak & \mui & \hi & \muo & \ho & \multirow{2}{*}{$p_{h}$} & \multirow{2}{*}{$p_{\mu_{0}}$} & \risoph & $R_{lim}$  \\
 & & & (º) & (\magarc) & (kpc) & (\magarc) & (kpc) & (\magarc) & (kpc) & & & (kpc) & (kpc)\\
(1) & (2) & (3) & (4) & (5) & (6) & (7) & (8) & (9) & (10) & (11) & (12) & (13) & (14) \vspace{0.1cm}\\\hline
\vspace{0.1cm}
\endfirsthead

\\ \hline
\vspace{-0.3cm}\\ 
\multirow{2}{*}{\#} & \multirow{2}{*}{ID}& \multirow{2}{*}{\thead{Profile \\ type}} & $i$ & \mubreak & \rbreak & \mui & \hi & \muo & \ho & \multirow{2}{*}{$p_{h}$} & \multirow{2}{*}{$p_{\mu_{0}}$} & \risoph & $R_{lim}$  \\
 & & & (º) & (\magarc) & (kpc) & (\magarc) & (kpc) & (\magarc) & (kpc) & & & (kpc) & (kpc)\\
(1) & (2) & (3) & (4) & (5) & (6) & (7) & (8) & (9) & (10) & (11) & (12) & (13) & (14) \vspace{0.1cm}\\\hline
\vspace{0.1cm}
\endhead
1&SHARDS10000327&III&54&$23.05^{+0.29}_{-0.68}$&$5.45^{+0.66}_{-0.95}$&$18.96^{+0.21}_{-0.27}$&$1.42^{+0.12}_{-0.15}$&$20.86^{+0.56}_{-0.42}$&$2.61^{+0.65}_{-0.26}$&$3.0\cdot10^{-4}$&$3.0\cdot10^{-4}$&$5.42^{+0.22}_{-0.18}$&10.34\\
2&SHARDS10000762&I&64&--&--&$19.19^{+0.13}_{-0.13}$&$4.50^{+0.16}_{-0.15}$&--&--&$2.0\cdot10^{-1}$&$2.0\cdot10^{-1}$&$16.06^{+0.39}_{-0.45}$&24.19\\
3&$\dagger$ SHARDS10000827&III&55&$22.44^{+1.31}_{-0.41}$&$8.08^{+4.44}_{-0.93}$&$18.63^{+0.18}_{-0.16}$&$2.29^{+0.21}_{-0.15}$&$20.23^{+1.23}_{-0.49}$&$3.92^{+1.98}_{-0.59}$&$2.4\cdot10^{-3}$&$2.4\cdot10^{-3}$&$10.18^{+0.42}_{-0.39}$&15.20\\
4&SHARDS10000840&III&48&$21.96^{+0.33}_{-0.52}$&$6.10^{+0.77}_{-0.84}$&$18.18^{+0.21}_{-0.35}$&$1.73^{+0.15}_{-0.21}$&$19.67^{+0.41}_{-0.26}$&$2.83^{+0.43}_{-0.20}$&$2.3\cdot10^{-4}$&$2.3\cdot10^{-4}$&$8.72^{+0.25}_{-0.27}$&13.72\\
5&SHARDS10000845&II&84&$19.71^{+0.29}_{-0.30}$&$8.86^{+0.75}_{-0.93}$&$16.68^{+0.21}_{-0.11}$&$3.15^{+0.38}_{-0.13}$&$15.41^{+0.23}_{-0.37}$&$2.23^{+0.11}_{-0.12}$&$3.4\cdot10^{-4}$&$2.4\cdot10^{-4}$&$15.89^{+0.62}_{-0.45}$&18.17\\
6&SHARDS10000849&III&69&$23.23^{+0.18}_{-0.36}$&$8.86^{+0.69}_{-0.60}$&$18.36^{+0.18}_{-0.22}$&$1.97^{+0.11}_{-0.12}$&$21.09^{+0.84}_{-0.82}$&$4.4^{+2.2}_{-1.1}$&$3.1\cdot10^{-4}$&$3.1\cdot10^{-4}$&$8.45^{+0.21}_{-0.18}$&12.10\\
7&SHARDS10001013&I&45&--&--&$17.356^{+0.060}_{-0.075}$&$1.132^{+0.018}_{-0.018}$&--&--&$2.4\cdot10^{-1}$&$2.4\cdot10^{-1}$&$5.87^{+0.15}_{-0.17}$&7.48\\
8&SHARDS10001269&I&67&--&--&$17.75^{+0.14}_{-0.13}$&$2.751^{+0.087}_{-0.076}$&--&--&$1.8\cdot10^{-2}$&$1.8\cdot10^{-2}$&$13.12^{+0.27}_{-0.29}$&18.54\\
9&SHARDS10001314&I&49&--&--&$20.12^{+0.13}_{-0.11}$&$1.843^{+0.068}_{-0.059}$&--&--&$3.9\cdot10^{-2}$&$3.9\cdot10^{-2}$&$4.60^{+0.14}_{-0.15}$&9.81\\
10&SHARDS10001344&III&54&$22.578^{+1.311}_{-0.050}$&$10.21^{+7.30}_{-0.43}$&$19.70^{+0.48}_{-0.27}$&$3.64^{+1.10}_{-0.32}$&$21.16^{+0.98}_{-0.17}$&$6.99^{+3.31}_{-0.43}$&$1.6\cdot10^{-3}$&$1.6\cdot10^{-3}$&$12.41^{+0.66}_{-0.81}$&24.22\\
11&SHARDS10001350&II&45&$21.64^{+0.42}_{-0.26}$&$10.21^{+1.22}_{-0.87}$&$18.89^{+0.21}_{-0.16}$&$4.02^{+0.59}_{-0.22}$&$17.40^{+0.26}_{-0.63}$&$2.61^{+0.16}_{-0.24}$&$5.8\cdot10^{-4}$&$5.8\cdot10^{-4}$&$13.52^{+0.21}_{-0.19}$&17.77\\
12&SHARDS10001648&III&72&$23.01^{+0.52}_{-0.53}$&$7.7^{+1.1}_{-1.1}$&$17.99^{+0.29}_{-0.38}$&$1.60^{+0.15}_{-0.17}$&$20.63^{+0.87}_{-0.50}$&$3.23^{+1.16}_{-0.34}$&$2.9\cdot10^{-4}$&$2.9\cdot10^{-4}$&$7.70^{+0.20}_{-0.24}$&13.90\\
13&SHARDS10001727&I&84&--&--&$17.447^{+0.052}_{-0.052}$&$2.104^{+0.029}_{-0.027}$&--&--&$1.6\cdot10^{-2}$&$1.6\cdot10^{-2}$&$11.08^{+0.14}_{-0.14}$&16.14\\
14&SHARDS10001847&II&75&$21.33^{+0.59}_{-0.37}$&$9.9^{+1.4}_{-1.0}$&$17.924^{+0.156}_{-0.092}$&$3.148^{+0.259}_{-0.095}$&$16.84^{+0.24}_{-0.69}$&$2.39^{+0.13}_{-0.23}$&$4.0\cdot10^{-5}$&$5.0\cdot10^{-5}$&$13.68^{+0.20}_{-0.25}$&21.34\\
15&SHARDS10002351&I&86&--&--&$16.95^{+0.13}_{-0.14}$&$1.290^{+0.048}_{-0.050}$&--&--&$4.0\cdot10^{-1}$&$4.2\cdot10^{-1}$&$7.08^{+0.60}_{-0.34}$&8.63\\
16&SHARDS10002730&III&53&$23.49^{+1.10}_{-0.85}$&$6.8^{+1.7}_{-1.2}$&$18.20^{+0.12}_{-0.20}$&$1.361^{+0.045}_{-0.098}$&$19.76^{+2.88}_{-0.64}$&$1.92^{+2.78}_{-0.23}$&$7.8\cdot10^{-3}$&$7.8\cdot10^{-3}$&$6.04^{+0.16}_{-0.15}$&9.99\\
17&SHARDS10002769&I&39&--&--&$20.41^{+0.15}_{-0.13}$&$3.79^{+0.19}_{-0.16}$&--&--&$9.1\cdot10^{-2}$&$9.1\cdot10^{-2}$&$11.28^{+0.84}_{-0.77}$&16.07\\
18&SHARDS10002942&III&21&$22.01^{+1.69}_{-0.34}$&$5.70^{+2.81}_{-0.57}$&$17.70^{+0.17}_{-0.22}$&$1.410^{+0.119}_{-0.096}$&$19.08^{+2.00}_{-0.31}$&$2.06^{+1.41}_{-0.15}$&$3.2\cdot10^{-3}$&$3.2\cdot10^{-3}$&$7.54^{+0.24}_{-0.26}$&10.90\\
19&SHARDS10003216&I&19&--&--&$18.66^{+0.17}_{-0.24}$&$1.049^{+0.060}_{-0.075}$&--&--&$4.2\cdot10^{-1}$&$4.2\cdot10^{-1}$&$4.16^{+0.15}_{-0.15}$&5.84\\
20&SHARDS10003312&III&38&$23.850^{+0.021}_{-0.799}$&$10.43^{+0.22}_{-2.02}$&$19.64^{+0.17}_{-0.31}$&$2.71^{+0.16}_{-0.33}$&$21.79^{+0.42}_{-0.50}$&$5.60^{+1.51}_{-0.68}$&$2.6\cdot10^{-4}$&$2.6\cdot10^{-4}$&$8.31^{+0.31}_{-0.32}$&17.90\\
21&SHARDS10003402&II&84&$20.09^{+0.19}_{-0.29}$&$3.96^{+0.25}_{-0.41}$&$17.78^{+0.14}_{-0.11}$&$1.88^{+0.25}_{-0.10}$&$16.36^{+0.13}_{-0.19}$&$1.161^{+0.029}_{-0.035}$&$1.1\cdot10^{-4}$&$8.0\cdot10^{-5}$&$7.067^{+0.117}_{-0.099}$&10.62\\
22&SHARDS10003647&III&66&$22.40^{+1.43}_{-0.36}$&$8.88^{+3.23}_{-0.95}$&$17.57^{+0.33}_{-0.24}$&$1.92^{+0.24}_{-0.13}$&$19.64^{+2.02}_{-0.37}$&$3.26^{+2.69}_{-0.24}$&$4.6\cdot10^{-4}$&$4.6\cdot10^{-4}$&$10.41^{+0.34}_{-0.35}$&16.23\\
23&SHARDS10004777&II&62&$24.854^{+0.076}_{-0.461}$&$12.92^{+0.37}_{-1.18}$&$19.939^{+0.064}_{-0.058}$&$2.875^{+0.064}_{-0.045}$&$18.37^{+0.44}_{-1.18}$&$2.11^{+0.19}_{-0.26}$&$<1.0\cdot10^{-5}$&$<1.0\cdot10^{-5}$&$8.22^{+0.14}_{-0.14}$&16.29\\

\hline\\
\pagebreak

24&SHARDS10009610&III&53&$22.54^{+0.31}_{-0.38}$&$10.37^{+1.21}_{-0.99}$&$17.80^{+0.23}_{-0.24}$&$2.32^{+0.17}_{-0.17}$&$20.03^{+0.47}_{-0.34}$&$4.31^{+0.66}_{-0.36}$&$1.2\cdot10^{-4}$&$1.2\cdot10^{-4}$&$12.11^{+0.29}_{-0.38}$&20.51\\
25&SHARDS20000593&III&46&$23.10^{+0.27}_{-0.21}$&$6.09^{+0.60}_{-0.60}$&$20.32^{+0.10}_{-0.15}$&$2.34^{+0.12}_{-0.17}$&$21.17^{+0.15}_{-0.11}$&$3.36^{+0.17}_{-0.12}$&$3.8\cdot10^{-3}$&$3.8\cdot10^{-3}$&$5.79^{+0.15}_{-0.16}$&17.23\\
26&SHARDS20000827&III&84&$22.79^{+1.31}_{-0.18}$&$10.90^{+3.45}_{-0.44}$&$17.83^{+0.30}_{-0.14}$&$2.333^{+0.260}_{-0.088}$&$20.17^{+2.29}_{-0.37}$&$4.34^{+6.24}_{-0.36}$&$1.5\cdot10^{-5}$&$1.5\cdot10^{-5}$&$11.67^{+0.39}_{-0.39}$&17.66\\
27&SHARDS20001051&I&27&--&--&$18.407^{+0.083}_{-0.084}$&$1.689^{+0.038}_{-0.036}$&--&--&$4.3\cdot10^{-1}$&$4.3\cdot10^{-1}$&$7.21^{+0.21}_{-0.22}$&10.45\\
28&SHARDS20001534&I&63&--&--&$18.507^{+0.058}_{-0.056}$&$1.257^{+0.019}_{-0.017}$&--&--&$4.9\cdot10^{-1}$&$4.9\cdot10^{-1}$&$5.24^{+0.17}_{-0.17}$&7.85\\
29&$\dagger$ SHARDS20002147&I&24&--&--&$18.63^{+0.15}_{-0.14}$&$1.312^{+0.056}_{-0.051}$&--&--&$8.8\cdot10^{-2}$&$8.8\cdot10^{-2}$&$5.25^{+0.19}_{-0.20}$&7.848\\
30&SHARDS20002235&I&31&--&--&$19.05^{+0.12}_{-0.10}$&$2.256^{+0.080}_{-0.077}$&--&--&$2.9\cdot10^{-2}$&$2.9\cdot10^{-2}$&$8.06^{+0.29}_{-0.24}$&11.70\\
31&SHARDS20002550&I&38&--&--&$19.22^{+0.16}_{-0.14}$&$4.74^{+0.21}_{-0.20}$&--&--&$4.9\cdot10^{-2}$&$4.9\cdot10^{-2}$&$16.34^{+0.47}_{-0.40}$&24.49\\
32&SHARDS20002889&I&17&--&--&$20.07^{+0.18}_{-0.14}$&$1.292^{+0.077}_{-0.062}$&--&--&$1.8\cdot10^{-1}$&$1.8\cdot10^{-1}$&$3.42^{+0.18}_{-0.18}$&5.35\\
33&SHARDS20002935&I&38&--&--&$19.36^{+0.13}_{-0.12}$&$2.92^{+0.11}_{-0.10}$&--&--&$1.6\cdot10^{-2}$&$1.6\cdot10^{-2}$&$9.59^{+0.24}_{-0.24}$&15.30\\
34&SHARDS20002966&I&47&--&--&$18.334^{+0.091}_{-0.096}$&$2.184^{+0.054}_{-0.048}$&--&--&$2.3\cdot10^{-1}$&$2.3\cdot10^{-1}$&$9.50^{+0.25}_{-0.25}$&13.70\\
35&SHARDS20002995&II&31&$22.67^{+0.48}_{-0.41}$&$10.7^{+1.2}_{-1.2}$&$19.25^{+0.14}_{-0.11}$&$3.40^{+0.22}_{-0.15}$&$17.66^{+0.62}_{-1.00}$&$2.32^{+0.27}_{-0.33}$&$6.0\cdot10^{-4}$&$5.8\cdot10^{-4}$&$11.50^{+0.25}_{-0.24}$&15.30\\
36&SHARDS20003134&II&78&$20.40^{+1.14}_{-0.97}$&$5.47^{+0.99}_{-1.11}$&$16.38^{+0.84}_{-0.54}$&$1.53^{+0.73}_{-0.24}$&$12.7^{+1.2}_{-1.5}$&$0.78^{+0.12}_{-0.10}$&$2.0\cdot10^{-2}$&$7.6\cdot10^{-4}$&$7.54^{+0.18}_{-0.20}$&8.52\\
37&SHARDS20003210&III&22&$22.06^{+0.35}_{-0.39}$&$5.69^{+0.68}_{-0.64}$&$17.95^{+0.27}_{-0.40}$&$1.46^{+0.15}_{-0.19}$&$20.21^{+0.32}_{-0.28}$&$3.13^{+0.36}_{-0.25}$&$9.9\cdot10^{-4}$&$9.9\cdot10^{-4}$&$8.06^{+0.26}_{-0.29}$&13.63\\
38&SHARDS20003217&I&28&--&--&$17.787^{+0.080}_{-0.090}$&$1.359^{+0.028}_{-0.030}$&--&--&$2.8\cdot10^{-1}$&$2.7\cdot10^{-1}$&$6.37^{+0.19}_{-0.19}$&8.56\\
39&$\dagger$ SHARDS20003377&I&68&--&--&$18.70^{+0.21}_{-0.18}$&$1.313^{+0.060}_{-0.059}$&--&--&$2.1\cdot10^{-3}$&$2.2\cdot10^{-2}$&$4.95^{+0.16}_{-0.14}$&8.93\\
40&SHARDS20003678&I&38&--&--&$19.56^{+0.25}_{-0.22}$&$1.392^{+0.088}_{-0.068}$&--&--&$1.8\cdot10^{-2}$&$1.8\cdot10^{-2}$&$4.11^{+0.22}_{-0.21}$&6.99\\
41&SHARDS20004359&I&40&--&--&$19.97^{+0.17}_{-0.16}$&$3.56^{+0.19}_{-0.17}$&--&--&$1.1\cdot10^{-2}$&$1.1\cdot10^{-2}$&$9.82^{+0.25}_{-0.30}$&17.80\\
42&SHARDS20004420&I&42&--&--&$19.81^{+0.17}_{-0.15}$&$4.13^{+0.22}_{-0.19}$&--&--&$8.2\cdot10^{-2}$&$8.2\cdot10^{-2}$&$11.84^{+0.36}_{-0.34}$&18.55\\
43&$\dagger$ SHARDS20004440&I&34&--&---&$18.13^{+0.15}_{-0.13}$&$1.668^{+0.067}_{-0.061}$&--&--&$1.3\cdot10^{-2}$&$1.3\cdot10^{-2}$&$7.46^{+0.27}_{-0.24}$&9.86\\
44&SHARDS20011817&I&82&--&--&$16.283^{+0.077}_{-0.076}$&$1.097^{+0.022}_{-0.020}$&--&--&$2.6\cdot10^{-1}$&$2.5\cdot10^{-1}$&$6.83^{+0.35}_{-0.31}$&8.42\\
\hline
\end{longtable}
\end{center}
\centering
\begin{minipage}[t]{25cm}
\emph{Columns}: (1) Numerical ID in our final sample of S0 and E/S0 galaxies at $0.2<z<0.6$. (2) ID SHARDS (release DR2 beta). (3) Profile type. (4) Galaxy inclination (degrees). (5) Surface brightness of the profile at the break radius \mubreak\ in \magarc. (6) Break radius \rbreak\ in kpc. (7) Central surface brightness of the inner profile \mui\ in \magarc. (8) Scale-length of the inner profile \hi\ in kpc. (9) Central surface brightness of the outer profile \muo\ in \magarc. (10) Scale-length of the outer profile \ho\ in kpc. (11) {\tt{Elbow}} $p$-value for the likelihood that \hi\ and \ho\ are equal. (12)  {\tt{Elbow}} $p$-value for for the likelihood that \mui\ and \muo\ are equal. (13) Isophotal radius for 23 \magarc\ of the object in the corrected Steidel $R$ band in kpc. (14) Limiting radius of the profile in kpc. The objects flagged with the $\dagger$ symbol after their ID SHARDS correspond to those objects in the AGN subsample (see Sect.\,\ref{Subsec:AGN}).  
\end{minipage}
}
\end{landscape}

\clearpage
\newpage
\onecolumn

\section{Specific comments on images, surface brightness profiles and disc profile analysis for the S0 and E/S0 galaxies at $z<0.6$}
\label{Sec:TypeIIIcomments}
Here we detail some of the most important characteristics of the S0 and E/S0 objects within our sample (see Table B.1), attending to their PSF-corrected photometric profiles. Only the first galaxy is provided in the printed edition. A colour version of the Appendix including all objects in Table B.1 is available in the online edition:\\
\clearpage
\newpage

\textbf{SHARDS10000327:} E/S0 galaxy with a Type-III profile, with no nearby galaxies, and medium inclination (see Table \ref{tab:fits_psforr}). The object shows a clear Type-III break at 5.4 kpc from the centre. The PDDs for $h$ and $\mu_{0}$ show two clearly separated peaks corresponding to the two sections of the surface brightness profile.

\begin{figure}[!h]
{\centering
\vspace{-0cm}

\begin{minipage}{.5\textwidth}
\hspace{1.2cm}
\begin{overpic}[width=0.8\textwidth]
{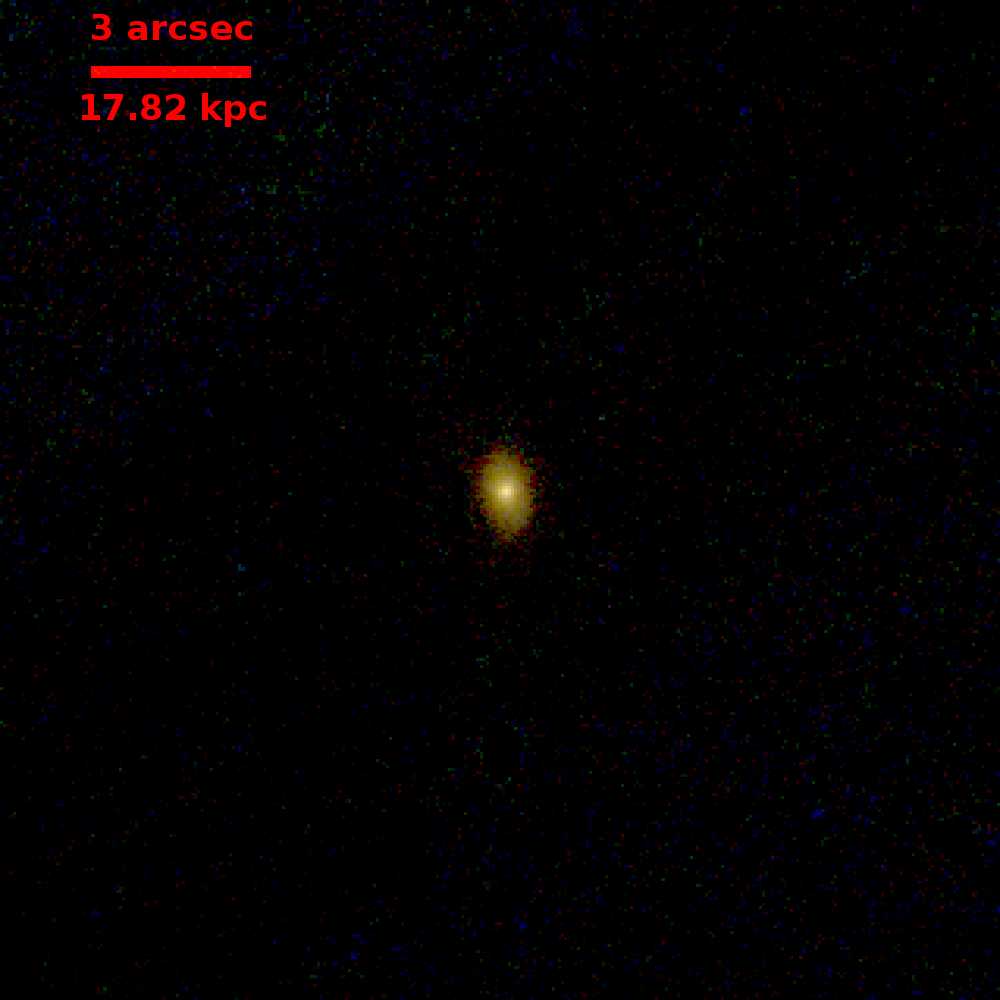}
\put(110,200){\color{yellow} \textbf{SHARDS10000327}}
\put(110,190){\color{yellow} \textbf{z=0.4755}}
\put(110,180){\color{yellow} \textbf{E/S0}}
\end{overpic}
\vspace{-1cm}
\end{minipage}%
\begin{minipage}{.5\textwidth}
\includegraphics[clip, trim=1cm 1cm 1.5cm 1.5cm, width=\textwidth]{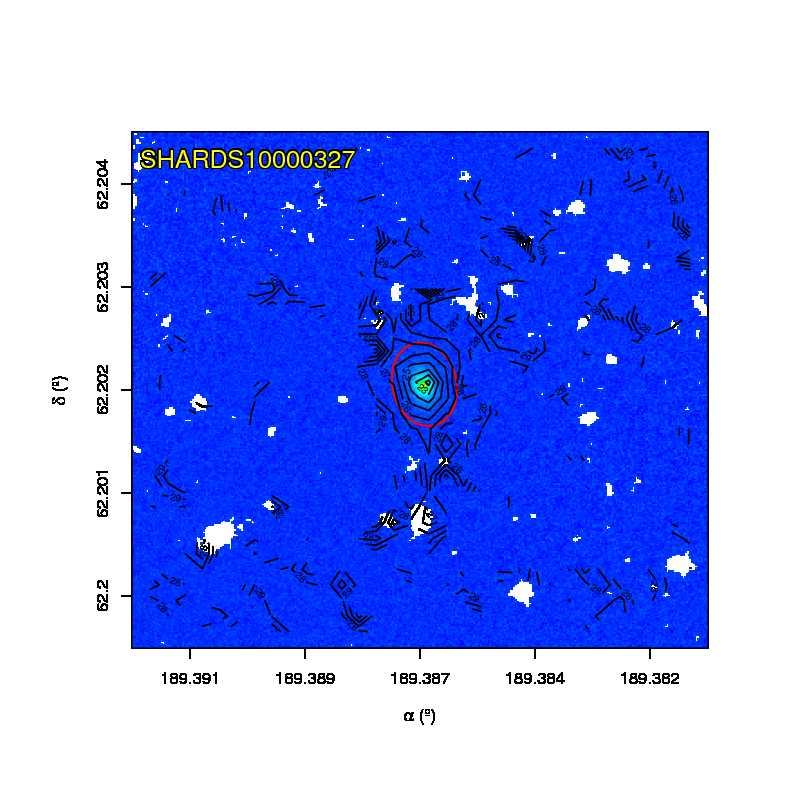}\vspace{-1cm}
\end{minipage}%

\begin{minipage}{.49\textwidth}
\includegraphics[clip, trim=0.1cm 0.1cm 0.1cm 0.1cm, width=\textwidth]{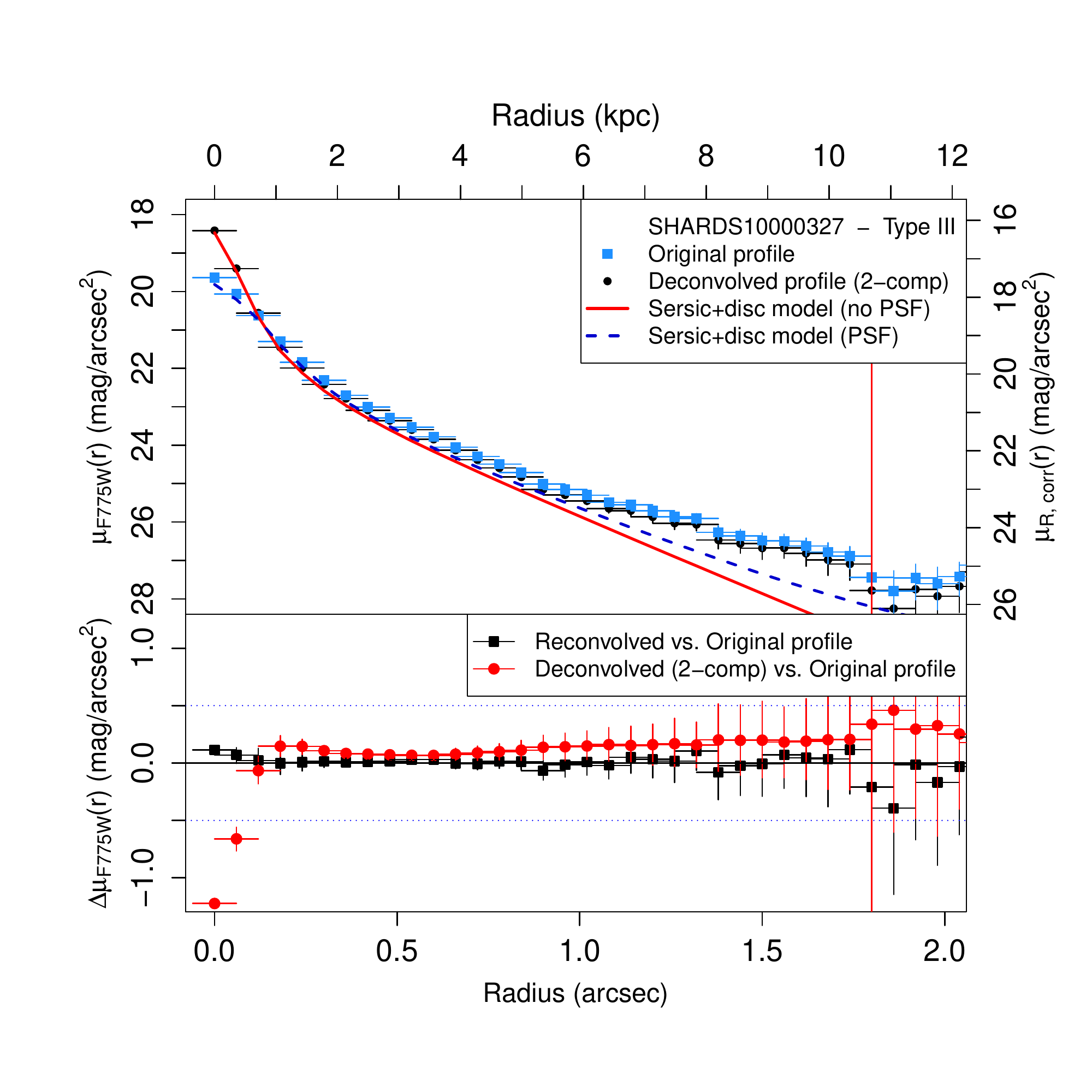}
\end{minipage}
\begin{minipage}{.49\textwidth}
\includegraphics[clip, trim=0.1cm 0.1cm 1cm 0.1cm, width=0.95\textwidth]{IMAGES/IMG_FINAL_SPLIT/SHARDS10000327.pdf}
\end{minipage}}

\vspace{-0.5cm}
\caption[]{\textbf{Upper row, left panel:} Masked false RGB image centred on the source (red: F775W, green: F606W, blue: F435W). The red segment represents 3 arcsec. \textbf{Upper row, right panel:} Deconvolved F775W image. The black lines represent the isophotal contours of the image in magnitudes. The white regions represent the masked areas. The red ellipse indicates the limiting radius. \textbf{Lower row, left panel:} Surface brightness profiles of the original image (blue) and the PSF-corrected image (black) for the observed F775W band (left axis) and for the rest-frame Steidel $R$ band (right axis). The red solid and blue dashed lines correspond to the models fitted during the deconvolution and used for checking the visual morphological selection (see the legend). The lower panel represents the differences between the original and the PSF-corrected profiles (red circles) and the difference between the original profile and the reconvolved PSF-corrected image profile (black squares). The vertical red line represents the limiting radius. \textbf{Lower row, right panel:} Surface brightness profile in the rest-frame Steidel $R$ band corrected for dust extinction, cosmological dimming and K-correction. The dashed lines and the shaded areas correspond to the exponential fittings of the inner and outer profiles in red and blue colours, respectively. The vertical and horizontal black dotted lines correspond to the peak of the PDD for the break radius and the surface brightness value at that location (\rbreak\ and \mubreak). The results of the break analysis are provided in the panel. [\emph{A colour version of this figure and the rest of figures in this Appendix (also in colour) are available in the online edition}.]}         
\label{fig:img_final}
\end{figure}
\clearpage
\newpage

\textbf{SHARDS10000762:} We classify with object as S0 galaxy with a Type-I profile. It presents a medium to high inclination (see Table \ref{tab:fits_psforr}). There are not any nearby galaxies. The inner profile shows an slightly but noticeable bump in the inner region, possibly a lens component. We apply masking to a small source of $\mu_{\mathrm{F775W}}\sim 23$ \magarc\ on the north direction along the semimajor axis. After masking the lens dominated part of the profile, the PDDs do not show any noticieable break.

\begin{figure}[!h]
{\centering
\vspace{-0cm}

\begin{minipage}{.5\textwidth}
\hspace{1.2cm}
\begin{overpic}[width=0.8\textwidth]
{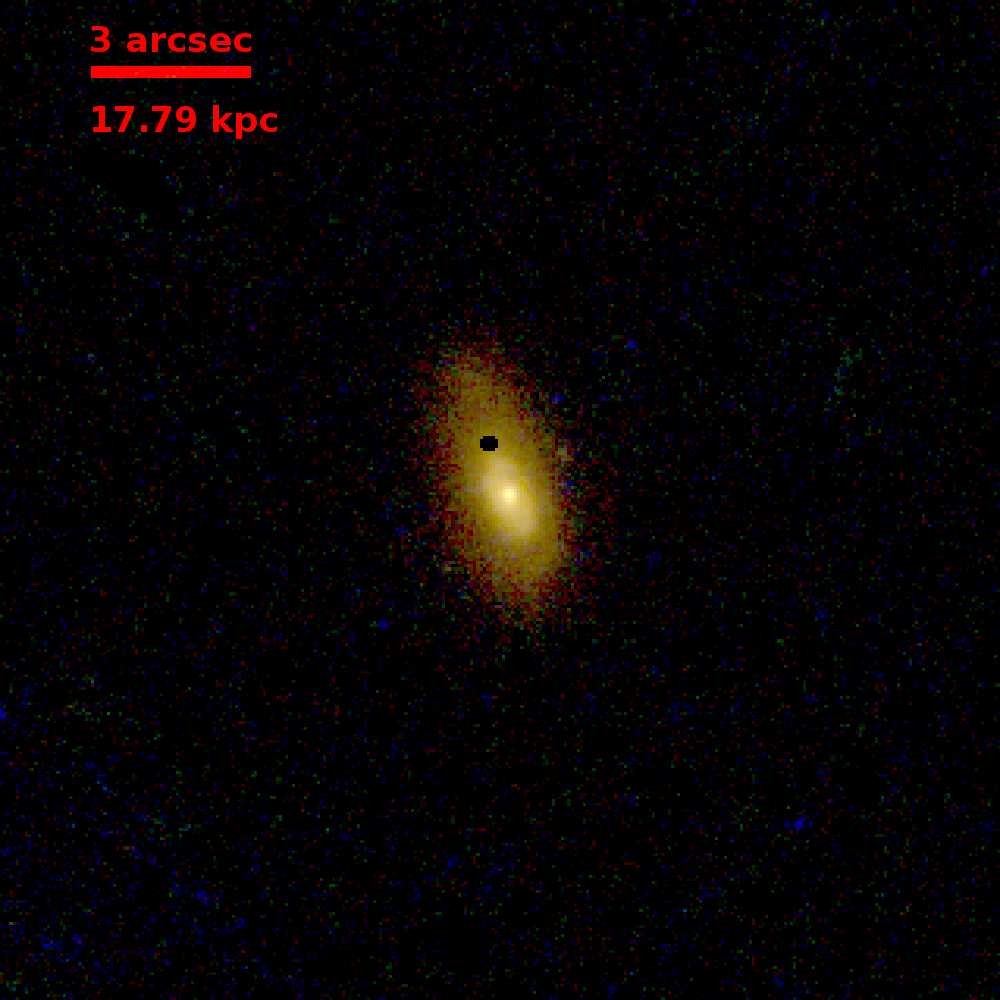}
\put(110,200){\color{yellow} \textbf{SHARDS10000762}}
\put(110,190){\color{yellow} \textbf{z=0.4742}}
\put(110,180){\color{yellow} \textbf{S0}}
\end{overpic}
\vspace{-1cm}
\end{minipage}%
\begin{minipage}{.5\textwidth}
\includegraphics[clip, trim=1cm 1cm 1.5cm 1.5cm, width=\textwidth]{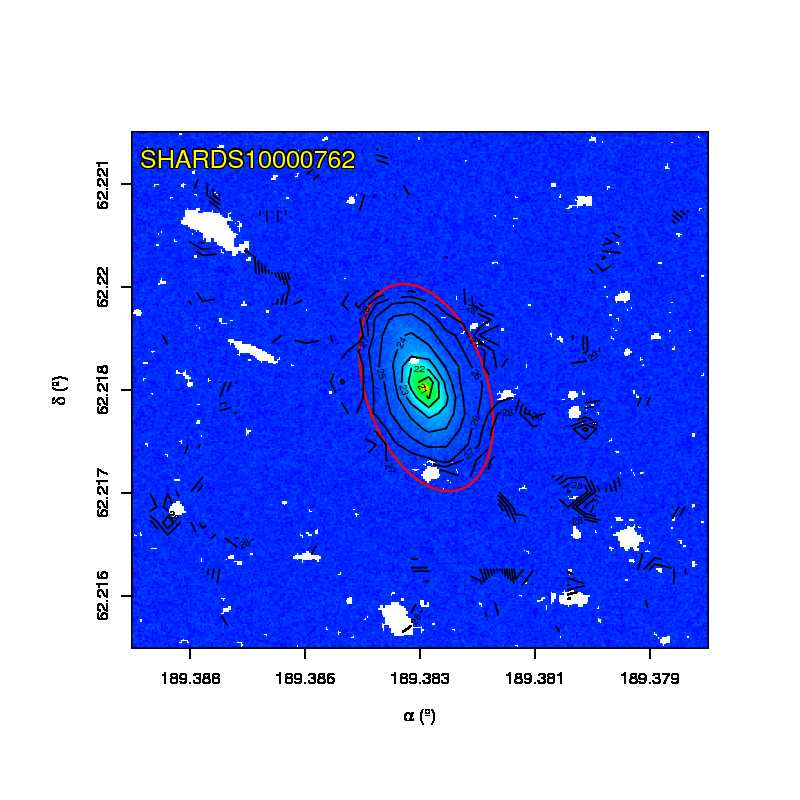}\vspace{-1cm}
\end{minipage}%

\begin{minipage}{.49\textwidth}
\includegraphics[clip, trim=0.1cm 0.1cm 0.1cm 0.1cm, width=\textwidth]{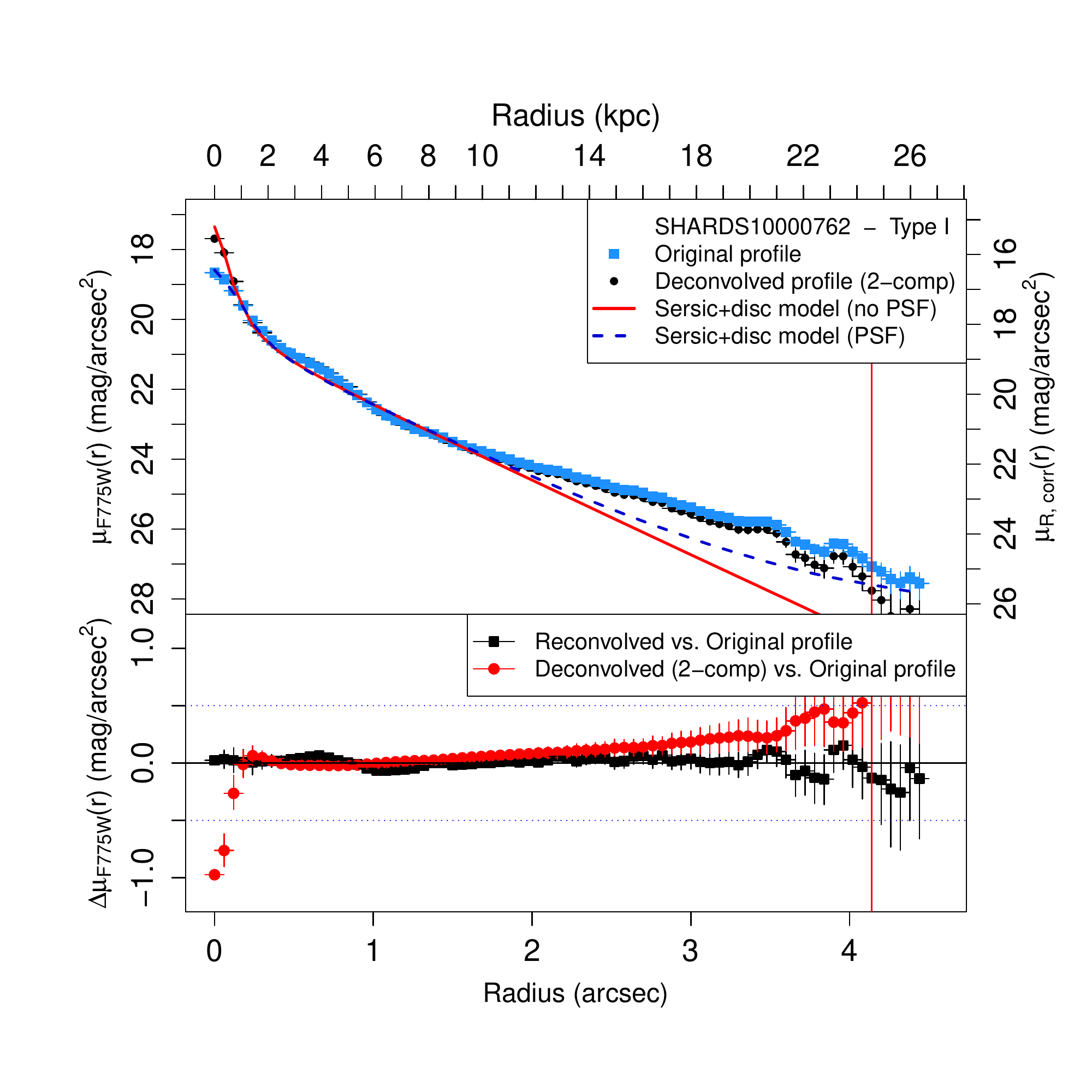}
\end{minipage}
\begin{minipage}{.49\textwidth}
\includegraphics[clip, trim=0.1cm 0.1cm 1cm 0.1cm, width=0.95\textwidth]{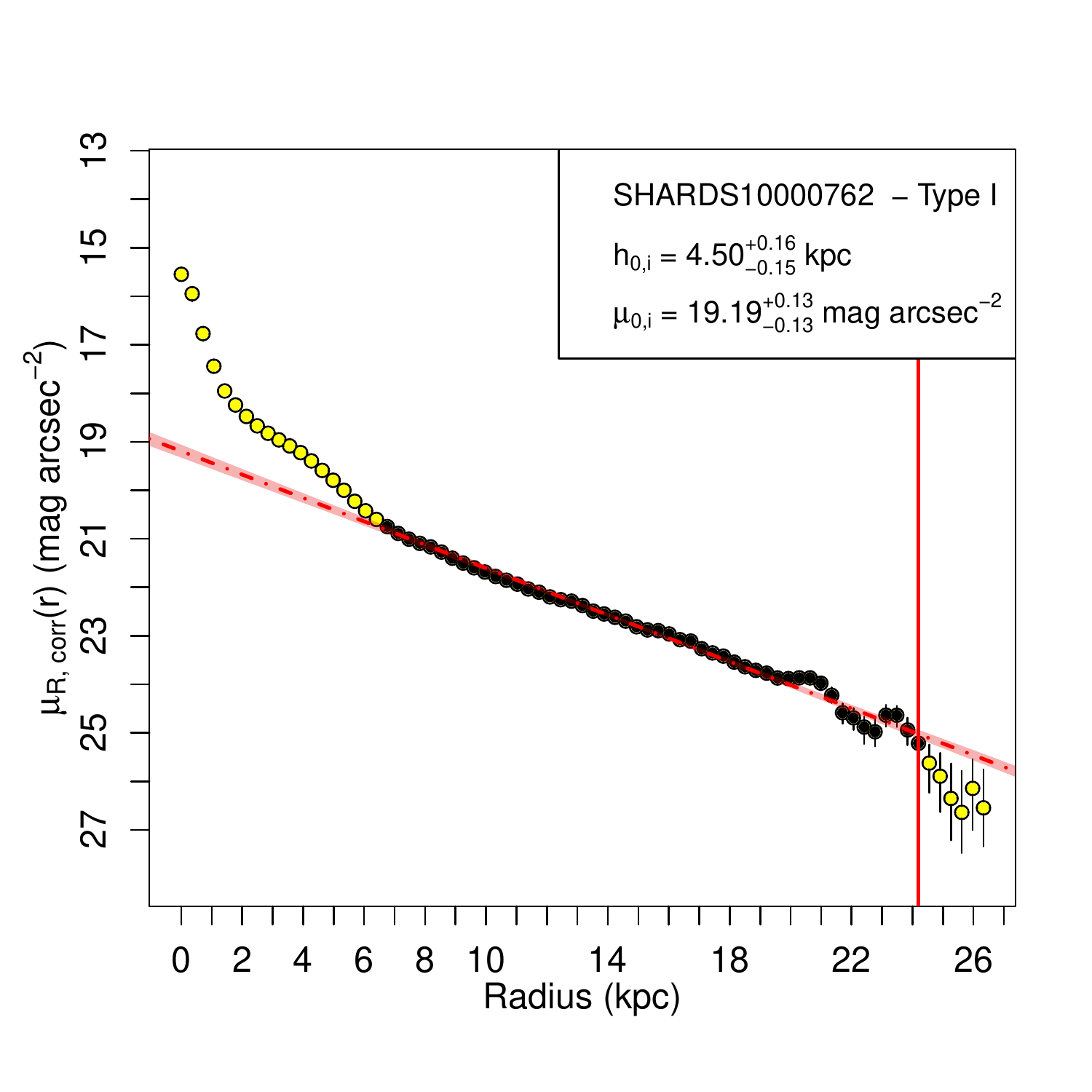}
\end{minipage}%

\vspace{-0.5cm}}
\caption[]{See caption of Fig.1. [\emph{Figure  available in the online edition}.]}         
\label{fig:img_final}
\end{figure}
\clearpage
\newpage

\textbf{SHARDS10000827:} S0 galaxy with Type-III profile. The object presents a medium inclination (see Table \ref{tab:fits_psforr}). It was flagged as an AGN source (see Sect.\,\ref{Subsec:AGN}). A high level of masking was needed due to a low surface brightness field object to the SW. To avoid any contamination, we calculated the surface brightness profile by using the opposite part of the galaxy and performing a manual aggressive masking to the whole FoV. The isophotal curves are centred around the main object, they do not present any large perturbations or deviations and show a position angle and ellipticity nearly constant. 

\begin{figure}[!h]
{\centering
\vspace{-0cm}

\begin{minipage}{.5\textwidth}
\hspace{1.2cm}
\begin{overpic}[width=0.8\textwidth]
{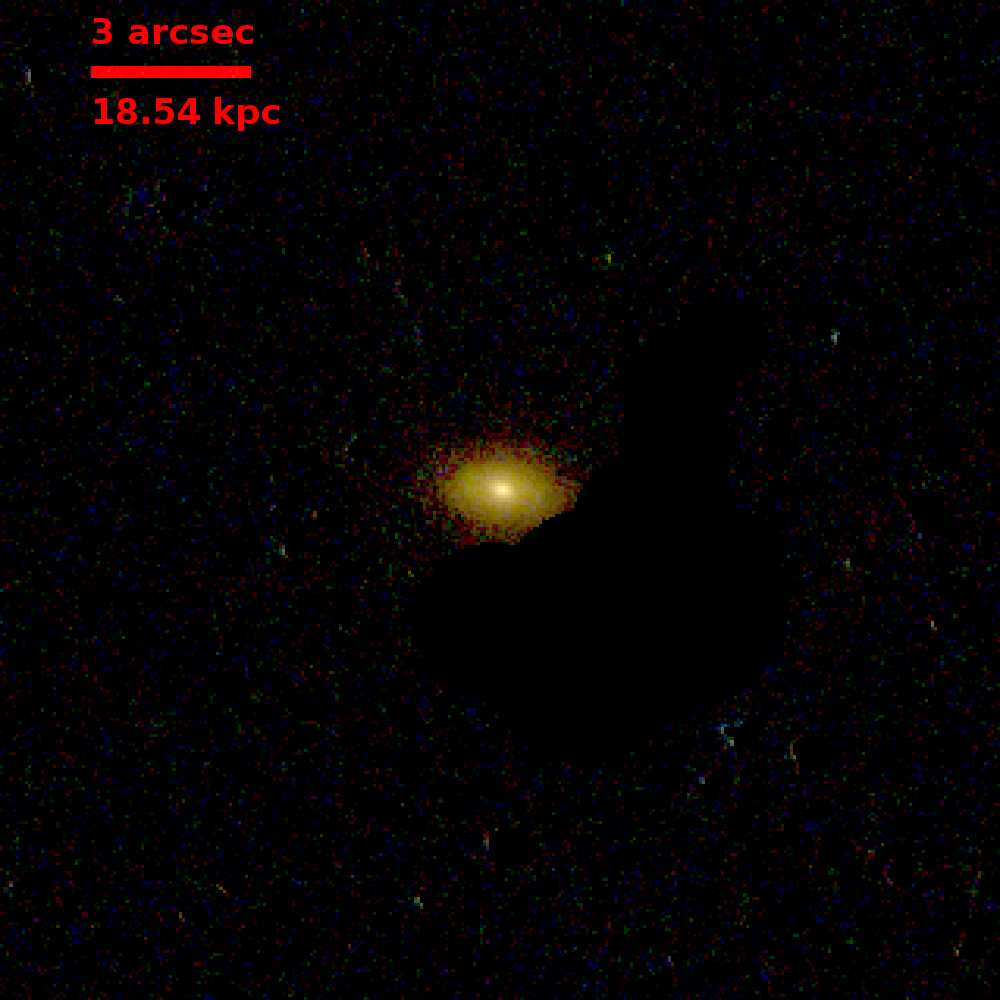}
\put(110,200){\color{yellow} \textbf{SHARDS10000827}}
\put(110,190){\color{yellow} \textbf{z=0.5116}}
\put(110,180){\color{yellow} \textbf{S0}}
\end{overpic}
\vspace{-1cm}
\end{minipage}%
\begin{minipage}{.5\textwidth}
\includegraphics[clip, trim=1cm 1cm 1.5cm 1.5cm, width=\textwidth]{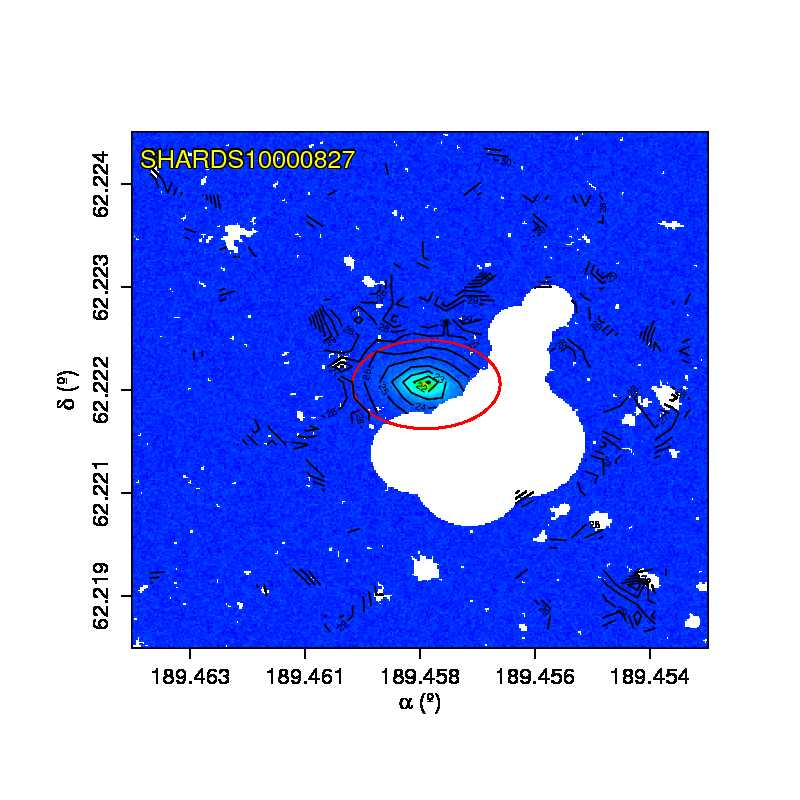}\vspace{-1cm}
\end{minipage}%

\begin{minipage}{.49\textwidth}
\includegraphics[clip, trim=0.1cm 0.1cm 0.1cm 0.1cm, width=\textwidth]{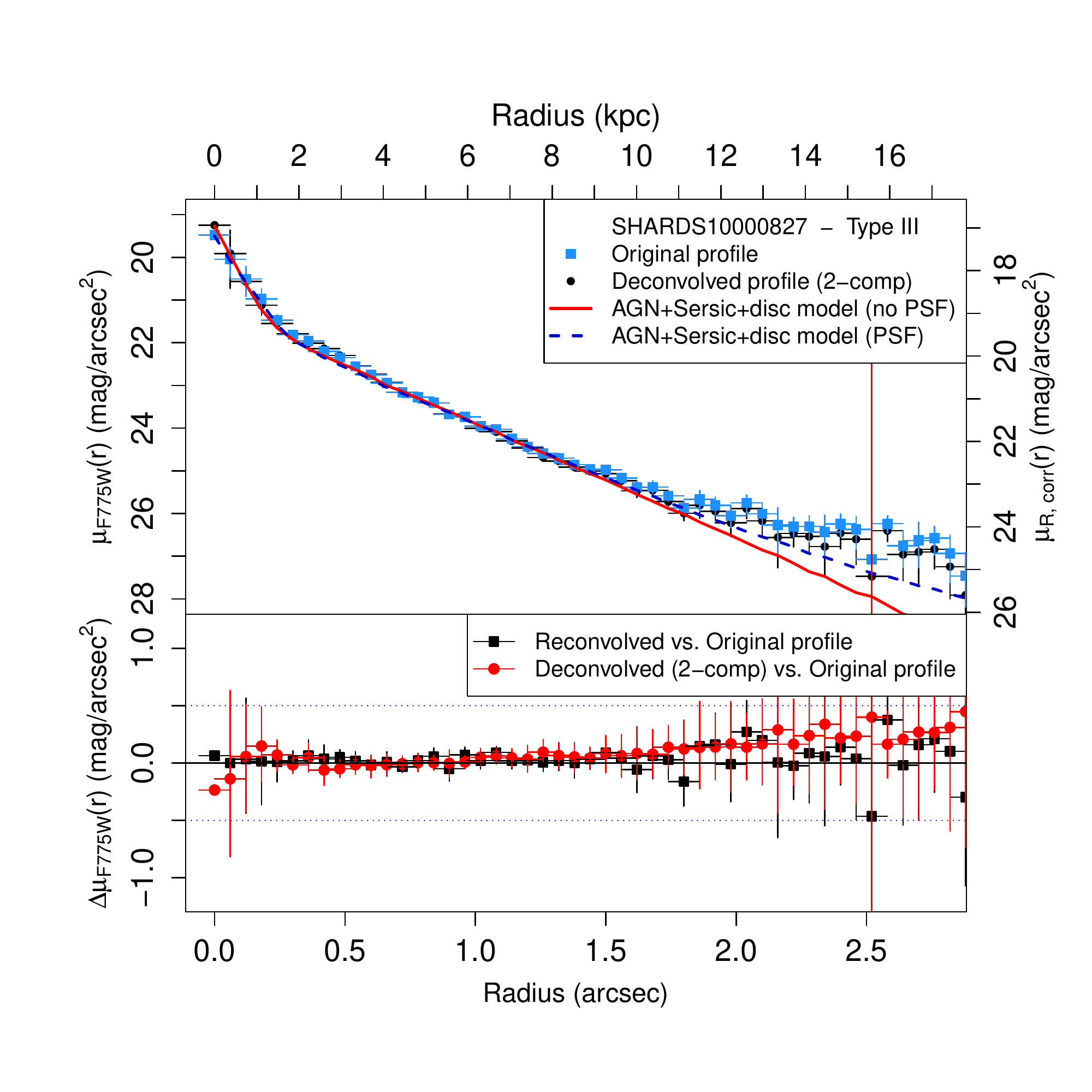}
\end{minipage}
\begin{minipage}{.49\textwidth}
\includegraphics[clip, trim=0.1cm 0.1cm 1cm 0.1cm, width=0.95\textwidth]{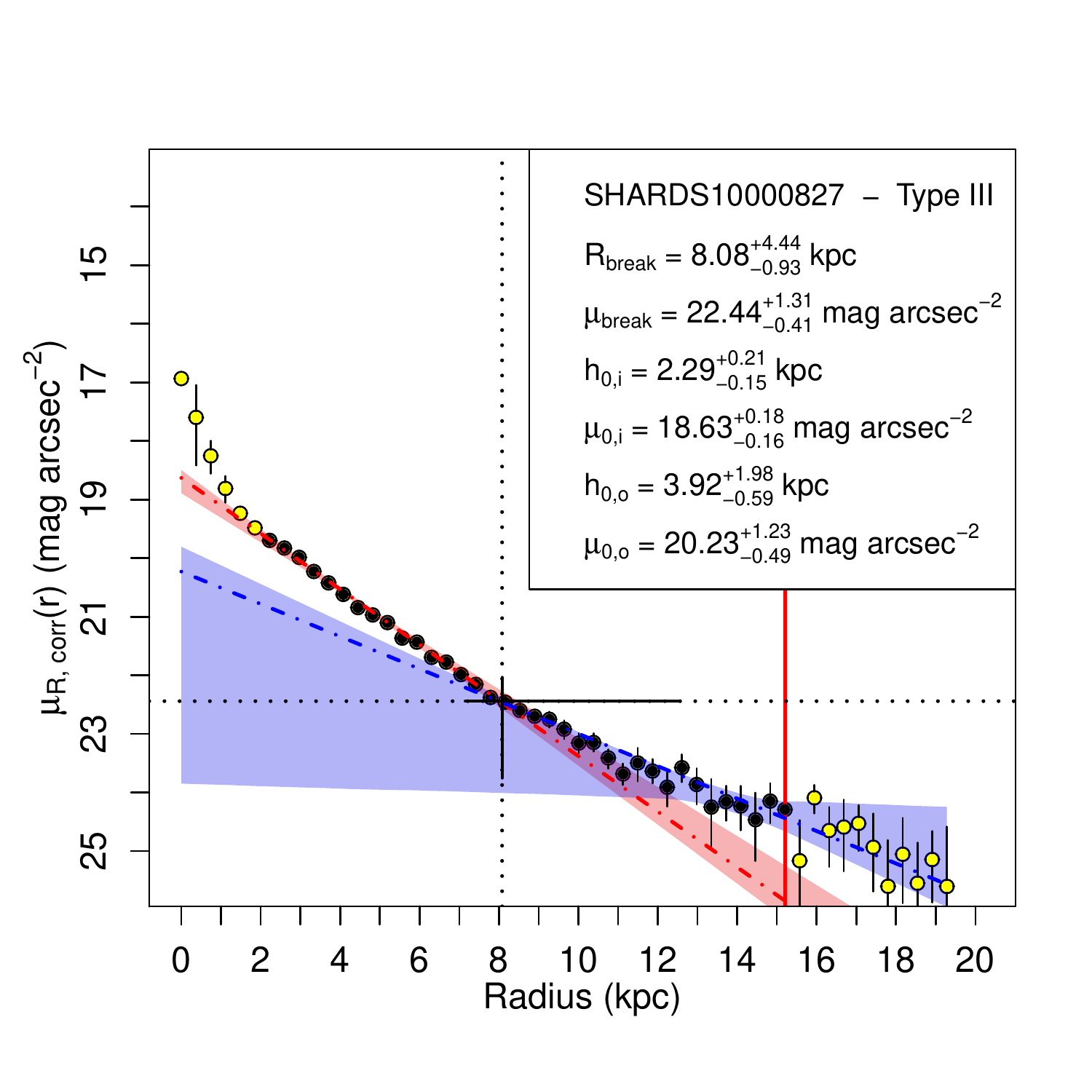}
\end{minipage}%

\vspace{-0.5cm}}
\caption[]{See caption of Fig.1. [\emph{Figure  available in the online edition}.]}         
\label{fig:img_final}
\end{figure}
\clearpage
\newpage

\textbf{SHARDS10000840:} E/S0 galaxy with Type-III profile. The object presents a medium to low inclination (see Table \ref{tab:fits_psforr}). The image needed an aggressive level of masking due to two nearby edge on field galaxies to the NE, including manual masking and revision of the whole FoV. We do not find any significant perturbations in the isophotes that could be due to the field objects. The PDDs for $h$ and $\mu_{0}$ show two clearly separated peaks corresponding to the two profiles.

\begin{figure}[!h]
{\centering
\vspace{-0cm}

\begin{minipage}{.5\textwidth}
\hspace{1.2cm}
\begin{overpic}[width=0.8\textwidth]
{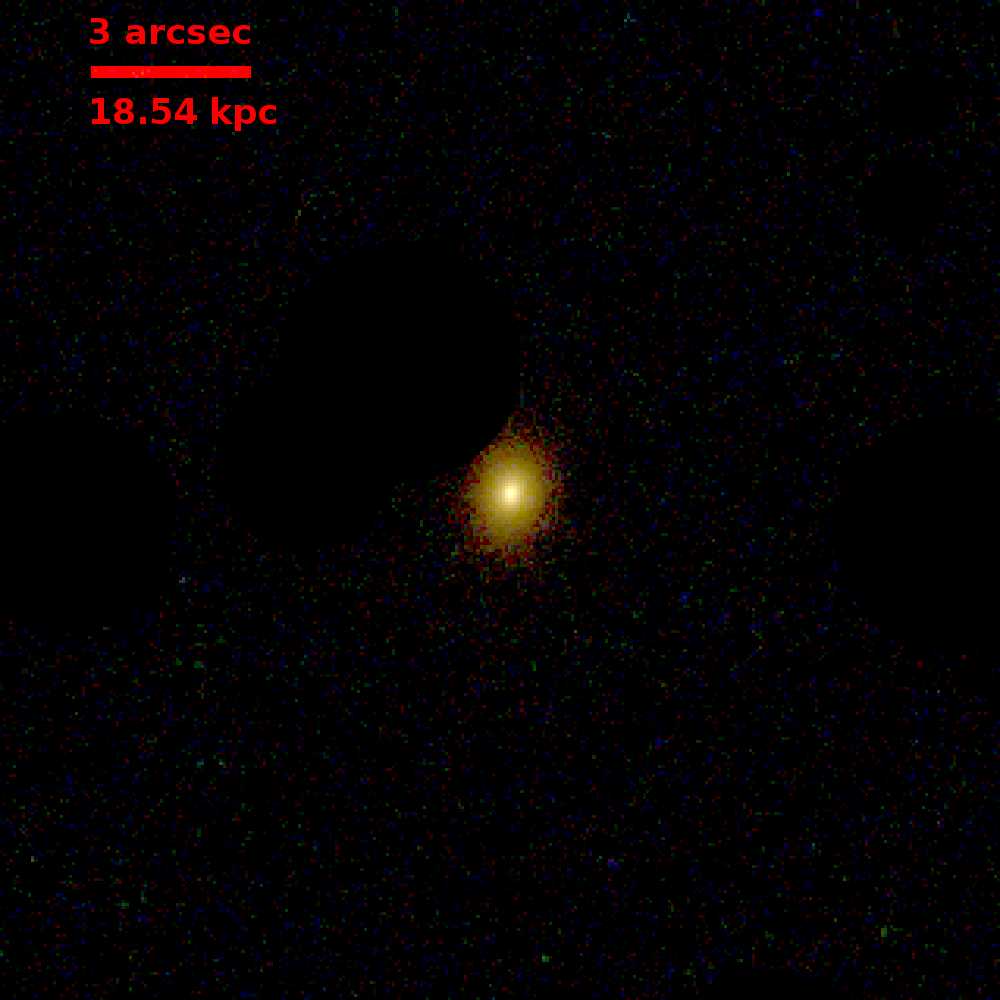}
\put(110,200){\color{yellow} \textbf{SHARDS10000840}}
\put(110,190){\color{yellow} \textbf{z=0.5118}}
\put(110,180){\color{yellow} \textbf{E/S0}}
\end{overpic}
\vspace{-1cm}
\end{minipage}%
\begin{minipage}{.5\textwidth}
\includegraphics[clip, trim=1cm 1cm 1.5cm 1.5cm, width=\textwidth]{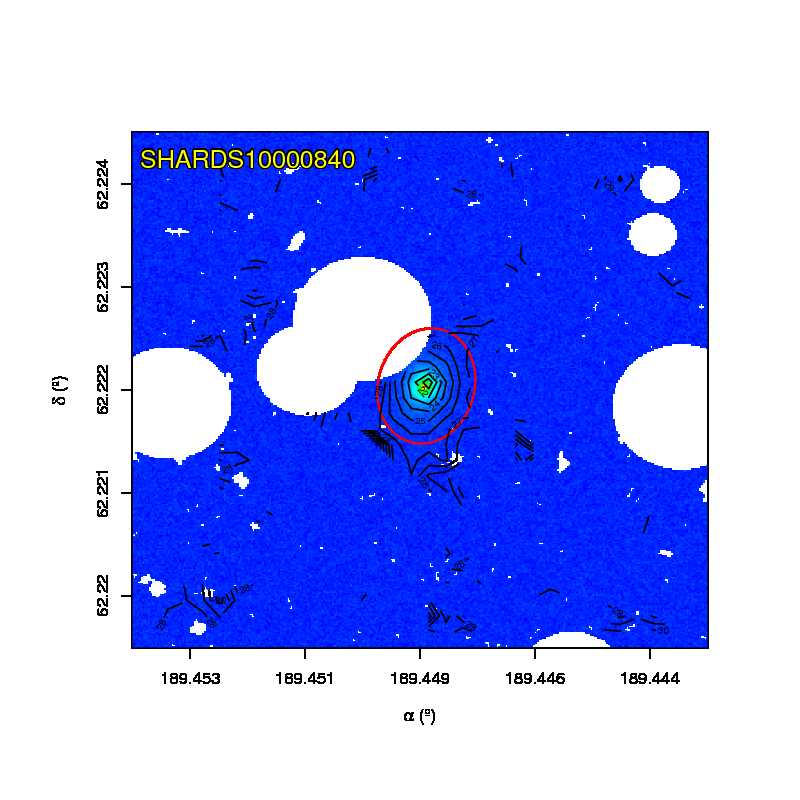}\vspace{-1cm}
\end{minipage}%

\begin{minipage}{.49\textwidth}
\includegraphics[clip, trim=0.1cm 0.1cm 0.1cm 0.1cm, width=\textwidth]{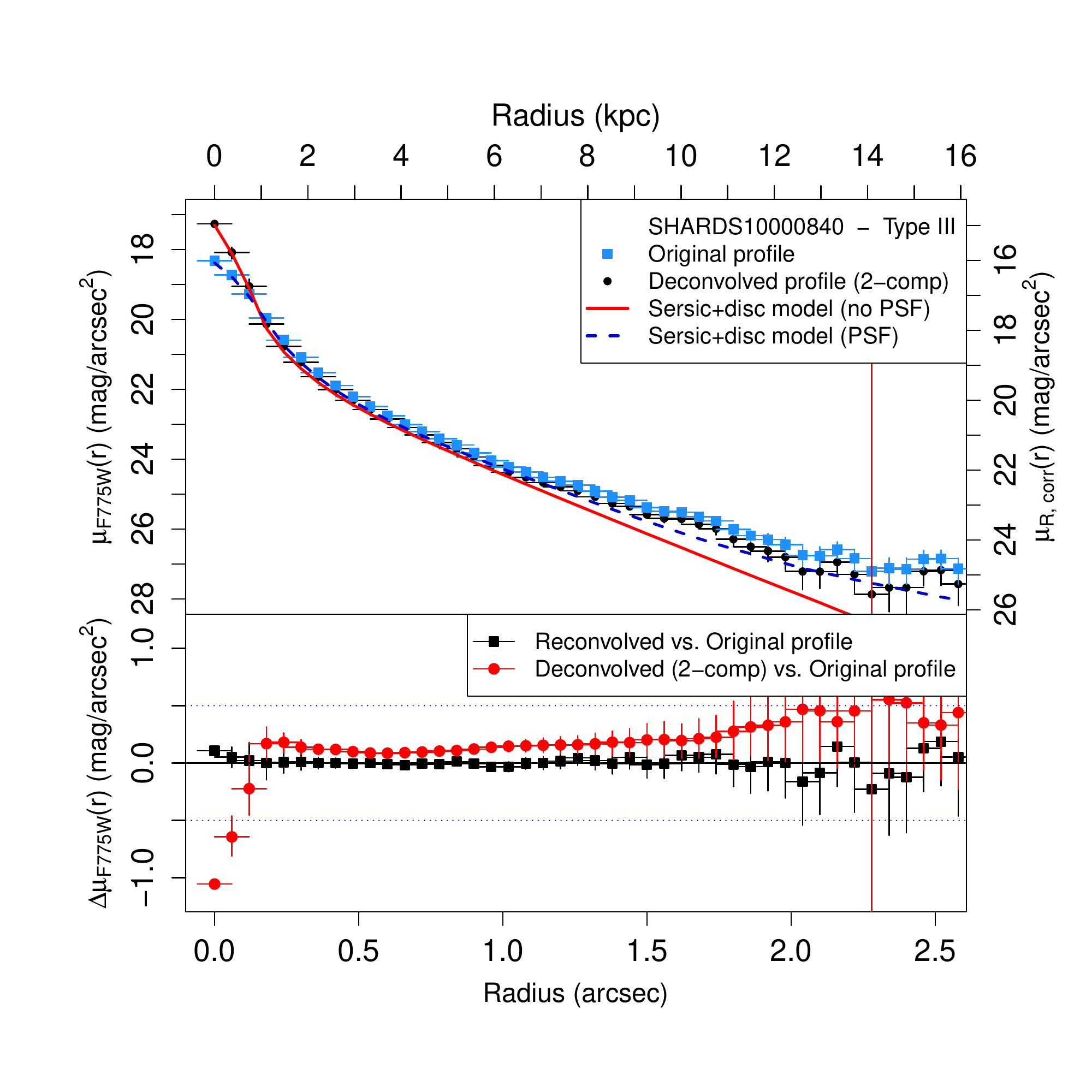}
\end{minipage}
\begin{minipage}{.49\textwidth}
\includegraphics[clip, trim=0.1cm 0.1cm 1cm 0.1cm, width=0.95\textwidth]{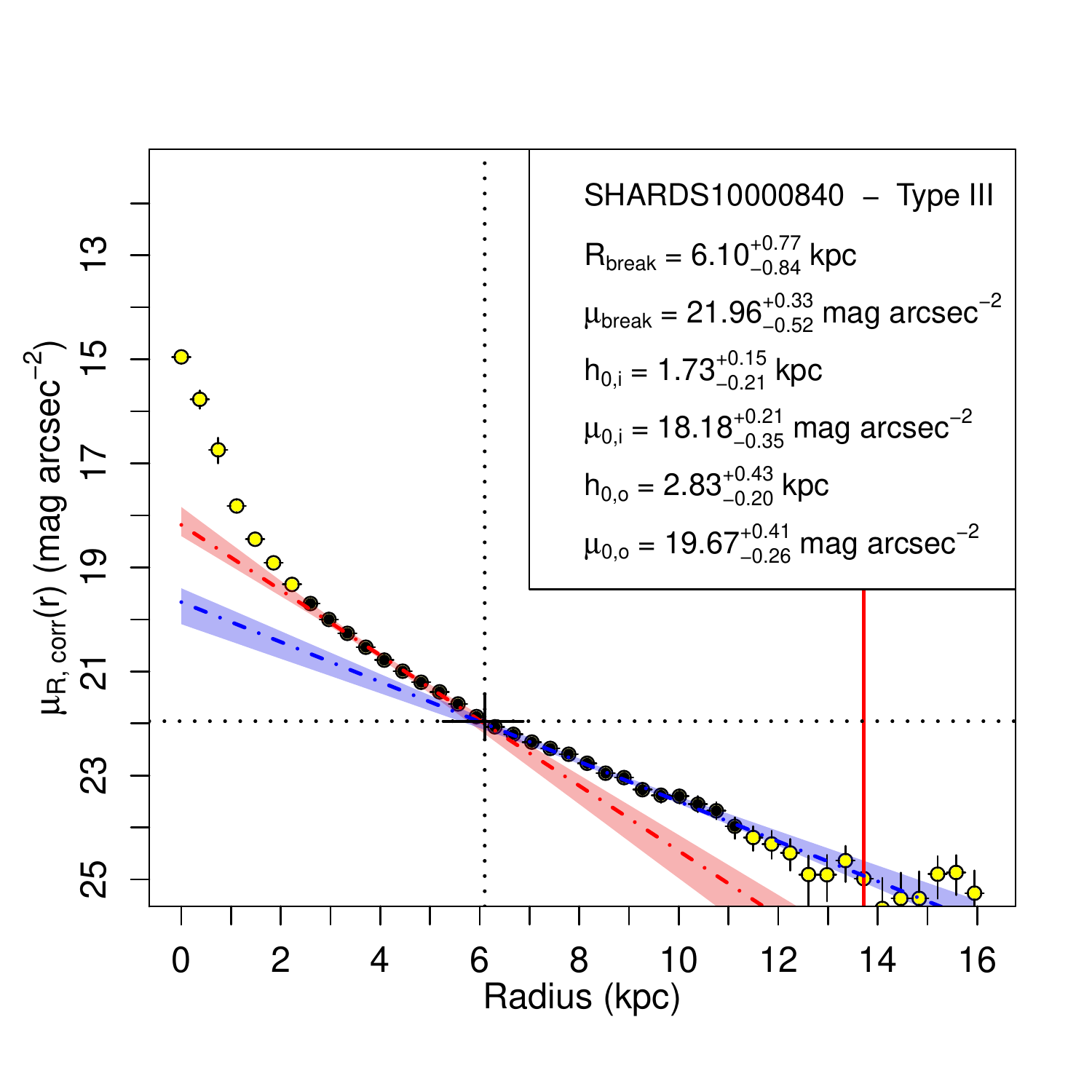}
\end{minipage}%

\vspace{-0.5cm}}
\caption[]{See caption of Fig.1. [\emph{Figure  available in the online edition}.]}         
\label{fig:img_final}
\end{figure}
\clearpage
\newpage

\textbf{SHARDS10000845:} S0 galaxy with Type-II profile. Its profile was generated by {\tt{ISOFIT}} instead of {\tt{ellipse}} due to its completely edge on orientation (see Table \ref{tab:fits_psforr}). The object appears to be isolated, therefore manual masking was not needed. The profile shows a clear and compact bulge component. The PDDs for $h$ and $\mu_{0}$ show two clearly separated peaks corresponding to the two profiles. The corresponding PDDs for \rbreak\ and \mubreak\ appear to be clearly gaussian and narrow. 

\begin{figure}[!h]
{\centering
\vspace{-0cm}

\begin{minipage}{.5\textwidth}
\hspace{1.2cm}
\begin{overpic}[width=0.8\textwidth]
{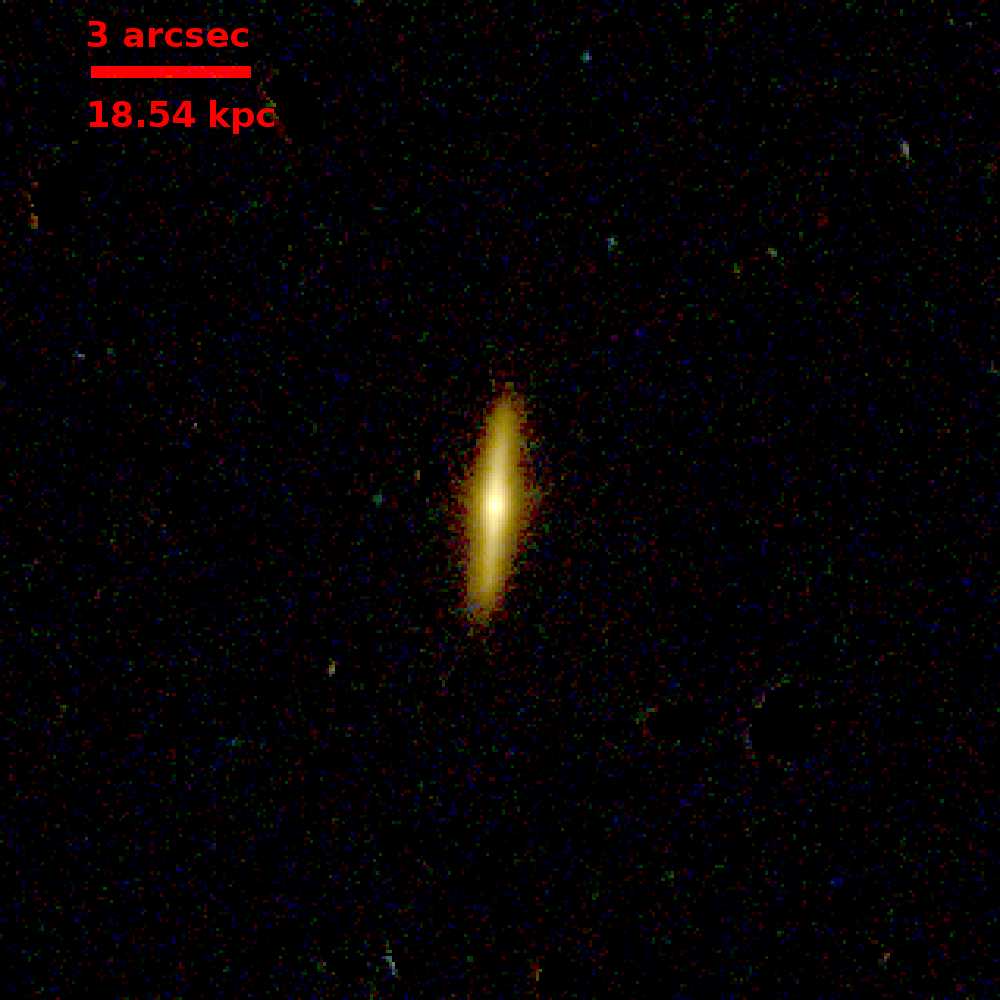}
\put(110,200){\color{yellow} \textbf{SHARDS10000845}}
\put(110,190){\color{yellow} \textbf{z=0.5123}}
\put(110,180){\color{yellow} \textbf{S0}}
\end{overpic}
\vspace{-1cm}
\end{minipage}%
\begin{minipage}{.5\textwidth}
\includegraphics[clip, trim=1cm 1cm 1.5cm 1.5cm, width=\textwidth]{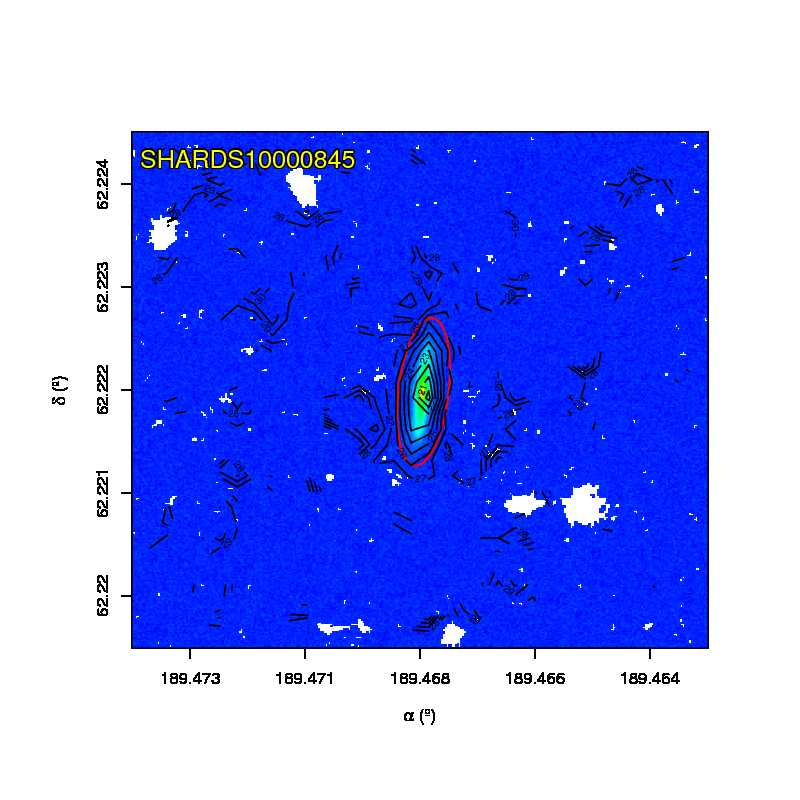}\vspace{-1cm}
\end{minipage}%

\begin{minipage}{.49\textwidth}
\includegraphics[clip, trim=0.1cm 0.1cm 0.1cm 0.1cm, width=\textwidth]{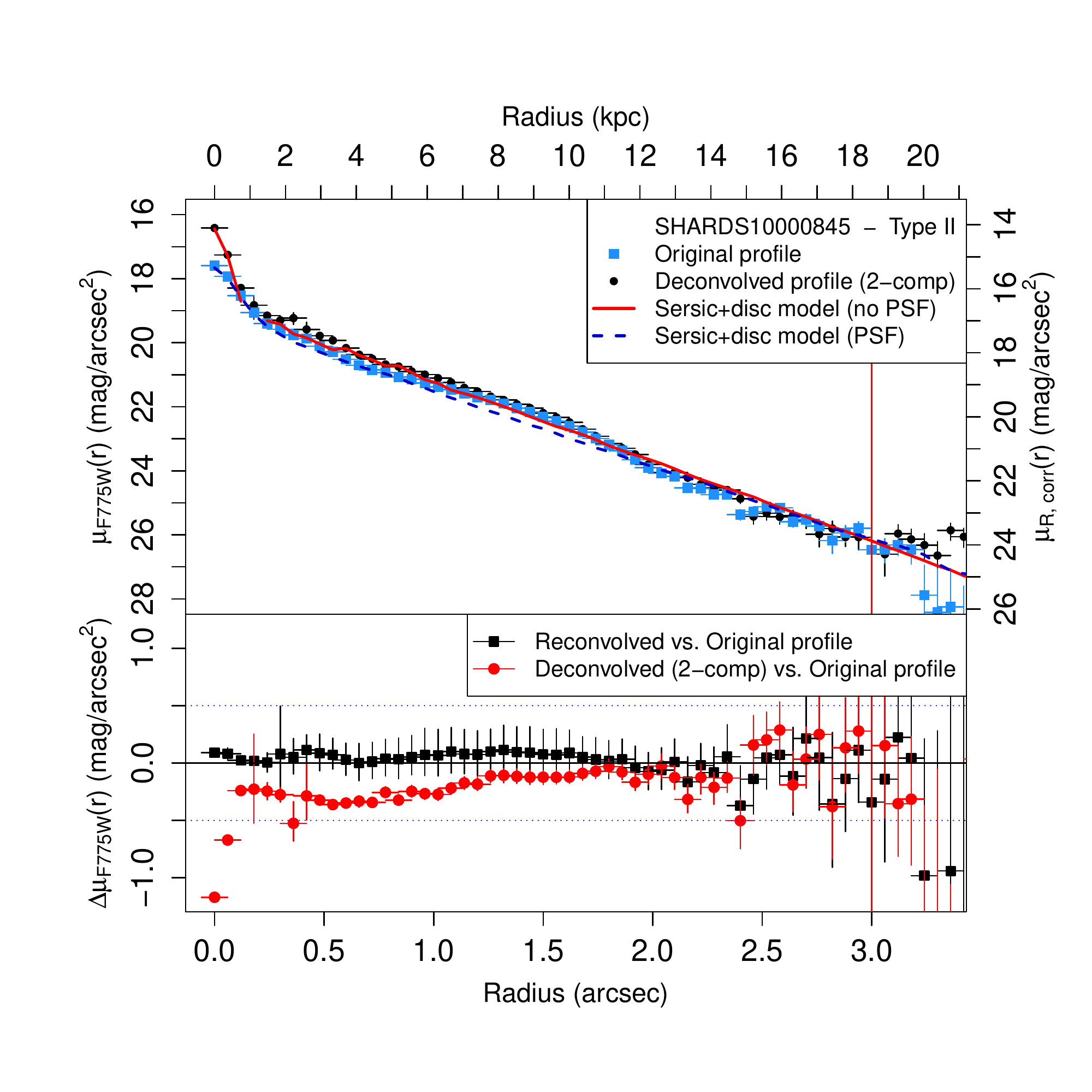}
\end{minipage}
\begin{minipage}{.49\textwidth}
\includegraphics[clip, trim=0.1cm 0.1cm 1cm 0.1cm, width=0.95\textwidth]{IMAGES/IMG_FINAL_SPLIT/SHARDS10000845.pdf}
\end{minipage}%

\vspace{-0.5cm}}
\caption[]{See caption of Fig.1. [\emph{Figure  available in the online edition}.]}         
\label{fig:img_final}
\end{figure}
\clearpage
\newpage

\textbf{SHARDS10000849:} S0 galaxy with Type-III profile. The object presents medium to high inclination (see Table \ref{tab:fits_psforr}). We have applied manual masking to two main sources in the FoV and a smaller one. Nevertheless, none of them appears to be close enough to distort the profile and any contribution to the outer parts of our profile was discarded. The PDDs for $h$ and $\mu_{0}$ show two clearly separated distributions, although the outer profile is somewhat skewed due to deviations from a  pure exponential profile.

\begin{figure}[!h]
{\centering
\vspace{-0cm}

\begin{minipage}{.5\textwidth}
\hspace{1.2cm}
\begin{overpic}[width=0.8\textwidth]
{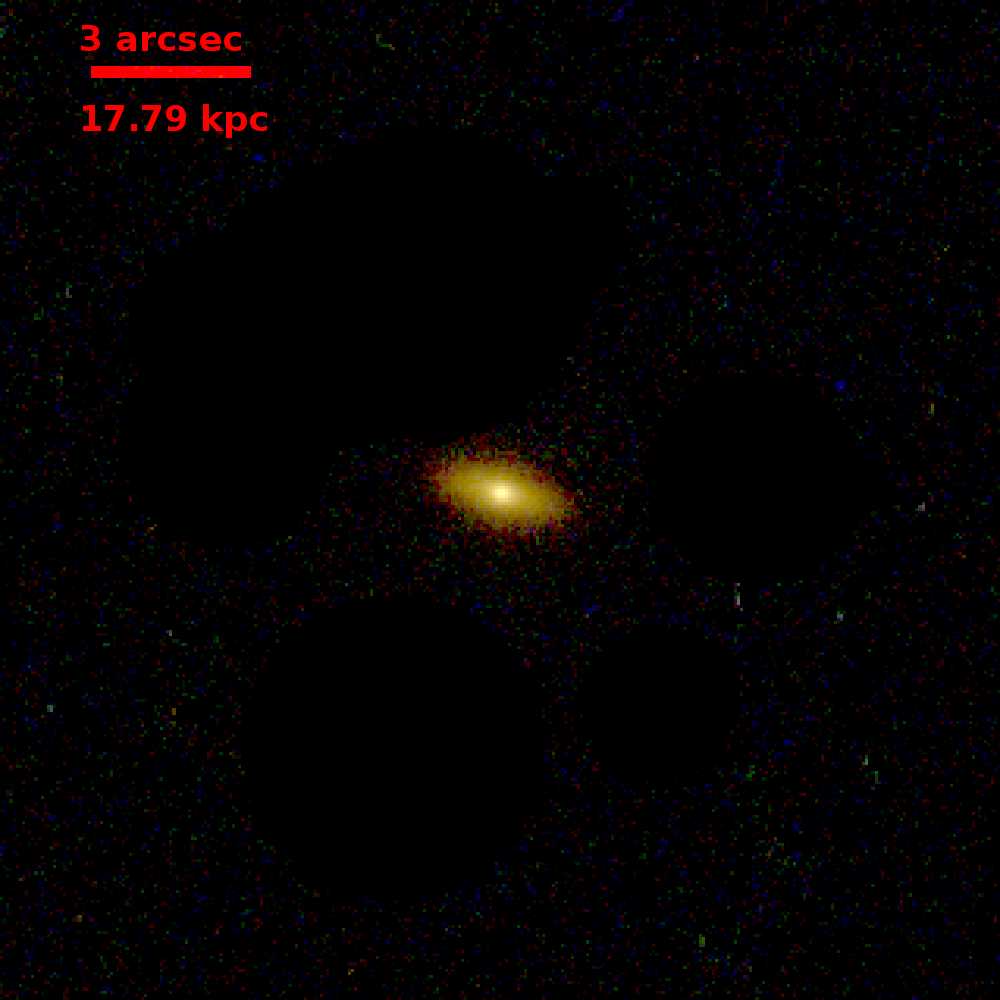}
\put(110,200){\color{yellow} \textbf{SHARDS10000849}}
\put(110,190){\color{yellow} \textbf{z=0.4746}}
\put(110,180){\color{yellow} \textbf{S0}}
\end{overpic}
\vspace{-1cm}
\end{minipage}%
\begin{minipage}{.5\textwidth}
\includegraphics[clip, trim=1cm 1cm 1.5cm 1.5cm, width=\textwidth]{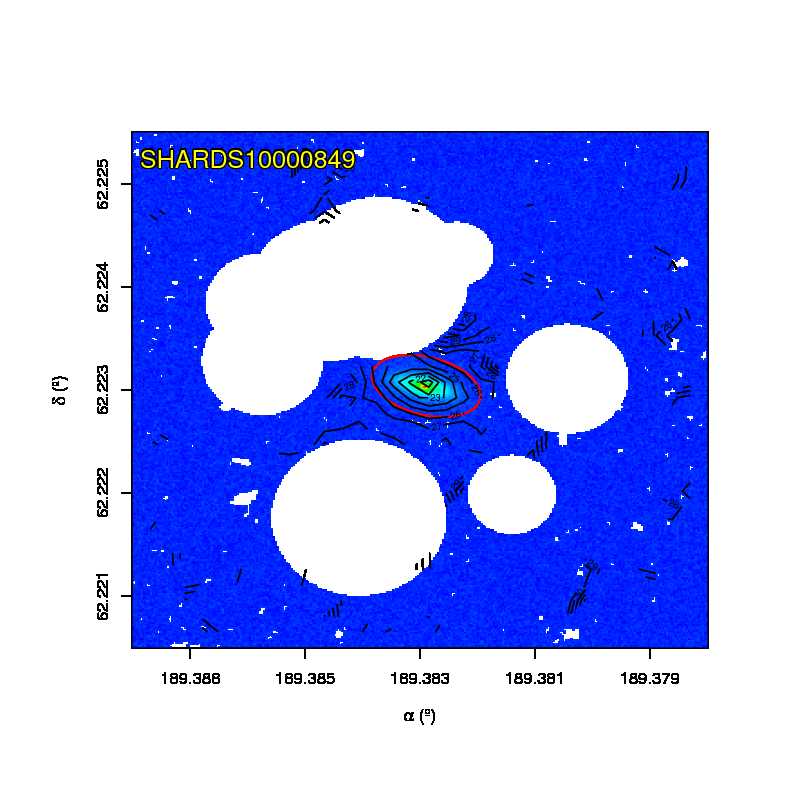}\vspace{-1cm}
\end{minipage}%

\begin{minipage}{.49\textwidth}
\includegraphics[clip, trim=0.1cm 0.1cm 0.1cm 0.1cm, width=\textwidth]{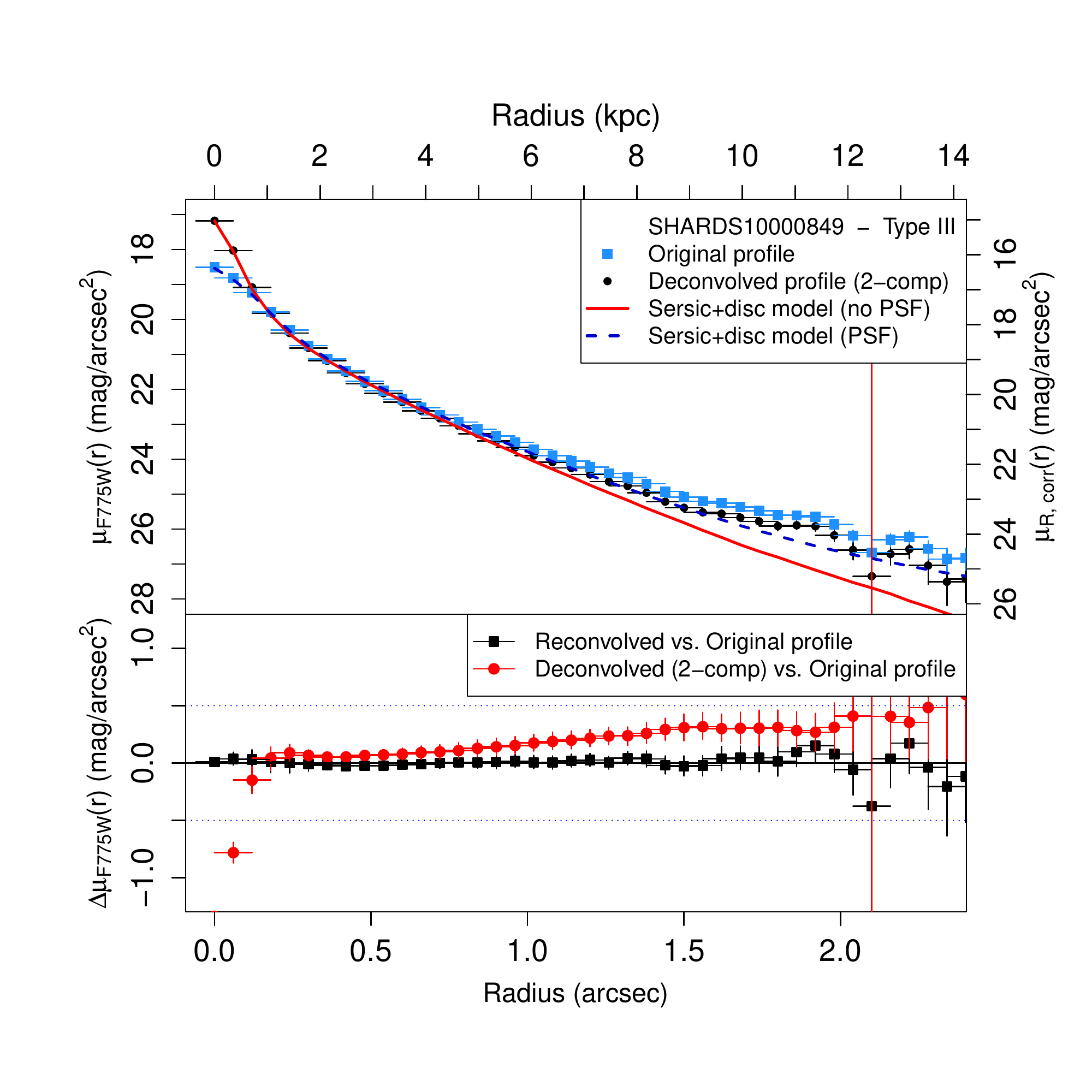}
\end{minipage}
\begin{minipage}{.49\textwidth}
\includegraphics[clip, trim=0.1cm 0.1cm 1cm 0.1cm, width=0.95\textwidth]{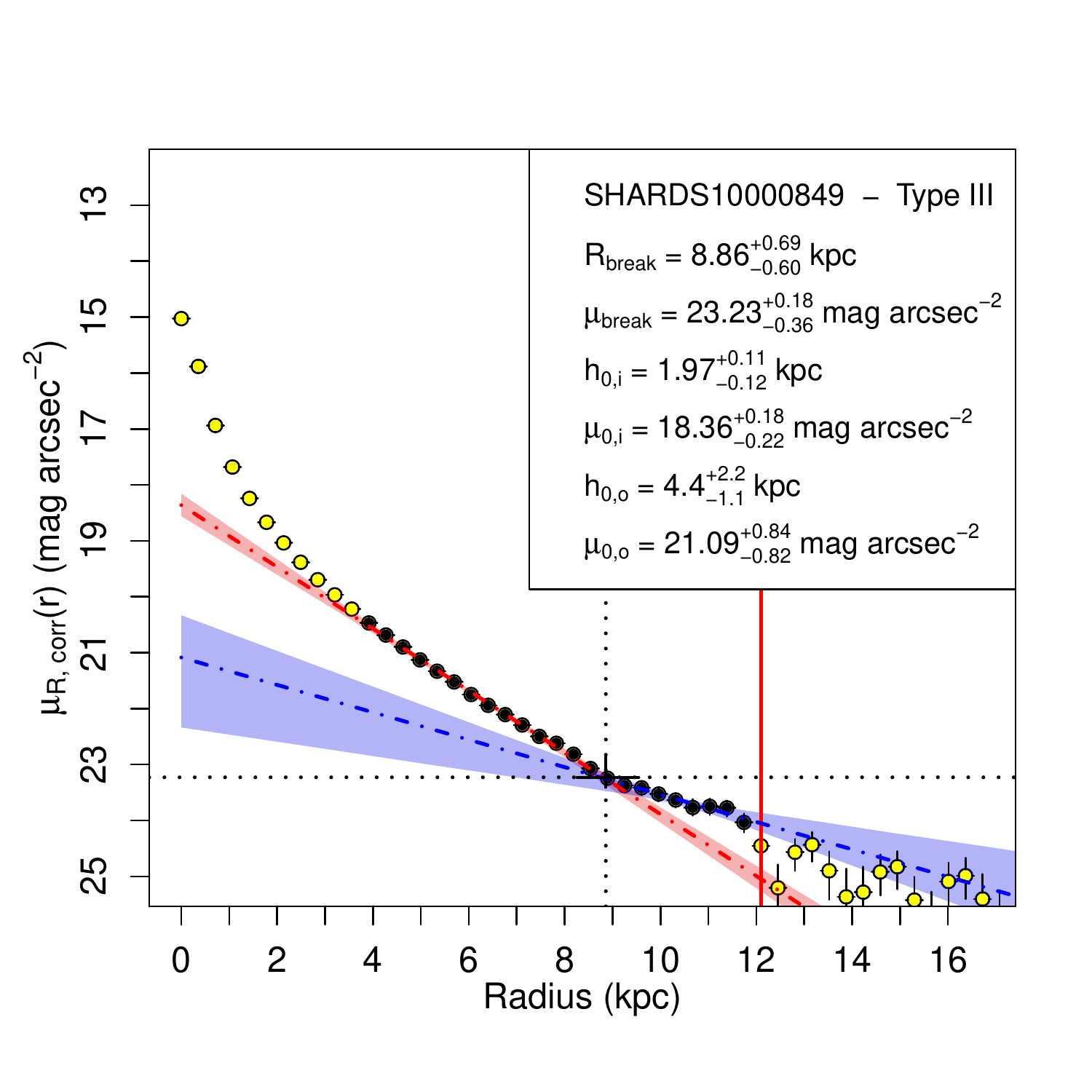}
\end{minipage}%

\vspace{-0.5cm}}
\caption[]{See caption of Fig.1. [\emph{Figure  available in the online edition}.]}         
\label{fig:img_final}
\end{figure}
\clearpage
\newpage

\textbf{SHARDS10001013:} Small S0 galaxy with Type-I profile. The original surface brightness profile appears to be almost bulgeless. It presents medium inclination and small apparent size ($\sim 1.4$ arcsec to $S/N =3$). There are not any nearby galaxies or FoV objects. The resolution is just not enough to resolve any possible detail apart from the disc itself. {\tt{Elbow}} does not reveal significant differences of any part on the disc.

\begin{figure}[!h]
{\centering
\vspace{-0cm}

\begin{minipage}{.5\textwidth}
\hspace{1.2cm}
\begin{overpic}[width=0.8\textwidth]
{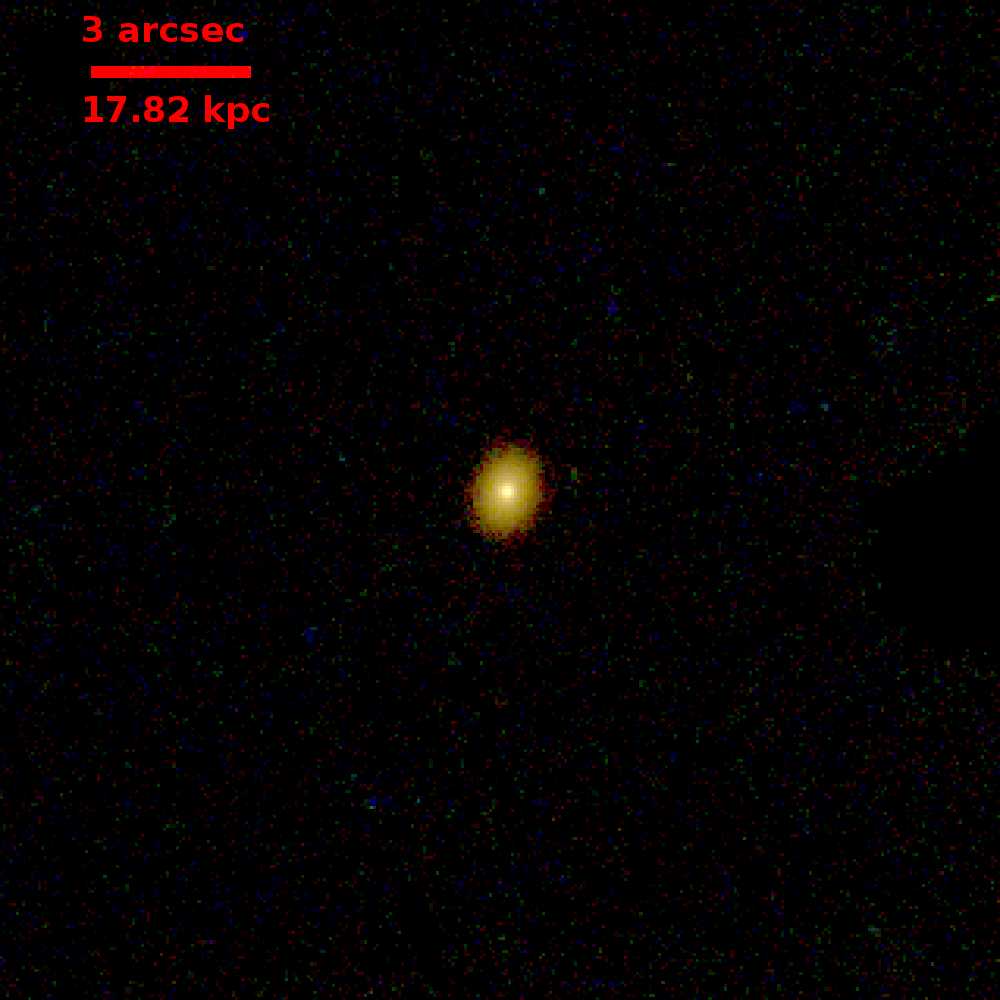}
\put(110,200){\color{yellow} \textbf{SHARDS10001013}}
\put(110,190){\color{yellow} \textbf{z=0.4753}}
\put(110,180){\color{yellow} \textbf{S0}}
\end{overpic}
\vspace{-1cm}
\end{minipage}%
\begin{minipage}{.5\textwidth}
\includegraphics[clip, trim=1cm 1cm 1.5cm 1.5cm, width=\textwidth]{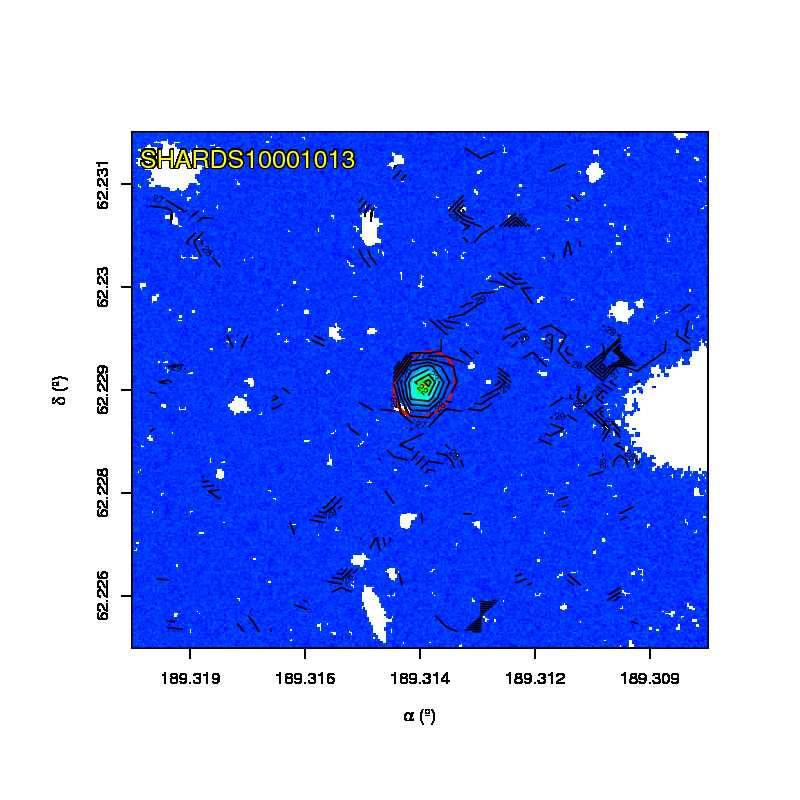}\vspace{-1cm}
\end{minipage}%

\begin{minipage}{.49\textwidth}
\includegraphics[clip, trim=0.1cm 0.1cm 0.1cm 0.1cm, width=\textwidth]{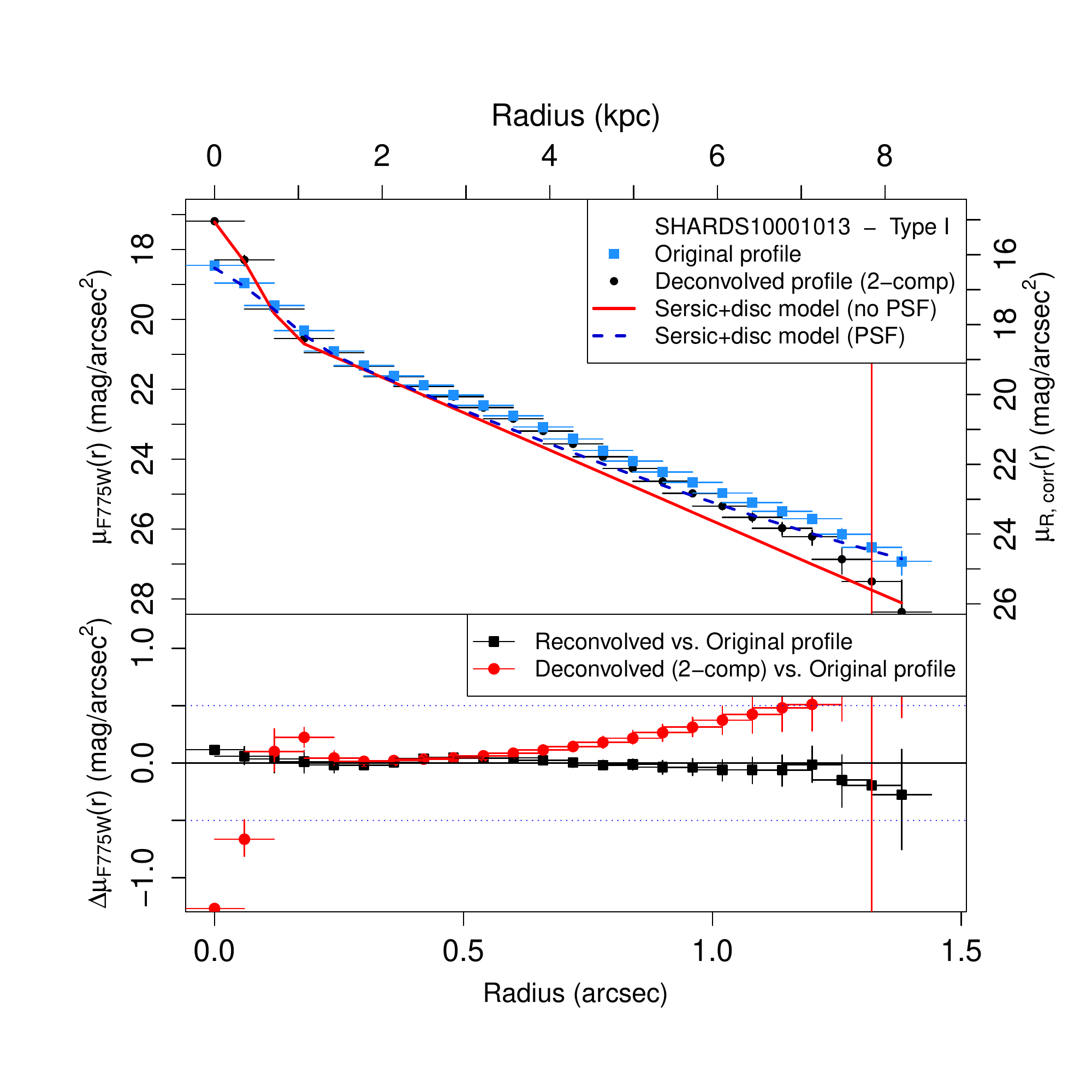}
\end{minipage}
\begin{minipage}{.49\textwidth}
\includegraphics[clip, trim=0.1cm 0.1cm 1cm 0.1cm, width=0.95\textwidth]{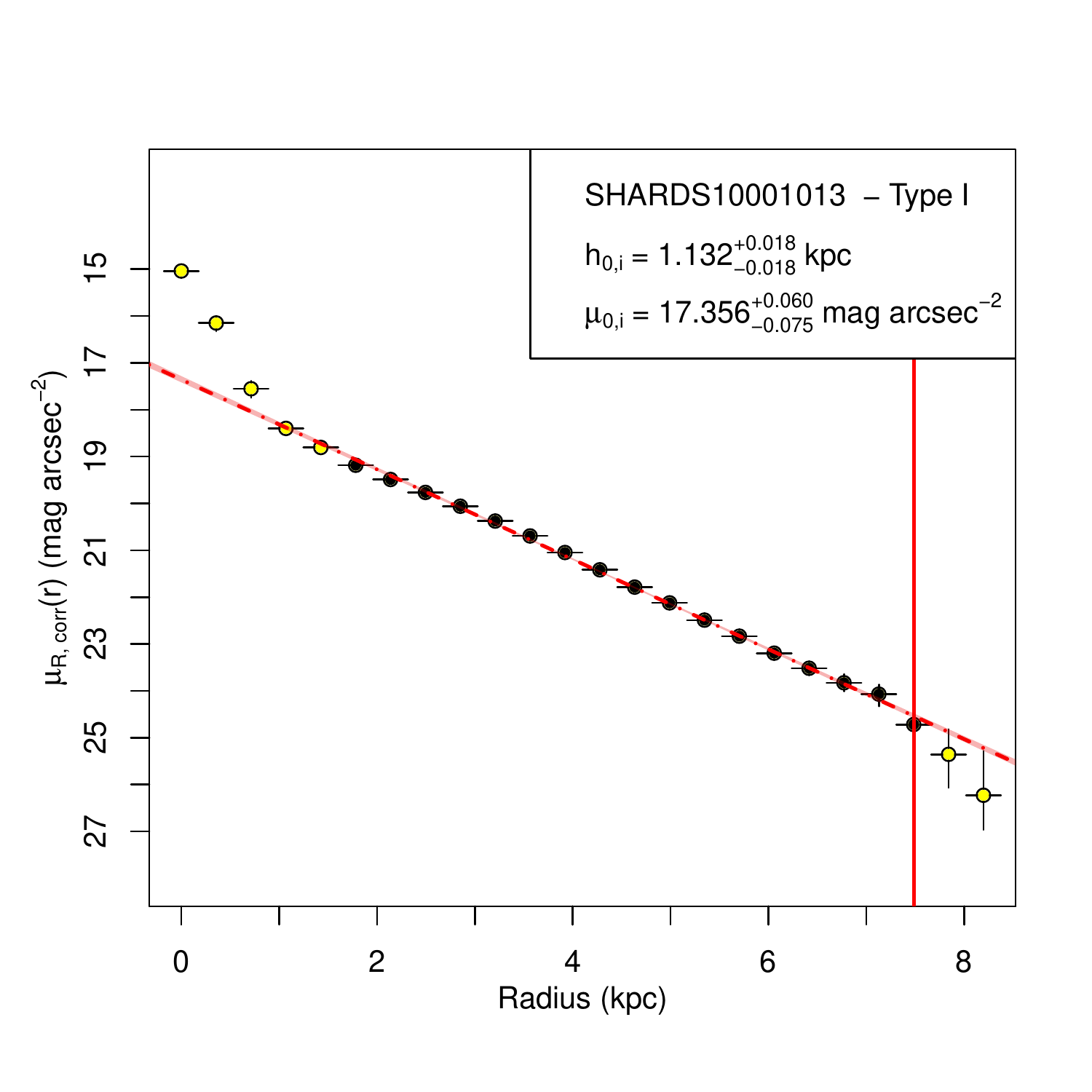}
\end{minipage}%

\vspace{-0.5cm}}
\caption[]{See caption of Fig.1. [\emph{Figure  available in the online edition}.]}         
\label{fig:img_final}
\end{figure}
\clearpage
\newpage

\textbf{SHARDS10001269:} S0 galaxy with Type-I profile. The object presents medium to high inclination (see Table \ref{tab:fits_psforr}). Some close field objects to the NW along the major axis required extensive manual masking, although only few masked pixels finally were within inside the fitting region. An apparent excess of light dominates at the very outer parts of the galaxy, but it is not significant after PSF subtraction ($p\sim 0.018$). This is a quite similar case to SHARDS10000845. The PDDs of $h$ and $\mu_{0}$ reveal a clear inner disc profile in contrast with the wider and skewed PDD of the outer profile.  
\begin{figure}[!h]
{\centering
\vspace{-0cm}

\begin{minipage}{.5\textwidth}
\hspace{1.2cm}
\begin{overpic}[width=0.8\textwidth]
{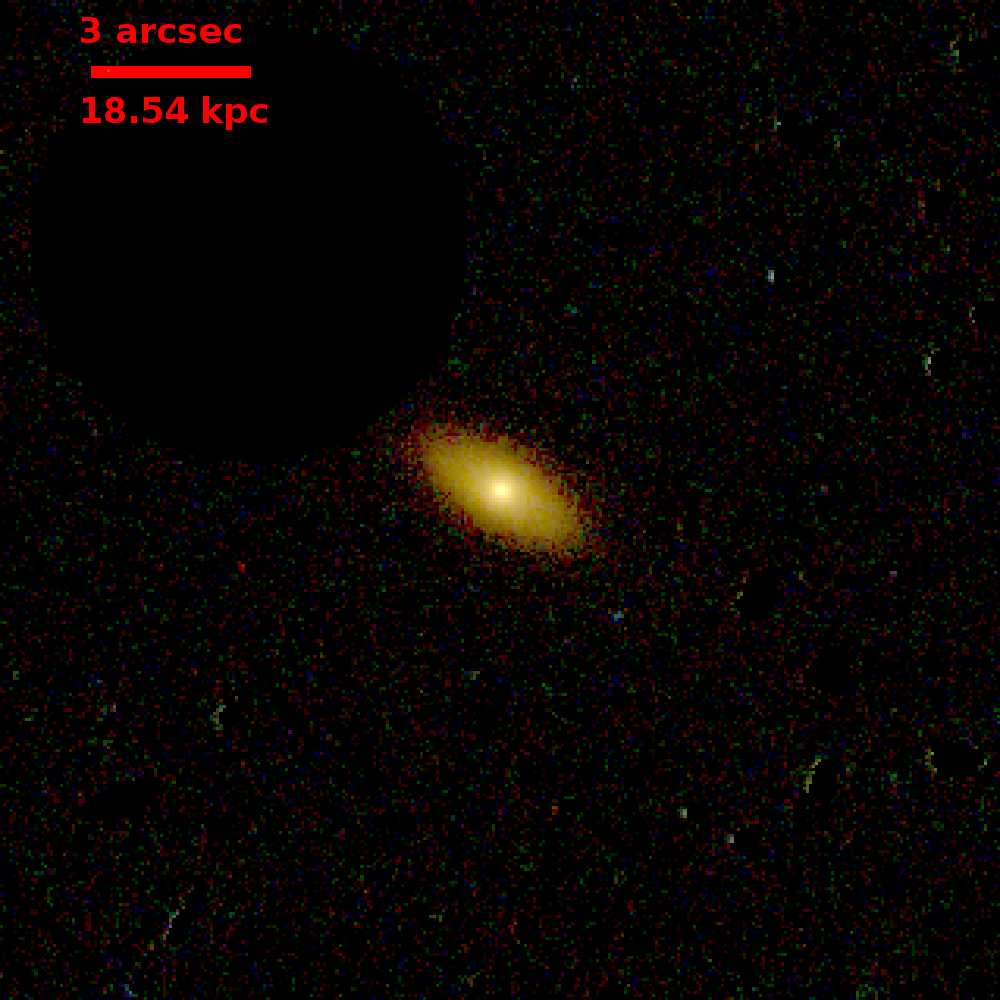}
\put(110,200){\color{yellow} \textbf{SHARDS10001269}}
\put(110,190){\color{yellow} \textbf{z=0.5116}}
\put(110,180){\color{yellow} \textbf{S0}}
\end{overpic}
\vspace{-1cm}
\end{minipage}%
\begin{minipage}{.5\textwidth}
\includegraphics[clip, trim=1cm 1cm 1.5cm 1.5cm, width=\textwidth]{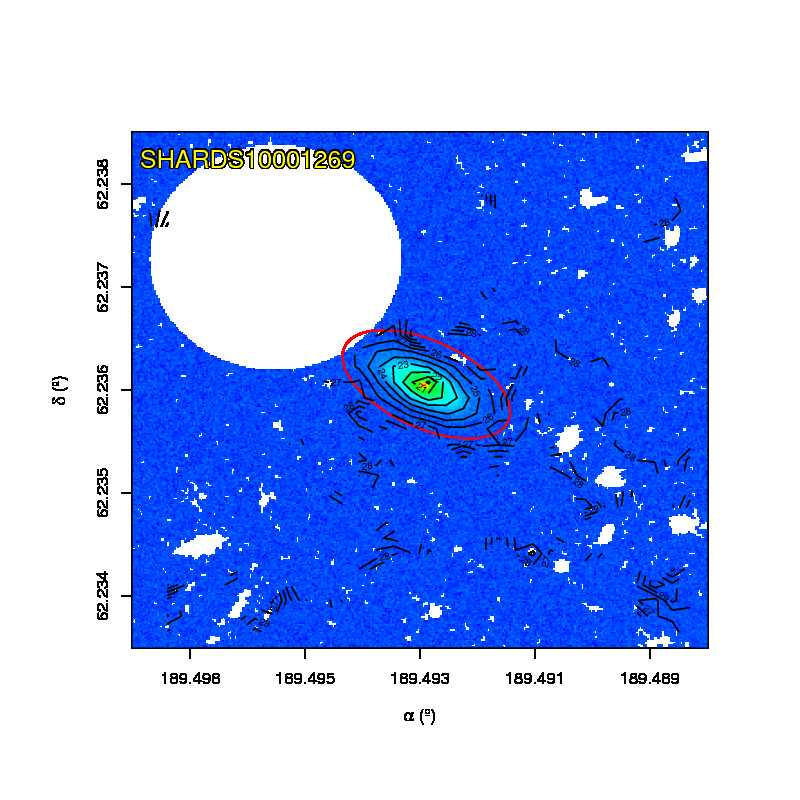}\vspace{-1cm}
\end{minipage}%

\begin{minipage}{.49\textwidth}
\includegraphics[clip, trim=0.1cm 0.1cm 0.1cm 0.1cm, width=\textwidth]{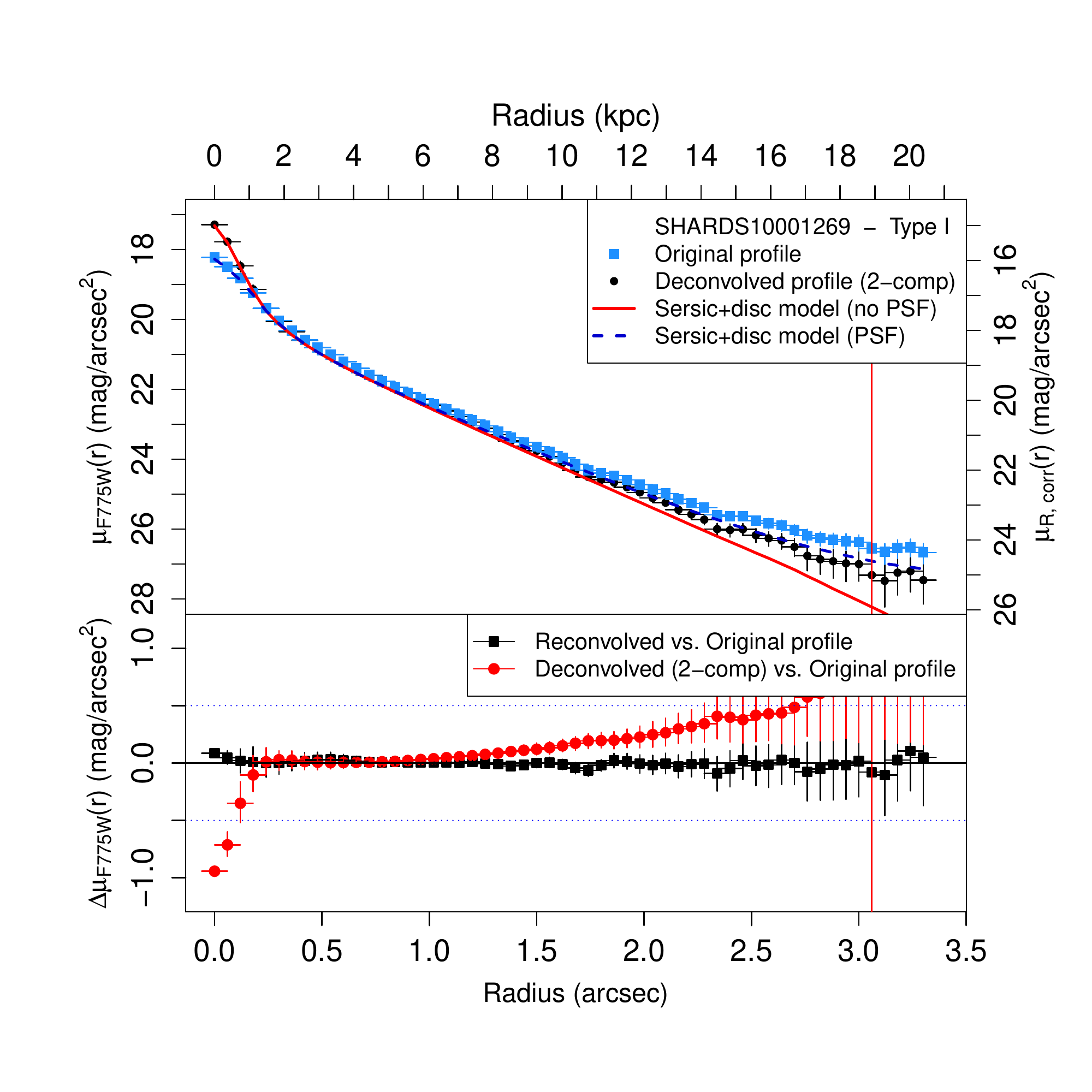}
\end{minipage}
\begin{minipage}{.49\textwidth}
\includegraphics[clip, trim=0.1cm 0.1cm 1cm 0.1cm, width=0.95\textwidth]{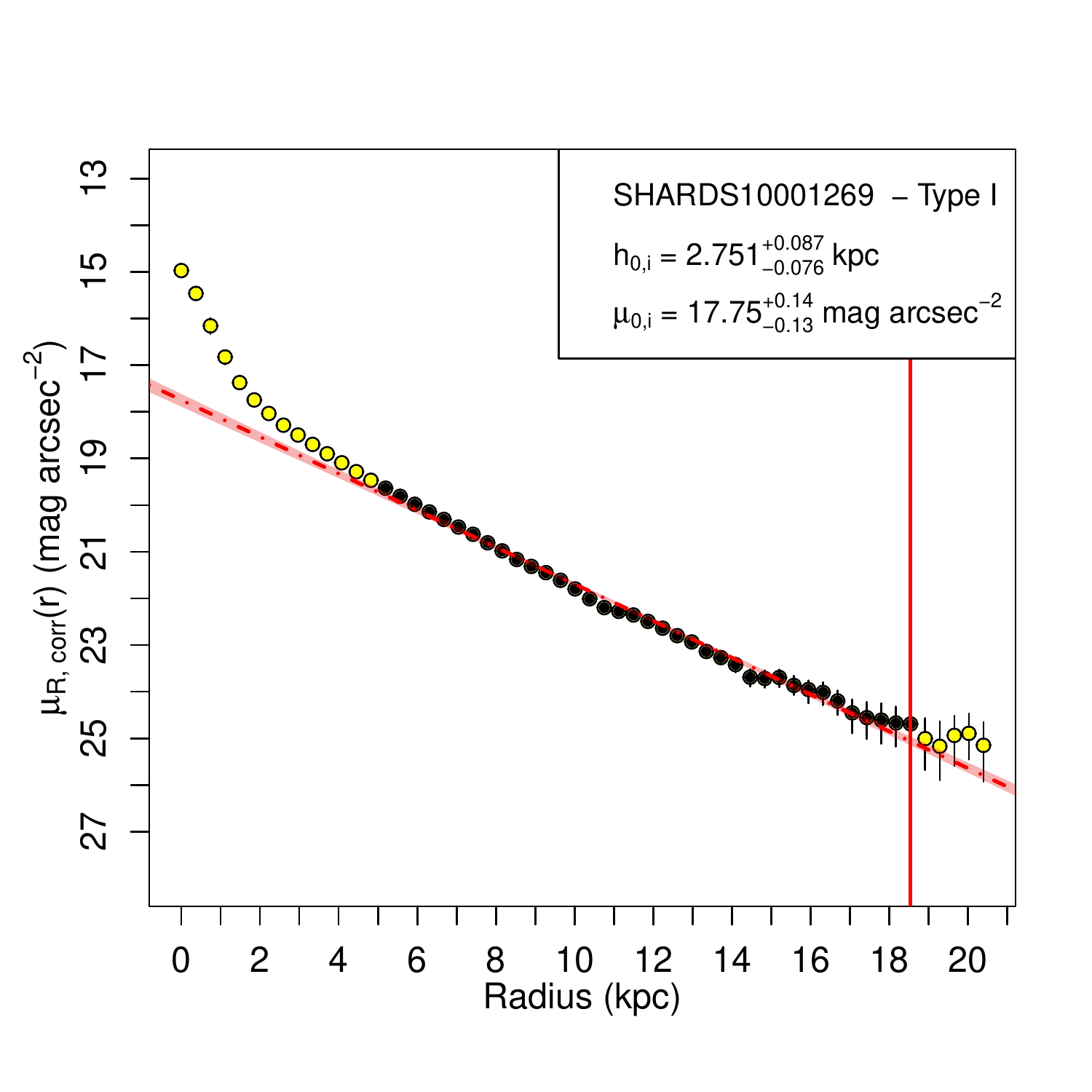}
\end{minipage}%

\vspace{-0.5cm}}
\caption[]{See caption of Fig.1. [\emph{Figure  available in the online edition}.]}         
\label{fig:img_final}
\end{figure}
\clearpage
\newpage

\textbf{SHARDS10001314:} S0 galaxy, apparently in interaction or at least overlapped with another source of similar size. The galaxy does not seem to be highly disturbed or asymmetric. High contamination from a close object required extensive masking of almost 30\% of the galaxy, leaving the other 70\% to analyse the surface brightness profile. The non-PSF corrected image shows an excess of light with almost constant surface brightness at $9-12$ kpc from the centre, which results negligible after PSF subtraction. The exponential profile appears to be slightly bended, but the break parameters present high uncertainities. The probabilities of the both distributions (inner and outer profile for being equal) are slightly lower than 5\% both in $\mu$ and $h$. We finally classified this object as Type I. 

\begin{figure}[!h]
{\centering
\vspace{-0cm}

\begin{minipage}{.5\textwidth}
\hspace{1.2cm}
\begin{overpic}[width=0.8\textwidth]
{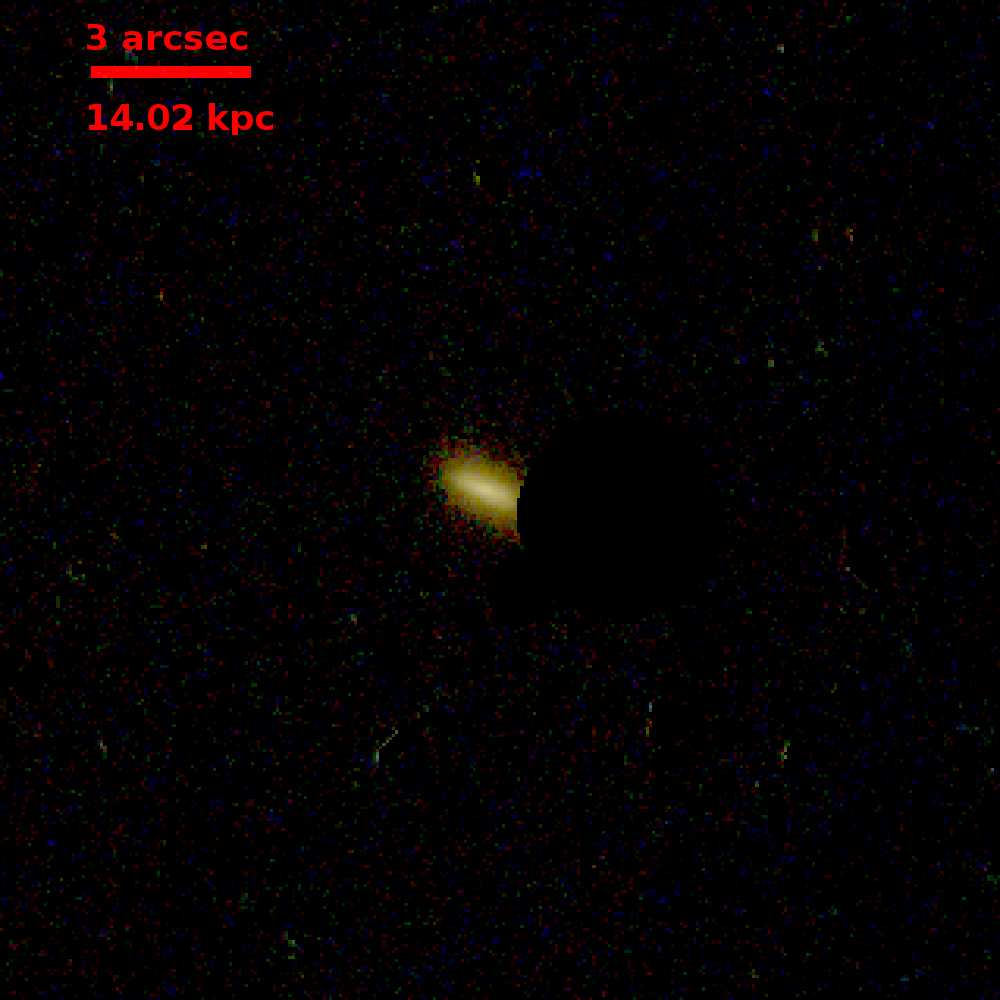}
\put(110,200){\color{yellow} \textbf{SHARDS10001314}}
\put(110,190){\color{yellow} \textbf{z=0.3215}}
\put(110,180){\color{yellow} \textbf{S0}}
\end{overpic}
\vspace{-1cm}
\end{minipage}%
\begin{minipage}{.5\textwidth}
\includegraphics[clip, trim=1cm 1cm 1.5cm 1.5cm, width=\textwidth]{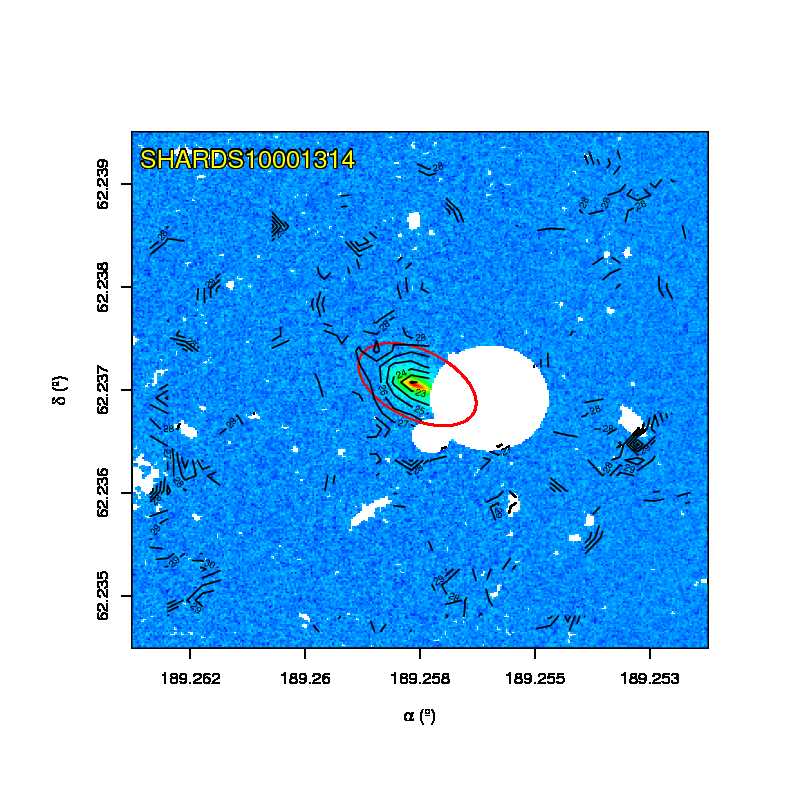}\vspace{-1cm}
\end{minipage}%

\begin{minipage}{.49\textwidth}
\includegraphics[clip, trim=0.1cm 0.1cm 0.1cm 0.1cm, width=\textwidth]{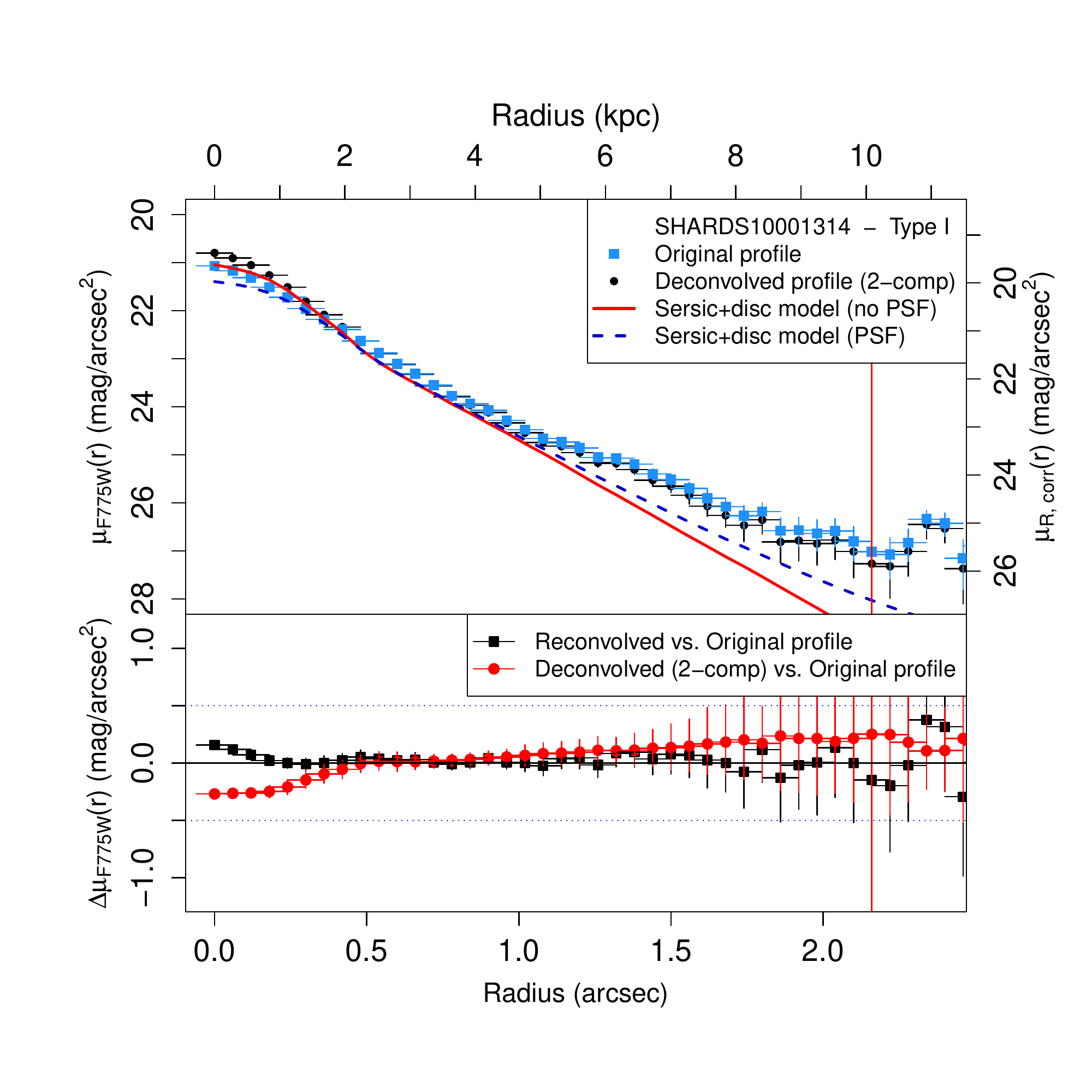}
\end{minipage}
\begin{minipage}{.49\textwidth}
\includegraphics[clip, trim=0.1cm 0.1cm 1cm 0.1cm, width=0.95\textwidth]{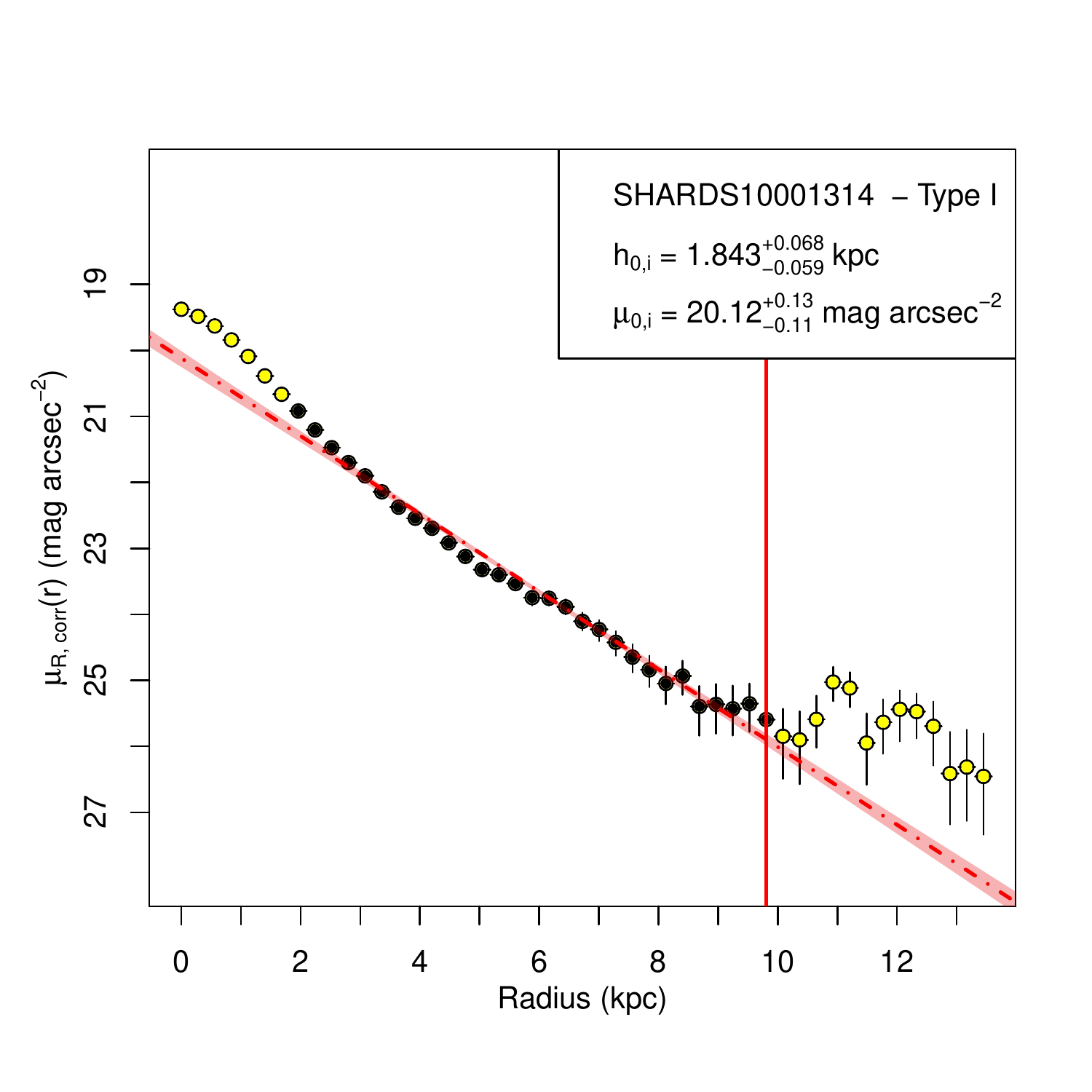}
\end{minipage}%

\vspace{-0.5cm}}
\caption[]{See caption of Fig.1. [\emph{Figure  available in the online edition}.]}         
\label{fig:img_final}
\end{figure}
\clearpage
\newpage

\textbf{SHARDS10001344:} S0 Type-III galaxy with medium inclination (see Table \ref{tab:fits_psforr}) and a very bright central component ($\mu_{R} = 16$ \magarc) in contrast with the low surface brightness of the disc structure. We applied an extensive masking to a nearby region to the N-EE and a small close source to the N-NW. The PDDs for $h$ and $\mu_{0}$ show two separated peaks corresponding to the two profiles. The PDDs of the inner disc profile appear with two peaks and distorted due to the low statistics that result from the aggresive masking performed to the central bulge to avoid contamination.

\begin{figure}[!h]
{\centering
\vspace{-0cm}

\begin{minipage}{.5\textwidth}
\hspace{1.2cm}
\begin{overpic}[width=0.8\textwidth]
{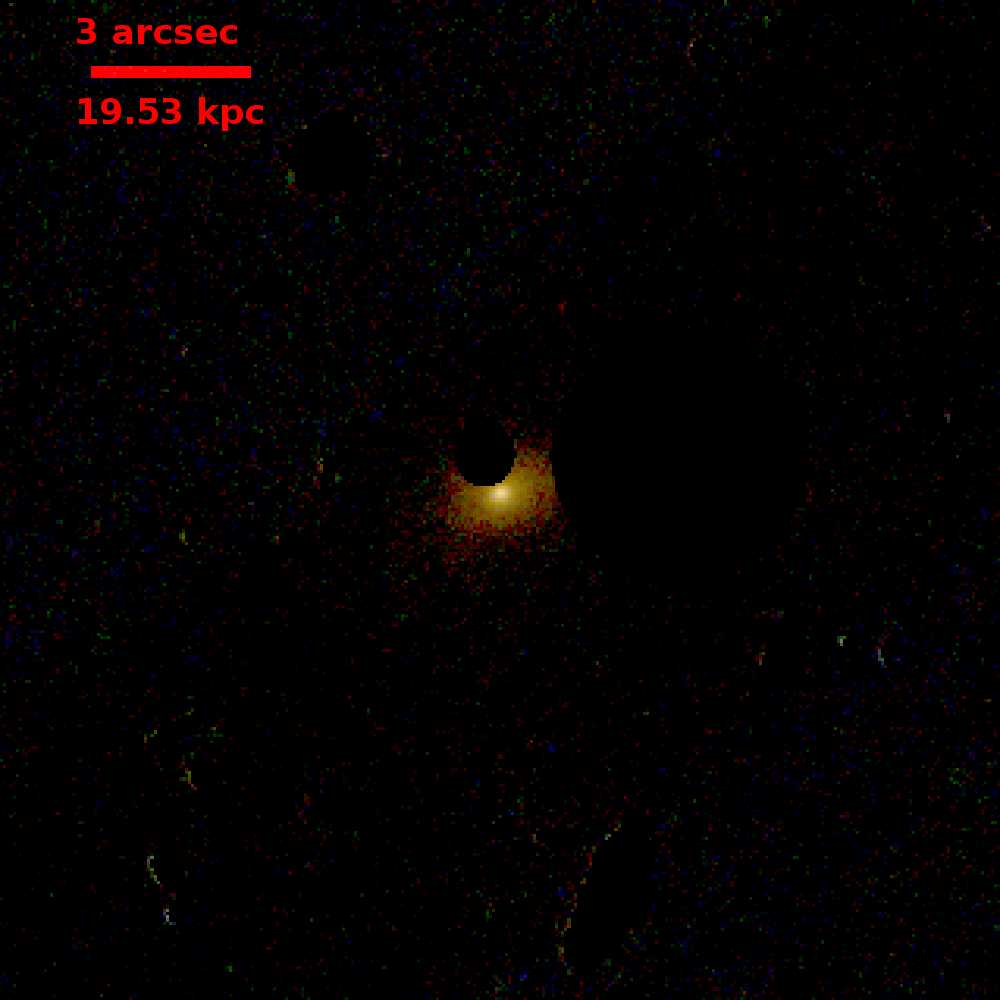}
\put(110,200){\color{yellow} \textbf{SHARDS10001344}}
\put(110,190){\color{yellow} \textbf{z=0.5673}}
\put(110,180){\color{yellow} \textbf{S0}}
\end{overpic}
\vspace{-1cm}
\end{minipage}%
\begin{minipage}{.5\textwidth}
\includegraphics[clip, trim=1cm 1cm 1.5cm 1.5cm, width=\textwidth]{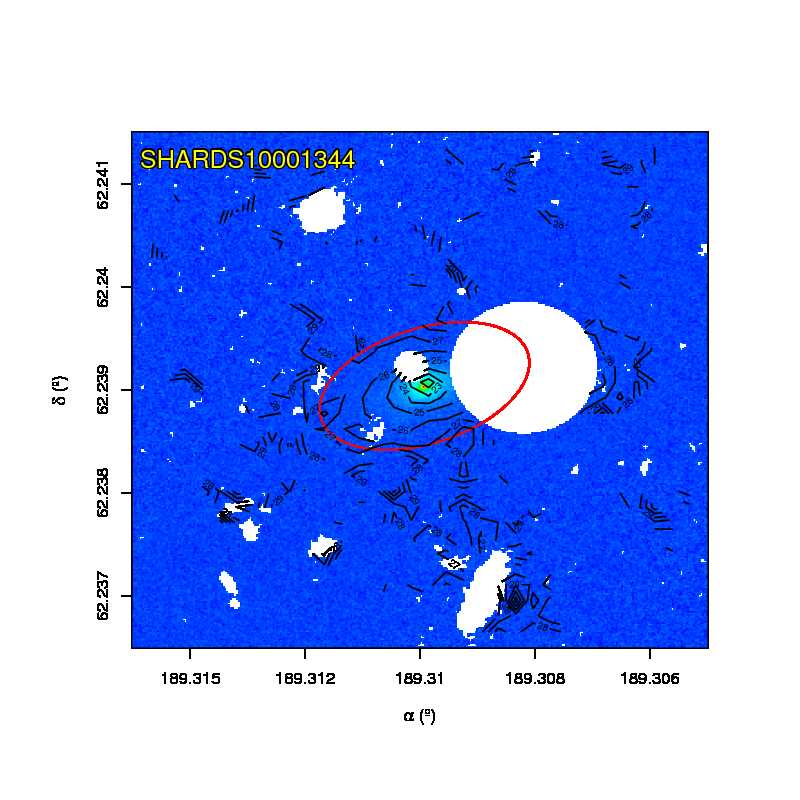}\vspace{-1cm}
\end{minipage}%

\begin{minipage}{.49\textwidth}
\includegraphics[clip, trim=0.1cm 0.1cm 0.1cm 0.1cm, width=\textwidth]{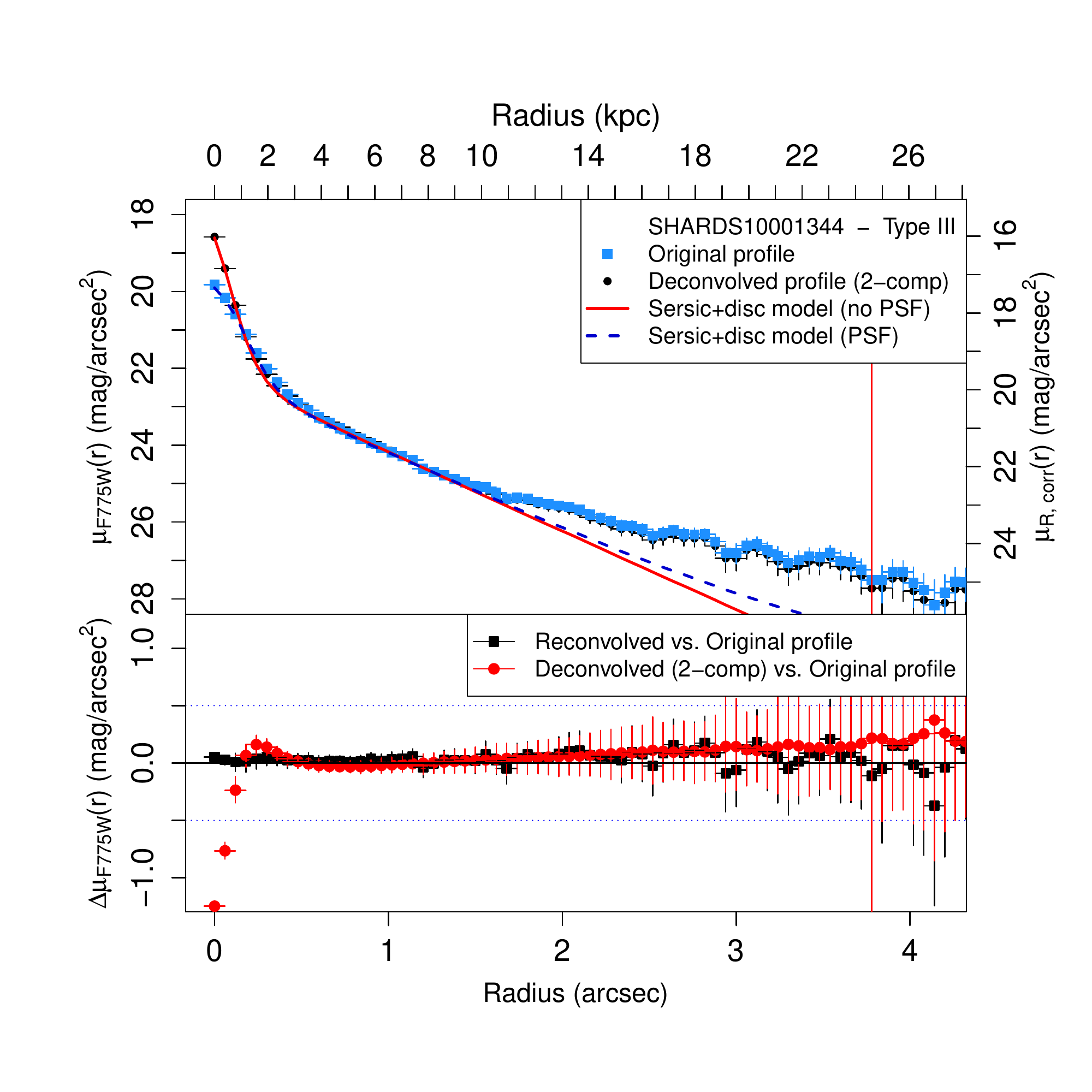}
\end{minipage}
\begin{minipage}{.49\textwidth}
\includegraphics[clip, trim=0.1cm 0.1cm 1cm 0.1cm, width=0.95\textwidth]{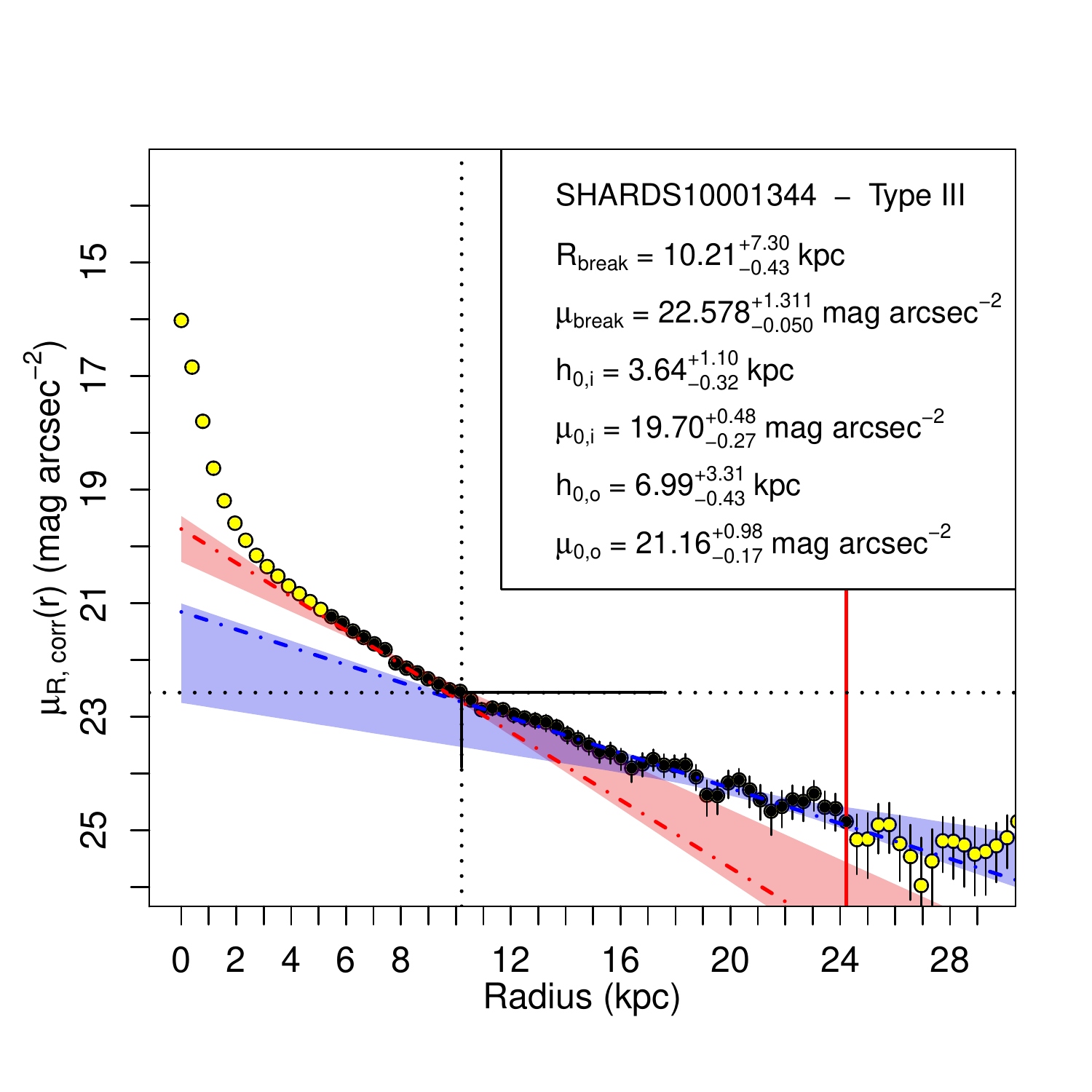}
\end{minipage}%

\vspace{-0.5cm}}
\caption[]{See caption of Fig.1. [\emph{Figure  available in the online edition}.]}         
\label{fig:img_final}
\end{figure}
\clearpage
\newpage

\textbf{SHARDS10001350:} S0 galaxy with a Type-II profile. The image shows a nearby galaxy to the main object, similar in size, that was carefully masked to prevent any flux contribution to the outer profile. Finally, the masked region was outside the limiting radius. We do not find any significant perturbations on the isophotes that could be due to the field objects, which are all beyond the fitting region. The PDDs for $h$ and $\mu_{0}$ show two clearly separated peaks corresponding to the two profiles. 

\begin{figure}[!h]
{\centering
\vspace{-0cm}

\begin{minipage}{.5\textwidth}
\hspace{1.2cm}
\begin{overpic}[width=0.8\textwidth]
{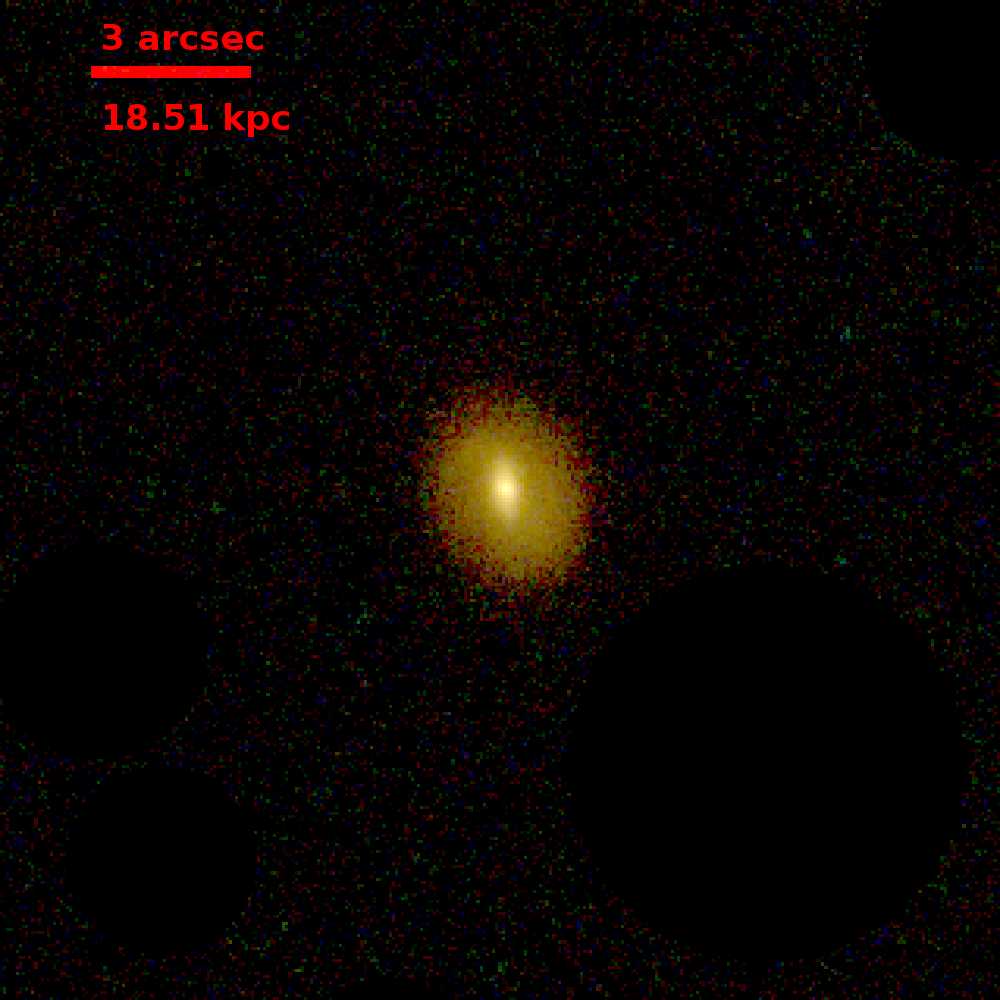}
\put(110,200){\color{yellow} \textbf{SHARDS10001350}}
\put(110,190){\color{yellow} \textbf{z=0.5106}}
\put(110,180){\color{yellow} \textbf{S0}}
\end{overpic}
\vspace{-1cm}
\end{minipage}%
\begin{minipage}{.5\textwidth}
\includegraphics[clip, trim=1cm 1cm 1.5cm 1.5cm, width=\textwidth]{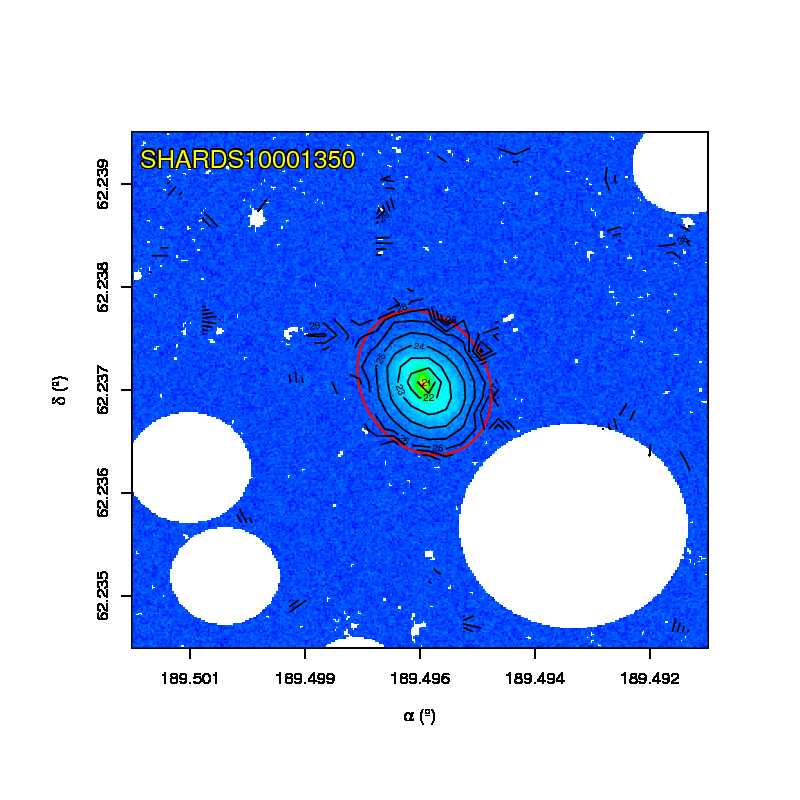}\vspace{-1cm}
\end{minipage}%

\begin{minipage}{.49\textwidth}
\includegraphics[clip, trim=0.1cm 0.1cm 0.1cm 0.1cm, width=\textwidth]{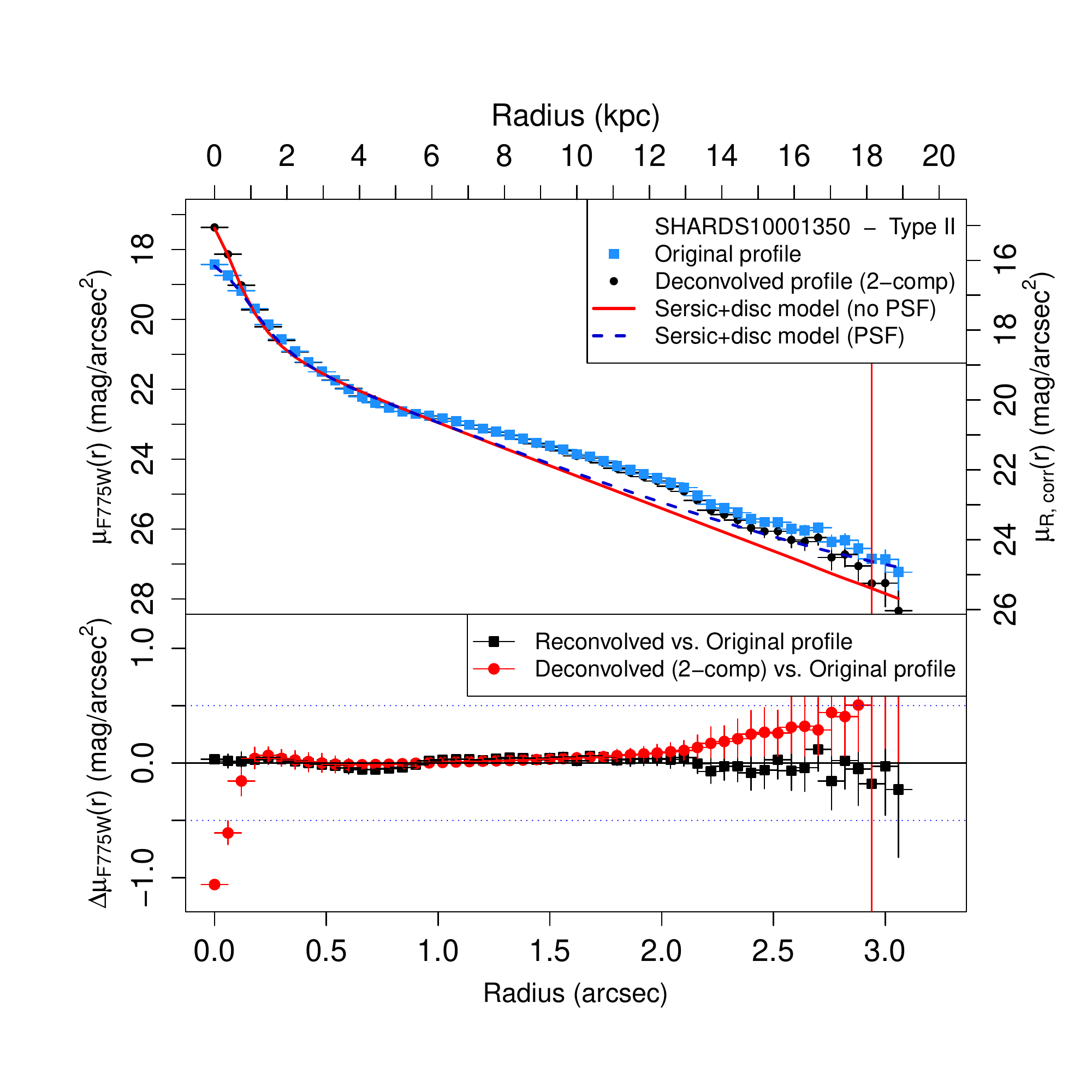}
\end{minipage}
\begin{minipage}{.49\textwidth}
\includegraphics[clip, trim=0.1cm 0.1cm 1cm 0.1cm, width=0.95\textwidth]{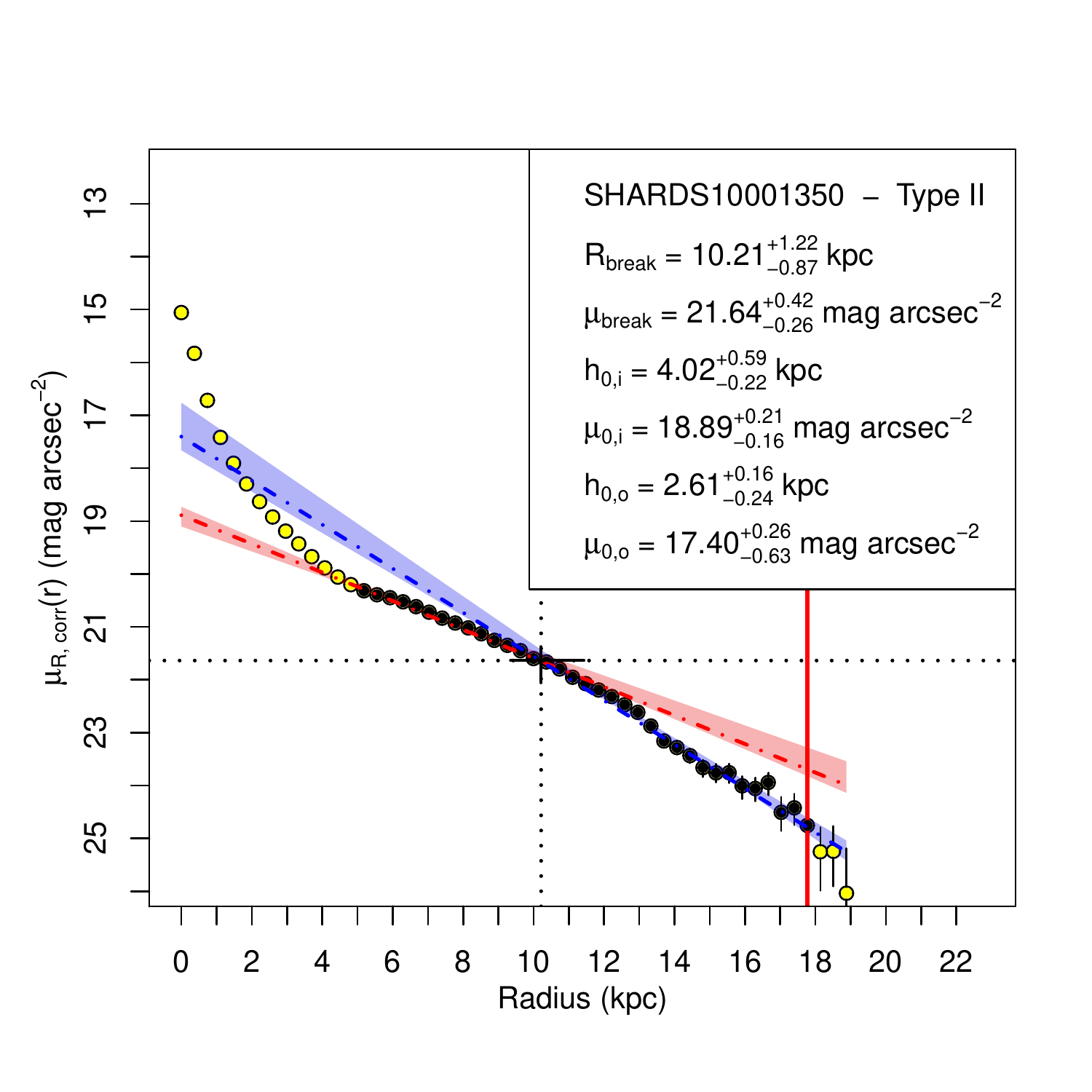}
\end{minipage}%

\vspace{-0.5cm}}
\caption[]{See caption of Fig.1. [\emph{Figure  available in the online edition}.]}         
\label{fig:img_final}
\end{figure}
\clearpage
\newpage


\textbf{SHARDS10001648:} Isolated Type-III S0 galaxy with medium inclination (see Table \ref{tab:fits_psforr}). The PDDs of $h$ and $\mu_{0}$ reveal that the excess of light found at \rbreak $=8-14$ kpc is significant and compatible with an outer exponential profile, different from the inner disc profile.

\begin{figure}[!h]
{\centering
\vspace{-0cm}

\begin{minipage}{.5\textwidth}
\hspace{1.2cm}
\begin{overpic}[width=0.8\textwidth]
{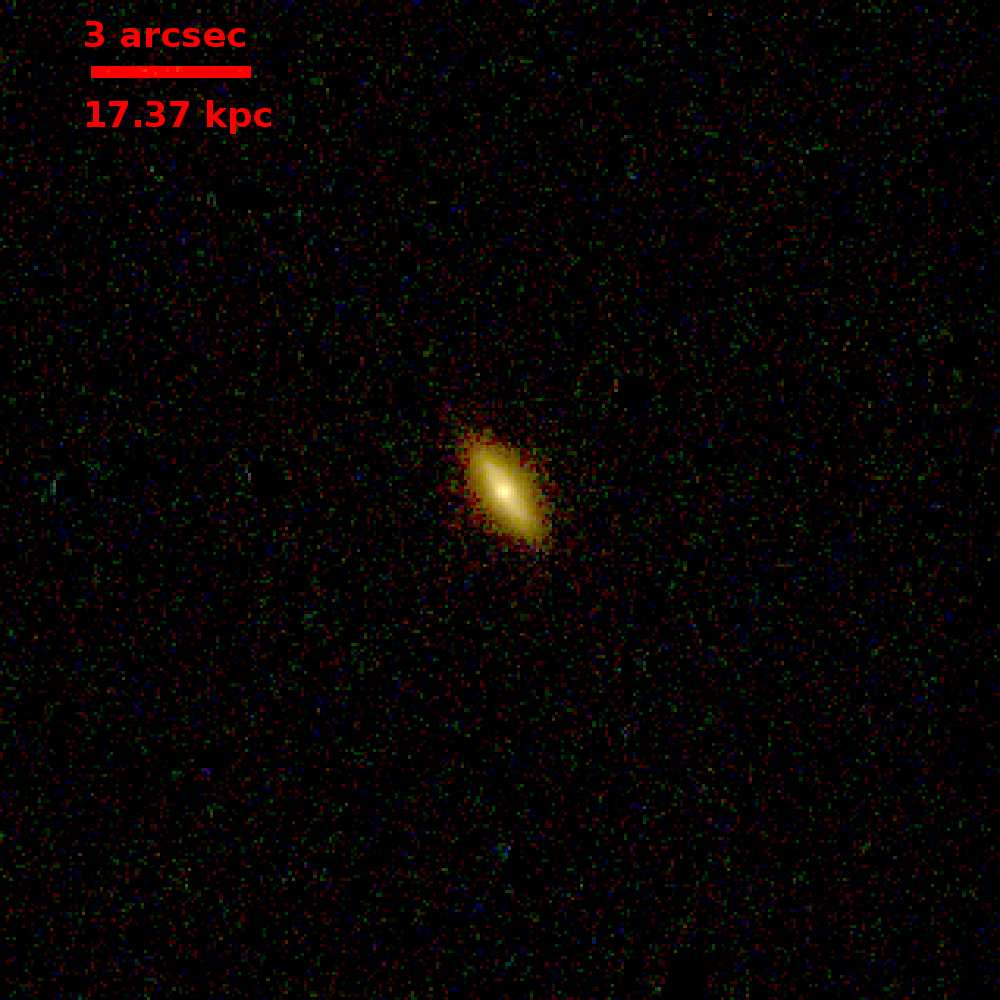}
\put(110,200){\color{yellow} \textbf{SHARDS10001648}}
\put(110,190){\color{yellow} \textbf{z=0.4540}}
\put(110,180){\color{yellow} \textbf{S0}}
\end{overpic}
\vspace{-1cm}
\end{minipage}%
\begin{minipage}{.5\textwidth}
\includegraphics[clip, trim=1cm 1cm 1.5cm 1.5cm, width=\textwidth]{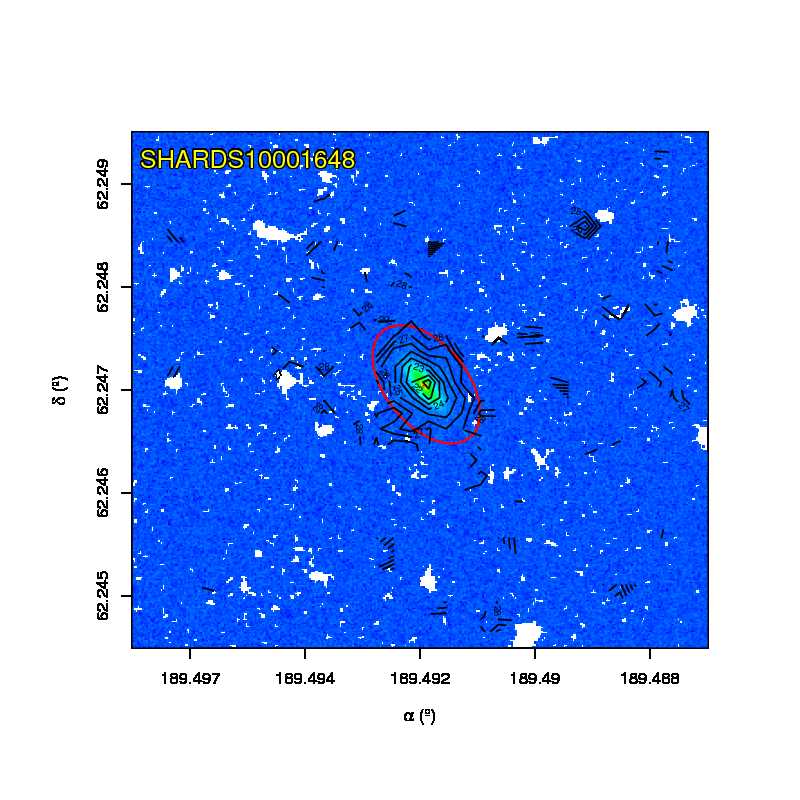}\vspace{-1cm}
\end{minipage}%

\begin{minipage}{.49\textwidth}
\includegraphics[clip, trim=0.1cm 0.1cm 0.1cm 0.1cm, width=\textwidth]{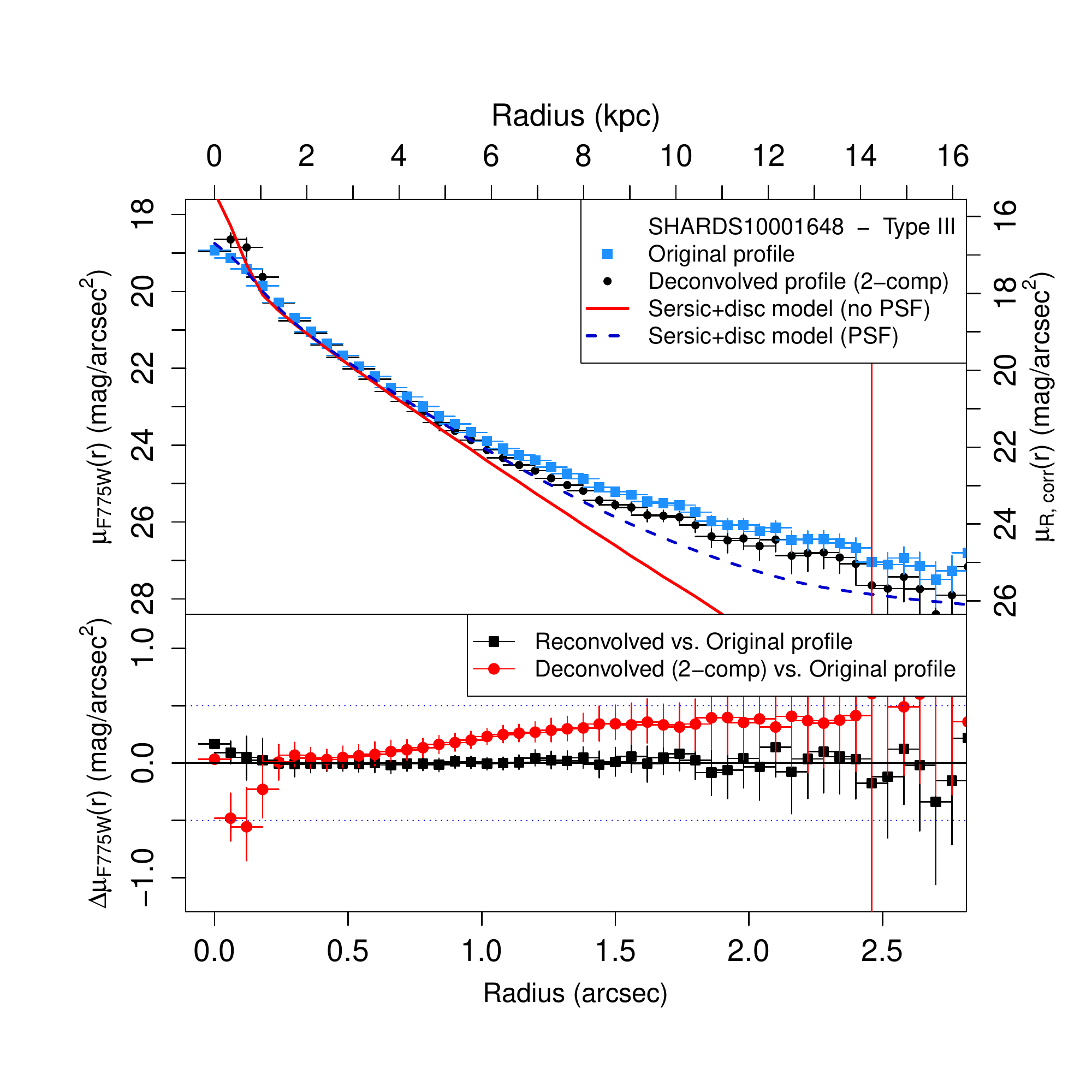}
\end{minipage}
\begin{minipage}{.49\textwidth}
\includegraphics[clip, trim=0.1cm 0.1cm 1cm 0.1cm, width=0.95\textwidth]{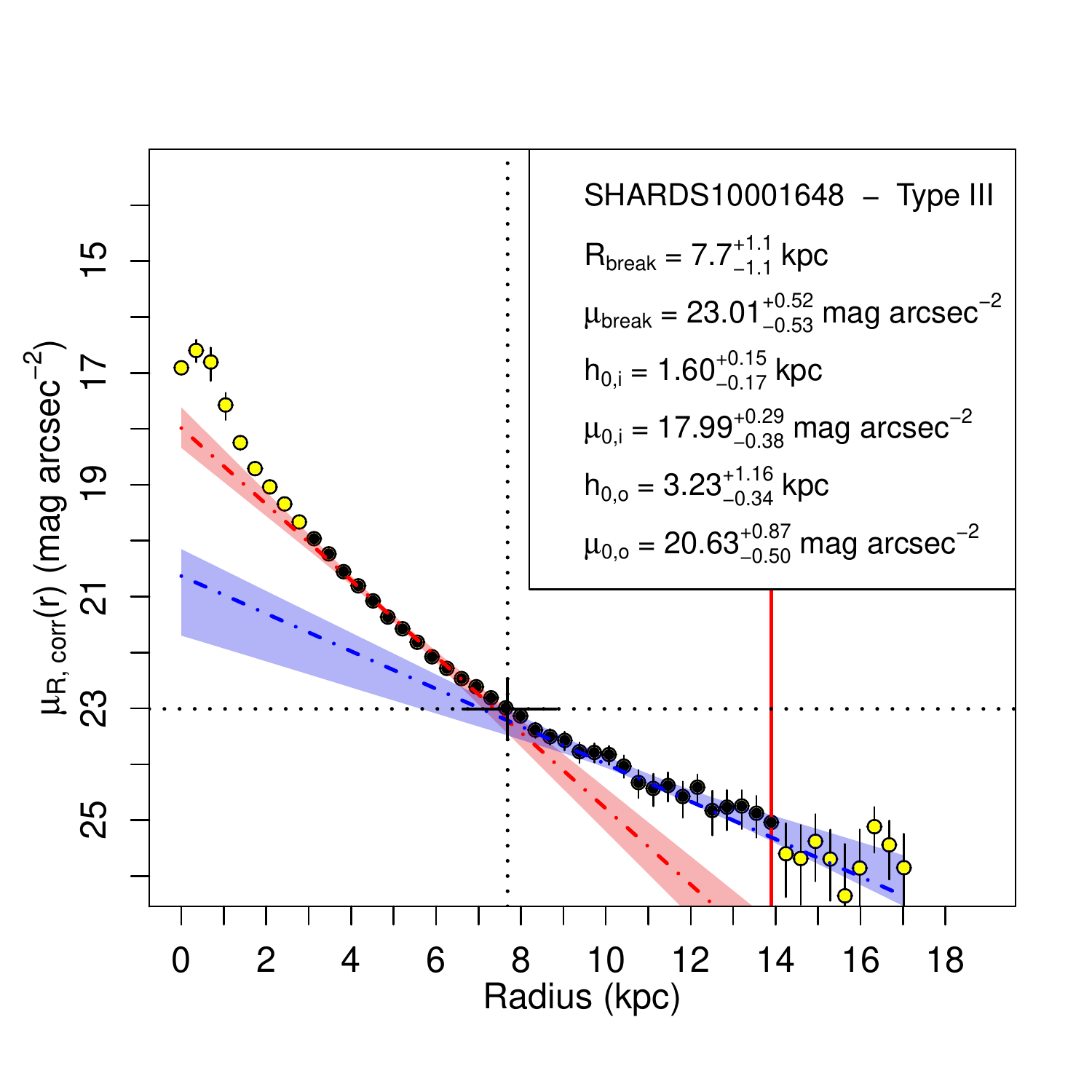}
\end{minipage}%

\vspace{-0.5cm}}
\caption[]{See caption of Fig.1. [\emph{Figure  available in the online edition}.]}         
\label{fig:img_final}
\end{figure}
\clearpage
\newpage


\textbf{SHARDS10001727:} S0 galaxy with Type-I profile. It was analysed by {\tt{ISOFIT}} instead of {\tt{ellipse}} due to its completely edge-on orientation (see Table \ref{tab:fits_psforr}). The PDDs of $h$ and $\mu_{0}$ do not show any significant break, and the general profile can be successfully modelled as a single exponential function. 

\begin{figure}[!h]
{\centering
\vspace{-0cm}

\begin{minipage}{.5\textwidth}
\hspace{1.2cm}
\begin{overpic}[width=0.8\textwidth]
{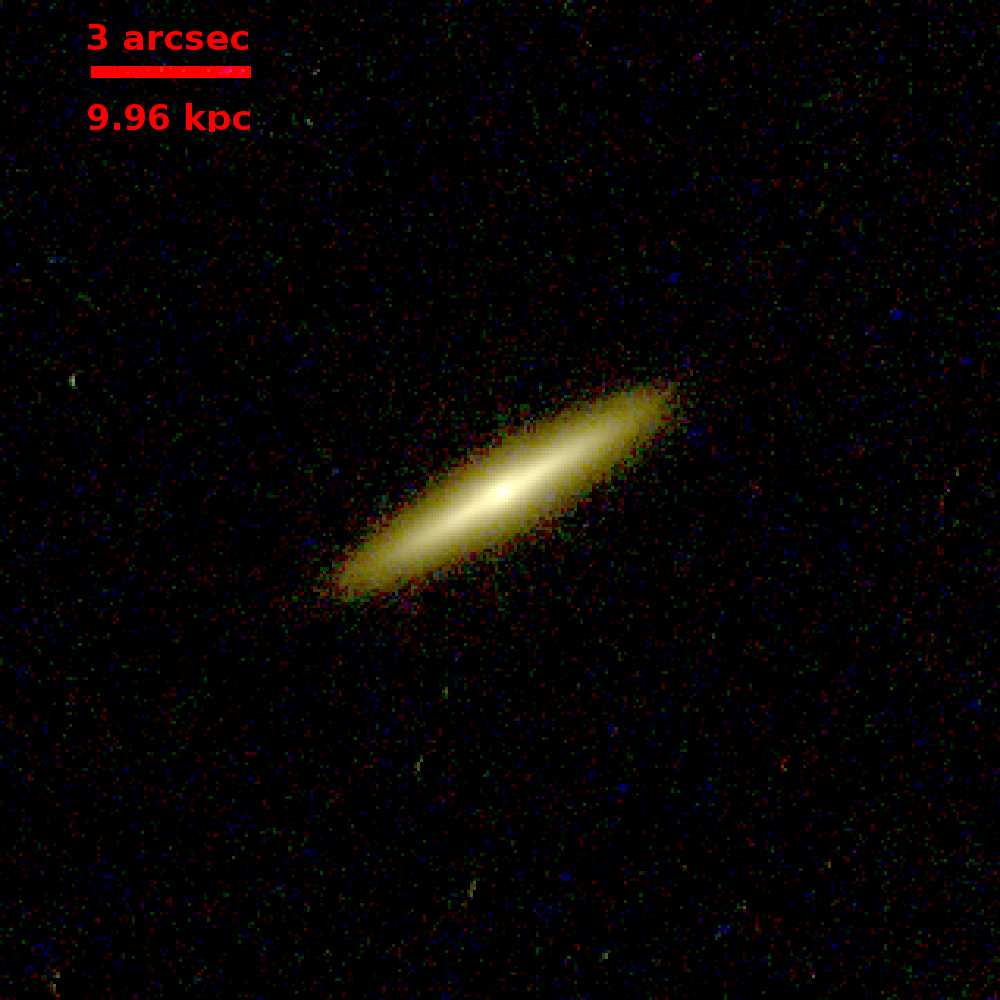}
\put(110,200){\color{yellow} \textbf{SHARDS10001727}}
\put(110,190){\color{yellow} \textbf{z=0.2013}}
\put(110,180){\color{yellow} \textbf{S0}}
\end{overpic}
\vspace{-1cm}
\end{minipage}%
\begin{minipage}{.5\textwidth}
\includegraphics[clip, trim=1cm 1cm 1.5cm 1.5cm, width=\textwidth]{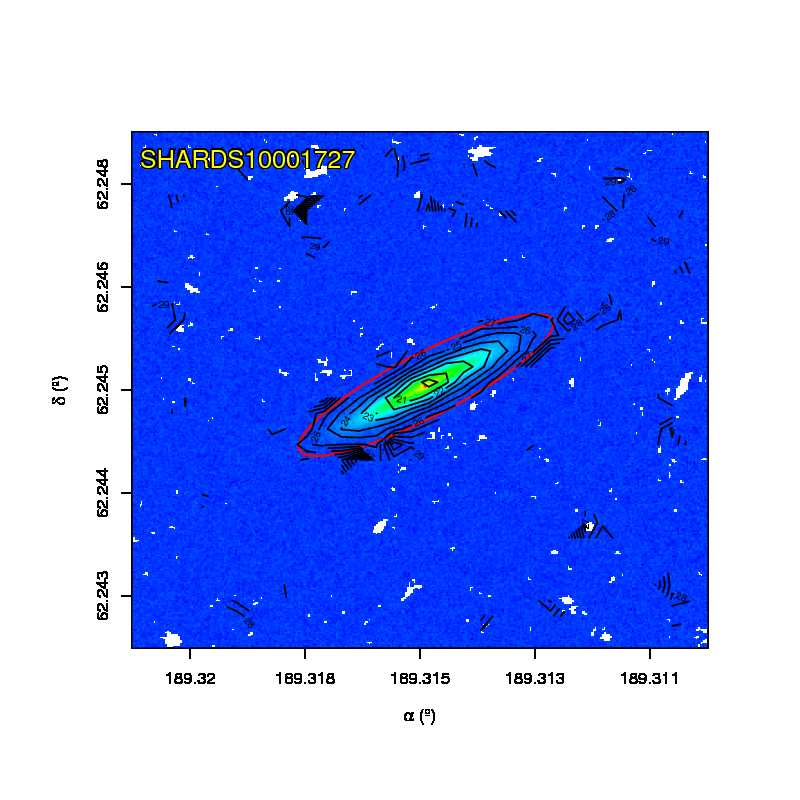}\vspace{-1cm}
\end{minipage}%

\begin{minipage}{.49\textwidth}
\includegraphics[clip, trim=0.1cm 0.1cm 0.1cm 0.1cm, width=\textwidth]{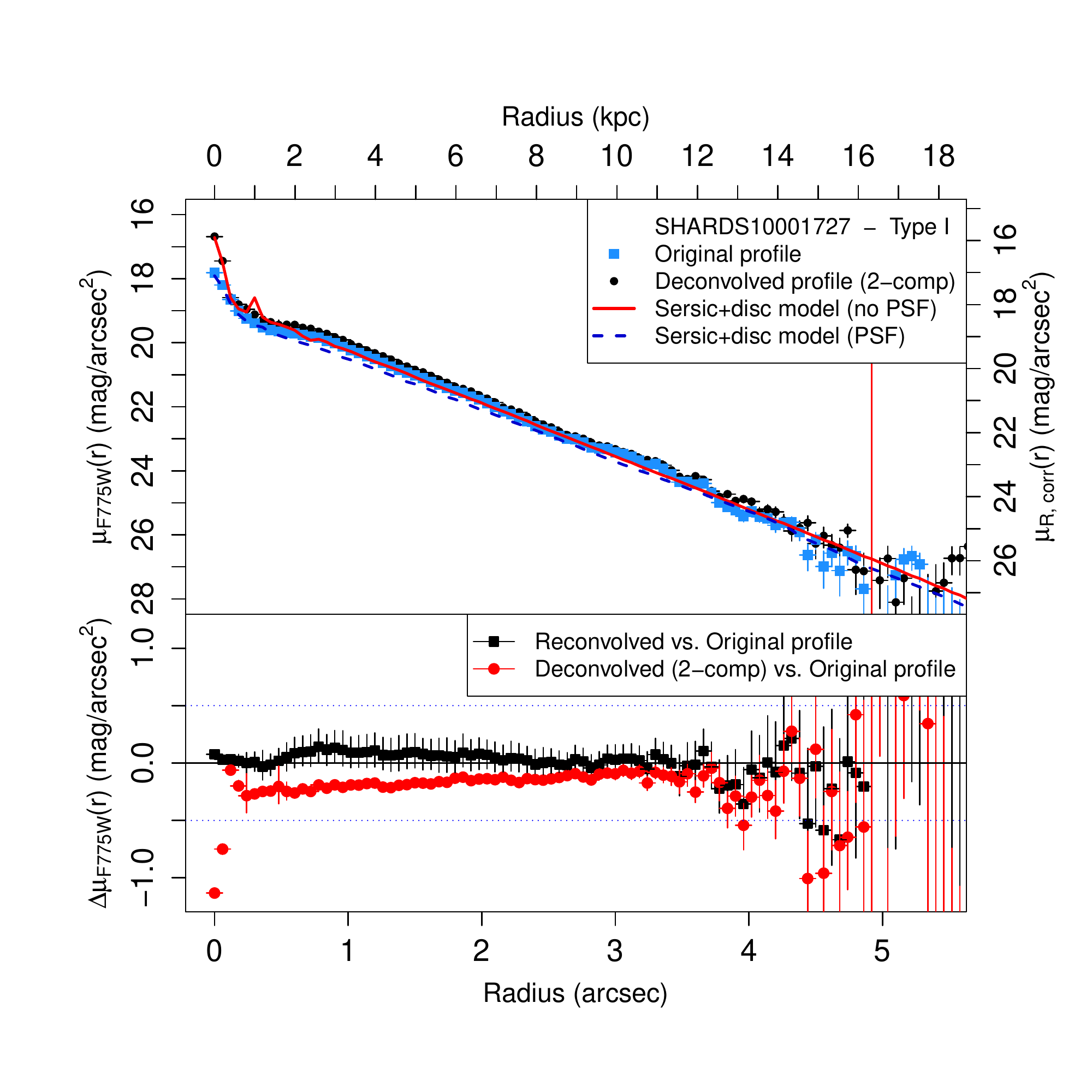}
\end{minipage}
\begin{minipage}{.49\textwidth}
\includegraphics[clip, trim=0.1cm 0.1cm 1cm 0.1cm, width=0.95\textwidth]{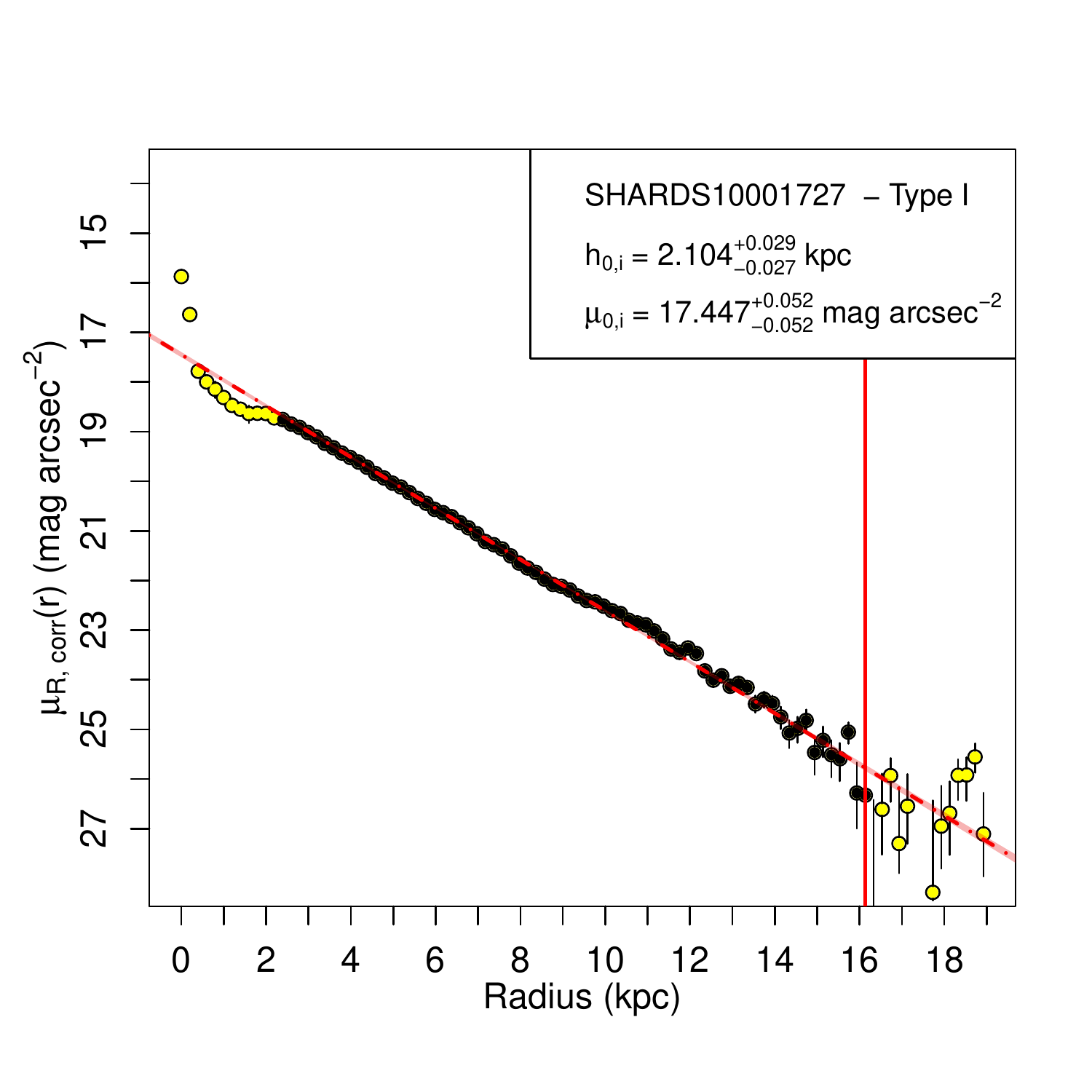}
\end{minipage}%

\vspace{-0.5cm}}
\caption[]{See caption of Fig.1. [\emph{Figure  available in the online edition}.]}         
\label{fig:img_final}
\end{figure}
\clearpage
\newpage


\textbf{SHARDS10001847:} S0 galaxy with Type II profile (see Table \ref{tab:fits_psforr}). The galaxy appears at a medium to high inclination. The limits for the profiles were chosen by hand, based several fittings performed varying the initial configurations, with the aim to reduce multiple peaks that appeared on the PDDs of $h$ and $\mu_{0}$. The profile show a Type-II break which is statistically significant. The outskirts present an apparent excess of light, but it is not statistically significant after PSF subtraction.

\begin{figure}[!h]
{\centering
\vspace{-0cm}

\begin{minipage}{.5\textwidth}
\hspace{1.2cm}
\begin{overpic}[width=0.8\textwidth]
{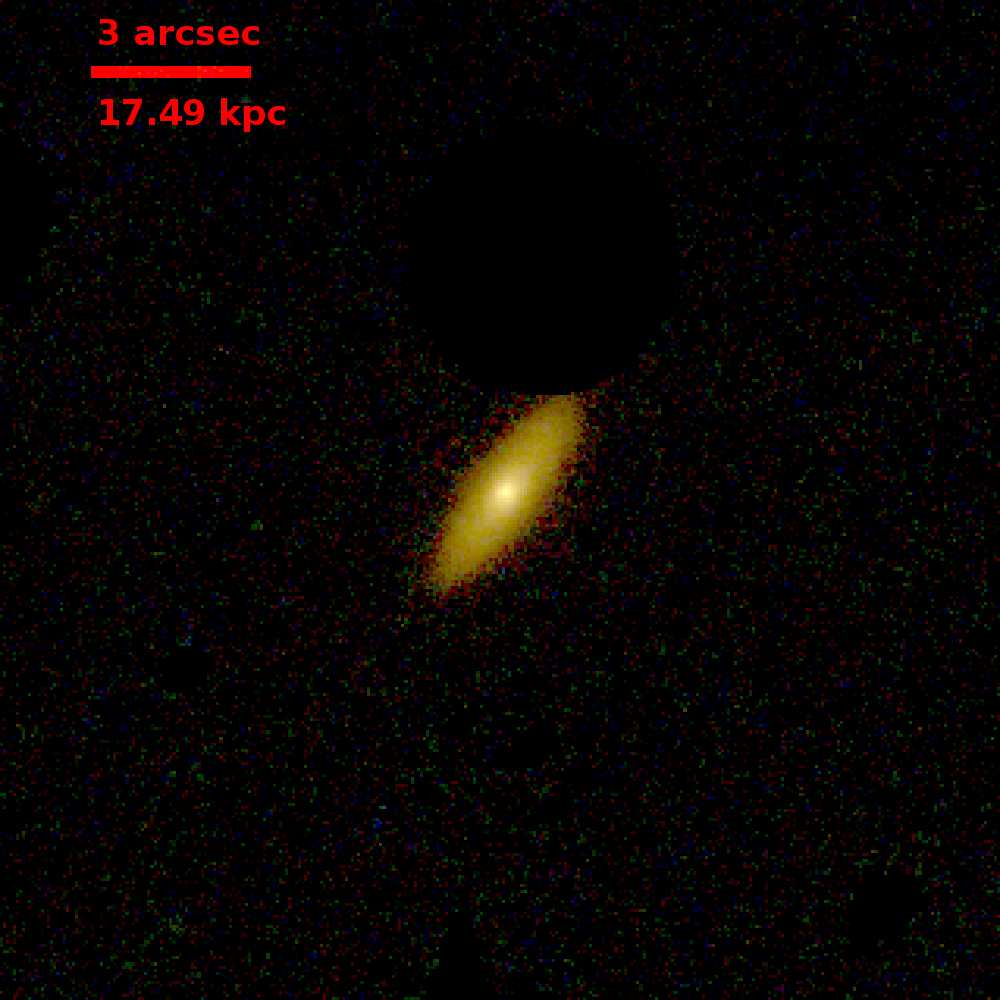}
\put(110,200){\color{yellow} \textbf{SHARDS10001847}}
\put(110,190){\color{yellow} \textbf{z=0.4598}}
\put(110,180){\color{yellow} \textbf{S0}}
\end{overpic}
\vspace{-1cm}
\end{minipage}%
\begin{minipage}{.5\textwidth}
\includegraphics[clip, trim=1cm 1cm 1.5cm 1.5cm, width=\textwidth]{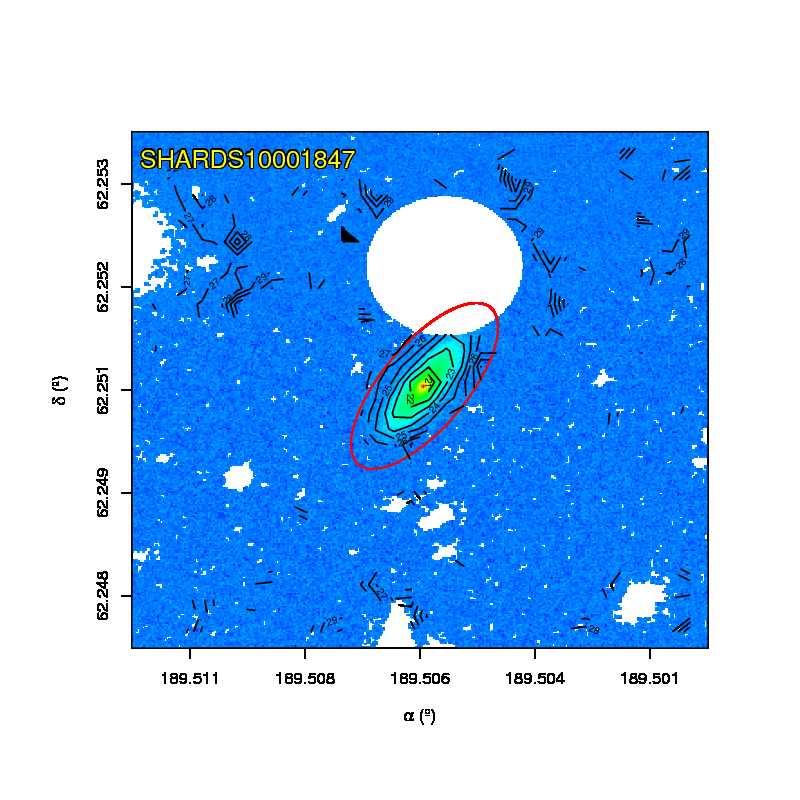}\vspace{-1cm}
\end{minipage}%

\begin{minipage}{.49\textwidth}
\includegraphics[clip, trim=0.1cm 0.1cm 0.1cm 0.1cm, width=\textwidth]{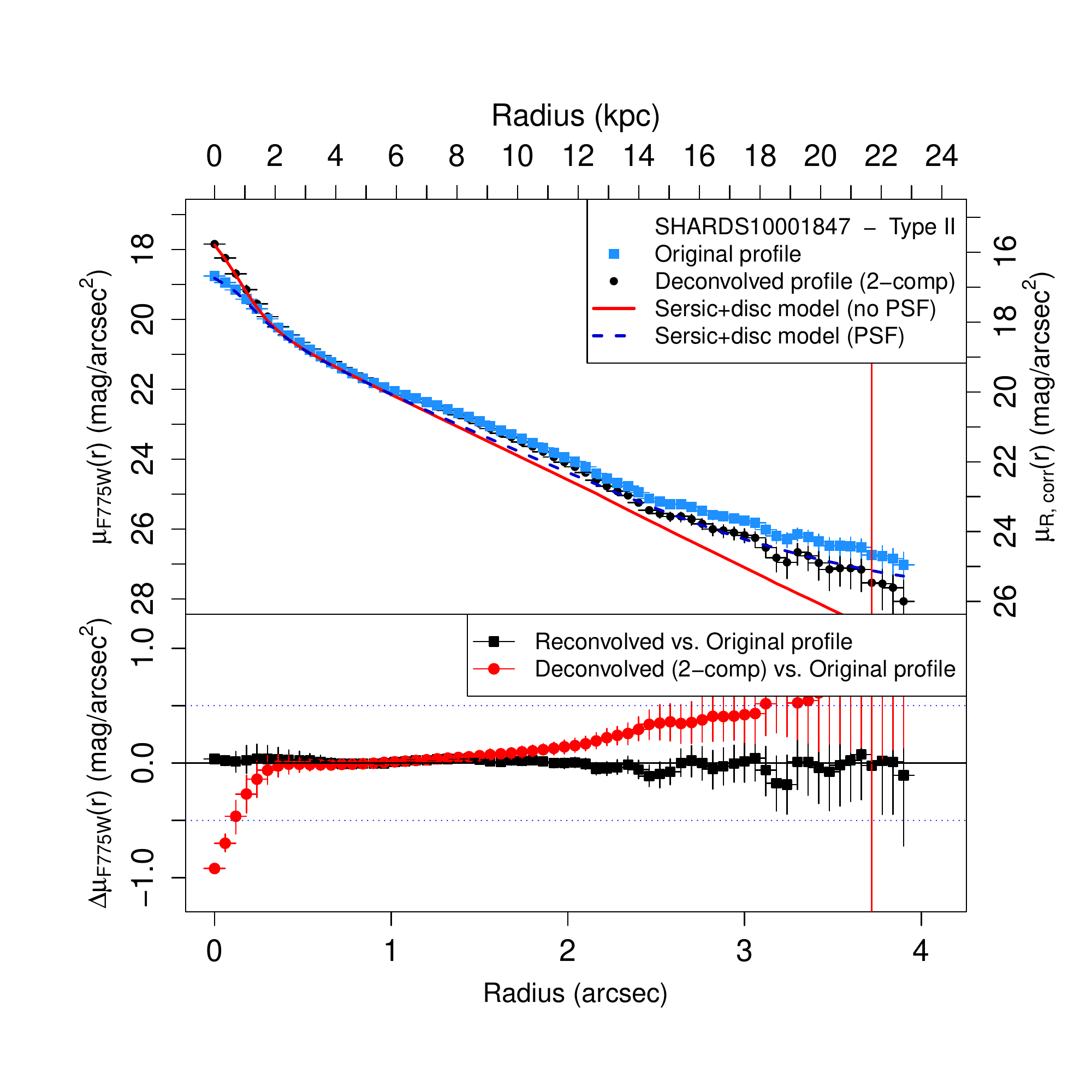}
\end{minipage}
\begin{minipage}{.49\textwidth}
\includegraphics[clip, trim=0.1cm 0.1cm 1cm 0.1cm, width=0.95\textwidth]{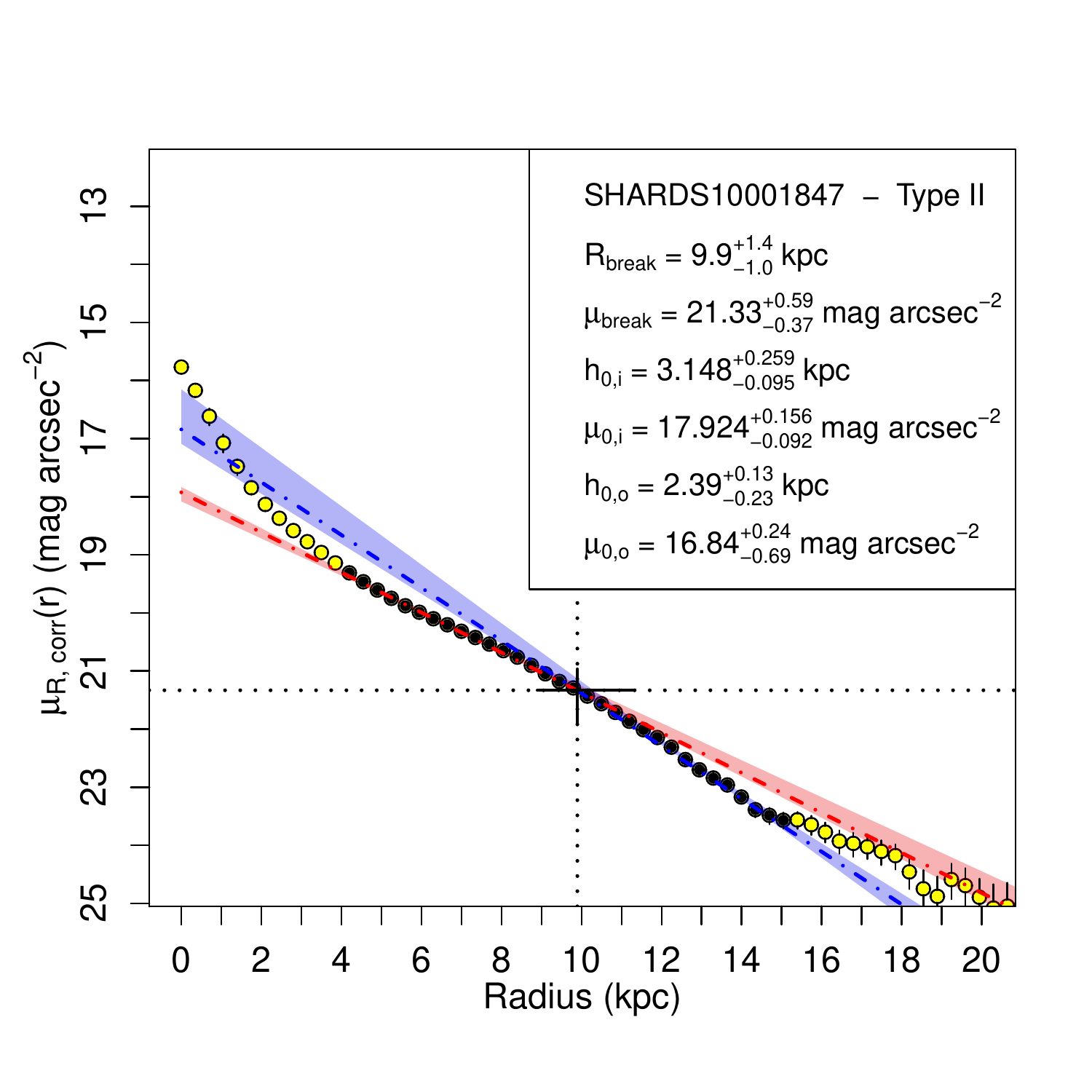}
\end{minipage}%

\vspace{-0.5cm}}
\caption[]{See caption of Fig.1. [\emph{Figure  available in the online edition}.]}         
\label{fig:img_final}
\end{figure}
\clearpage
\newpage


\textbf{SHARDS10002351:} Small S0 galaxy with pure exponential profile (Type I). It was analysed by {\tt{ISOFIT}} instead of {\tt{ellipse}} due to its edge on-orientation. The object is almost isolated. The PDDs of $h$ and $\mu_{0}$ show no statistically significant differences between any parts of the disc.   

\begin{figure}[!h]
{\centering
\vspace{-0cm}

\begin{minipage}{.5\textwidth}
\hspace{1.2cm}
\begin{overpic}[width=0.8\textwidth]
{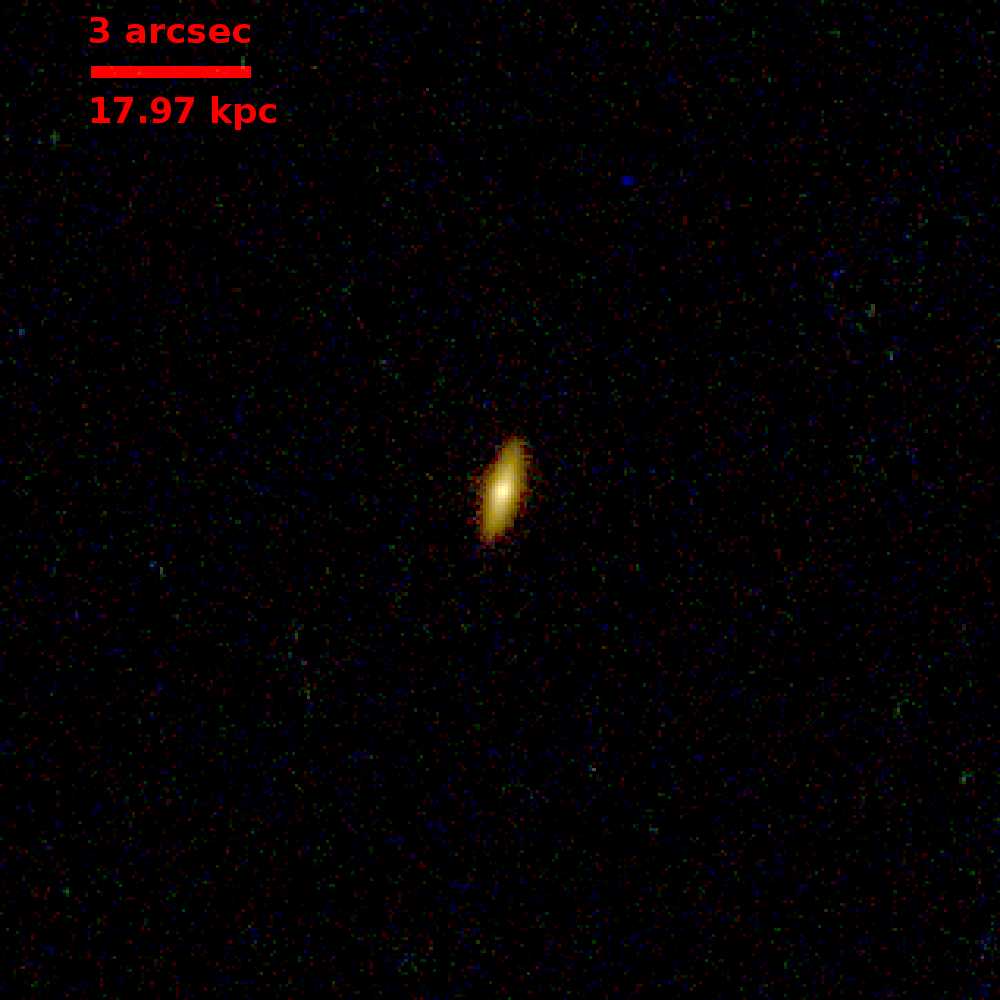}
\put(110,200){\color{yellow} \textbf{SHARDS10002351}}
\put(110,190){\color{yellow} \textbf{z=0.4834}}
\put(110,180){\color{yellow} \textbf{S0}}
\end{overpic}
\vspace{-1cm}
\end{minipage}%
\begin{minipage}{.5\textwidth}
\includegraphics[clip, trim=1cm 1cm 1.5cm 1.5cm, width=\textwidth]{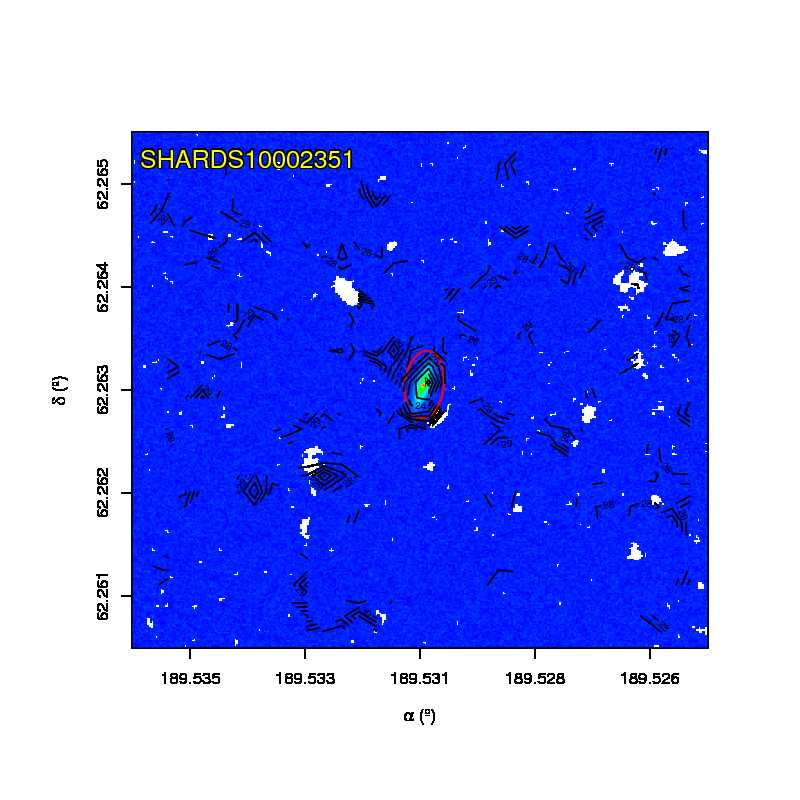}\vspace{-1cm}
\end{minipage}%

\begin{minipage}{.49\textwidth}
\includegraphics[clip, trim=0.1cm 0.1cm 0.1cm 0.1cm, width=\textwidth]{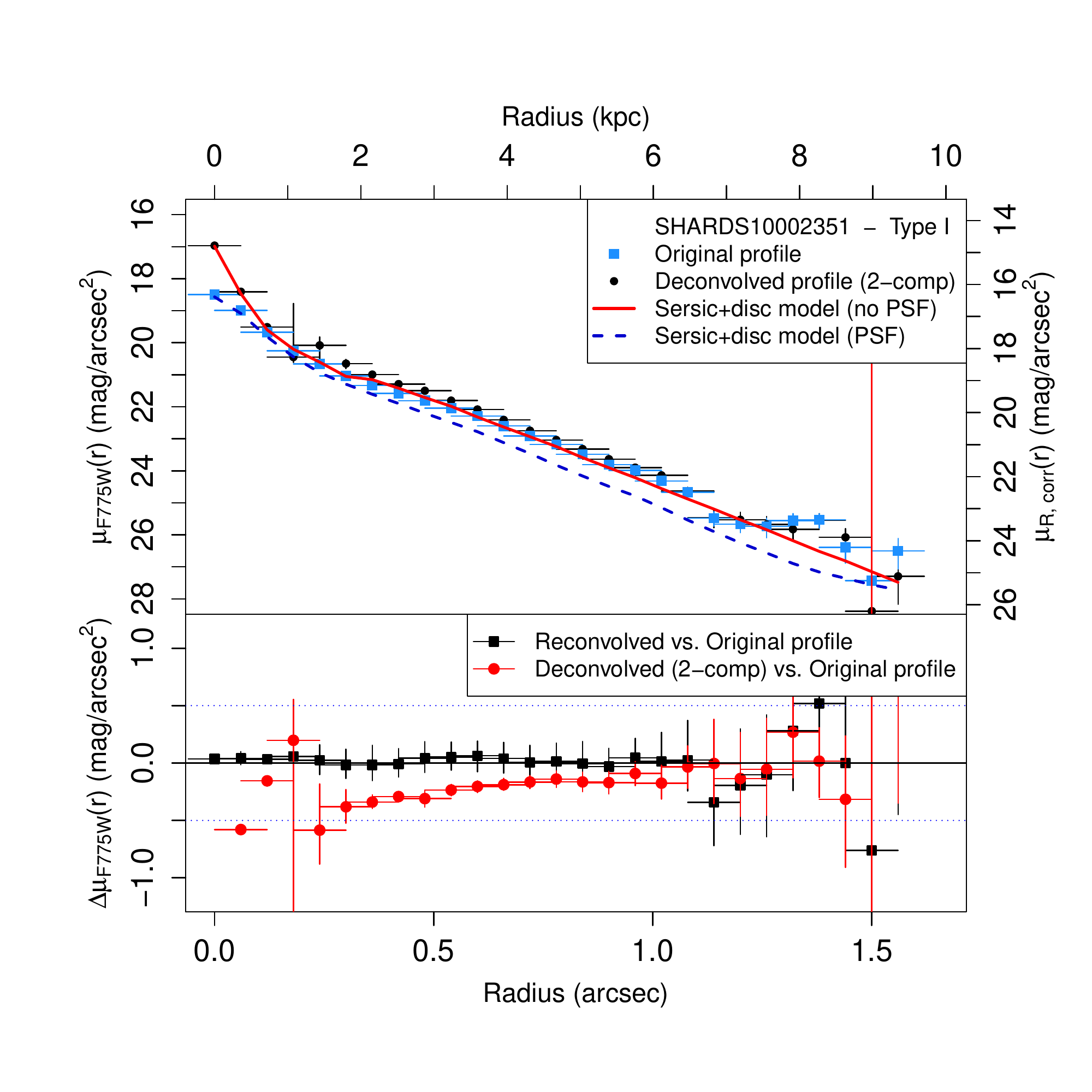}
\end{minipage}
\begin{minipage}{.49\textwidth}
\includegraphics[clip, trim=0.1cm 0.1cm 1cm 0.1cm, width=0.95\textwidth]{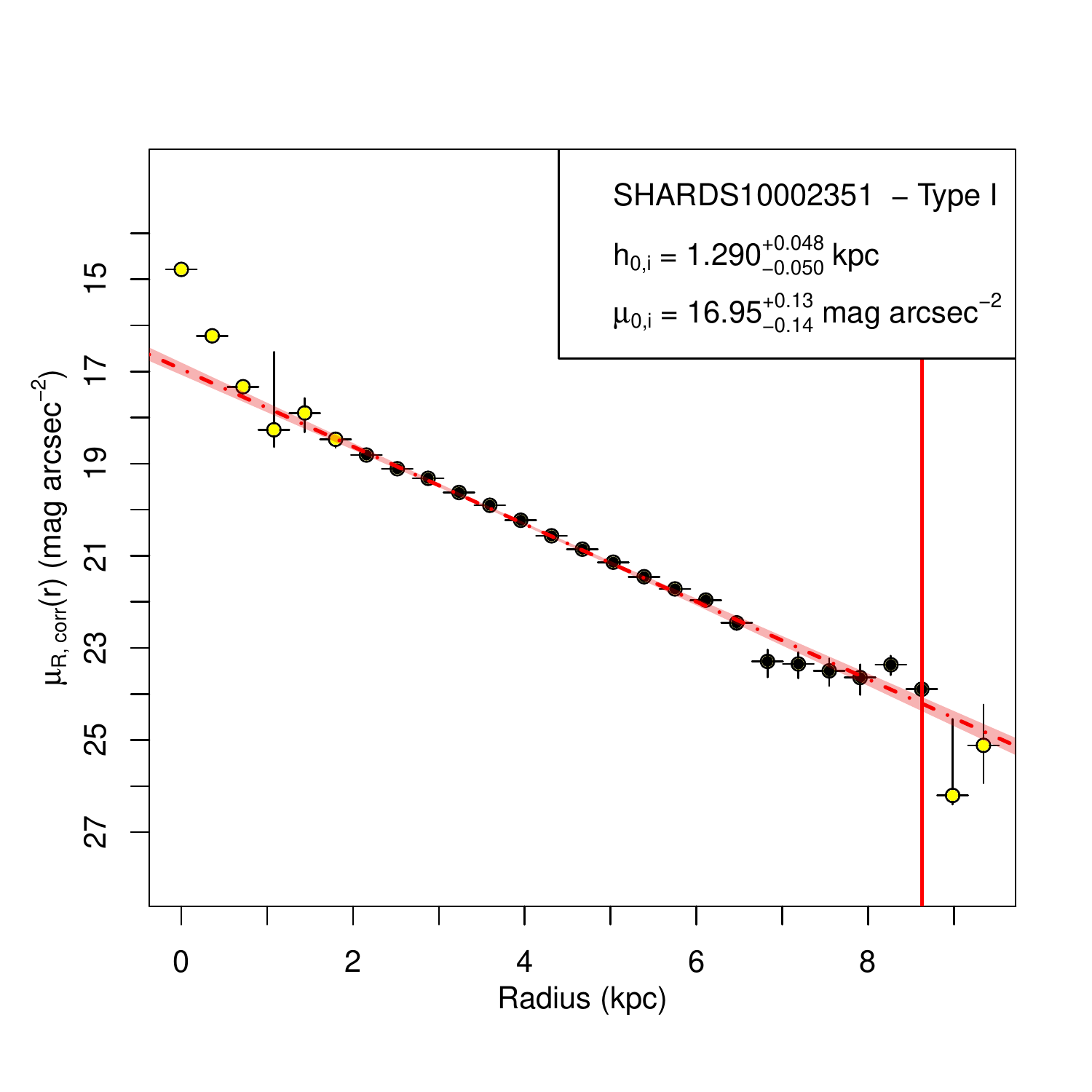}
\end{minipage}%

\vspace{-0.5cm}}
\caption[]{See caption of Fig.1. [\emph{Figure  available in the online edition}.]}         
\label{fig:img_final}
\end{figure}
\clearpage
\newpage


\textbf{SHARDS10002730:} Small S0 galaxy with Type-III profile. It has a medium inclination (see Table \ref{tab:fits_psforr}). It required extensive masking, but no masked pixel finally was inside the fitting region. The automated break analysis reveals that the excess of light shown at the outermost part of the galaxy is statistically significant as a Type-III break despite the PSF subtraction, although the the probability of being a Type I is $p\sim0.008$.

\begin{figure}[!h]
{\centering
\vspace{-0cm}

\begin{minipage}{.5\textwidth}
\hspace{1.2cm}
\begin{overpic}[width=0.8\textwidth]
{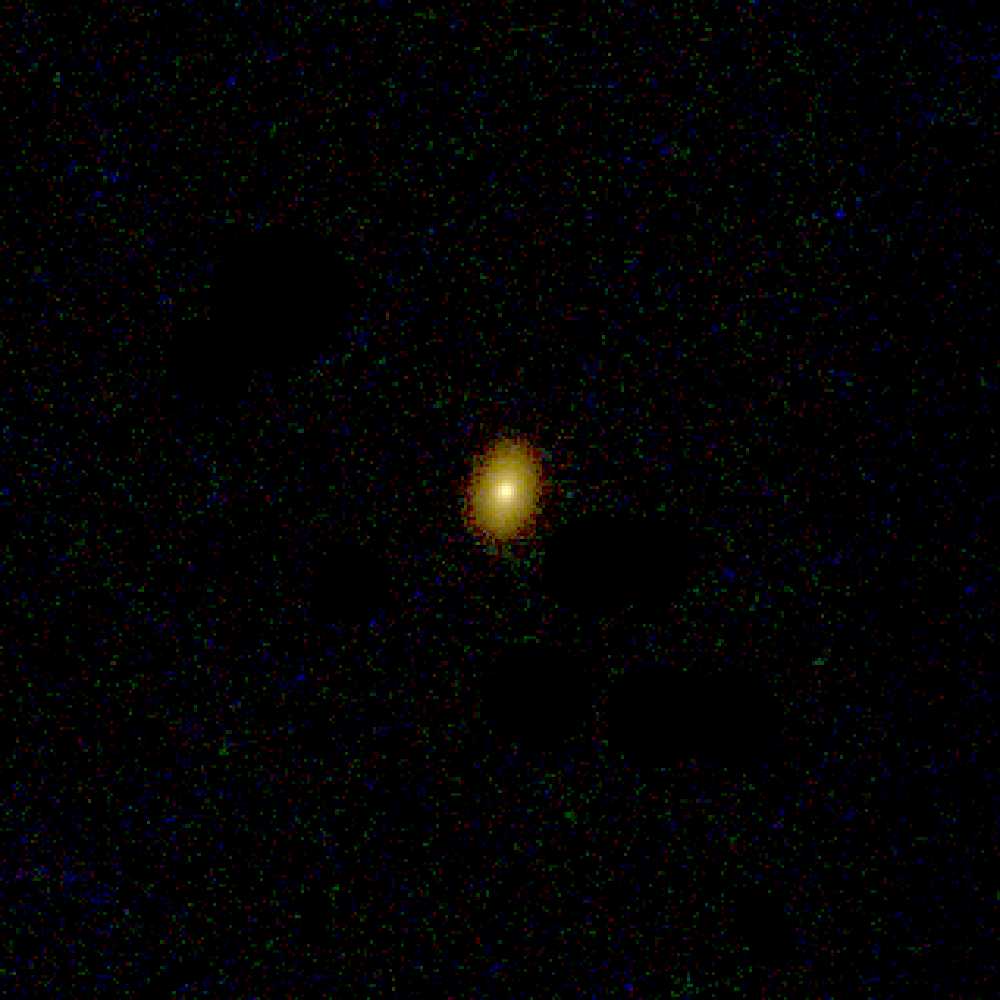}
\put(110,200){\color{yellow} \textbf{SHARDS10002730}}
\put(110,190){\color{yellow} \textbf{z=0.4472}}
\put(110,180){\color{yellow} \textbf{S0}}
\end{overpic}
\vspace{-1cm}
\end{minipage}%
\begin{minipage}{.5\textwidth}
\includegraphics[clip, trim=1cm 1cm 1.5cm 1.5cm, width=\textwidth]{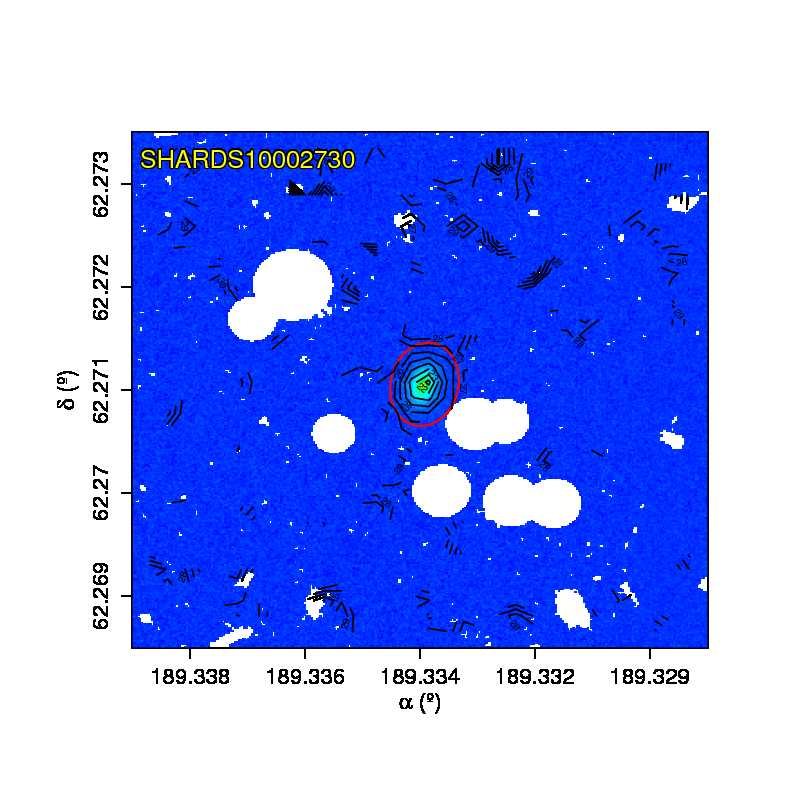}\vspace{-1cm}
\end{minipage}%

\begin{minipage}{.49\textwidth}
\includegraphics[clip, trim=0.1cm 0.1cm 0.1cm 0.1cm, width=\textwidth]{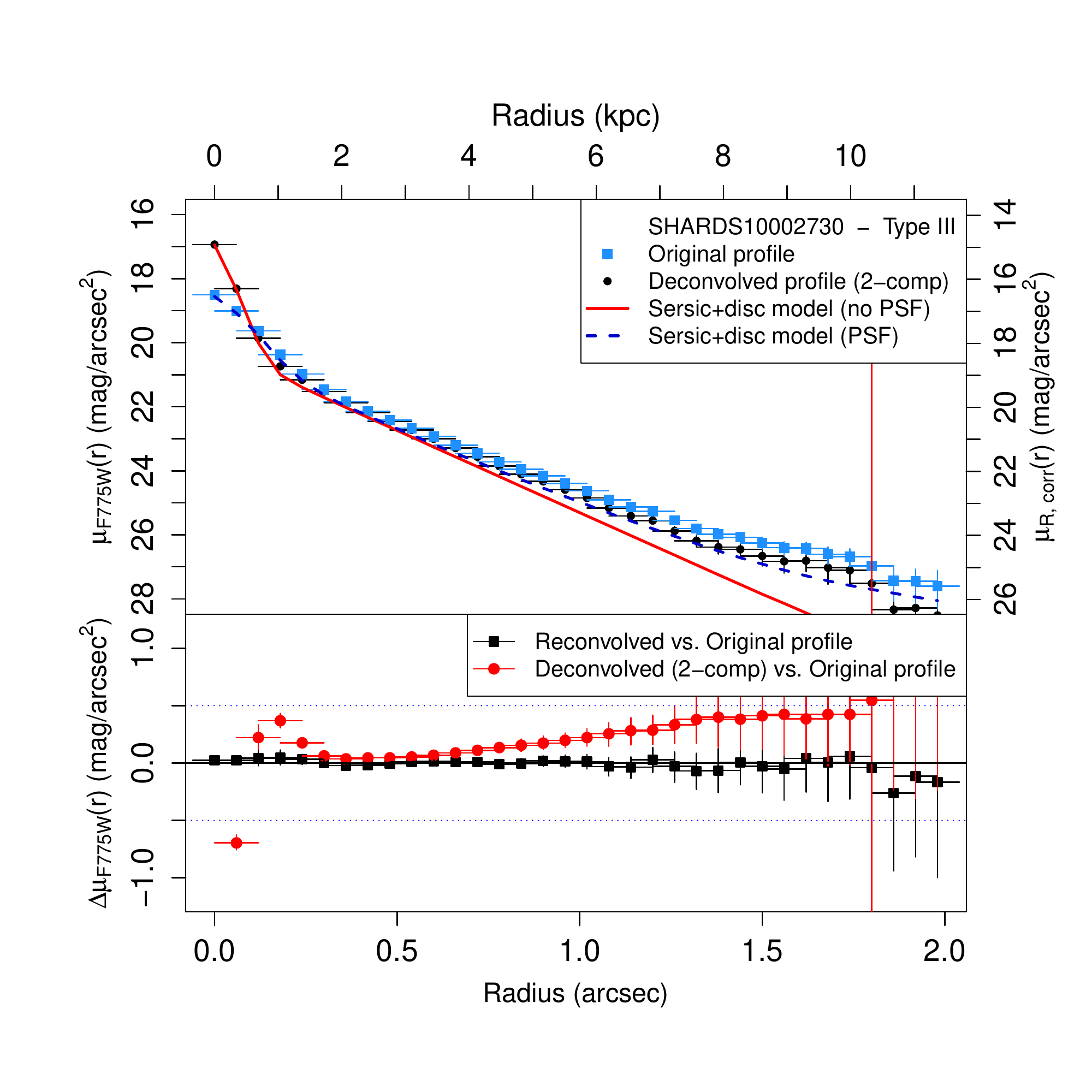}
\end{minipage}
\begin{minipage}{.49\textwidth}
\includegraphics[clip, trim=0.1cm 0.1cm 1cm 0.1cm, width=0.95\textwidth]{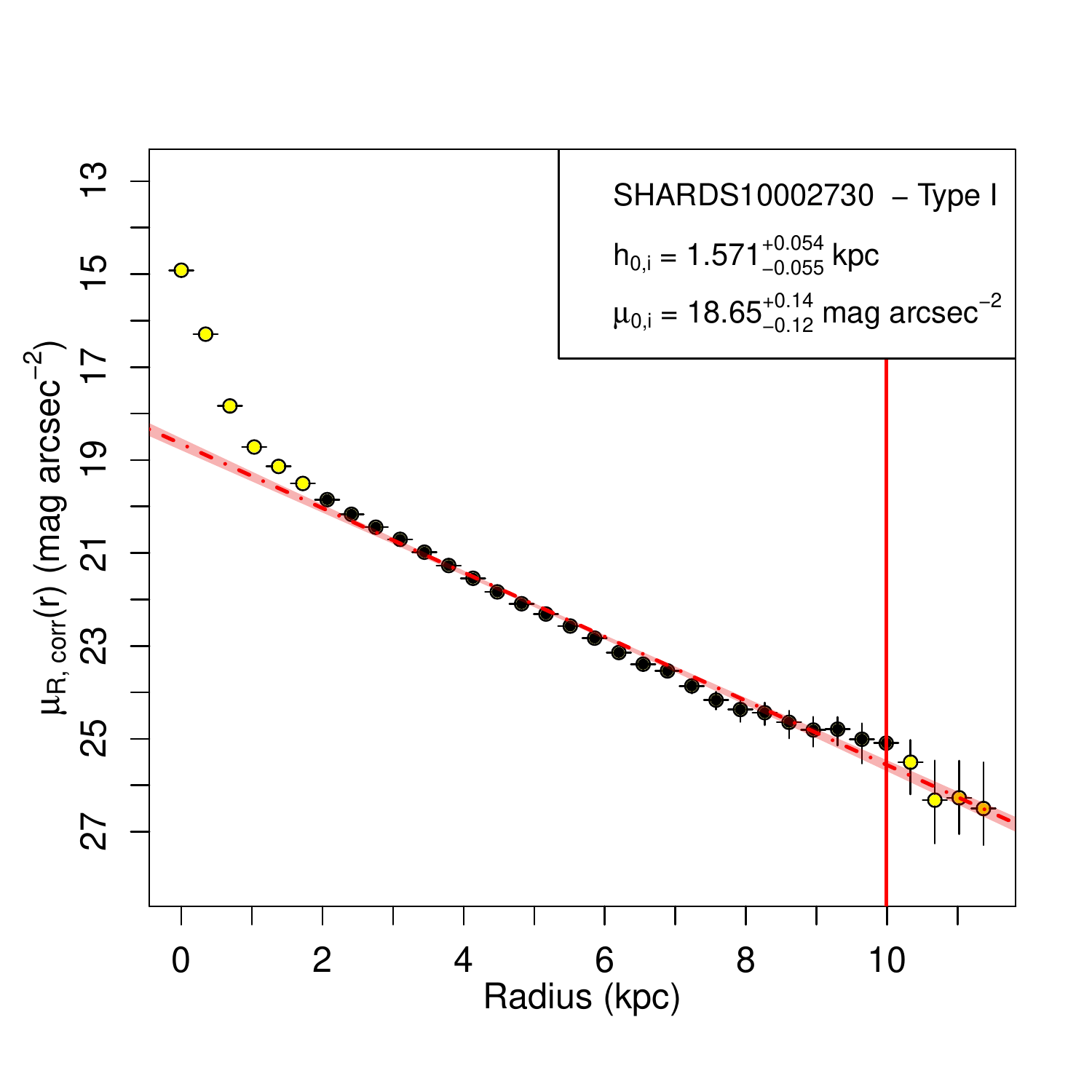}
\end{minipage}%

\vspace{-0.5cm}}
\caption[]{See caption of Fig.1. [\emph{Figure  available in the online edition}.]}         
\label{fig:img_final}
\end{figure}
\clearpage
\newpage


\textbf{SHARDS10002769} E/S0 galaxy with Type-I profile and a low to medium inclination (see Table \ref{tab:fits_psforr}). Manual and extensive masking was required due to multiple close field objects. The profile shows a very bright bulge and a wavy exponential profile. The PDDs of $h$ and $\mu_{0}$ reveal multiple Gaussian peaked distributions, due to the irregularities along the radius. None of them is compatible with a double exponential profile. 

\begin{figure}[!h]
{\centering
\vspace{-0cm}

\begin{minipage}{.5\textwidth}
\hspace{1.2cm}
\begin{overpic}[width=0.8\textwidth]
{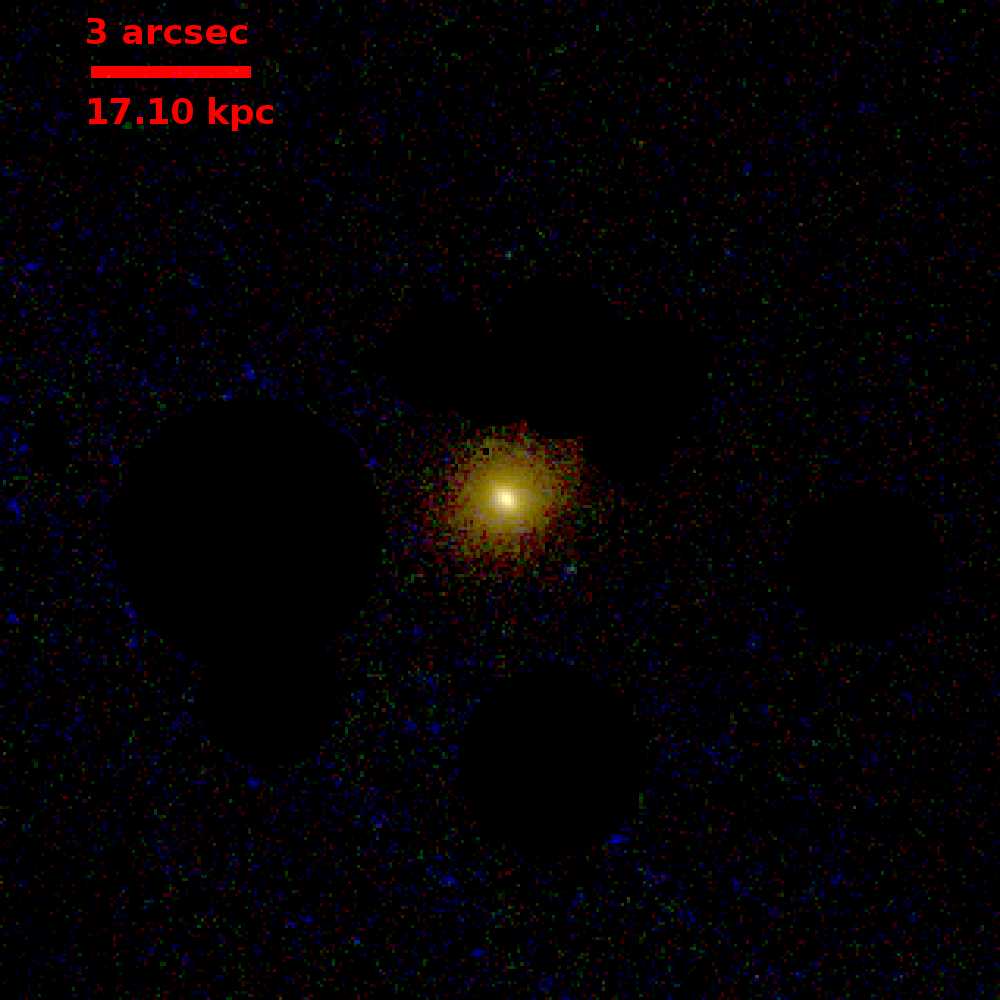}
\put(110,200){\color{yellow} \textbf{SHARDS10002769}}
\put(110,190){\color{yellow} \textbf{z=0.4417}}
\put(110,180){\color{yellow} \textbf{E/S0}}
\end{overpic}
\vspace{-1cm}
\end{minipage}%
\begin{minipage}{.5\textwidth}
\includegraphics[clip, trim=1cm 1cm 1.5cm 1.5cm, width=\textwidth]{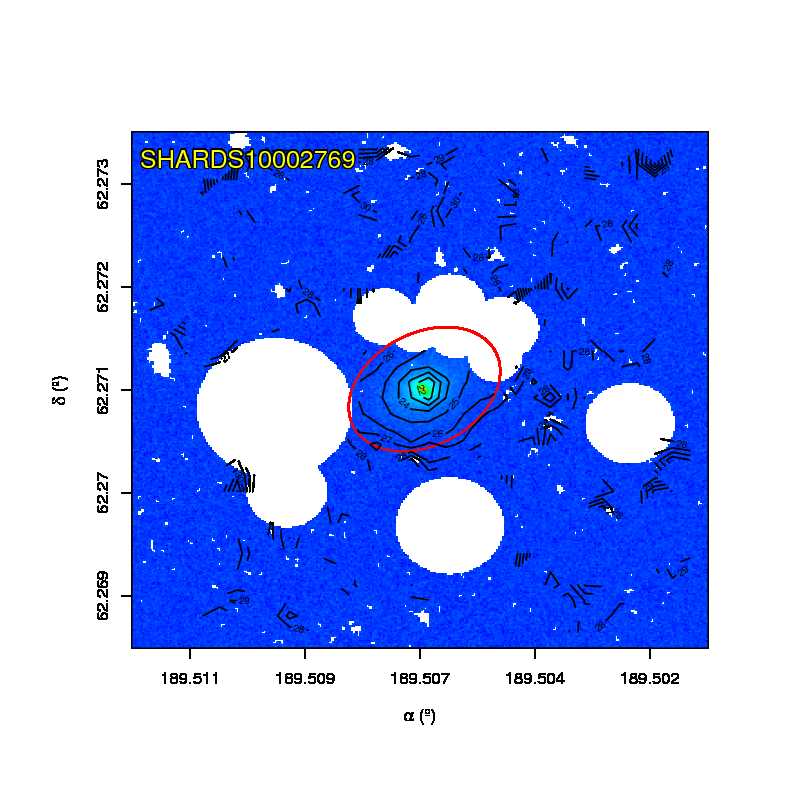}\vspace{-1cm}
\end{minipage}%

\begin{minipage}{.49\textwidth}
\includegraphics[clip, trim=0.1cm 0.1cm 0.1cm 0.1cm, width=\textwidth]{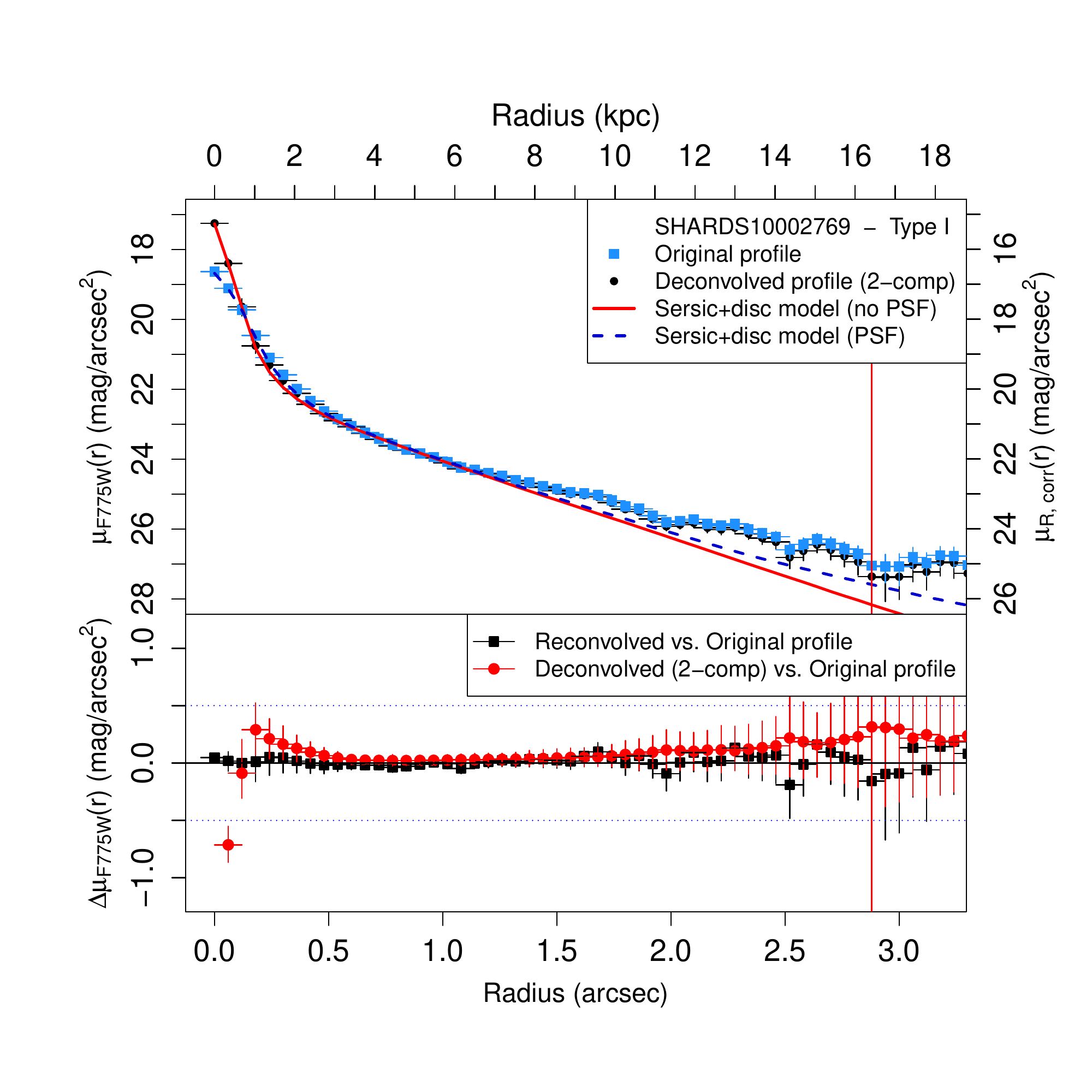}
\end{minipage}
\begin{minipage}{.49\textwidth}
\includegraphics[clip, trim=0.1cm 0.1cm 1cm 0.1cm, width=0.95\textwidth]{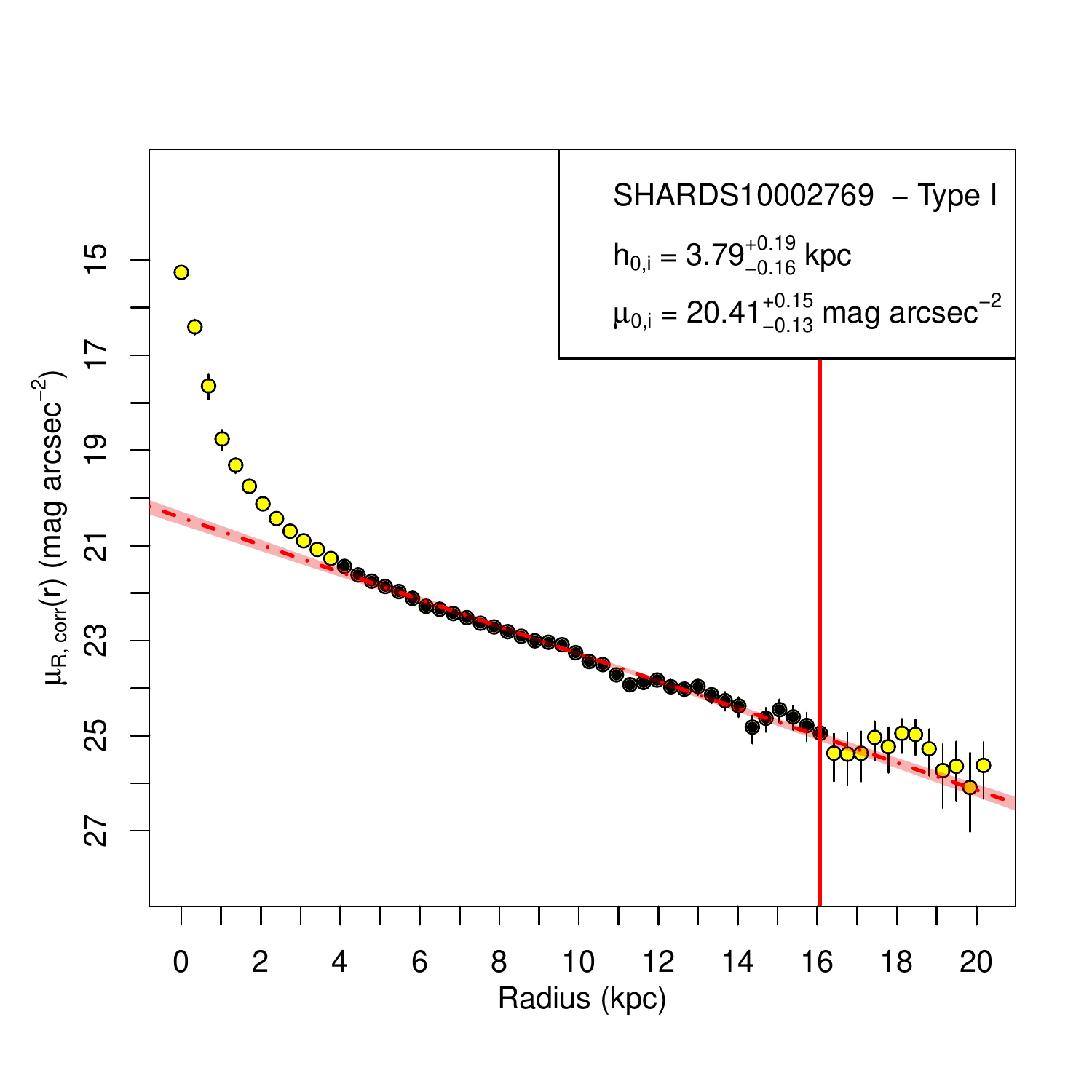}
\end{minipage}%

\vspace{-0.5cm}}
\caption[]{See caption of Fig.1. [\emph{Figure  available in the online edition}.]}         
\label{fig:img_final}
\end{figure}
\clearpage
\newpage

\textbf{SHARDS10002942:} Type-III S0 galaxy with face on orientation (see Table \ref{tab:fits_psforr}). Multiple and extensive masking was required due to the presence of nearby objects. The profile shows a bulge and Type-III disc composition, as the small excess of light at the outermost part of the galaxy is proved to be statistically significant by the automatic break analysis after the PSF subtraction.

\begin{figure}[!h]
{\centering
\vspace{-0cm}

\begin{minipage}{.5\textwidth}
\hspace{1.2cm}
\begin{overpic}[width=0.8\textwidth]
{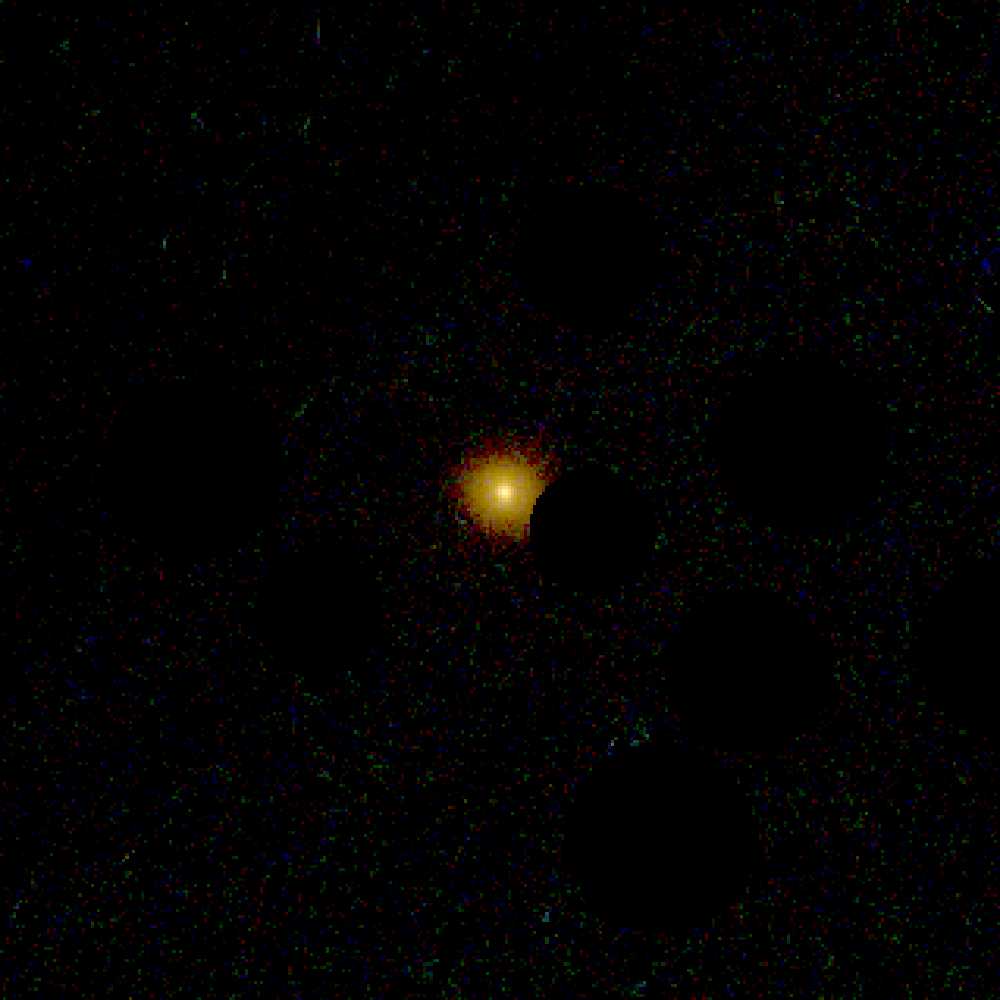}
\put(110,200){\color{yellow} \textbf{SHARDS10002942}}
\put(110,190){\color{yellow} \textbf{z=0.5644}}
\put(110,180){\color{yellow} \textbf{S0}}
\end{overpic}
\vspace{-1cm}
\end{minipage}%
\begin{minipage}{.5\textwidth}
\includegraphics[clip, trim=1cm 1cm 1.5cm 1.5cm, width=\textwidth]{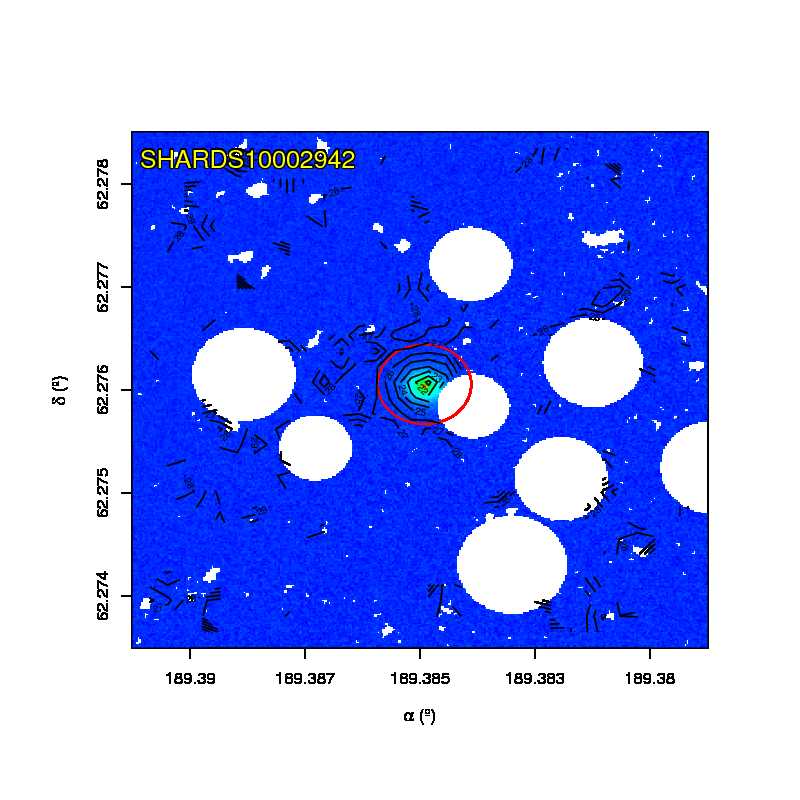}\vspace{-1cm}
\end{minipage}%

\begin{minipage}{.49\textwidth}
\includegraphics[clip, trim=0.1cm 0.1cm 0.1cm 0.1cm, width=\textwidth]{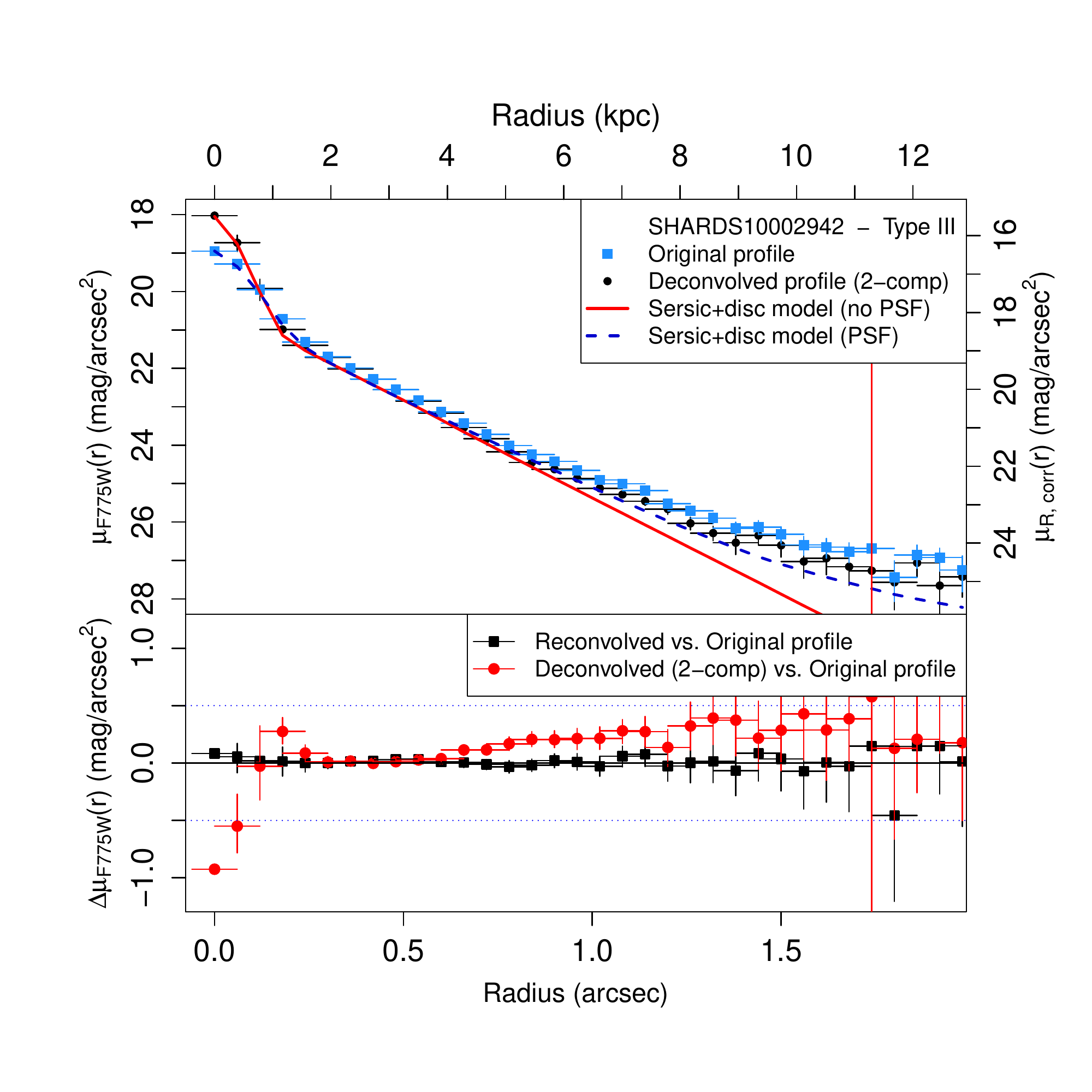}
\end{minipage}
\begin{minipage}{.49\textwidth}
\includegraphics[clip, trim=0.1cm 0.1cm 1cm 0.1cm, width=0.95\textwidth]{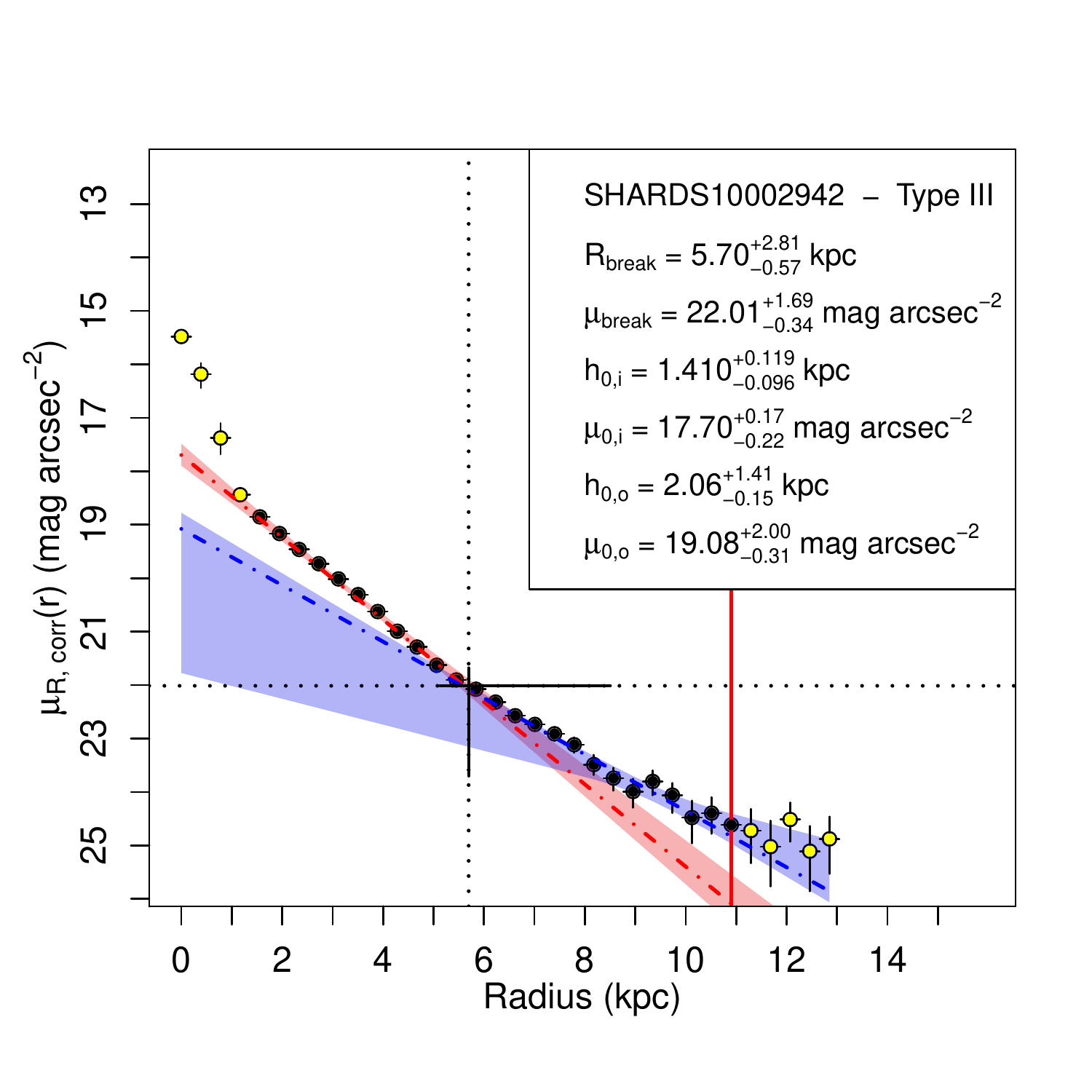}
\end{minipage}%

\vspace{-0.5cm}}
\caption[]{See caption of Fig.1. [\emph{Figure  available in the online edition}.]}         
\label{fig:img_final}
\end{figure}
\clearpage
\newpage


\textbf{SHARDS10003216:} E/S0 galaxy with very small size. The profile seems to be almost featureless, without any bulge or break visible, so it was classified as a Type I. 

\begin{figure}[!h]
{\centering
\vspace{-0cm}

\begin{minipage}{.5\textwidth}
\hspace{1.2cm}
\begin{overpic}[width=0.8\textwidth]
{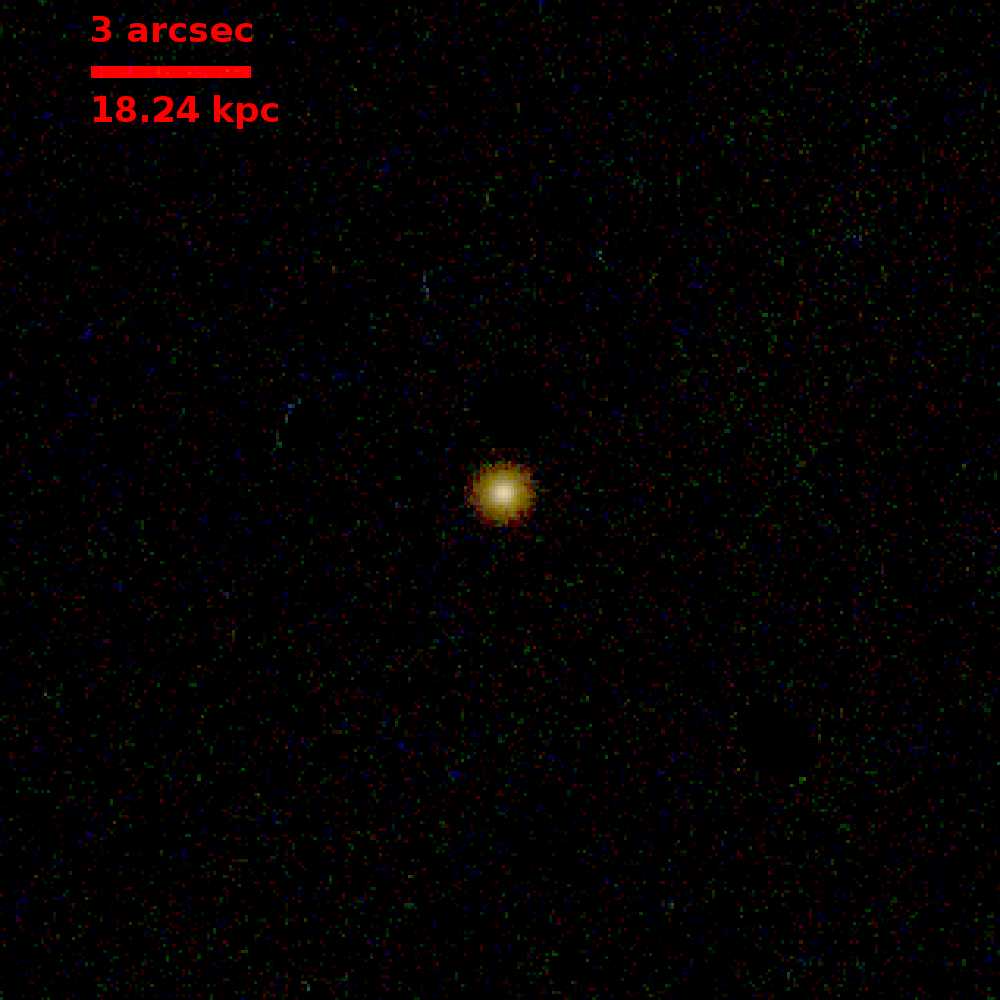}
\put(110,200){\color{yellow} \textbf{SHARDS10003216}}
\put(110,190){\color{yellow} \textbf{z=0.4970}}
\put(110,180){\color{yellow} \textbf{E/S0}}
\end{overpic}
\vspace{-1cm}
\end{minipage}%
\begin{minipage}{.5\textwidth}
\includegraphics[clip, trim=1cm 1cm 1.5cm 1.5cm, width=\textwidth]{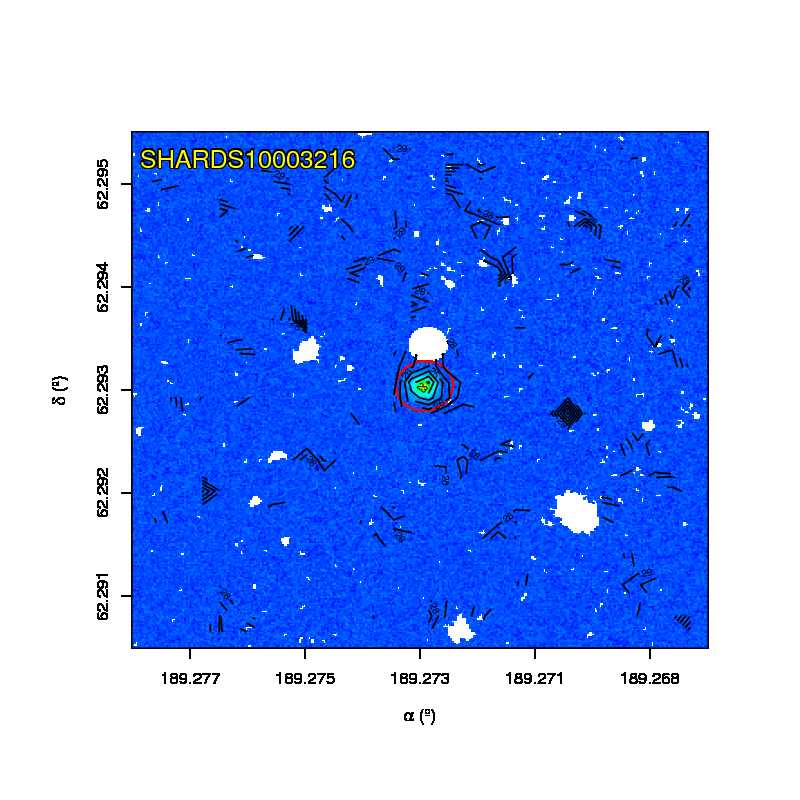}\vspace{-1cm}
\end{minipage}%

\begin{minipage}{.49\textwidth}
\includegraphics[clip, trim=0.1cm 0.1cm 0.1cm 0.1cm, width=\textwidth]{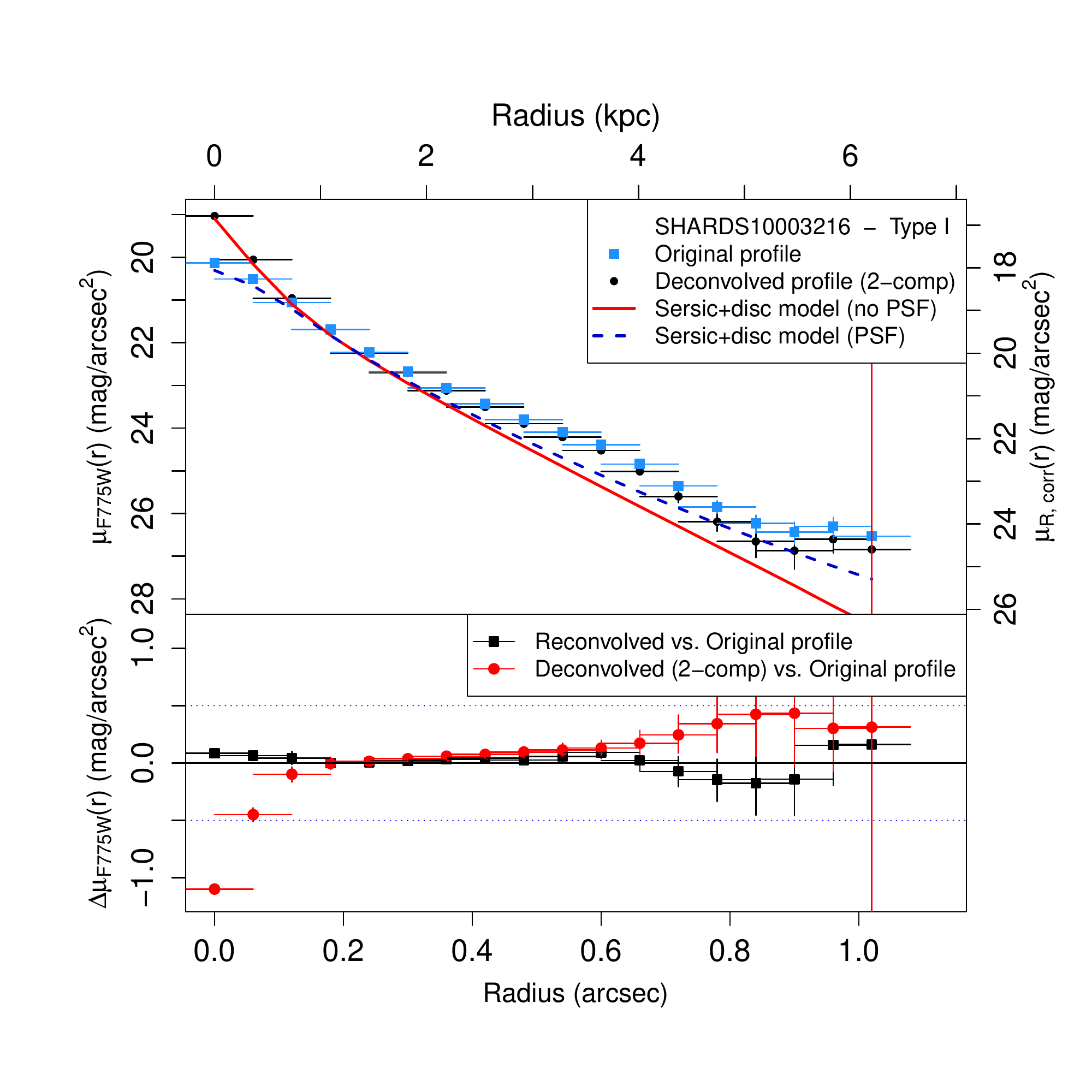}
\end{minipage}
\begin{minipage}{.49\textwidth}
\includegraphics[clip, trim=0.1cm 0.1cm 1cm 0.1cm, width=0.95\textwidth]{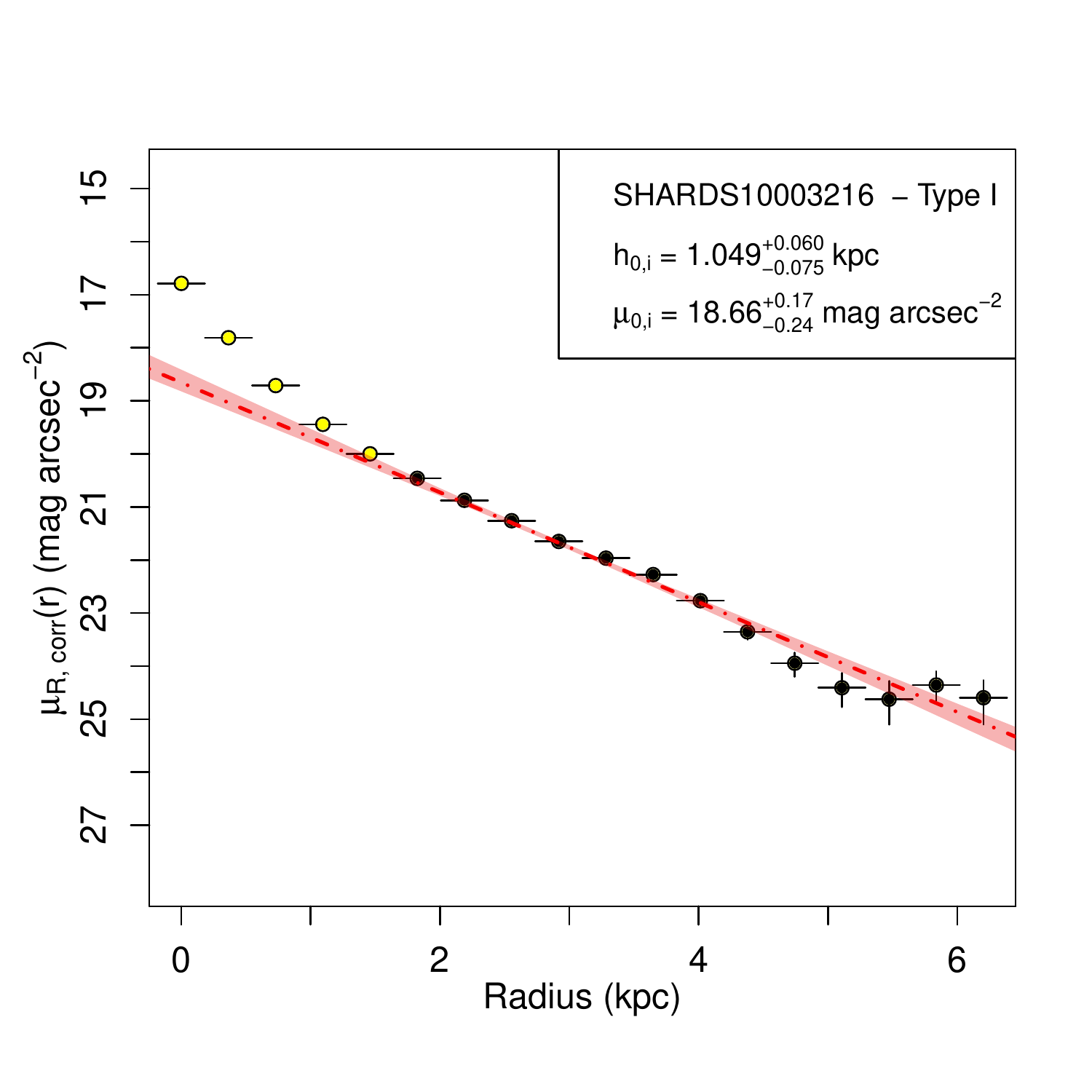}
\end{minipage}%

\vspace{-0.5cm}}
\caption[]{See caption of Fig.1. [\emph{Figure  available in the online edition}.]}         
\label{fig:img_final}
\end{figure}
\clearpage
\newpage


\textbf{SHARDS10003312:} E/S0 galaxy with Type-III profile. It has a medium inclination (see Table \ref{tab:fits_psforr}) and it is almost isolated, so no manual masking was needed. The automated break analysis successfully detected two exponential profiles, with narrow and separated PDDs, although the distributions associated with the outer profile presented higher dispersions due to the irregularities found in the profile. 

\begin{figure}[!h]
{\centering
\vspace{-0cm}

\begin{minipage}{.5\textwidth}
\hspace{1.2cm}
\begin{overpic}[width=0.8\textwidth]
{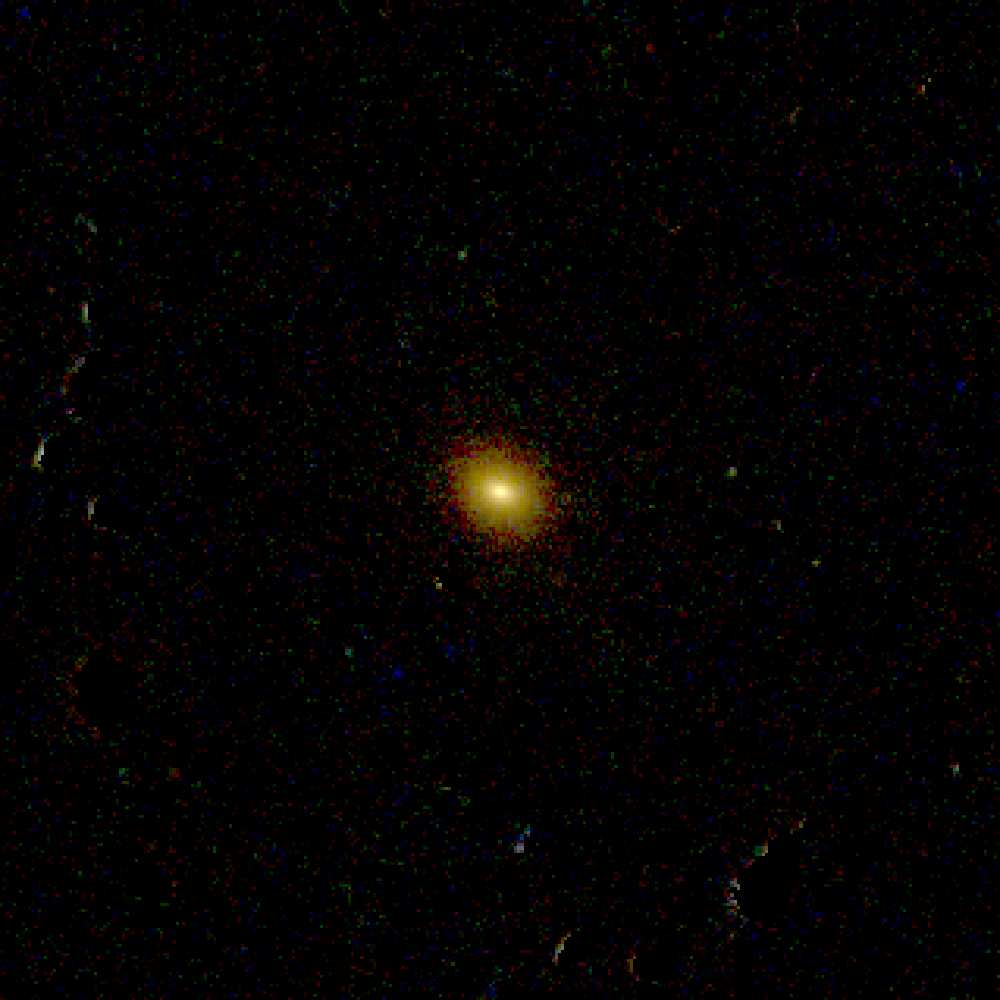}
\put(110,200){\color{yellow} \textbf{SHARDS10003312}}
\put(110,190){\color{yellow} \textbf{z=0.4973}}
\put(110,180){\color{yellow} \textbf{E/S0}}
\end{overpic}
\vspace{-1cm}
\end{minipage}%
\begin{minipage}{.5\textwidth}
\includegraphics[clip, trim=1cm 1cm 1.5cm 1.5cm, width=\textwidth]{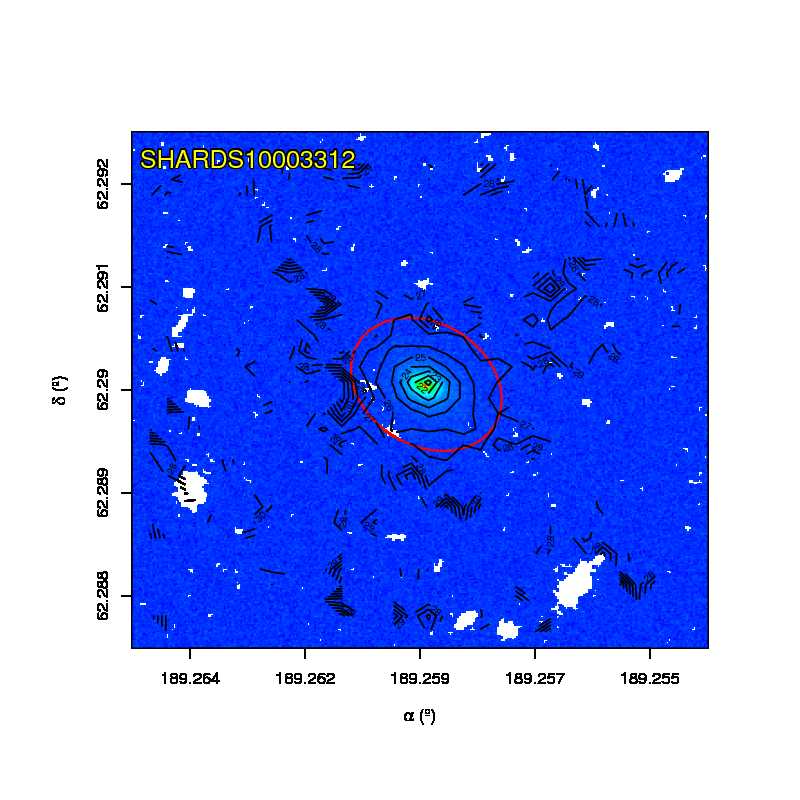}\vspace{-1cm}
\end{minipage}%

\begin{minipage}{.49\textwidth}
\includegraphics[clip, trim=0.1cm 0.1cm 0.1cm 0.1cm, width=\textwidth]{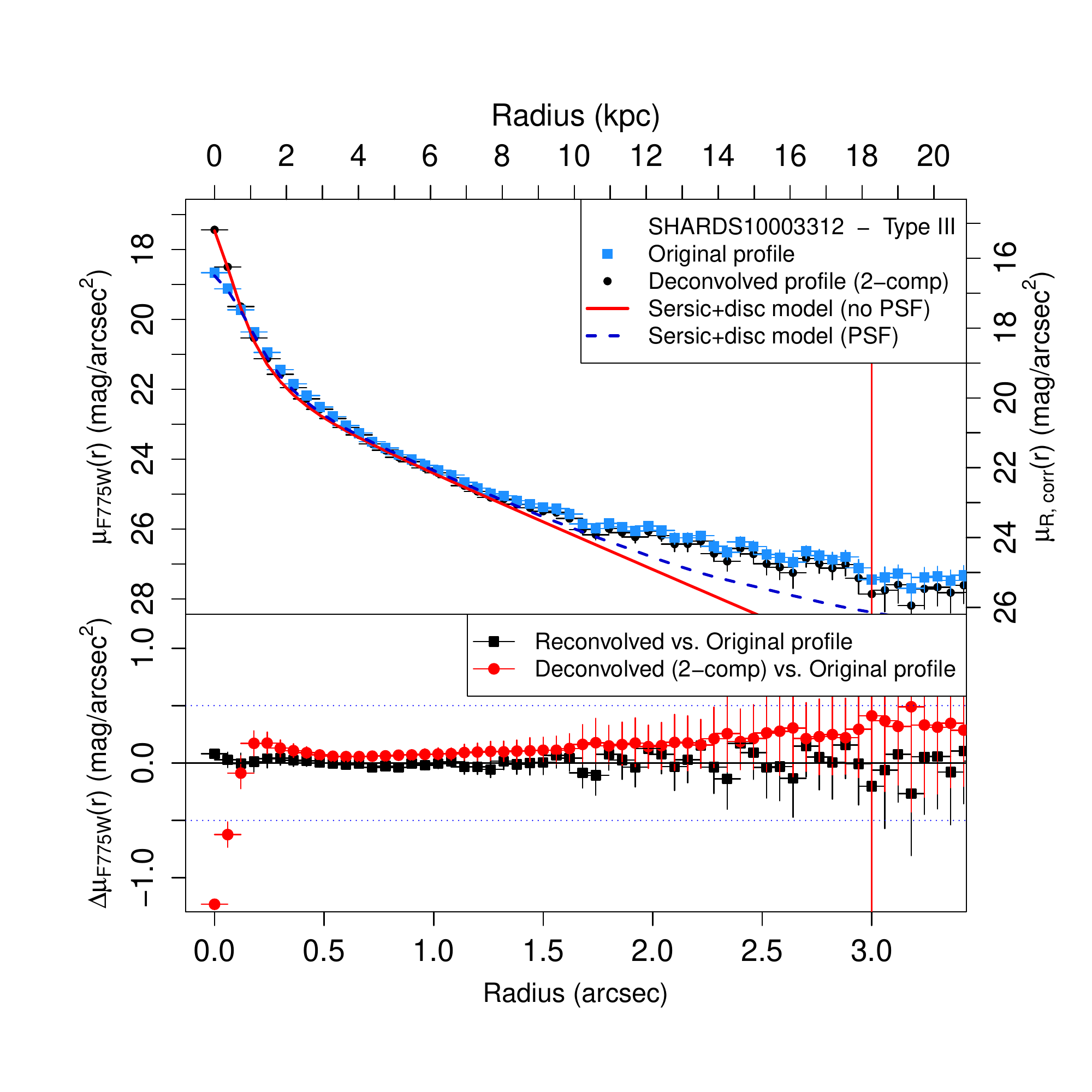}
\end{minipage}
\begin{minipage}{.49\textwidth}
\includegraphics[clip, trim=0.1cm 0.1cm 1cm 0.1cm, width=0.95\textwidth]{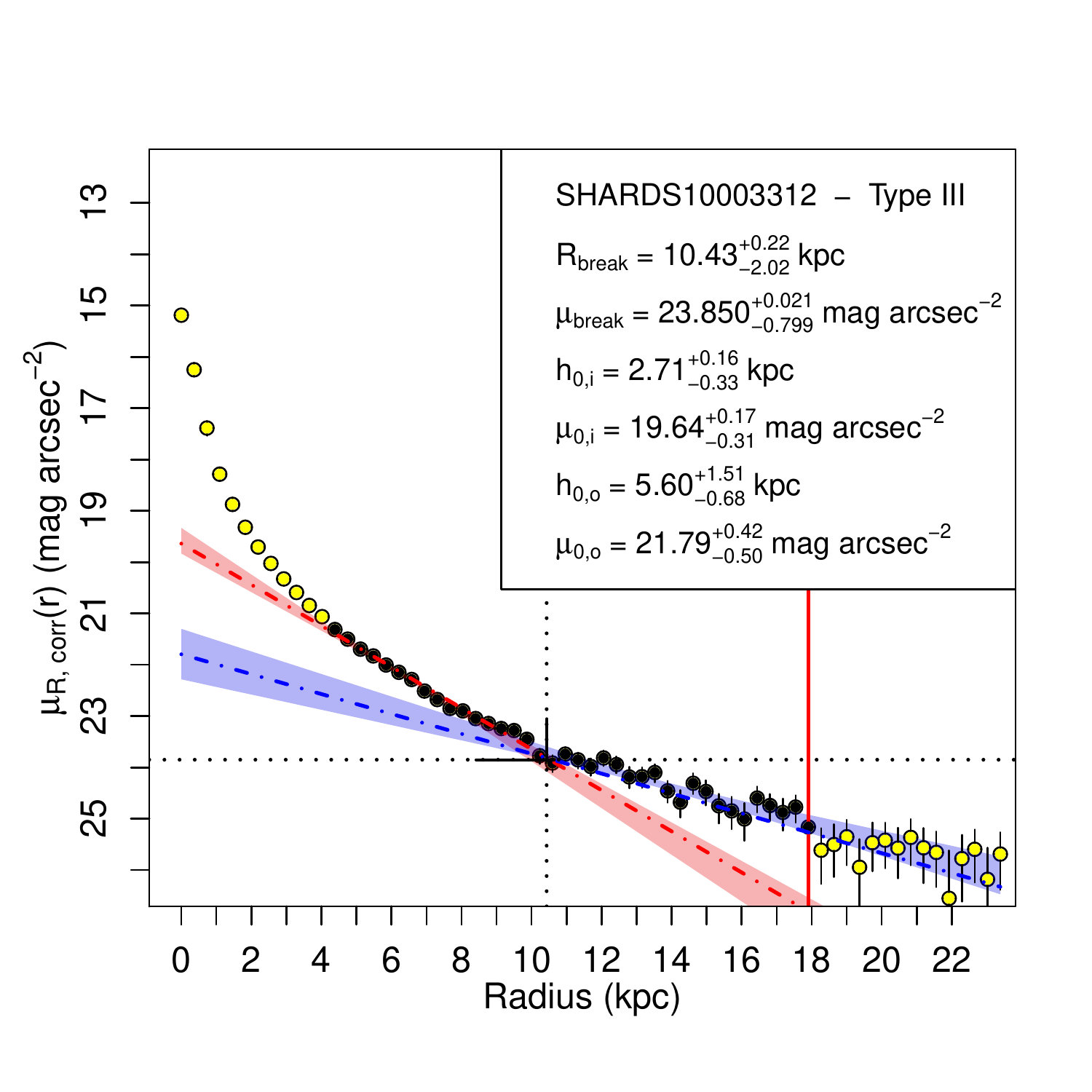}
\end{minipage}%

\vspace{-0.5cm}}
\caption[]{See caption of Fig.1. [\emph{Figure  available in the online edition}.]}         
\label{fig:img_final}
\end{figure}
\clearpage
\newpage


\textbf{SHARDS10003402:} S0 galaxy with a Type-II profile. Manual masking was required due to the presence of multiple field objects, none of them inside the fitting region. It was analysed by {\tt{ISOFIT}} instead of {\tt{ellipse}} due to its completely edge-on orientation (see Table \ref{tab:fits_psforr}). The outer region ($R > 9$ kpc) was removed from the profile analysis due to the presence of several irregularities. The PDDs show two close, but clearly separated distributions for $h$ and $\mu_{0}$, thus the disc is well represented by a Type-II profile.

\begin{figure}[!h]
{\centering
\vspace{-0cm}

\begin{minipage}{.5\textwidth}
\hspace{1.2cm}
\begin{overpic}[width=0.8\textwidth]
{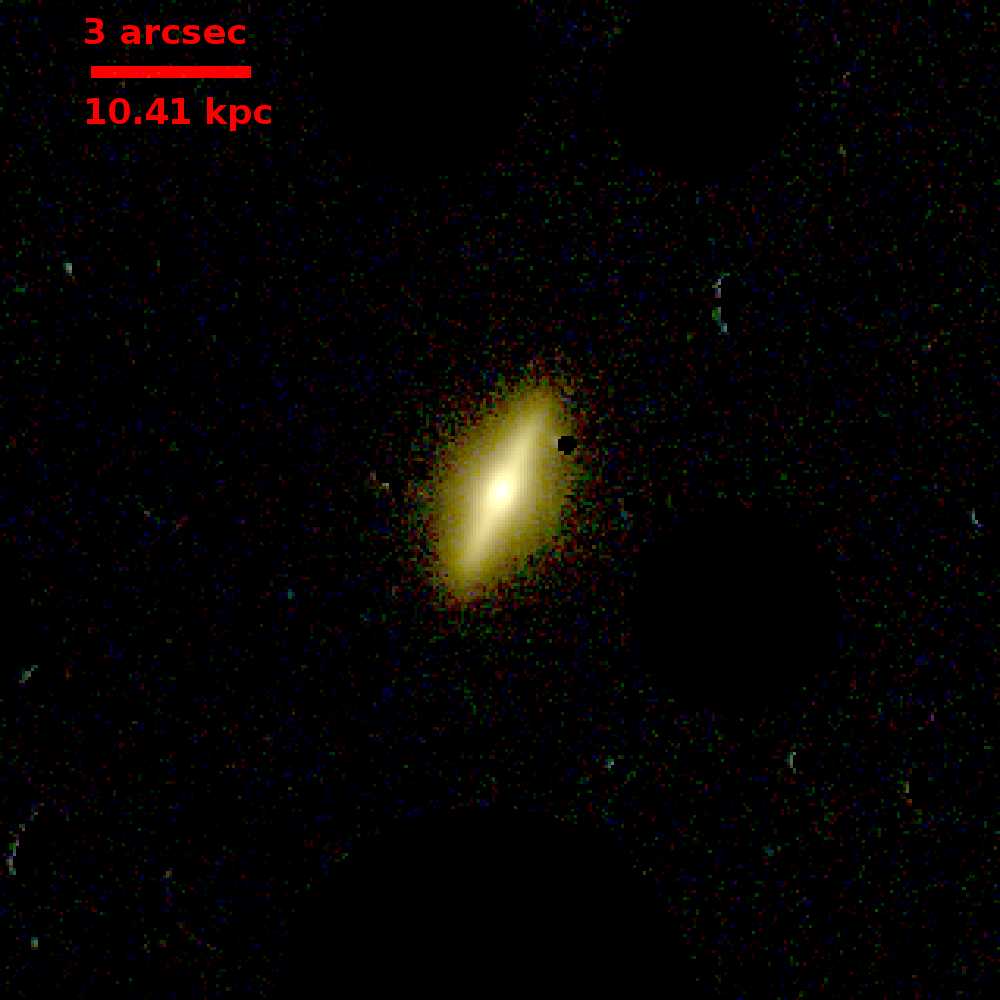}
\put(110,200){\color{yellow} \textbf{SHARDS10003402}}
\put(110,190){\color{yellow} \textbf{z=0.2133}}
\put(110,180){\color{yellow} \textbf{S0}}
\end{overpic}
\vspace{-1cm}
\end{minipage}%
\begin{minipage}{.5\textwidth}
\includegraphics[clip, trim=1cm 1cm 1.5cm 1.5cm, width=\textwidth]{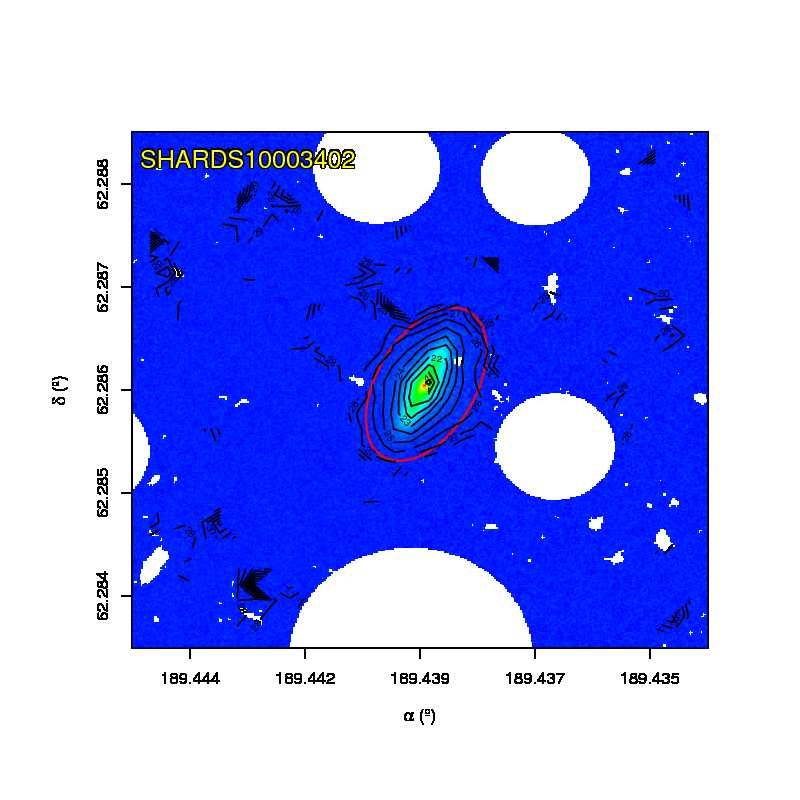}\vspace{-1cm}
\end{minipage}%

\begin{minipage}{.49\textwidth}
\includegraphics[clip, trim=0.1cm 0.1cm 0.1cm 0.1cm, width=\textwidth]{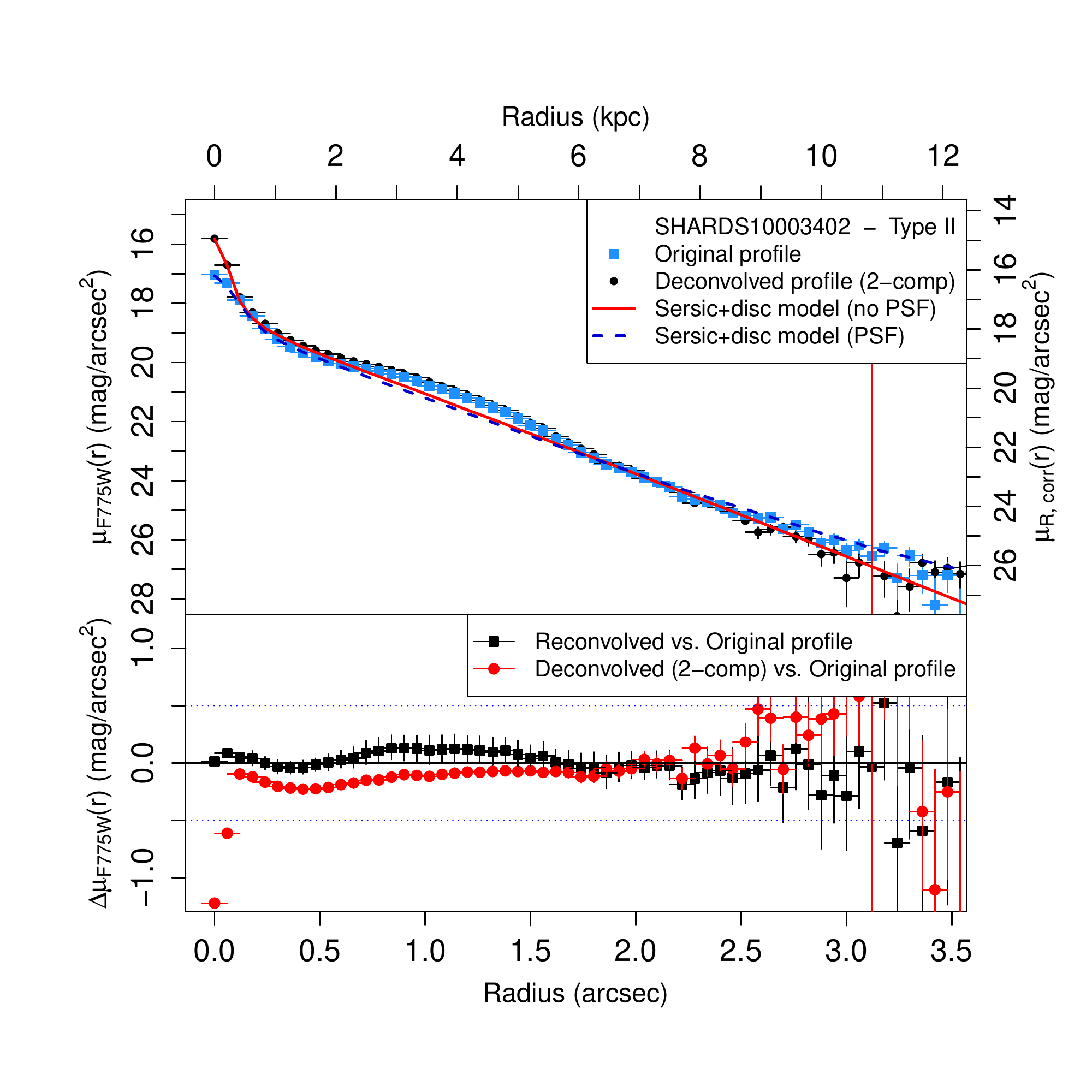}
\end{minipage}
\begin{minipage}{.49\textwidth}
\includegraphics[clip, trim=0.1cm 0.1cm 1cm 0.1cm, width=0.95\textwidth]{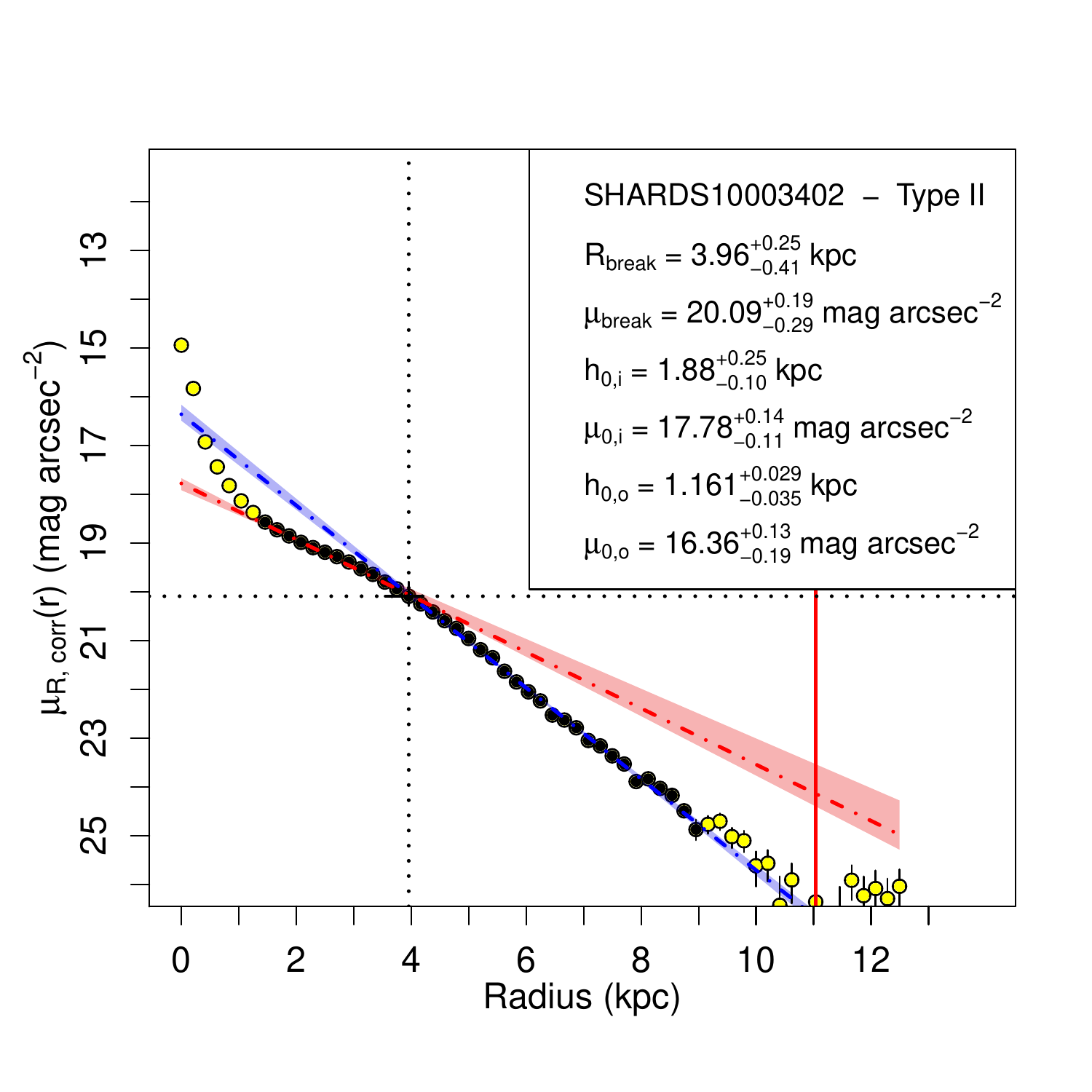}
\end{minipage}%

\vspace{-0.5cm}}
\caption[]{See caption of Fig.1. [\emph{Figure  available in the online edition}.]}         
\label{fig:img_final}
\end{figure}
\clearpage
\newpage


\textbf{SHARDS10003647:} S0 galaxy with a Type-III profile. It shows a medium inclination (see Table \ref{tab:fits_psforr}). Manual masking was required to avoid contamination from small field objects. The outermost region of the disc profile seems to be wavy, distorting the PDDs. The PDD of $h$ and $\mu_{0}$ appear to be clearly separated for the inner and outer profiles, although the outer profile is clearly skewed and wider than the inner one. The object present a significant Type-III profile despite the wide distribution of the outer profile.

\begin{figure}[!h]
{\centering
\vspace{-0cm}

\begin{minipage}{.5\textwidth}
\hspace{1.2cm}
\begin{overpic}[width=0.8\textwidth]
{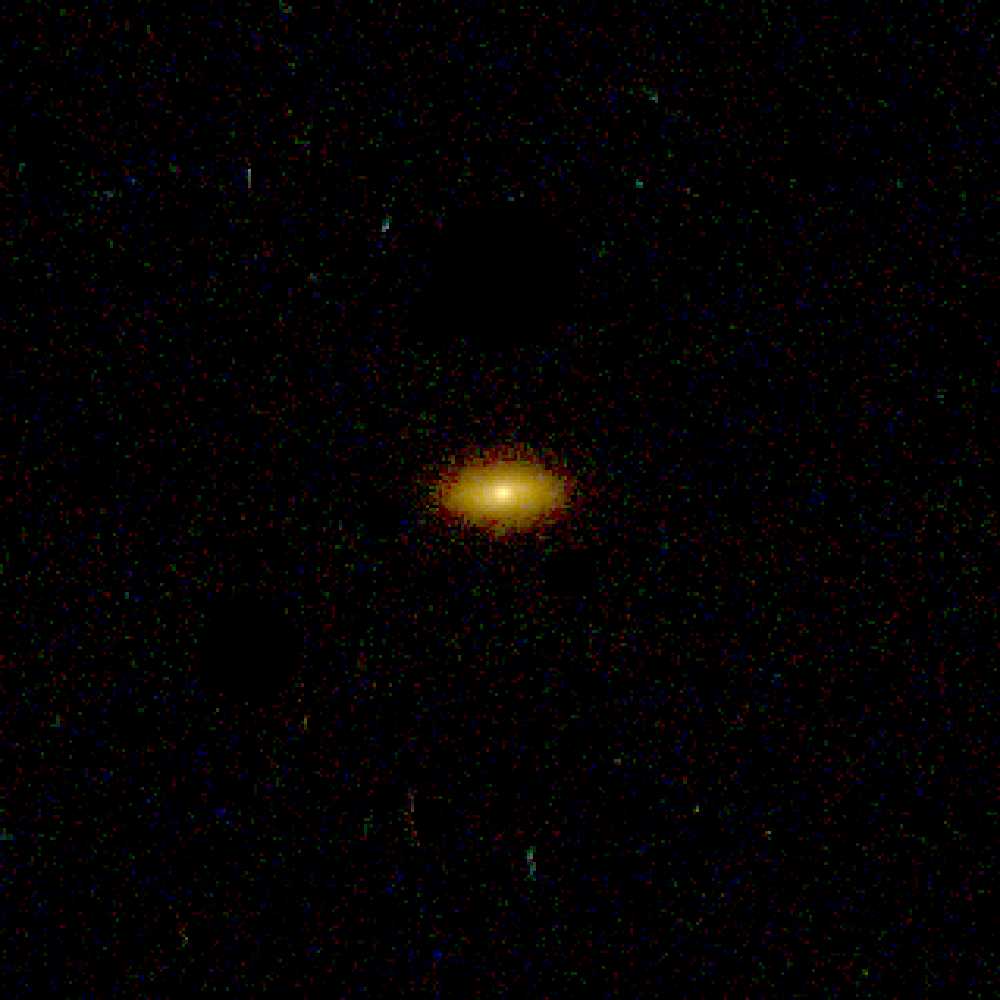}
\put(110,200){\color{yellow} \textbf{SHARDS10003647}}
\put(110,190){\color{yellow} \textbf{z=0.5553}}
\put(110,180){\color{yellow} \textbf{S0}}
\end{overpic}
\vspace{-1cm}
\end{minipage}%
\begin{minipage}{.5\textwidth}
\includegraphics[clip, trim=1cm 1cm 1.5cm 1.5cm, width=\textwidth]{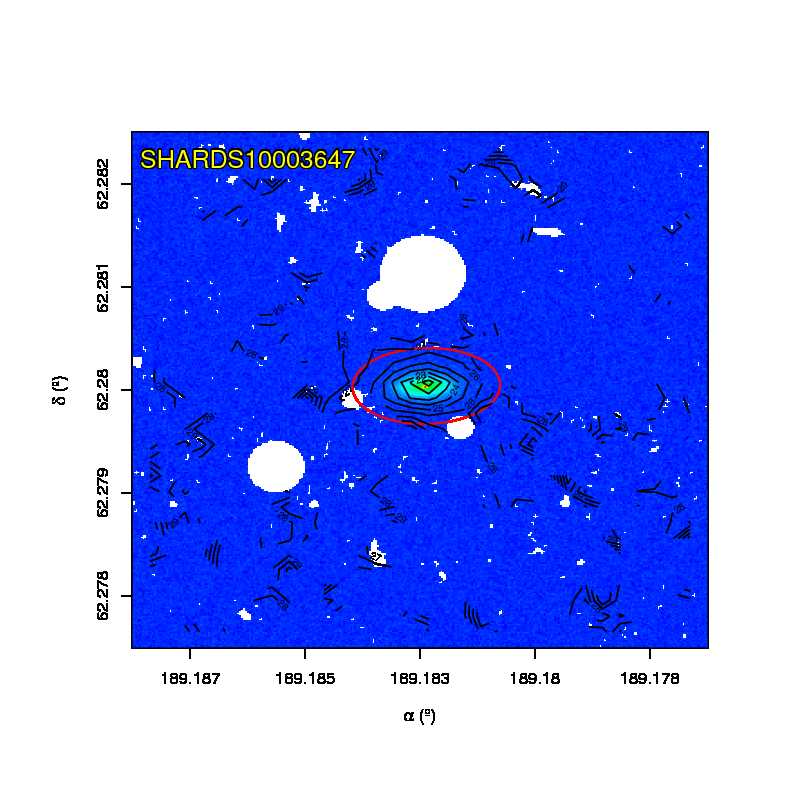}\vspace{-1cm}
\end{minipage}%

\begin{minipage}{.49\textwidth}
\includegraphics[clip, trim=0.1cm 0.1cm 0.1cm 0.1cm, width=\textwidth]{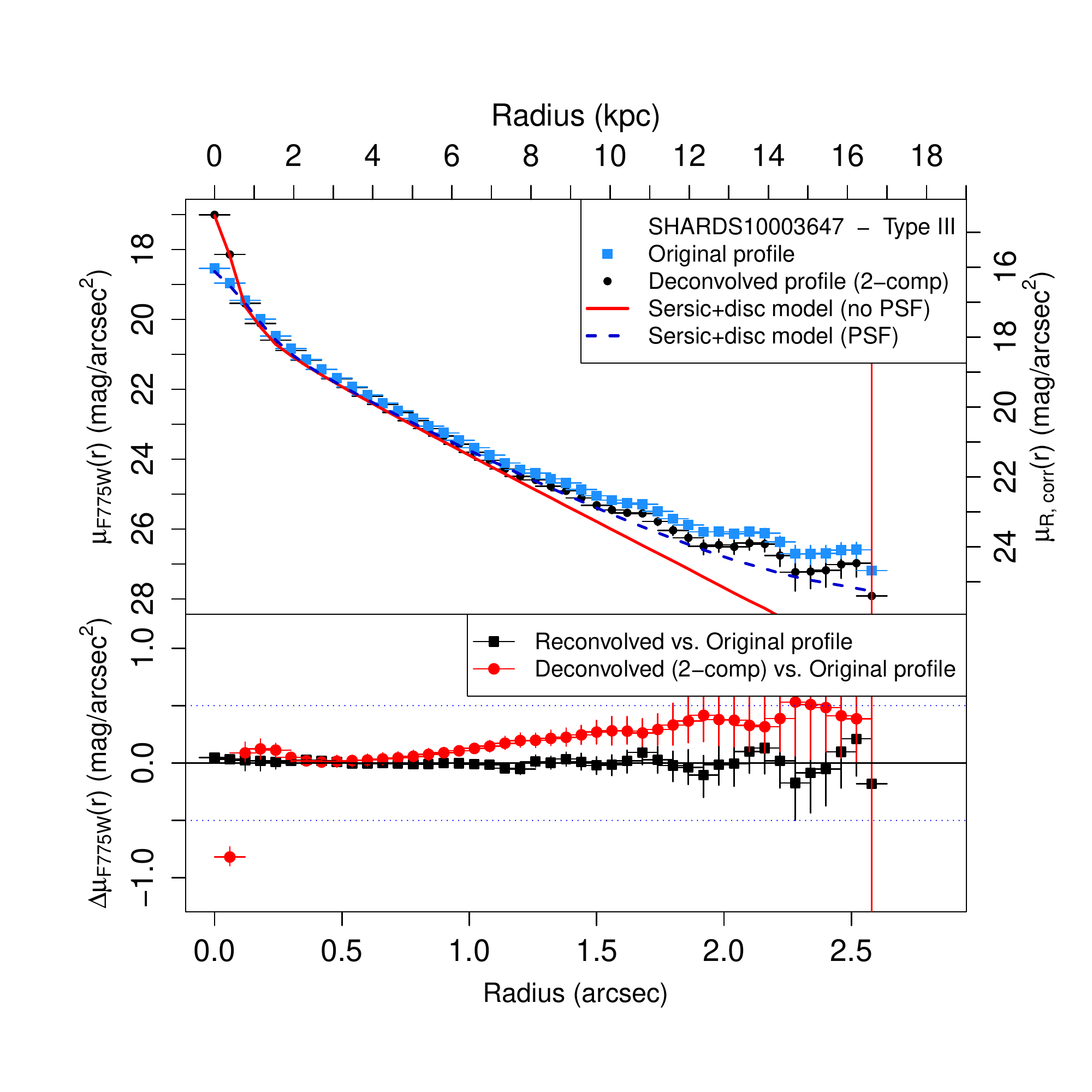}
\end{minipage}
\begin{minipage}{.49\textwidth}
\includegraphics[clip, trim=0.1cm 0.1cm 1cm 0.1cm, width=0.95\textwidth]{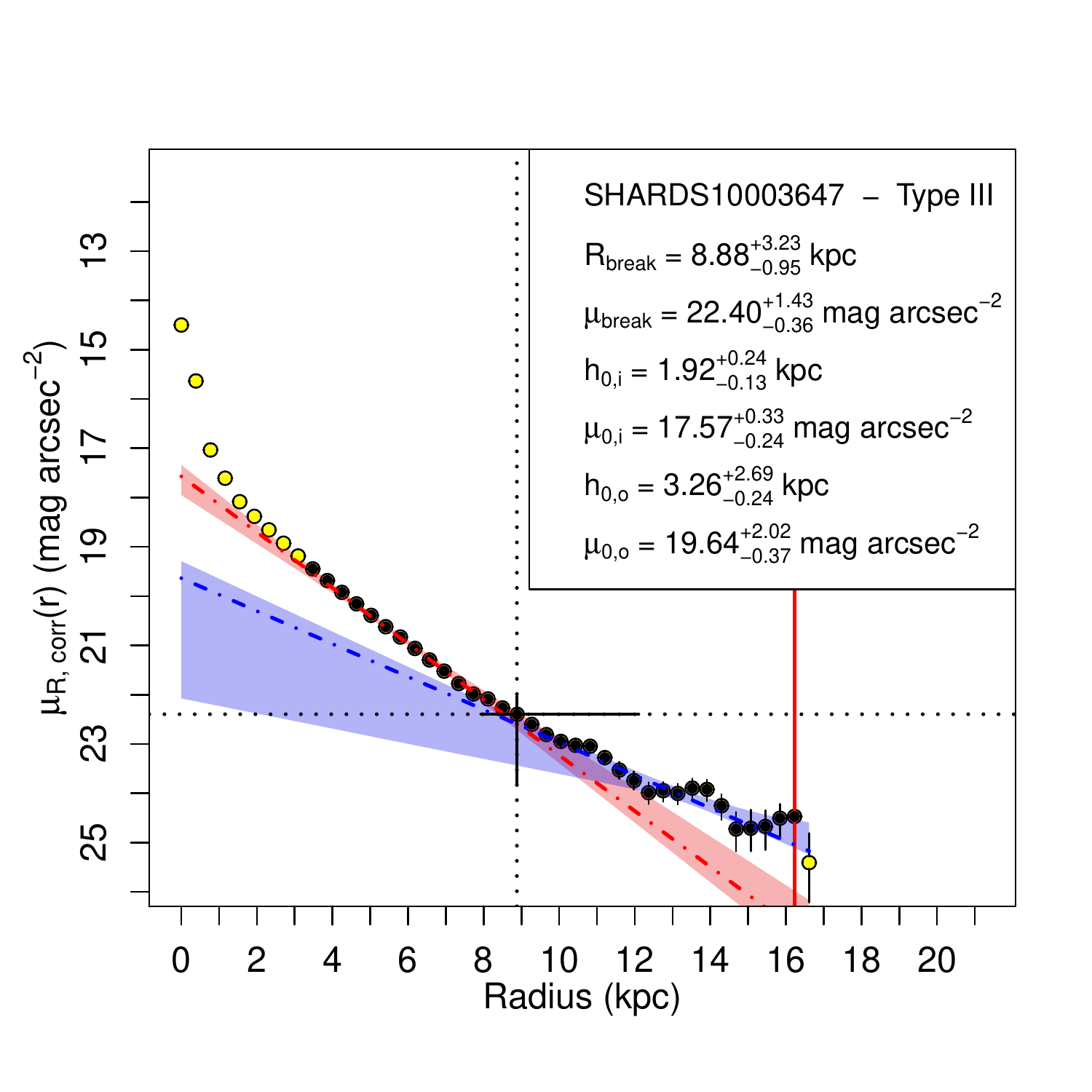}
\end{minipage}%

\vspace{-0.5cm}}
\caption[]{See caption of Fig.1. [\emph{Figure  available in the online edition}.]}         
\label{fig:img_final}
\end{figure}
\clearpage
\newpage


\textbf{SHARDS10004777:} S0 galaxy with a Type-II profile. The galaxy presents a medium inclination (see Table \ref{tab:fits_psforr}). Manual masking was applied to small sources at the outermost part (SE). We avoided the inner bump in the fit, because it probably traces a lens component, rather than a disc. The PDDs for $h$ and $\mu_{0}$ show clearly separated peaks corresponding to the outermost Type-II break.

\begin{figure}[!h]
{\centering
\vspace{-0cm}

\begin{minipage}{.5\textwidth}
\hspace{1.2cm}
\begin{overpic}[width=0.8\textwidth]
{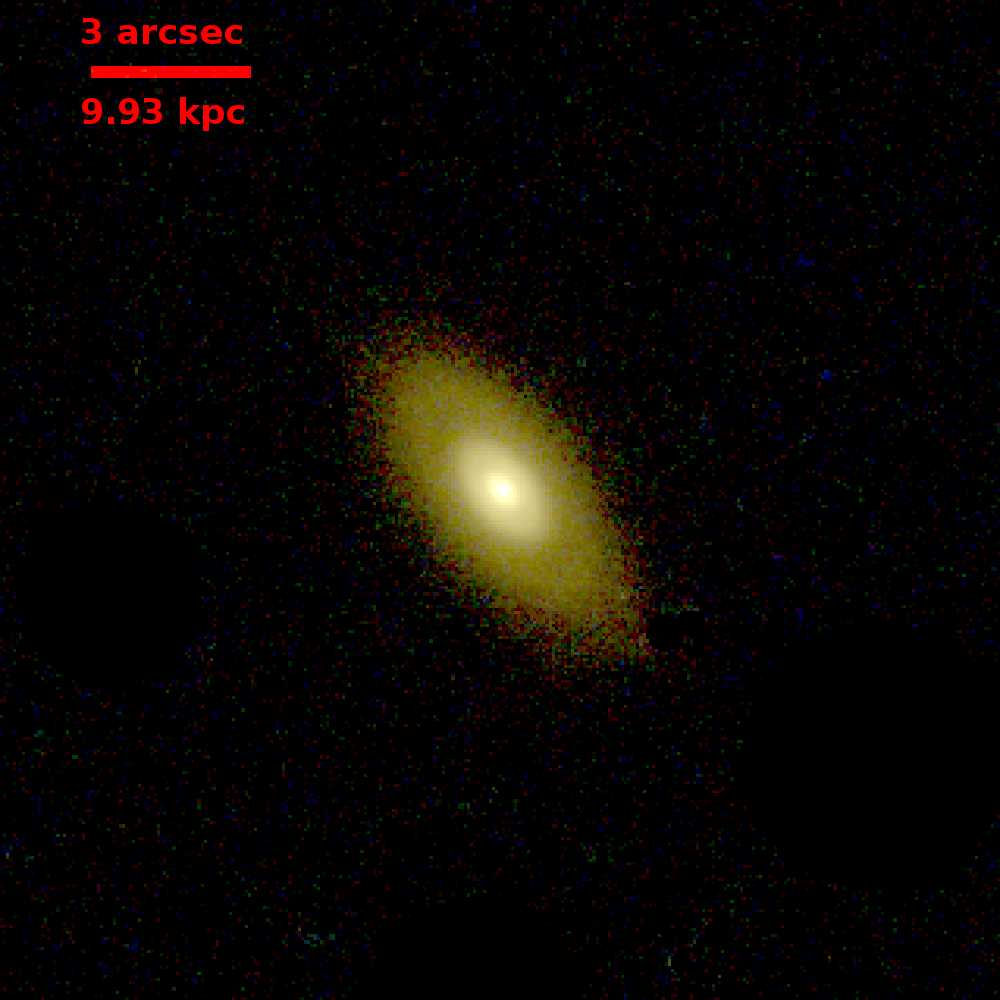}
\put(110,200){\color{yellow} \textbf{SHARDS10004777}}
\put(110,190){\color{yellow} \textbf{z=0.2010}}
\put(110,180){\color{yellow} \textbf{S0}}
\end{overpic}
\vspace{-1cm}
\end{minipage}%
\begin{minipage}{.5\textwidth}
\includegraphics[clip, trim=1cm 1cm 1.5cm 1.5cm, width=\textwidth]{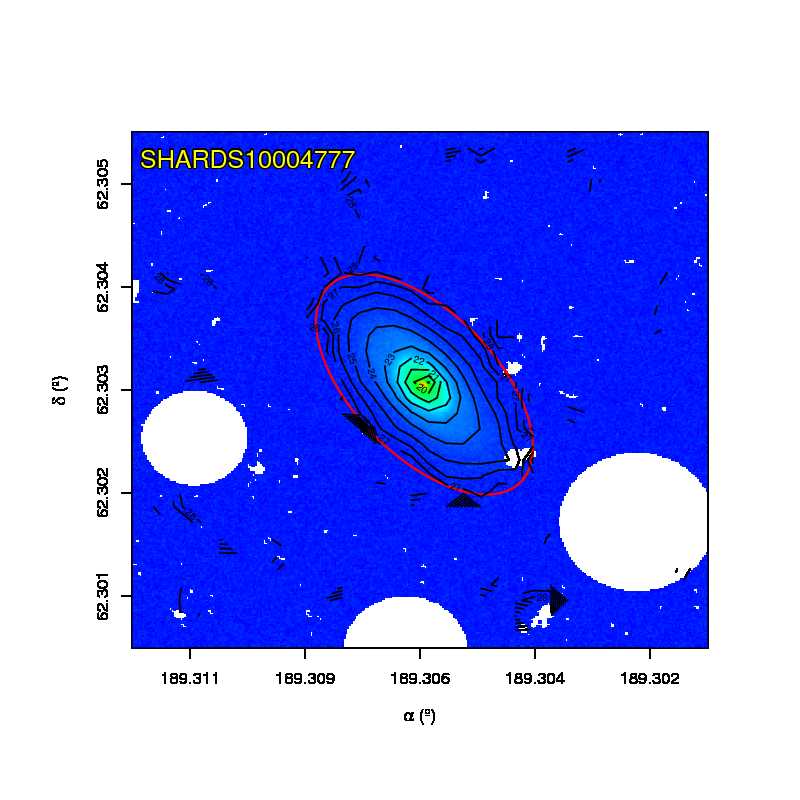}\vspace{-1cm}
\end{minipage}%

\begin{minipage}{.49\textwidth}
\includegraphics[clip, trim=0.1cm 0.1cm 0.1cm 0.1cm, width=\textwidth]{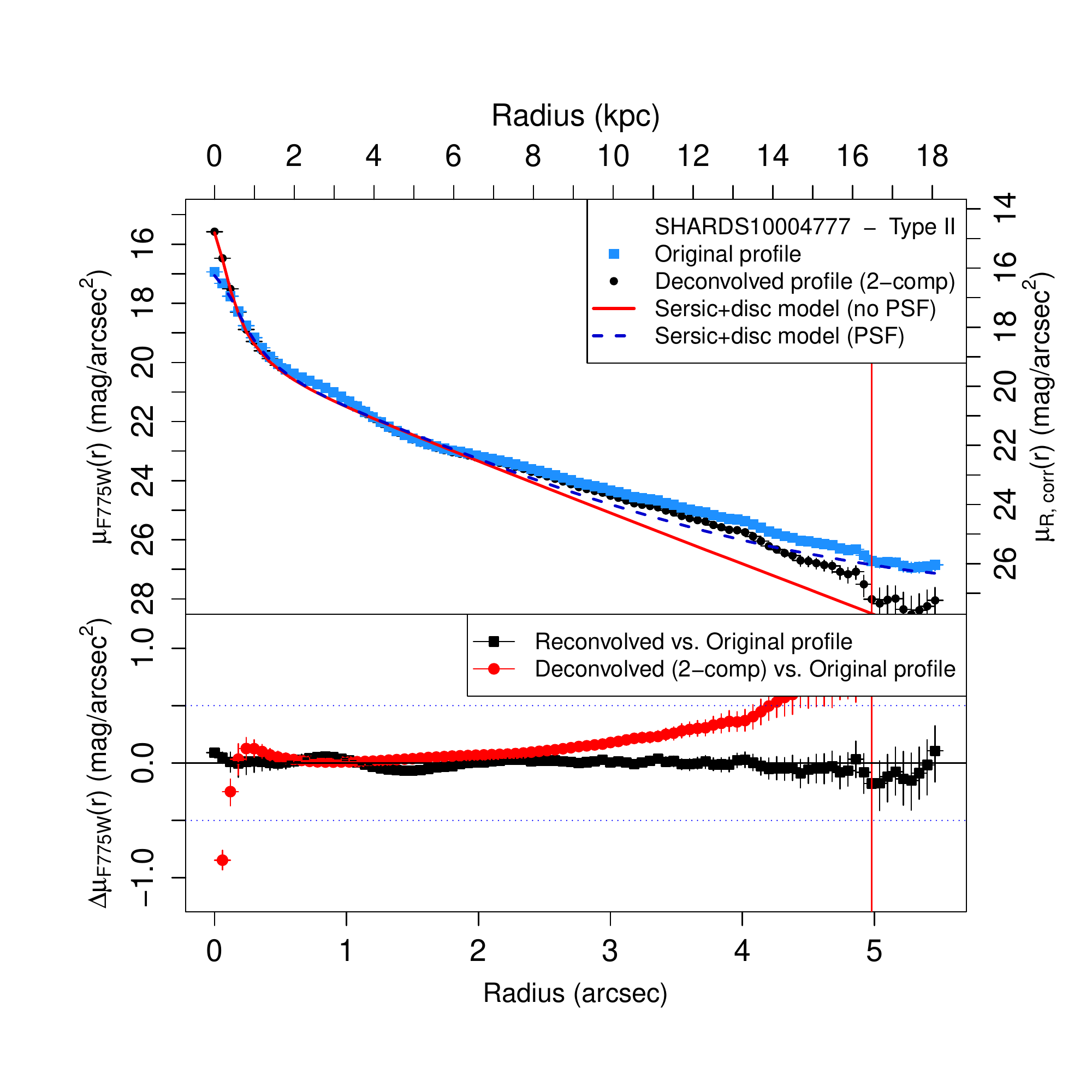}
\end{minipage}
\begin{minipage}{.49\textwidth}
\includegraphics[clip, trim=0.1cm 0.1cm 1cm 0.1cm, width=0.95\textwidth]{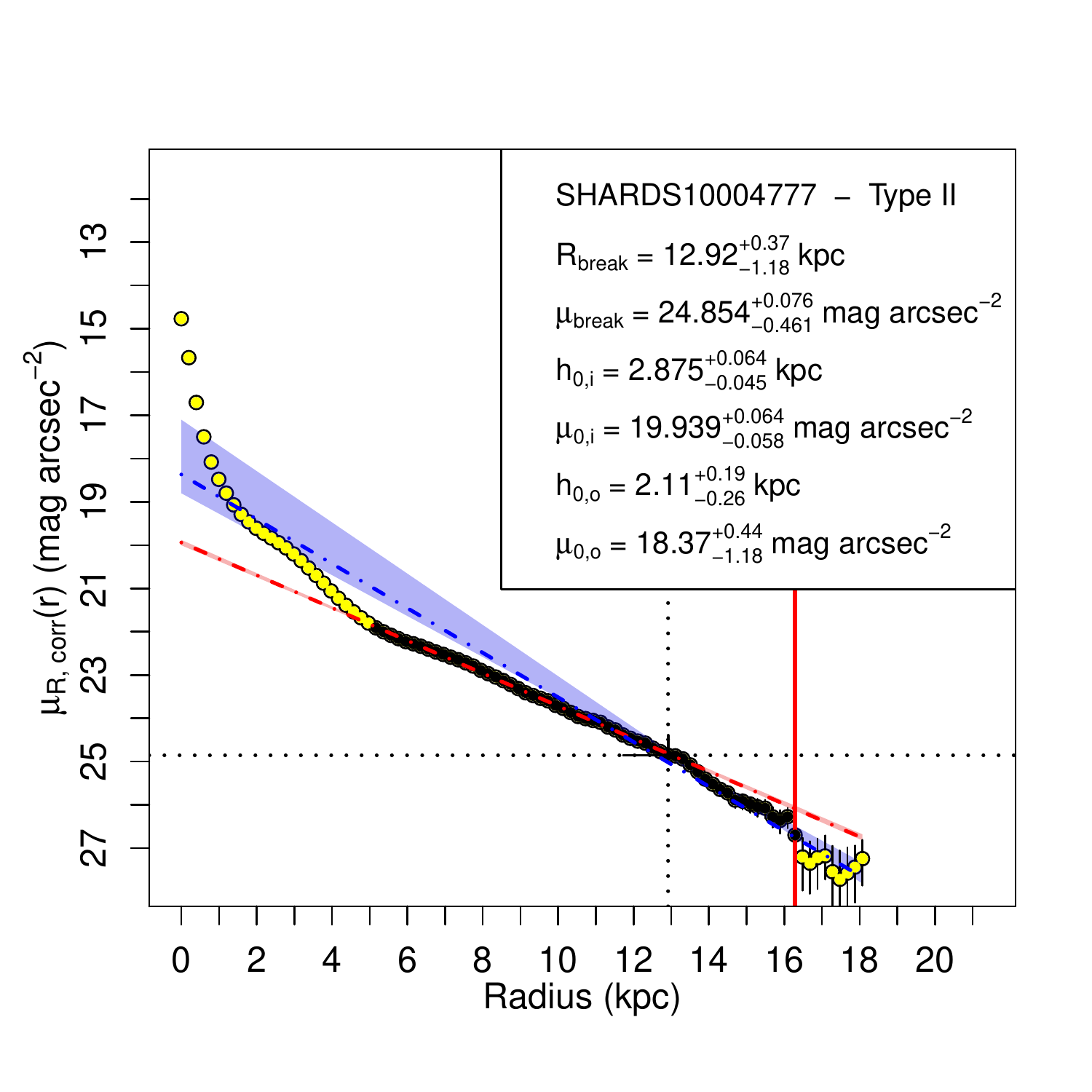}
\end{minipage}%

\vspace{-0.5cm}}
\caption[]{See caption of Fig.1. [\emph{Figure  available in the online edition}.]}         
\label{fig:img_final}
\end{figure}
\clearpage
\newpage

\textbf{SHARDS10009610:} S0 galaxy with a Type-III profile. Manual masking was applied to multiple sources, but none of them finally laid within in the final fitting region. The PDDs for $h$ and $\mu_{0}$ show two clearly separated peaks corresponding to the inner and outer profiles. The outer profile PDD seems to be slightly wider than the inner one due to little distortions at high radius.
(see Table \ref{tab:fits_psforr})

\begin{figure}[!h]
{\centering
\vspace{-0cm}

\begin{minipage}{.5\textwidth}
\hspace{1.2cm}
\begin{overpic}[width=0.8\textwidth]
{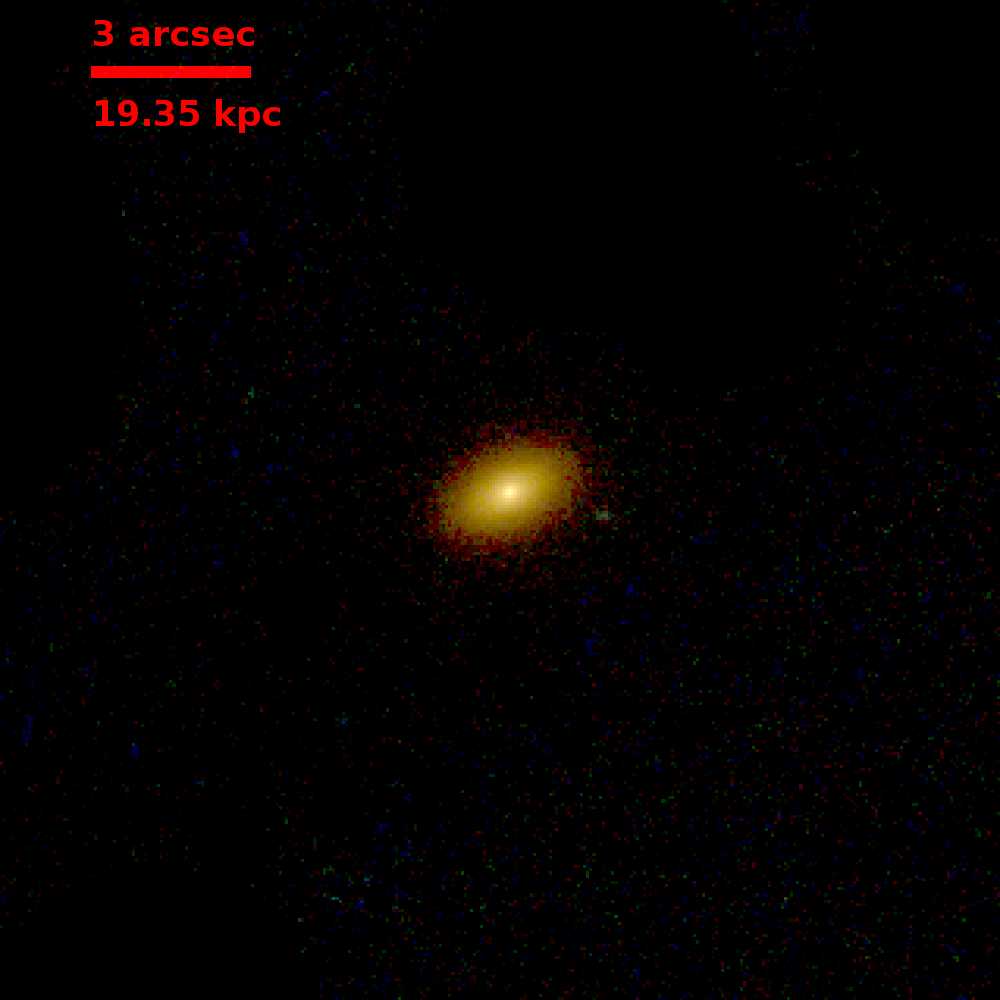}
\put(110,200){\color{yellow} \textbf{SHARDS10009610}}
\put(110,190){\color{yellow} \textbf{z=0.5568}}
\put(110,180){\color{yellow} \textbf{S0}}
\end{overpic}
\vspace{-1cm}
\end{minipage}%
\begin{minipage}{.5\textwidth}
\includegraphics[clip, trim=1cm 1cm 1.5cm 1.5cm, width=\textwidth]{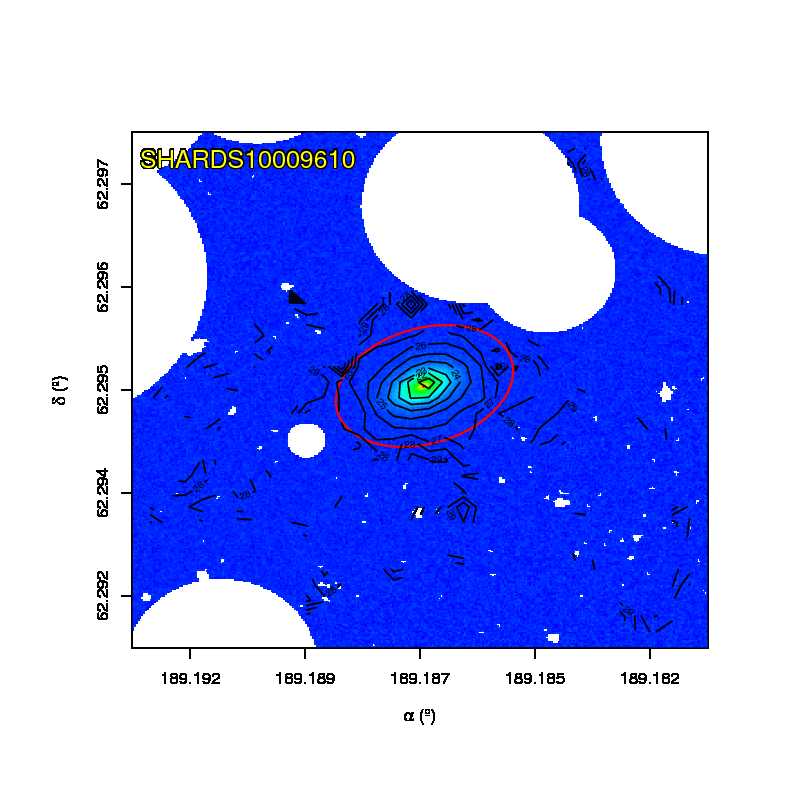}\vspace{-1cm}
\end{minipage}%

\begin{minipage}{.49\textwidth}
\includegraphics[clip, trim=0.1cm 0.1cm 0.1cm 0.1cm, width=\textwidth]{IMAGES/DECOMPOSITIONS/SHARDS10009610_plots_1.pdf}
\end{minipage}
\begin{minipage}{.49\textwidth}
\includegraphics[clip, trim=0.1cm 0.1cm 1cm 0.1cm, width=0.95\textwidth]{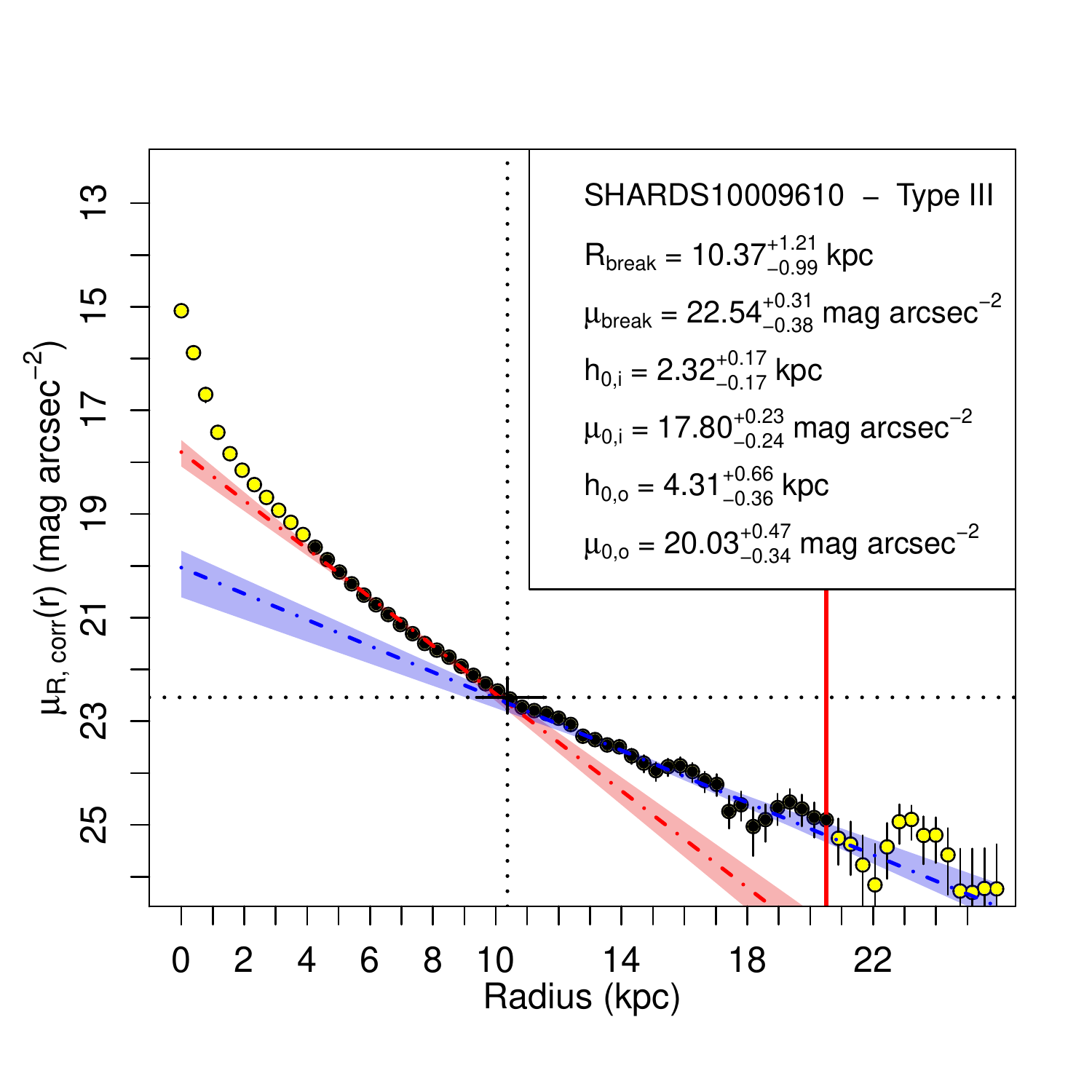}
\end{minipage}%

\vspace{-0.5cm}}
\caption[]{See caption of Fig.1. [\emph{Figure  available in the online edition}.]}         
\label{fig:img_final}
\end{figure}
\clearpage
\newpage

\textbf{SHARDS20000593:} Type-III S0 galaxy. The central bulge is very bright, and the ellipticity is low, but there is a prominent exponential component in the surface brightness profile that led us to classify it as S0, instead of elliptical. Extensive masking in the low emission regions. The disc profile presents a signficative excess of emission beyond $R\sim6$ kpc  with respect to the inner exponential section of the disc that cannot be explained by PSF contribution. The PDDs for $h$ and $\mu_{0}$ show two clearly separated peaks corresponding to the two profiles. 

\begin{figure}[!h]
{\centering
\vspace{-0cm}

\begin{minipage}{.5\textwidth}
\hspace{1.2cm}
\begin{overpic}[width=0.8\textwidth]
{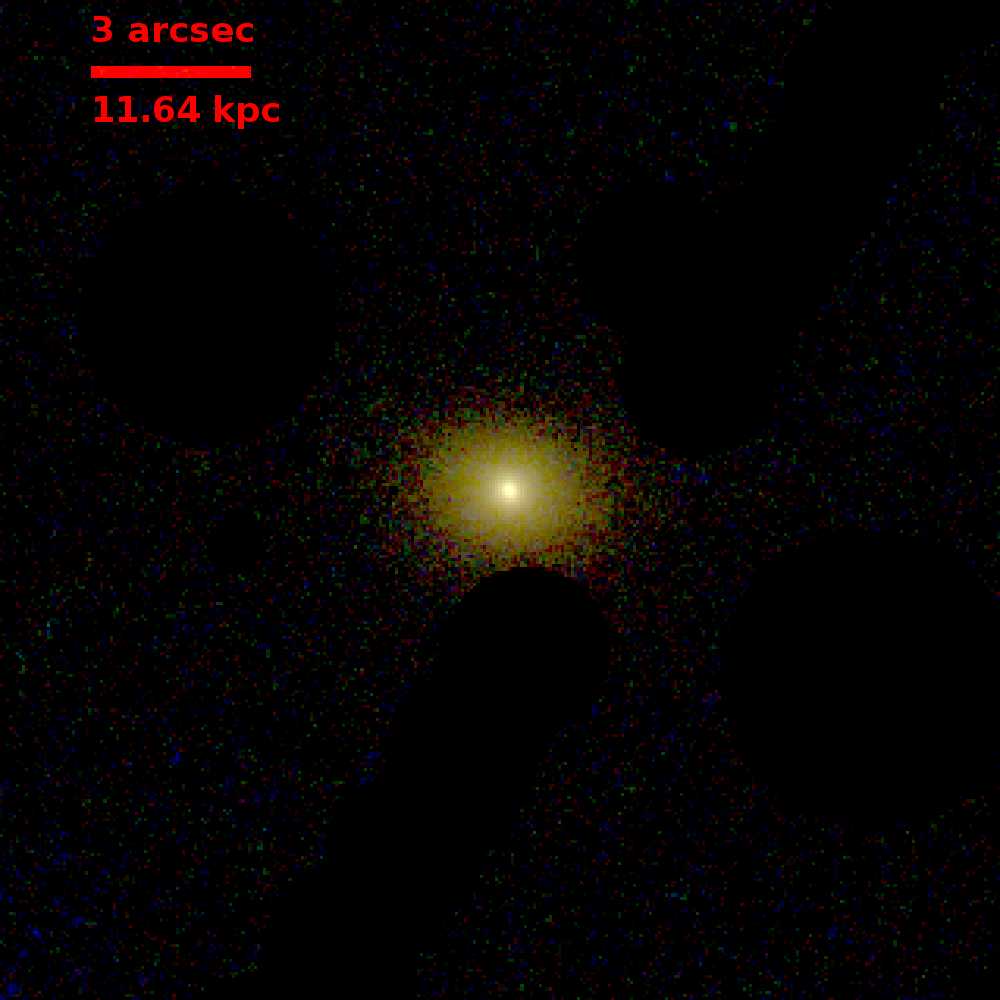}
\put(110,200){\color{yellow} \textbf{SHARDS20000593}}
\put(110,190){\color{yellow} \textbf{z=0.2470}}
\put(110,180){\color{yellow} \textbf{S0}}
\end{overpic}
\vspace{-1cm}
\end{minipage}%
\begin{minipage}{.5\textwidth}
\includegraphics[clip, trim=1cm 1cm 1.5cm 1.5cm, width=\textwidth]{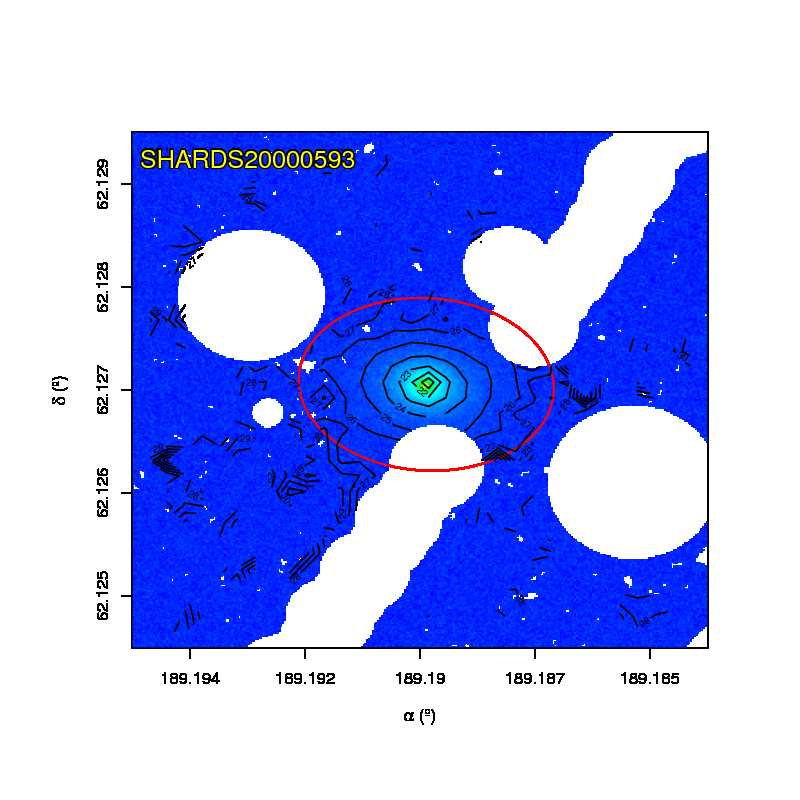}\vspace{-1cm}
\end{minipage}%

\begin{minipage}{.49\textwidth}
\includegraphics[clip, trim=0.1cm 0.1cm 0.1cm 0.1cm, width=\textwidth]{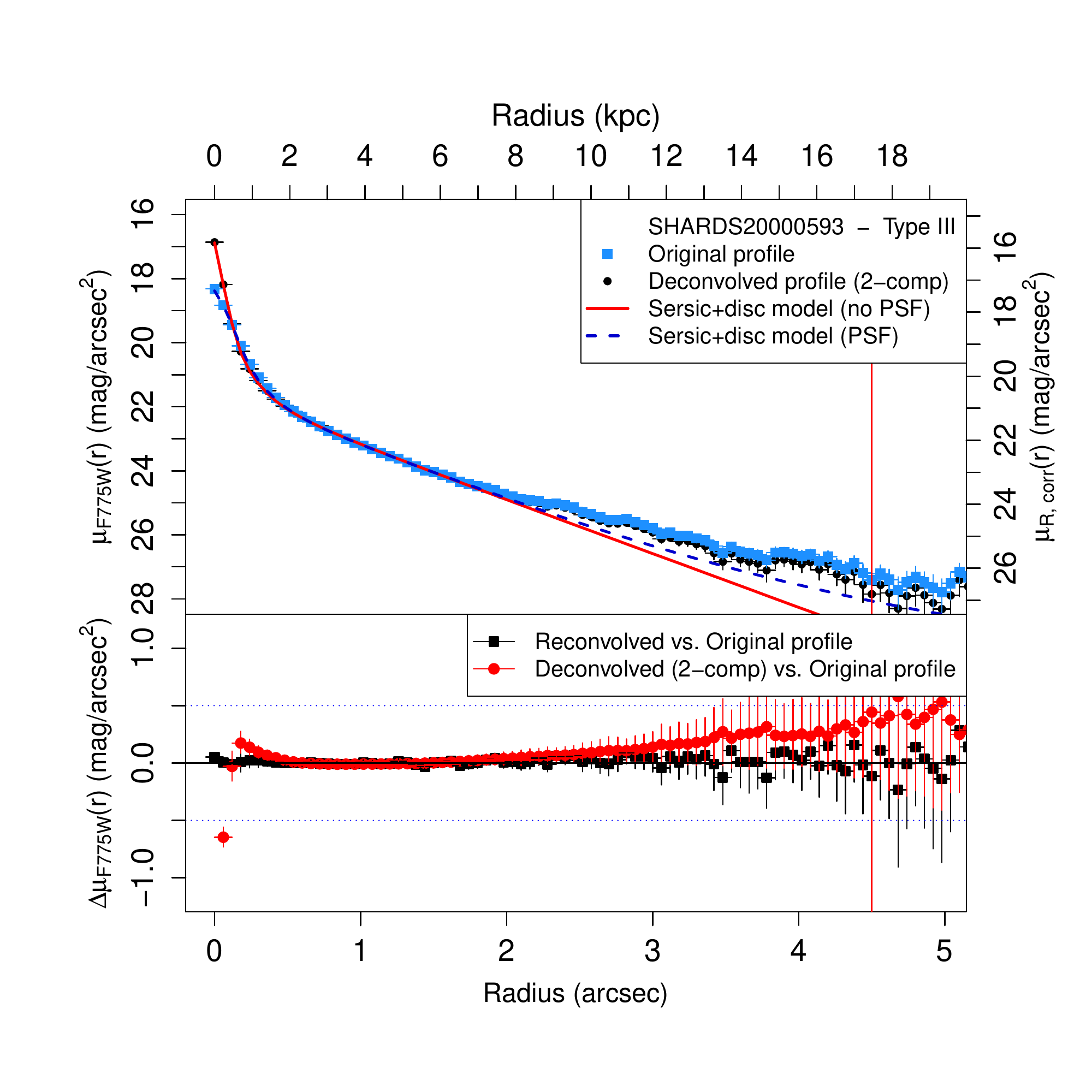}
\end{minipage}
\begin{minipage}{.49\textwidth}
\includegraphics[clip, trim=0.1cm 0.1cm 1cm 0.1cm, width=0.95\textwidth]{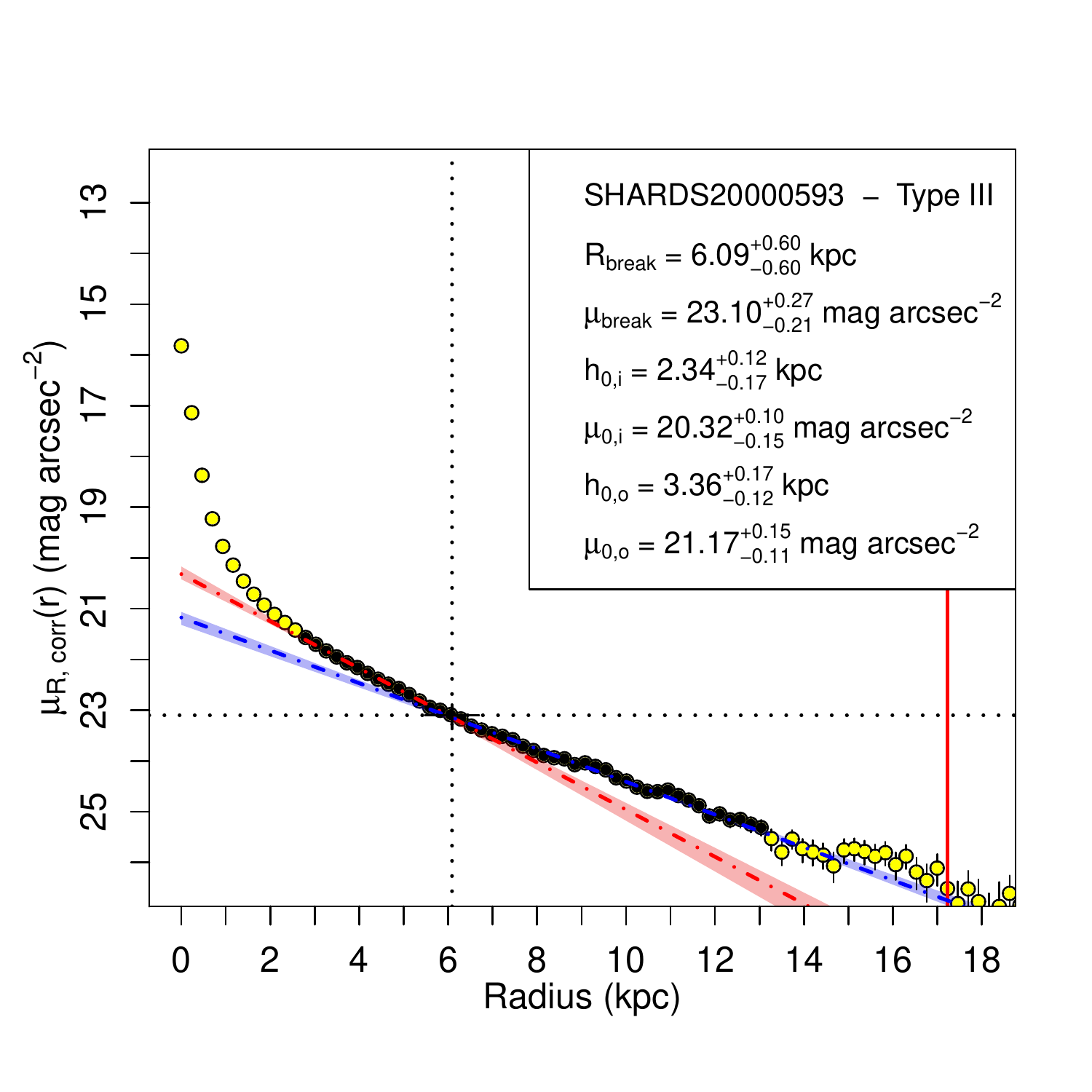}
\end{minipage}%

\vspace{-0.5cm}}
\caption[]{See caption of Fig.1. [\emph{Figure  available in the online edition}.]}         
\label{fig:img_final}
\end{figure}
\clearpage
\newpage

\textbf{SHARDS20000827:} Type-III S0 galaxy. It is highly inclined (see Table \ref{tab:fits_psforr})
so it was analysed with {\tt{ISOFIT}} instead of {\tt{Ellipse}}. The profile presents a signficative excess of emission since $R\sim10$ kpc that cannot be explained by PSF contribution. The PDDs for $h$ and $\mu_{0}$ show two clearly separated peaks corresponding to the two profiles, although they present large uncertainities for the outer profile fit. 

\begin{figure}[!h]
{\centering
\vspace{-0cm}

\begin{minipage}{.5\textwidth}
\hspace{1.2cm}
\begin{overpic}[width=0.8\textwidth]
{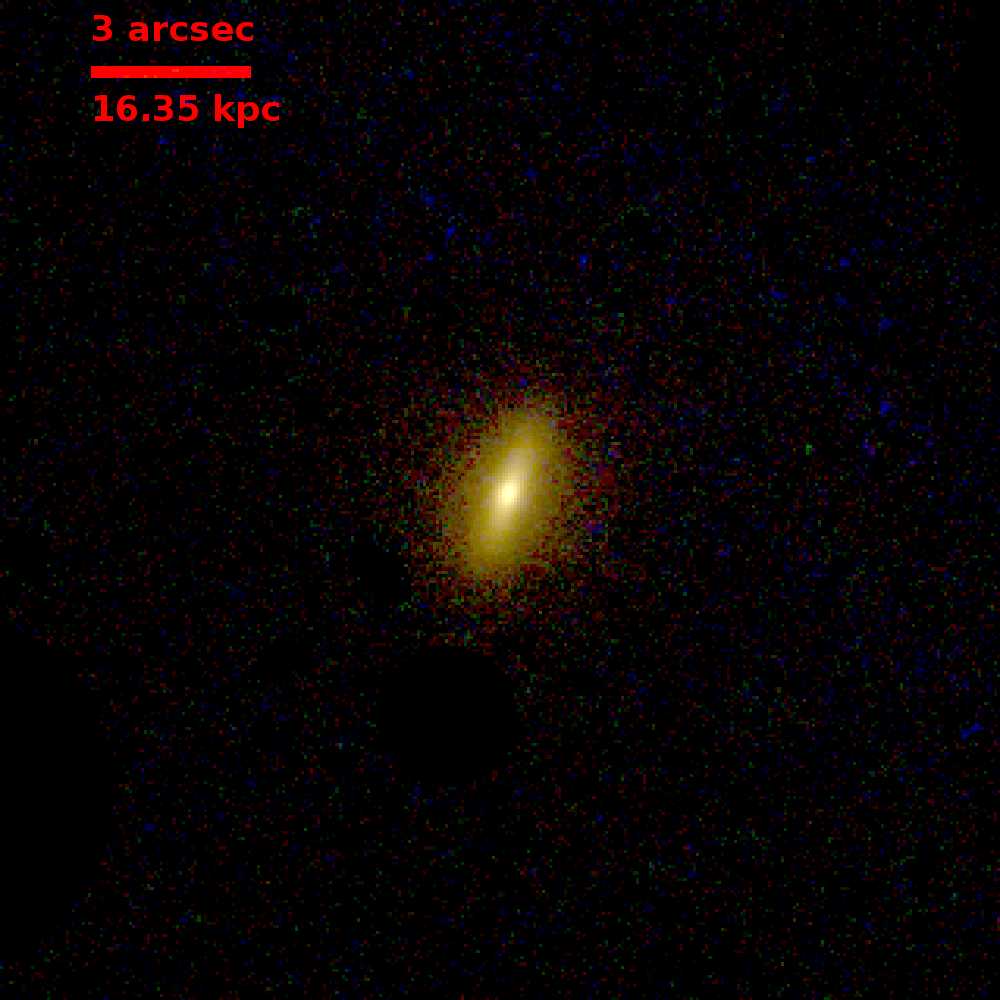}
\put(110,200){\color{yellow} \textbf{SHARDS20000827}}
\put(110,190){\color{yellow} \textbf{z=0.4090}}
\put(110,180){\color{yellow} \textbf{S0}}
\end{overpic}
\vspace{-1cm}
\end{minipage}%
\begin{minipage}{.5\textwidth}
\includegraphics[clip, trim=1cm 1cm 1.5cm 1.5cm, width=\textwidth]{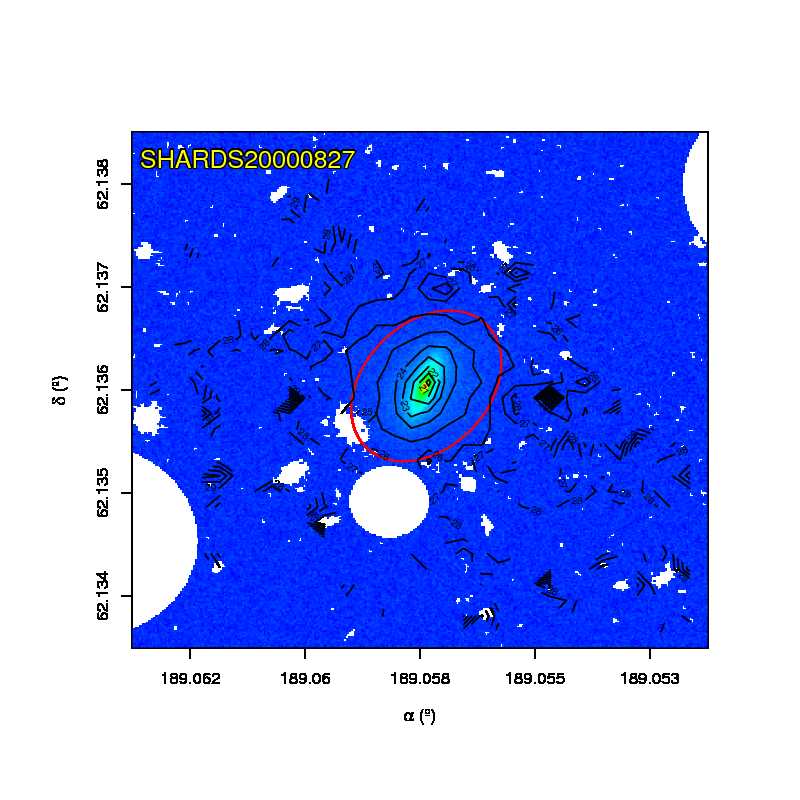}\vspace{-1cm}
\end{minipage}%

\begin{minipage}{.49\textwidth}
\includegraphics[clip, trim=0.1cm 0.1cm 0.1cm 0.1cm, width=\textwidth]{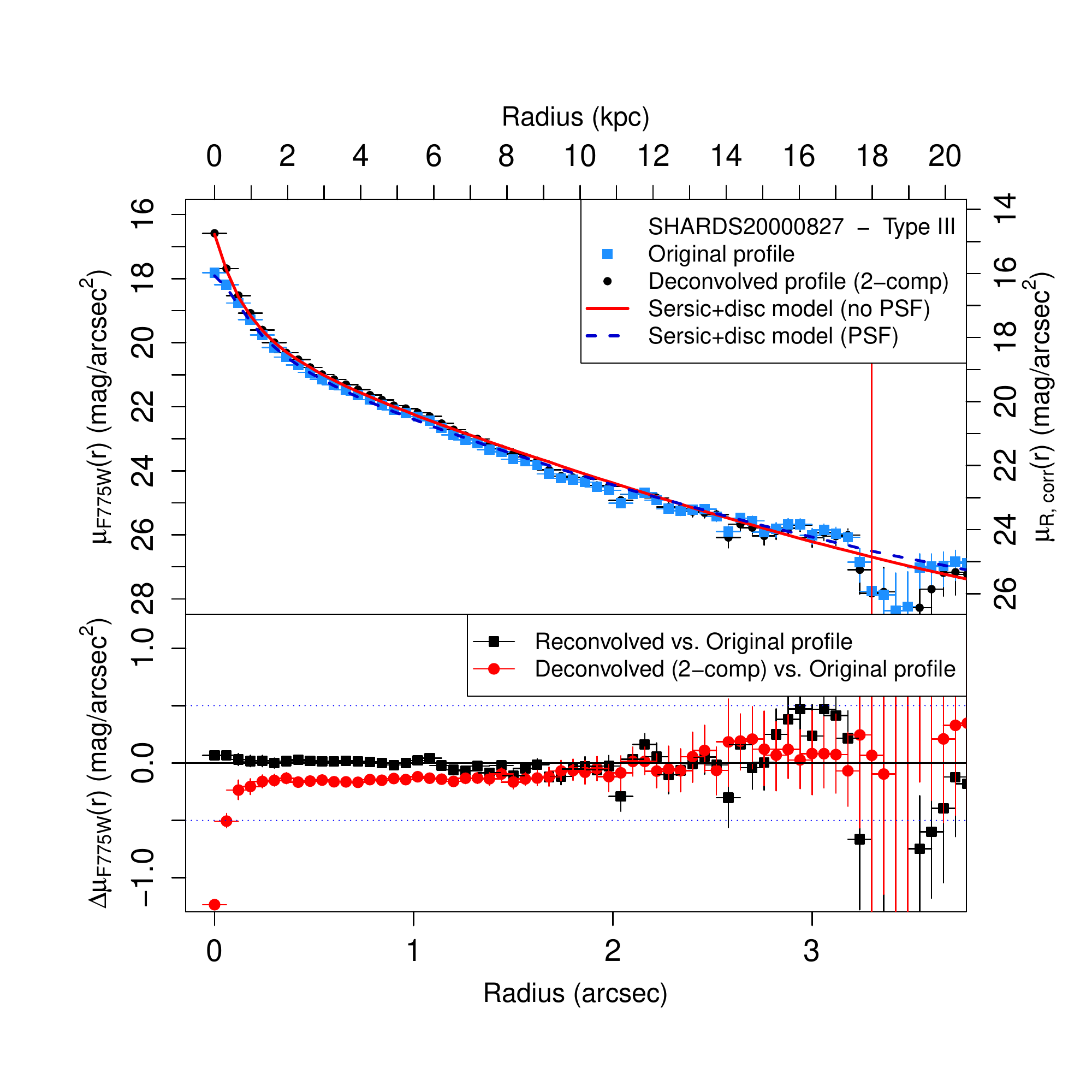}
\end{minipage}
\begin{minipage}{.49\textwidth}
\includegraphics[clip, trim=0.1cm 0.1cm 1cm 0.1cm, width=0.95\textwidth]{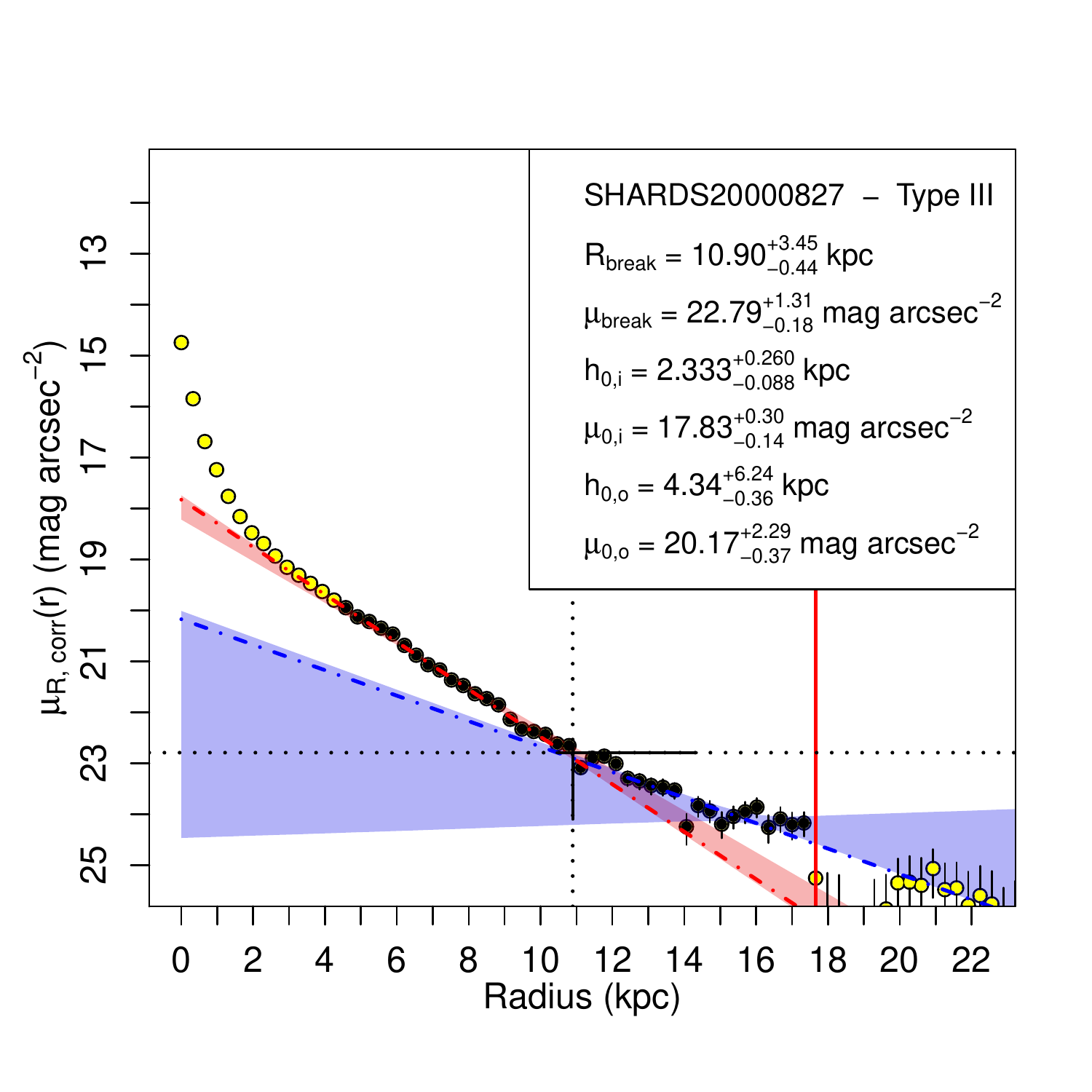}
\end{minipage}%

\vspace{-0.5cm}}
\caption[]{See caption of Fig.1. [\emph{Figure  available in the online edition}.]}         
\label{fig:img_final}
\end{figure}
\clearpage
\newpage

\textbf{SHARDS20001051:} Small S0 galaxy with a Type-I profile. The object appears to be completely face on (see Table \ref{tab:fits_psforr}), but inspection of the surface brightness profile reveals a typical bulge + exponential disc distribution. The disc appears to be featureless, and {\tt{Elbow}} does not reveal any significant breaks within the observed disc region. 

\begin{figure}[!h]
{\centering
\vspace{-0cm}

\begin{minipage}{.5\textwidth}
\hspace{1.2cm}
\begin{overpic}[width=0.8\textwidth]
{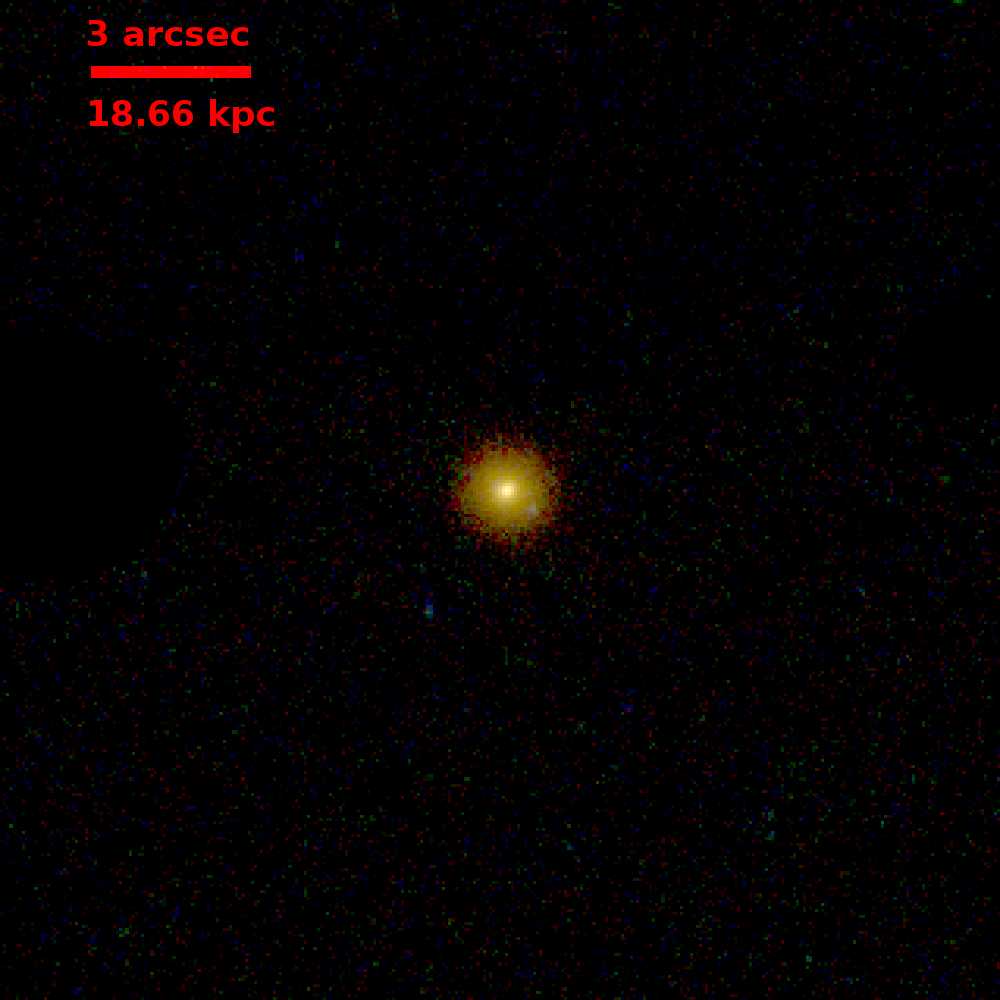}
\put(110,200){\color{yellow} \textbf{SHARDS20001051}}
\put(110,190){\color{yellow} \textbf{z=0.5183}}
\put(110,180){\color{yellow} \textbf{S0}}
\end{overpic}
\vspace{-1cm}
\end{minipage}%
\begin{minipage}{.5\textwidth}
\includegraphics[clip, trim=1cm 1cm 1.5cm 1.5cm, width=\textwidth]{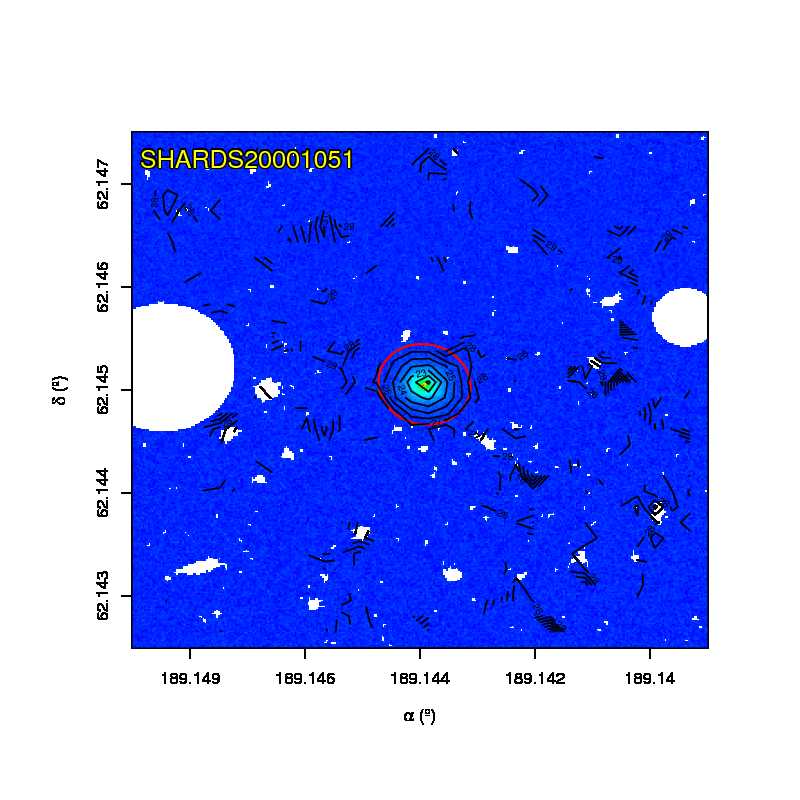}\vspace{-1cm}
\end{minipage}%

\begin{minipage}{.49\textwidth}
\includegraphics[clip, trim=0.1cm 0.1cm 0.1cm 0.1cm, width=\textwidth]{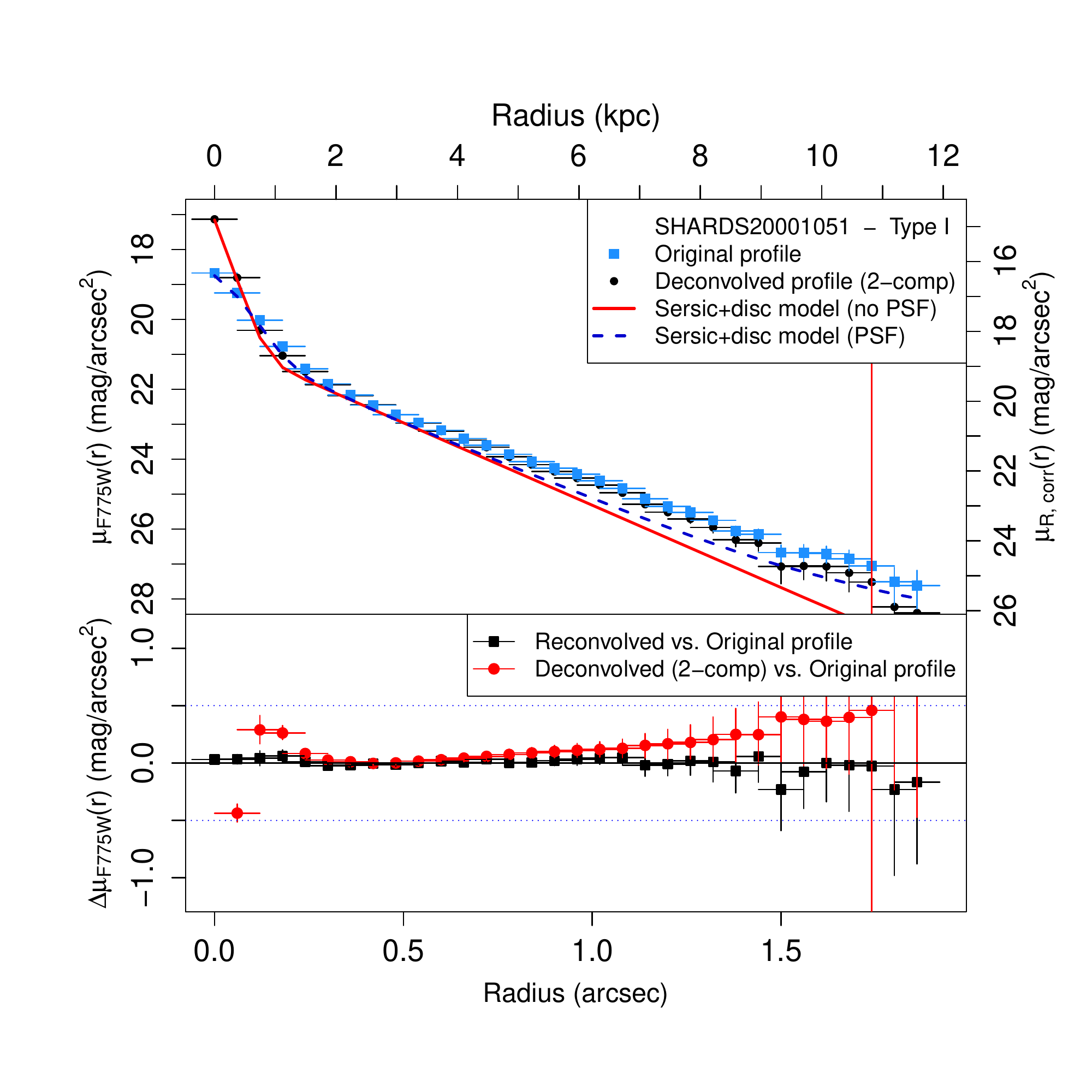}
\end{minipage}
\begin{minipage}{.49\textwidth}
\includegraphics[clip, trim=0.1cm 0.1cm 1cm 0.1cm, width=0.95\textwidth]{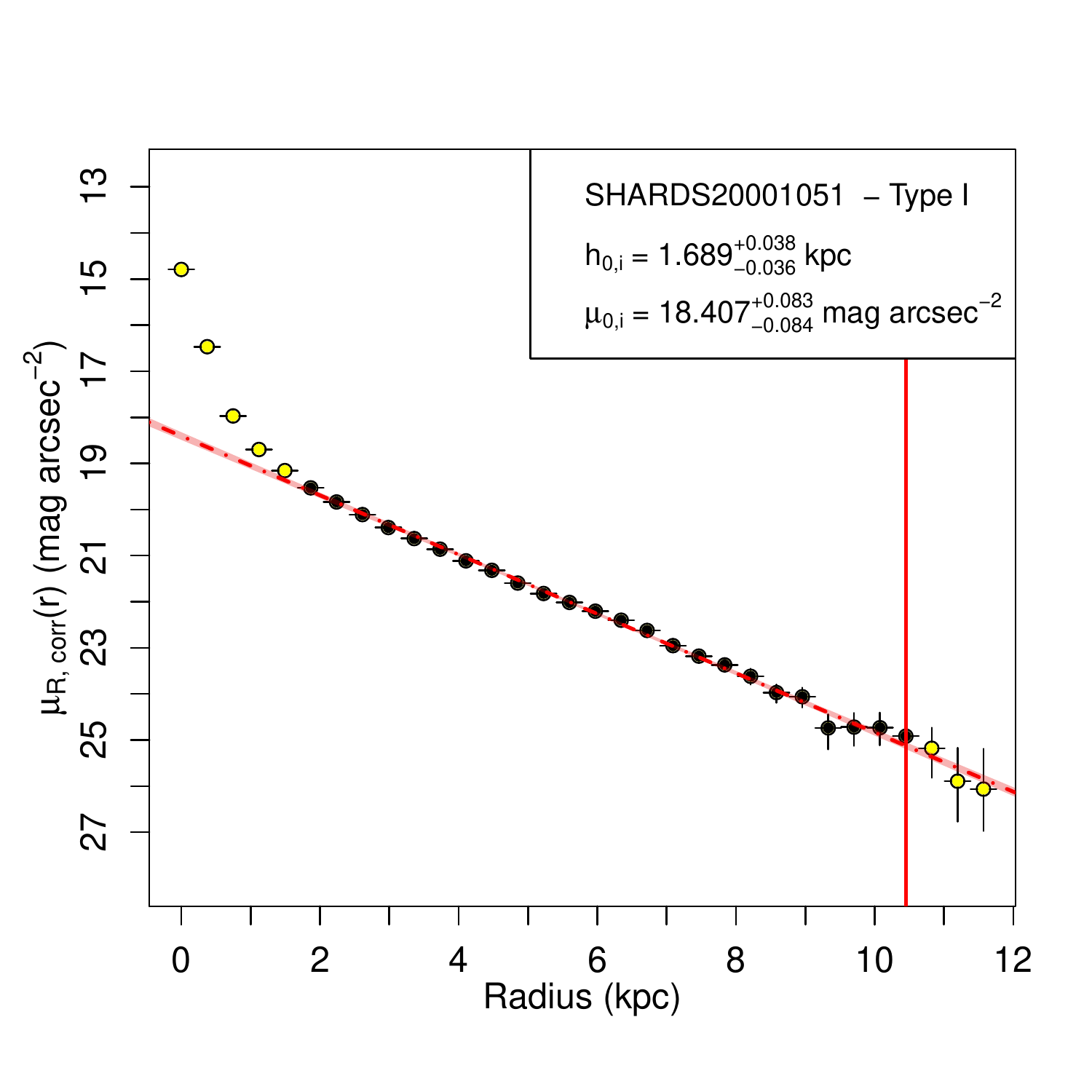}
\end{minipage}%

\vspace{-0.5cm}}
\caption[]{See caption of Fig.1. [\emph{Figure  available in the online edition}.]}         
\label{fig:img_final}
\end{figure}
\clearpage
\newpage

\textbf{SHARDS20001534:} Small S0 galaxy with a Type-I profile. It has a very similar profile to SHARDS20001051, but in this case, the object reveals a clear disc structure with medium to high inclination (see Table \ref{tab:fits_psforr}). After inspection of the surface brightness profile, we detected a typical bulge + exponential disc distribution. The disc appears to be featureless, and the automatic break analysis does not reveal any significant breaks within the limiting radius.

\begin{figure}[!h]
{\centering
\vspace{-0cm}

\begin{minipage}{.5\textwidth}
\hspace{1.2cm}
\begin{overpic}[width=0.8\textwidth]
{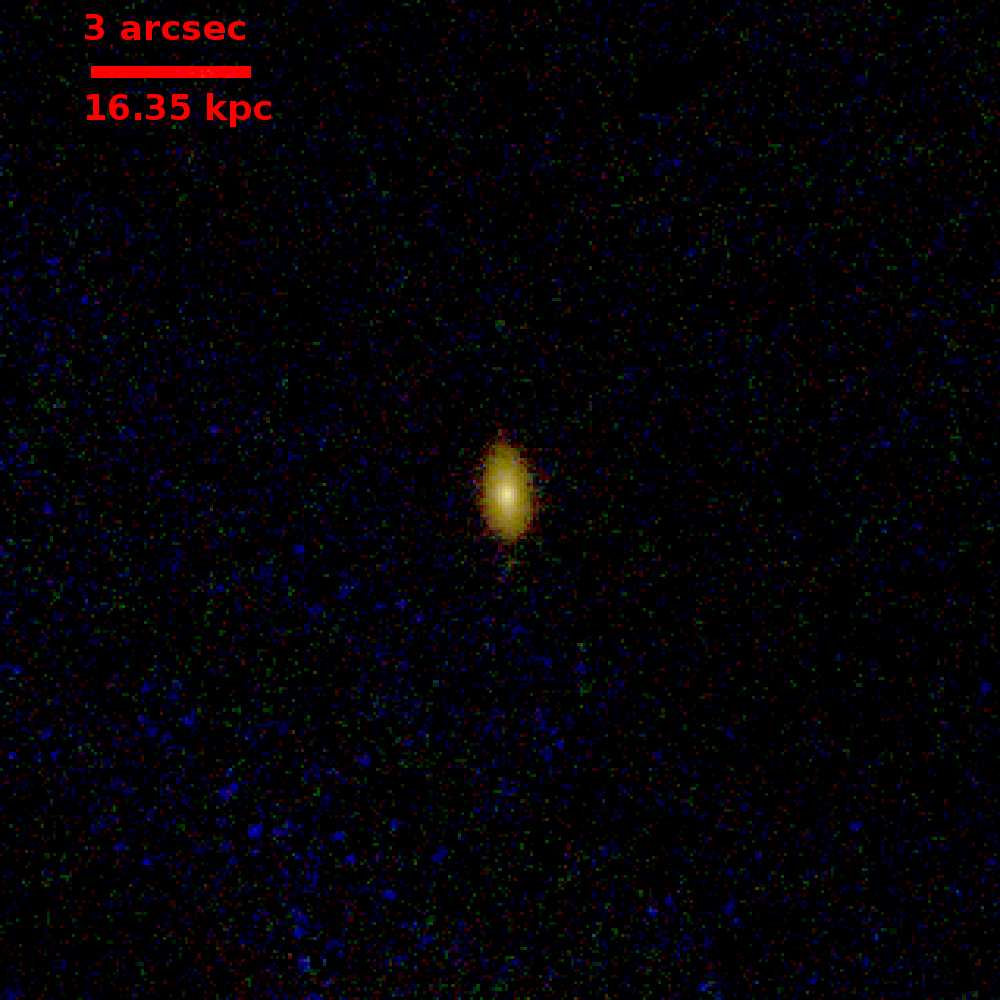}
\put(110,200){\color{yellow} \textbf{SHARDS20001534}}
\put(110,190){\color{yellow} \textbf{z=0.4101}}
\put(110,180){\color{yellow} \textbf{S0}}
\end{overpic}
\vspace{-1cm}
\end{minipage}%
\begin{minipage}{.5\textwidth}
\includegraphics[clip, trim=1cm 1cm 1.5cm 1.5cm, width=\textwidth]{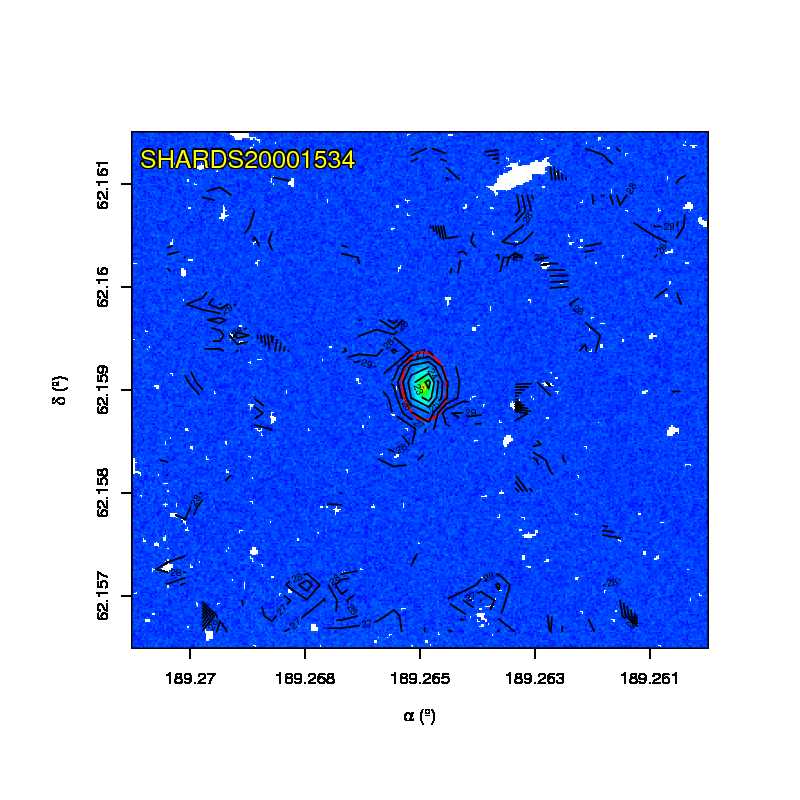}\vspace{-1cm}
\end{minipage}%

\begin{minipage}{.49\textwidth}
\includegraphics[clip, trim=0.1cm 0.1cm 0.1cm 0.1cm, width=\textwidth]{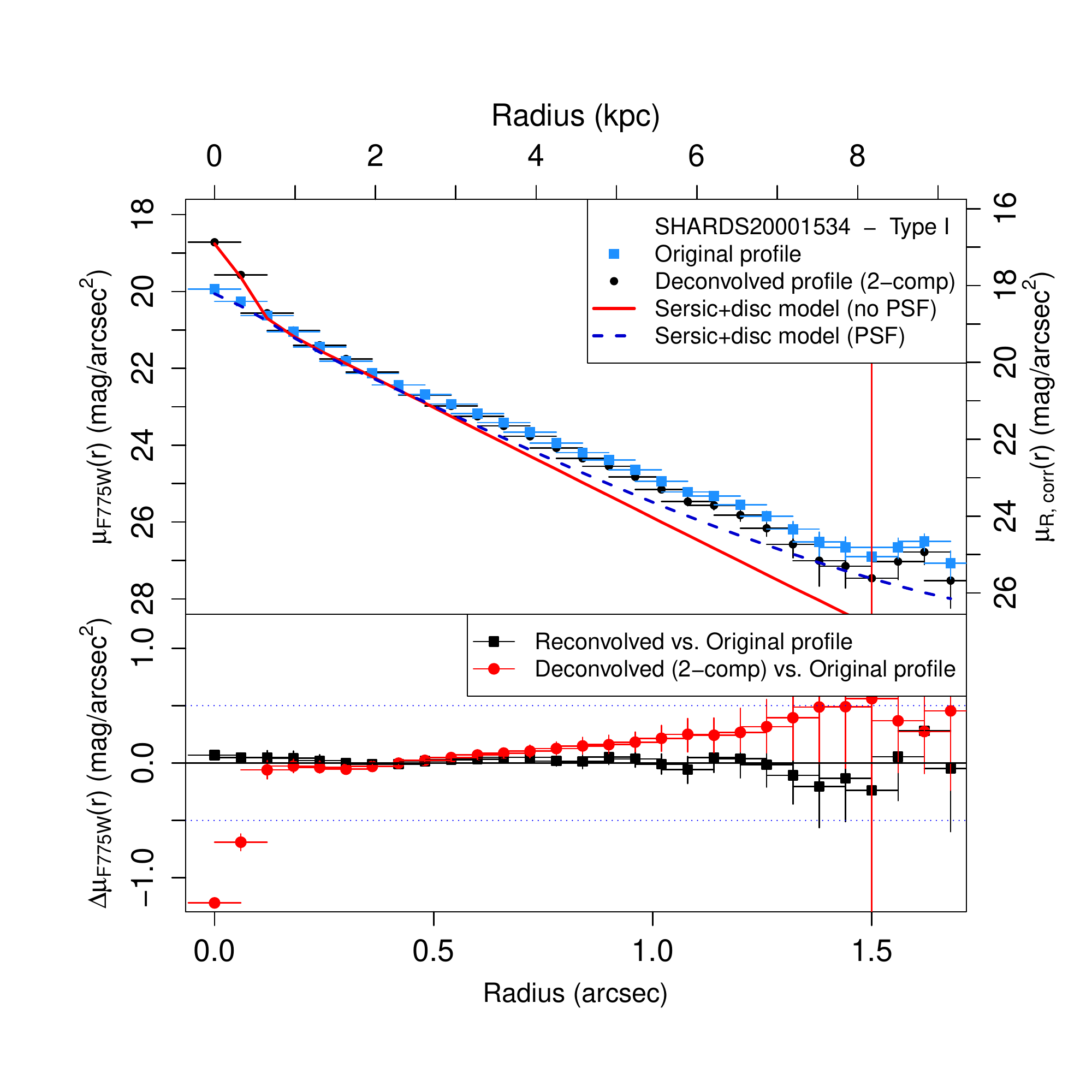}
\end{minipage}
\begin{minipage}{.49\textwidth}
\includegraphics[clip, trim=0.1cm 0.1cm 1cm 0.1cm, width=0.95\textwidth]{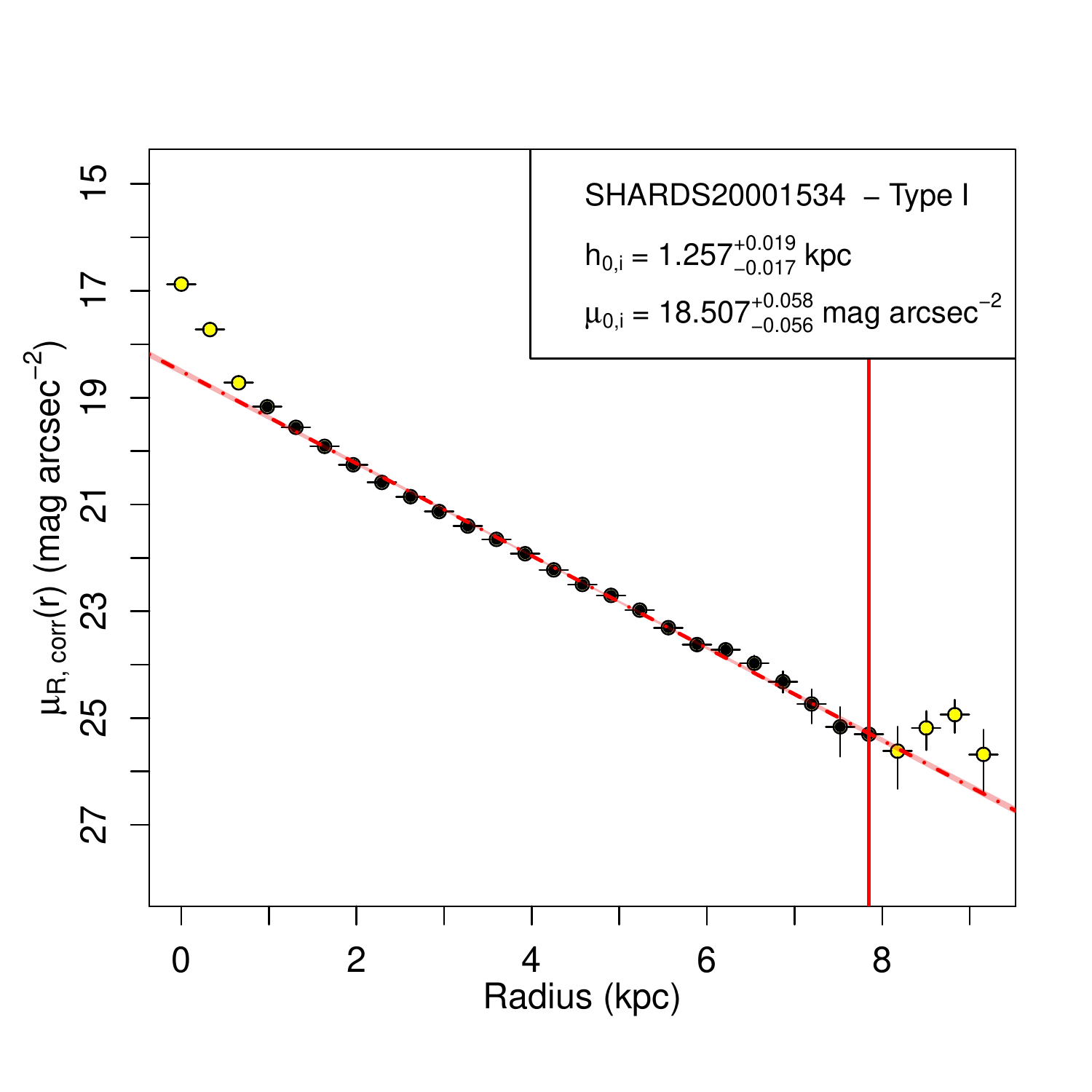}
\end{minipage}%

\vspace{-0.5cm}}
\caption[]{See caption of Fig.1. [\emph{Figure  available in the online edition}.]}         
\label{fig:img_final}
\end{figure}
\clearpage
\newpage

\textbf{SHARDS20002147:} Small S0 galaxy with a Type-I disc and with very similar morphology to SHARDS20001051. It was flagged as an AGN source (see Sect.\,\ref{Subsec:AGN}). The image presents some small isophotal irregularities, but they all are outside the fitting region. Masking was applied to two sources (to the N and SW of the main object). The surface brightness profile presents typical of bulge + exponential disc shape. The automatic break analysis reveals that, despite of the large uncertainities associated with the break parameters, the probability for this object of being compatible with a Type-I profile is $p=0.022$, thus we classified it as Type I. 

\begin{figure}[!h]
{\centering
\vspace{-0cm}

\begin{minipage}{.5\textwidth}
\hspace{1.2cm}
\begin{overpic}[width=0.8\textwidth]
{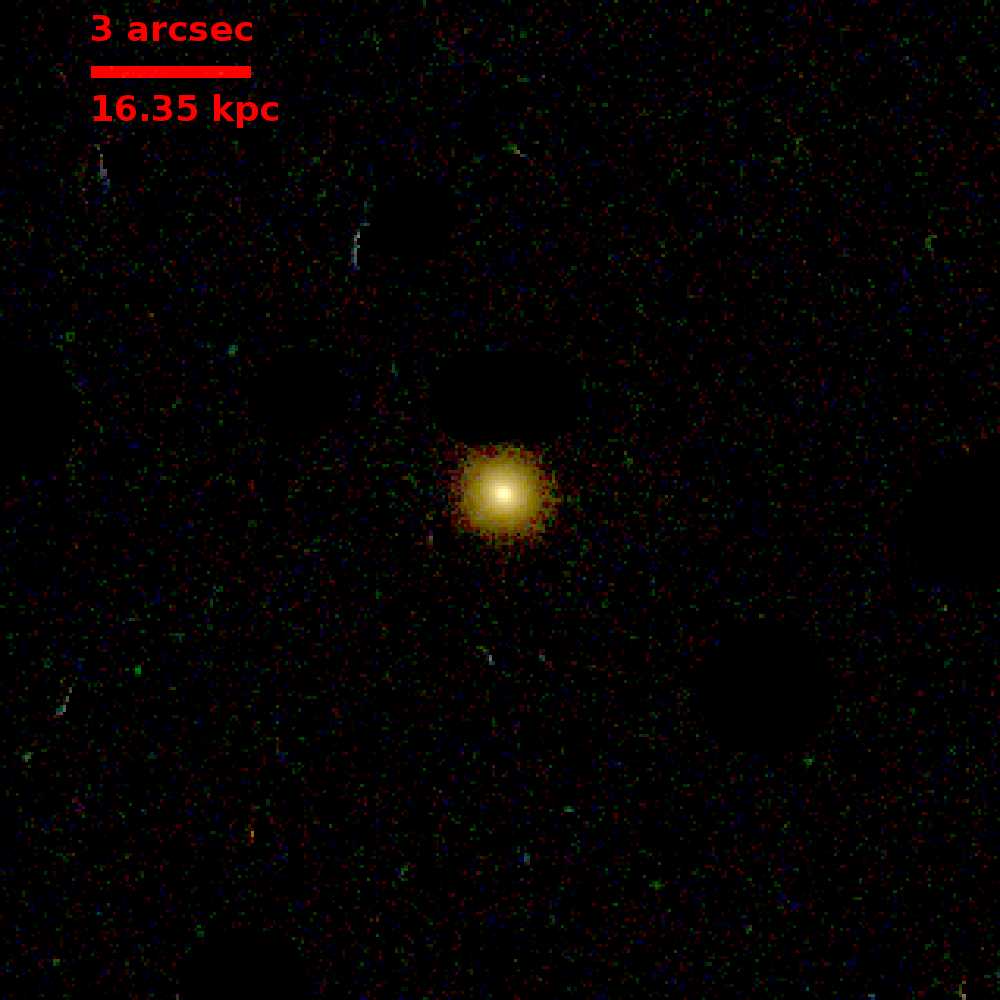}
\put(110,200){\color{yellow} \textbf{SHARDS20002147}}
\put(110,190){\color{yellow} \textbf{z=0.4101}}
\put(110,180){\color{yellow} \textbf{S0}}
\end{overpic}
\vspace{-1cm}
\end{minipage}%
\begin{minipage}{.5\textwidth}
\includegraphics[clip, trim=1cm 1cm 1.5cm 1.5cm, width=\textwidth]{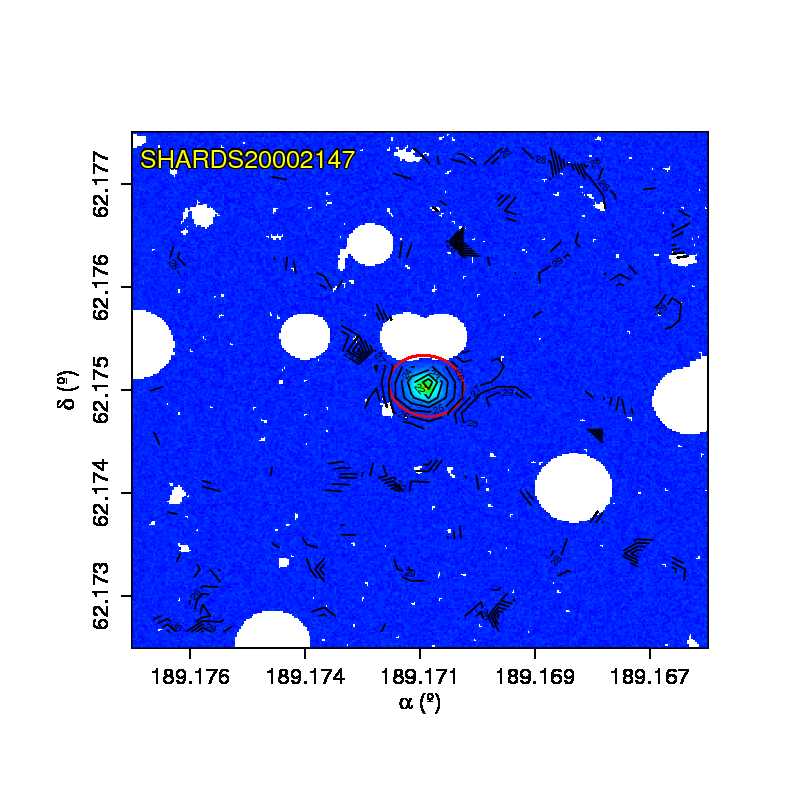}\vspace{-1cm}
\end{minipage}%

\begin{minipage}{.49\textwidth}
\includegraphics[clip, trim=0.1cm 0.1cm 0.1cm 0.1cm, width=\textwidth]{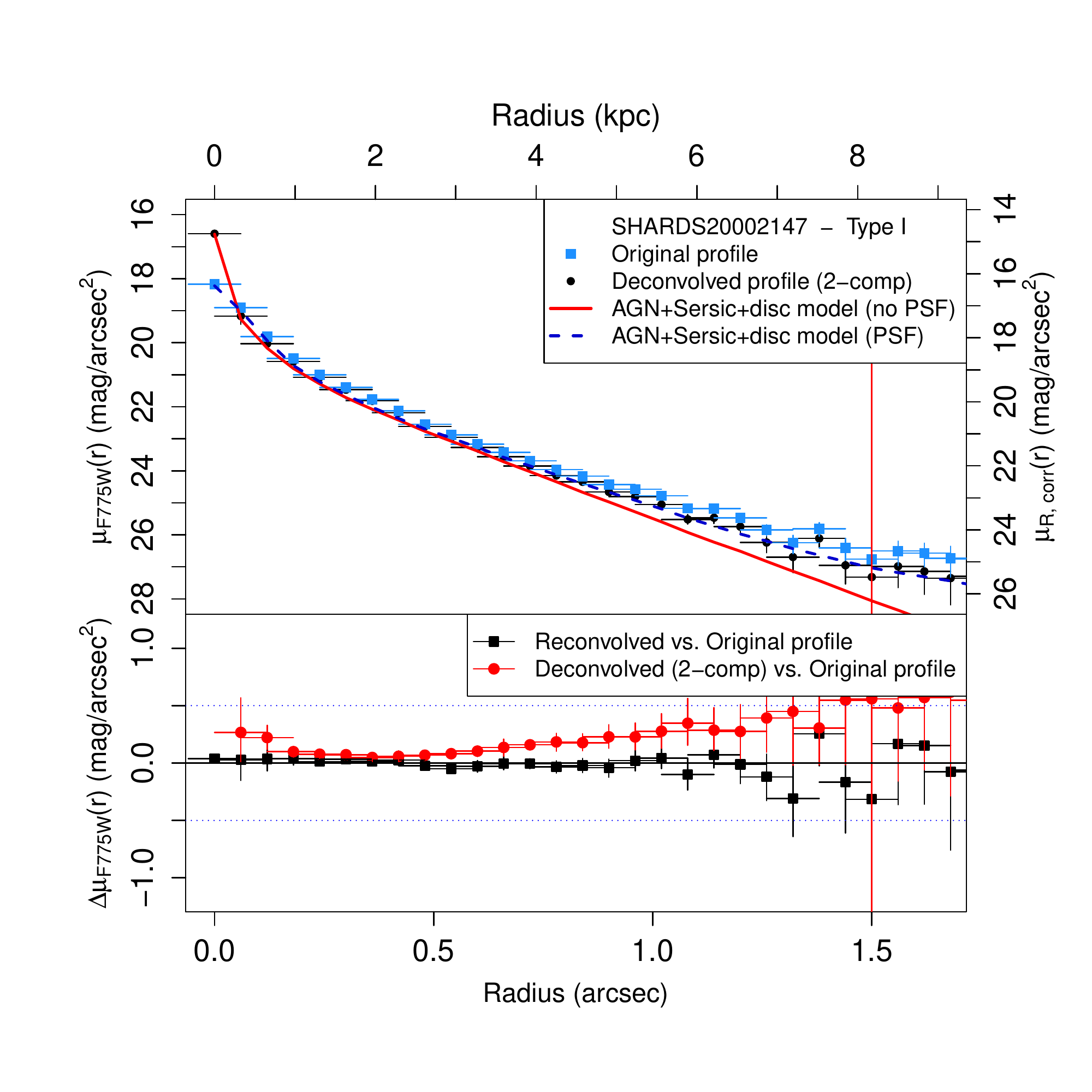}
\end{minipage}
\begin{minipage}{.49\textwidth}
\includegraphics[clip, trim=0.1cm 0.1cm 1cm 0.1cm, width=0.95\textwidth]{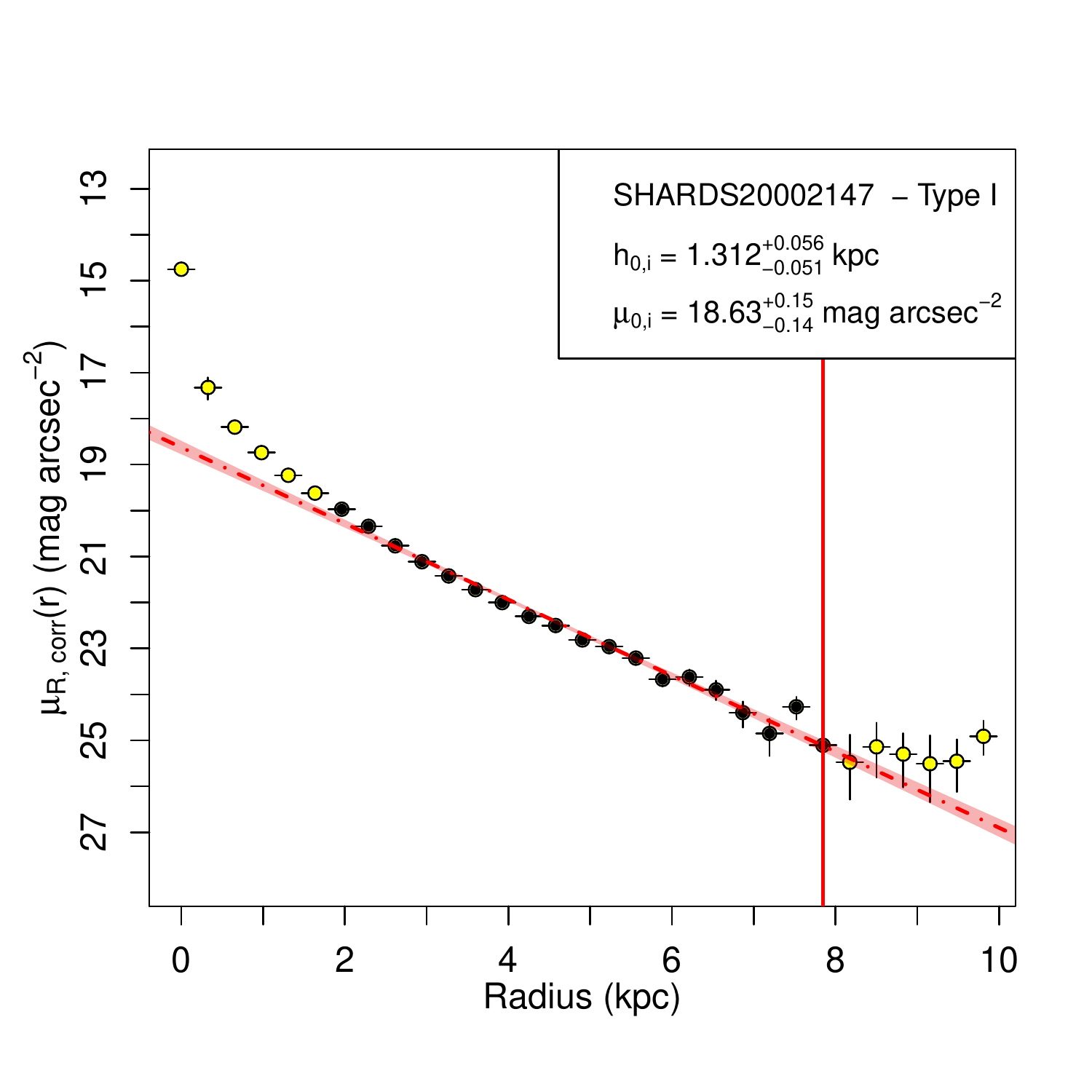}
\end{minipage}%

\vspace{-0.5cm}}
\caption[]{See caption of Fig.1. [\emph{Figure  available in the online edition}.]}         
\label{fig:img_final}
\end{figure}
\clearpage
\newpage

\textbf{SHARDS20002235:} E/S0 galaxy of Type-I disc partially overlapped with a galaxy of similar apparent size. Extensive manual masking has been applied to this object in order to extract the surface brightness profile from the farthest regions of the companion. The inspection of the surface brightness profile reveals a typical bulge + exponential disc distribution. The disc appears to be featureless, and the automatic break analysis does not detect any significant breaks within the limiting radius.

\begin{figure}[!h]
{\centering
\vspace{-0cm}

\begin{minipage}{.5\textwidth}
\hspace{1.2cm}
\begin{overpic}[width=0.8\textwidth]
{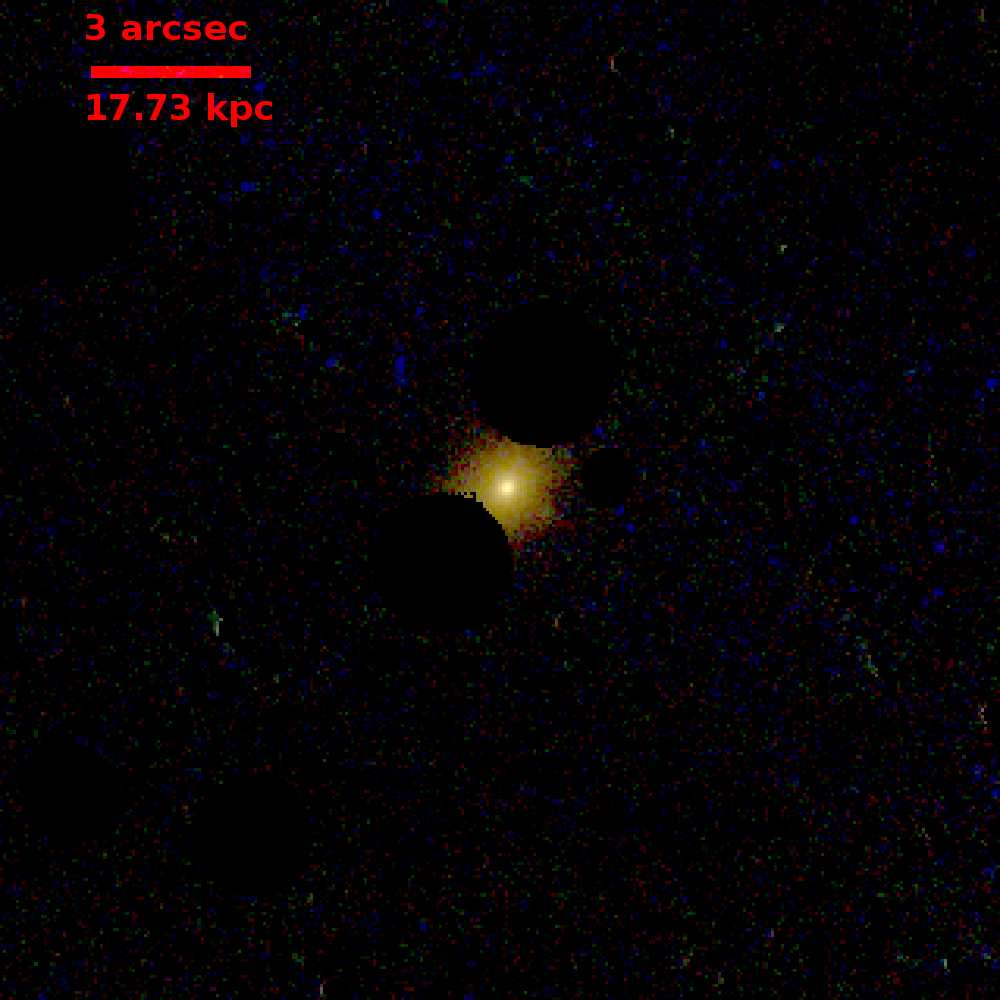}
\put(110,200){\color{yellow} \textbf{SHARDS20002235}}
\put(110,190){\color{yellow} \textbf{z=0.4714}}
\put(110,180){\color{yellow} \textbf{E/S0}}
\end{overpic}
\vspace{-1cm}
\end{minipage}%
\begin{minipage}{.5\textwidth}
\includegraphics[clip, trim=1cm 1cm 1.5cm 1.5cm, width=\textwidth]{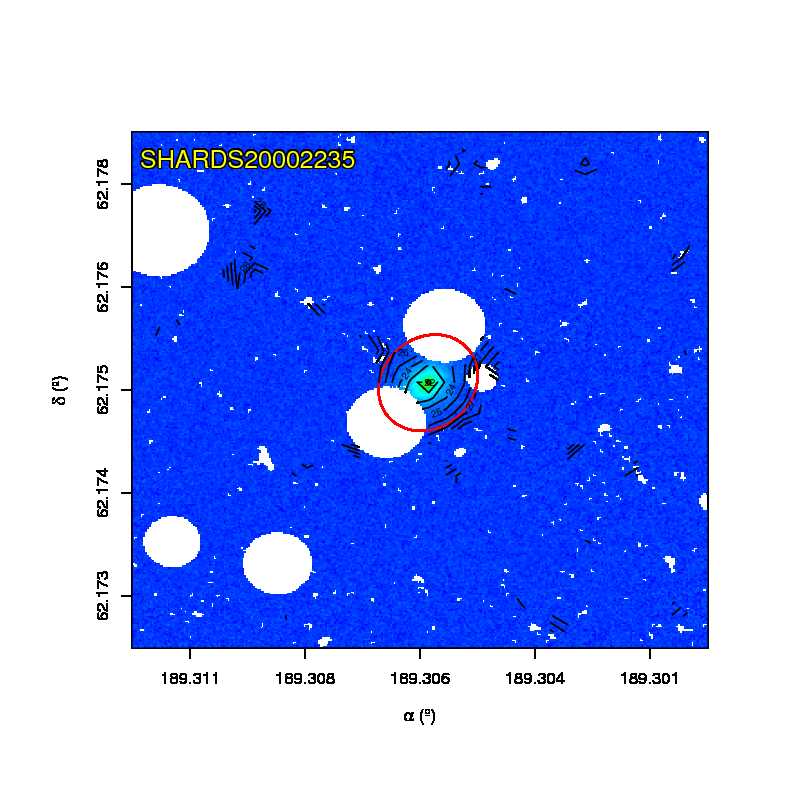}\vspace{-1cm}
\end{minipage}%

\begin{minipage}{.49\textwidth}
\includegraphics[clip, trim=0.1cm 0.1cm 0.1cm 0.1cm, width=\textwidth]{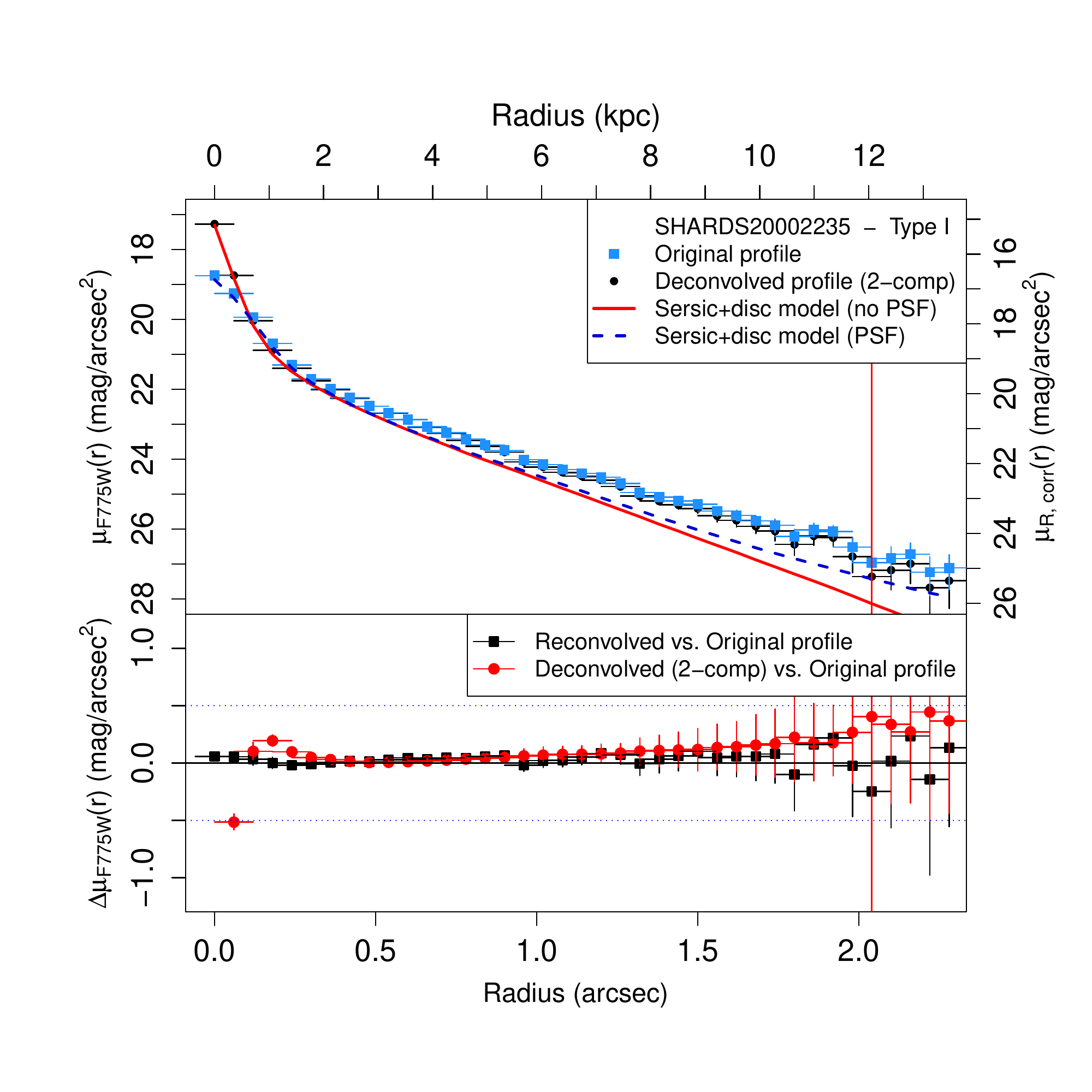}
\end{minipage}
\begin{minipage}{.49\textwidth}
\includegraphics[clip, trim=0.1cm 0.1cm 1cm 0.1cm, width=0.95\textwidth]{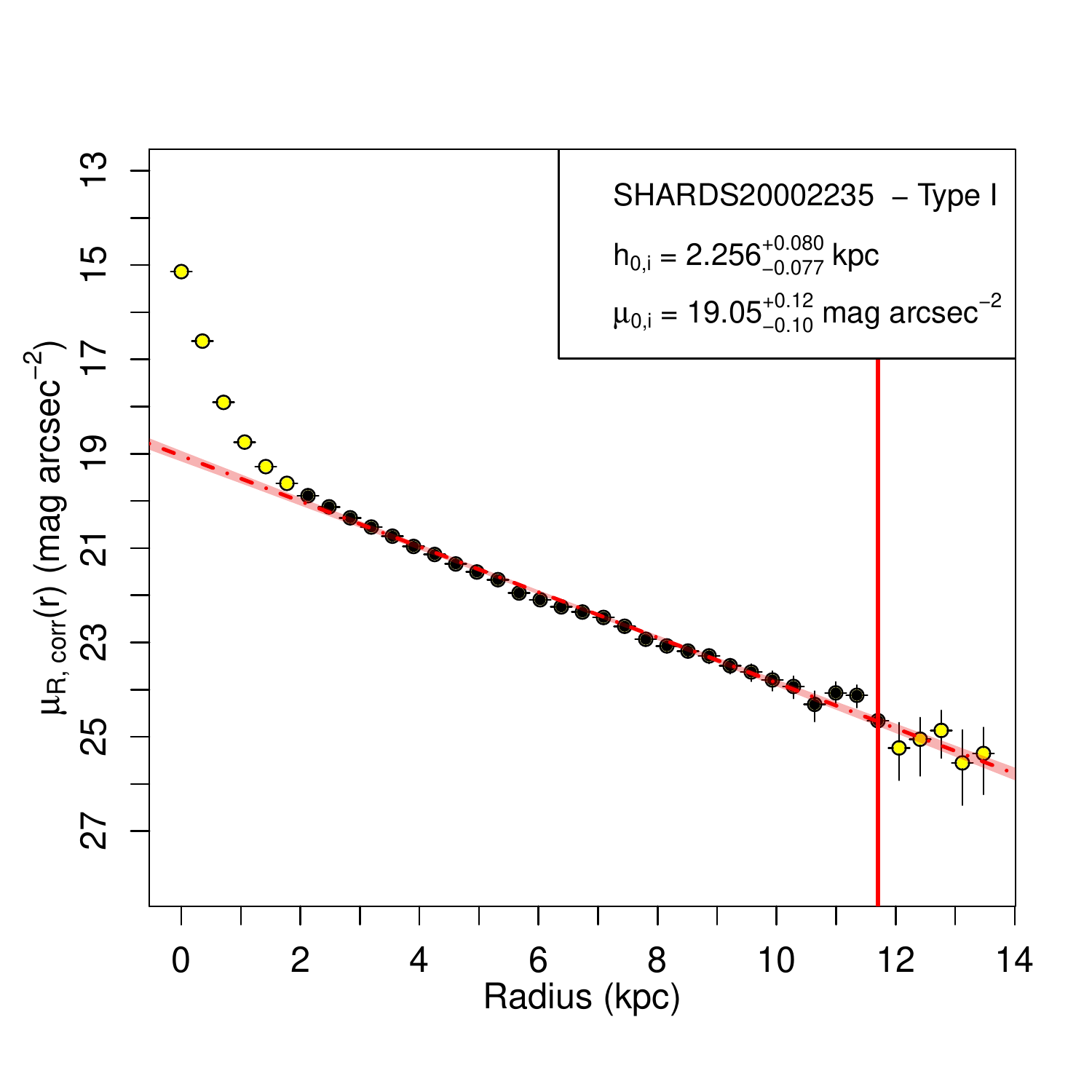}
\end{minipage}%

\vspace{-0.5cm}}
\caption[]{See caption of Fig.1. [\emph{Figure  available in the online edition}.]}         
\label{fig:img_final}
\end{figure}
\clearpage
\newpage

\textbf{SHARDS20002550:} E/S0 galaxy with a Type-I profile. The image required extensive masking due to the presence of a nearby saturated star. The masked area lies outside the final fitting region. It shows a clear bulge + exponential surface brightness profile. The automated break analysis revealed a noticeable change of profile in the transition zone between the bulge and the main disc, but we discarded because of being too close to the centre, in order to avoid bulge contributions. 
(see Table \ref{tab:fits_psforr})

\begin{figure}[!h]
{\centering
\vspace{-0cm}

\begin{minipage}{.5\textwidth}
\hspace{1.2cm}
\begin{overpic}[width=0.8\textwidth]
{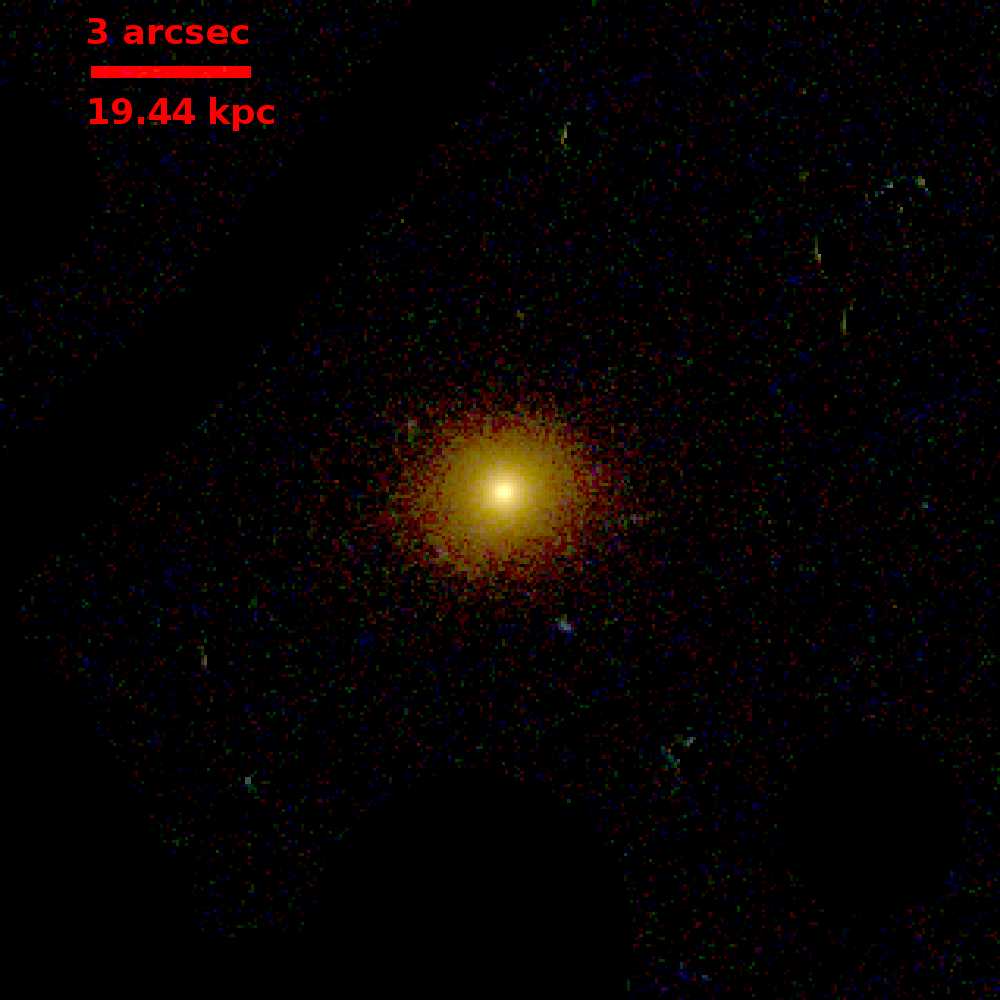}
\put(110,200){\color{yellow} \textbf{SHARDS20002550}}
\put(110,190){\color{yellow} \textbf{z=0.5616}}
\put(110,180){\color{yellow} \textbf{E/S0}}
\end{overpic}
\vspace{-1cm}
\end{minipage}%
\begin{minipage}{.5\textwidth}
\includegraphics[clip, trim=1cm 1cm 1.5cm 1.5cm, width=\textwidth]{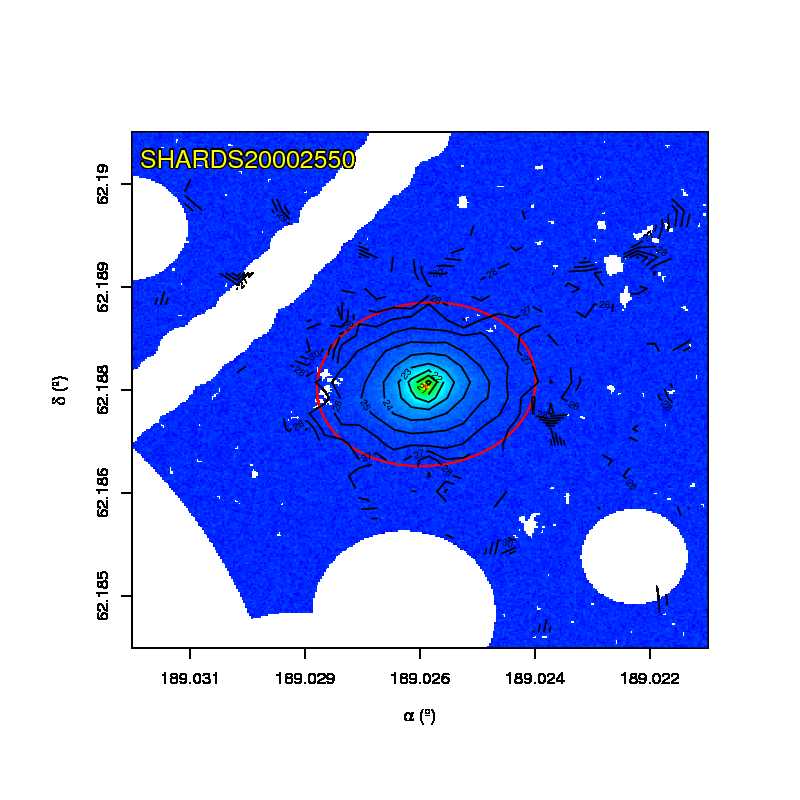}\vspace{-1cm}
\end{minipage}%

\begin{minipage}{.49\textwidth}
\includegraphics[clip, trim=0.1cm 0.1cm 0.1cm 0.1cm, width=\textwidth]{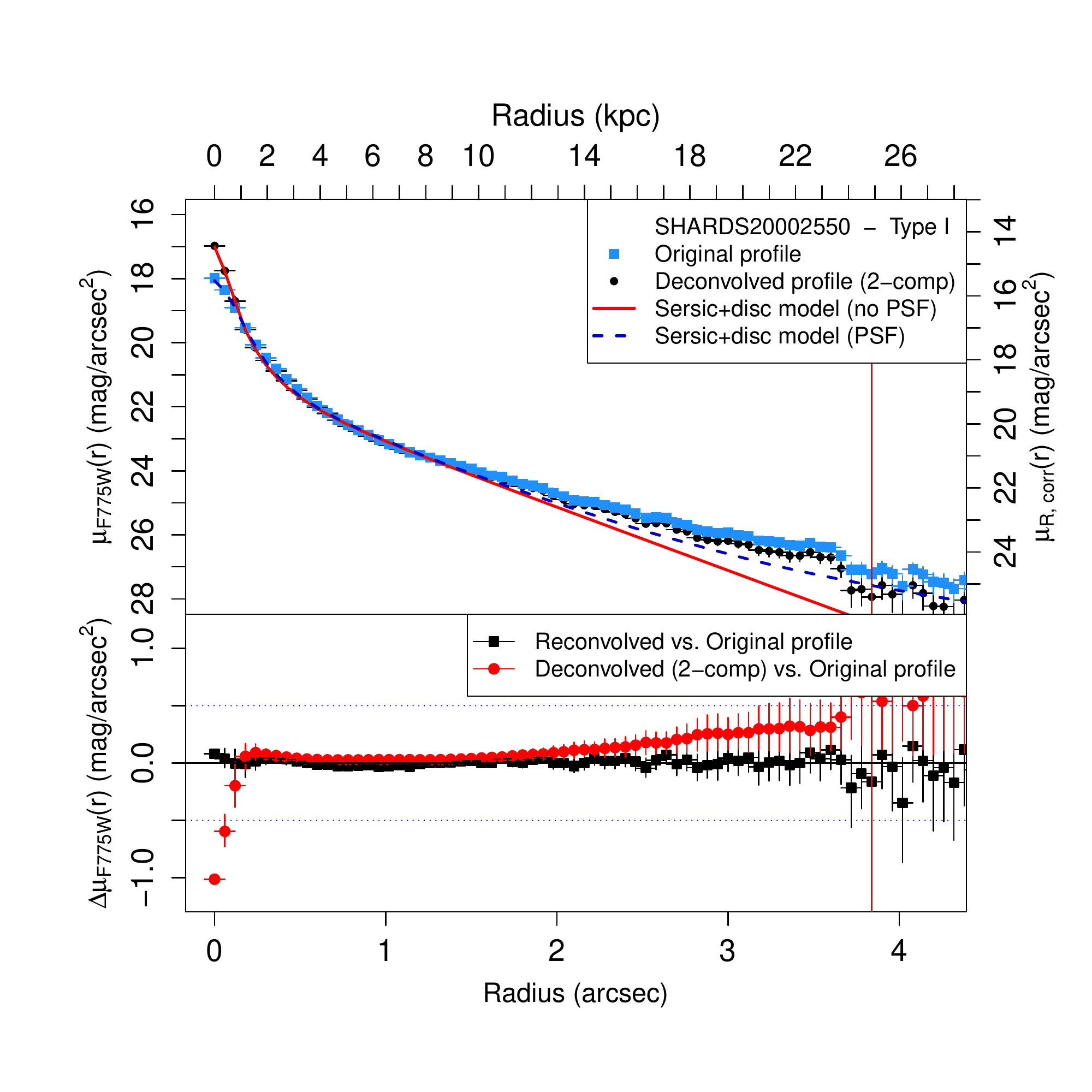}
\end{minipage}
\begin{minipage}{.49\textwidth}
\includegraphics[clip, trim=0.1cm 0.1cm 1cm 0.1cm, width=0.95\textwidth]{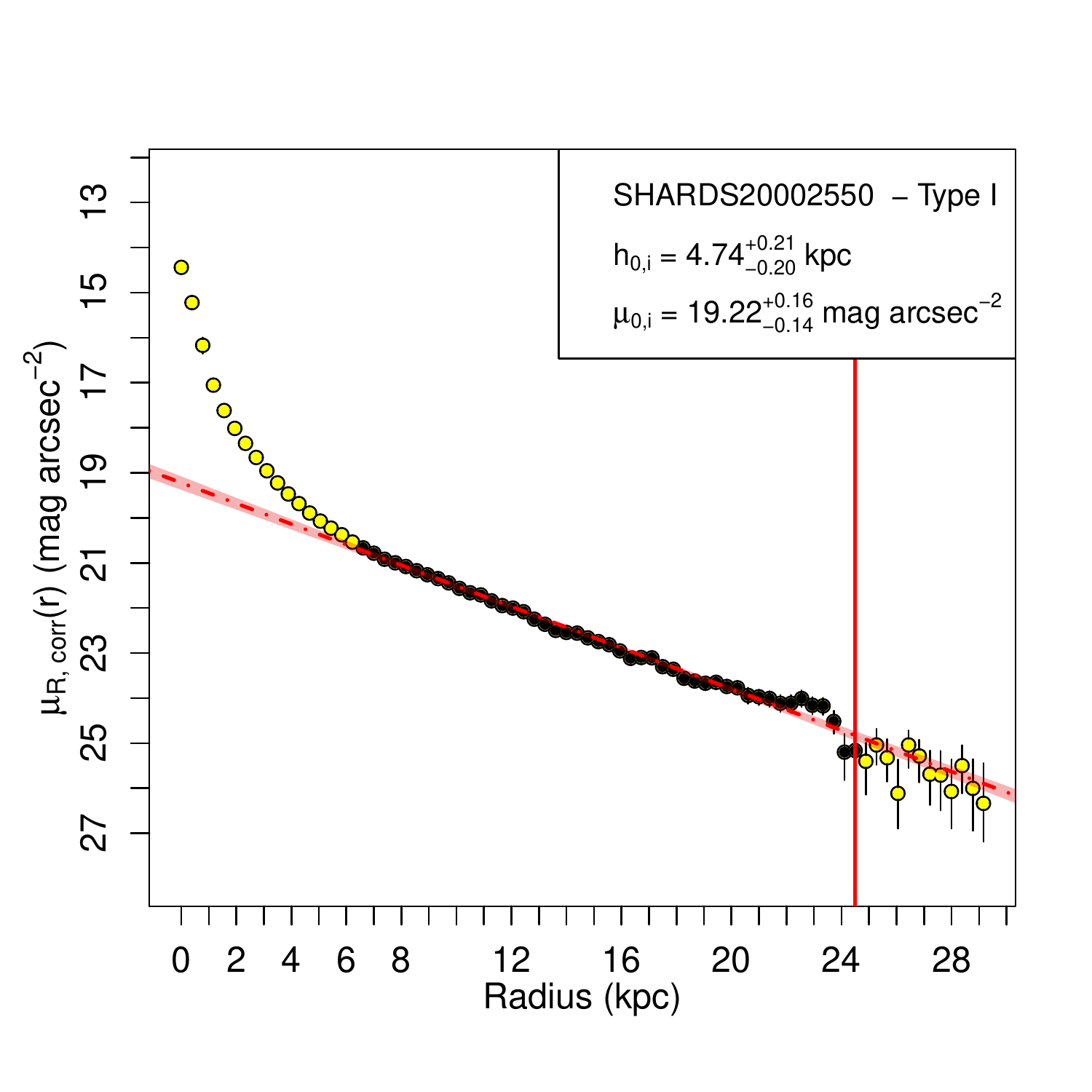}
\end{minipage}%

\vspace{-0.5cm}}
\caption[]{See caption of Fig.1. [\emph{Figure  available in the online edition}.]}         
\label{fig:img_final}
\end{figure}
\clearpage
\newpage

\textbf{SHARDS20002889:} Small S0 galaxy with Type-I profile. The original surface brightness profile appears to be almost bulgeless. It has a medium inclination and small apparent size. It has no nearby galaxies or field objects. The resolution is not enough to resolve any possible detail besides the disc itself. Automatic break analysis does not reveal significant differences of any part of the disc.

\begin{figure}[!h]
{\centering
\vspace{-0cm}

\begin{minipage}{.5\textwidth}
\hspace{1.2cm}
\begin{overpic}[width=0.8\textwidth]
{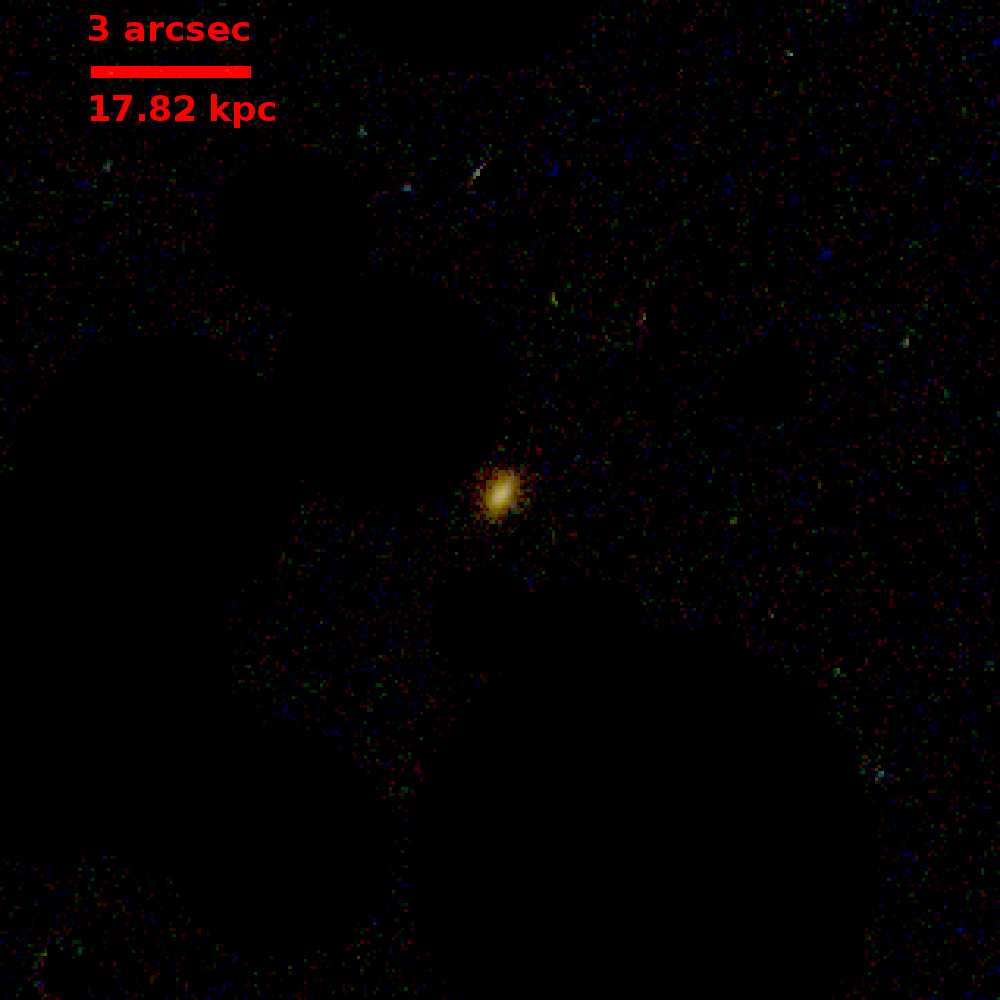}
\put(110,200){\color{yellow} \textbf{SHARDS20002889}}
\put(110,190){\color{yellow} \textbf{z=0.4760}}
\put(110,180){\color{yellow} \textbf{S0}}
\end{overpic}
\vspace{-1cm}
\end{minipage}%
\begin{minipage}{.5\textwidth}
\includegraphics[clip, trim=1cm 1cm 1.5cm 1.5cm, width=\textwidth]{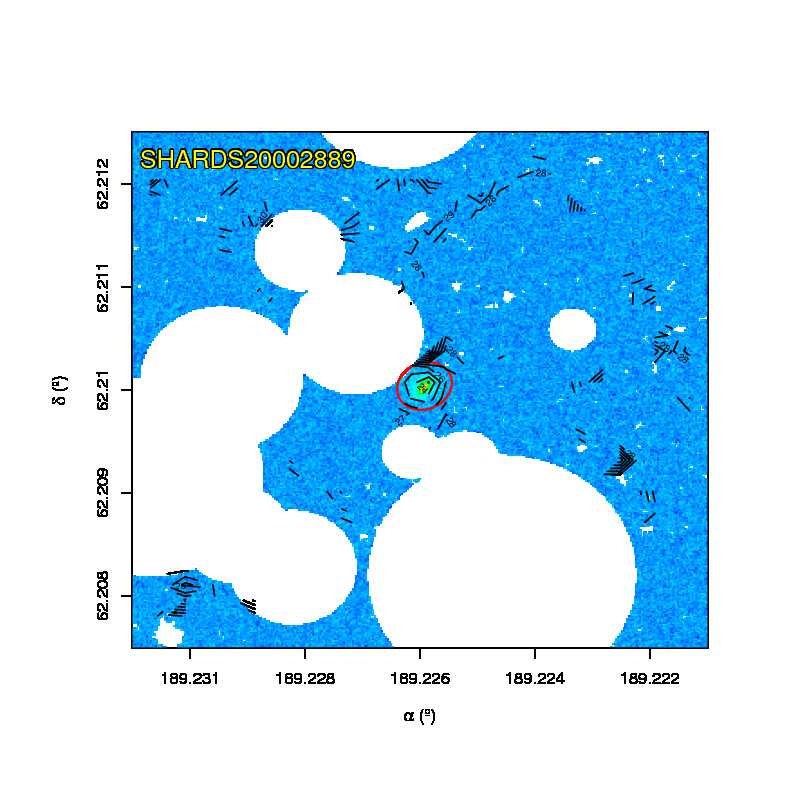}\vspace{-1cm}
\end{minipage}%

\begin{minipage}{.49\textwidth}
\includegraphics[clip, trim=0.1cm 0.1cm 0.1cm 0.1cm, width=\textwidth]{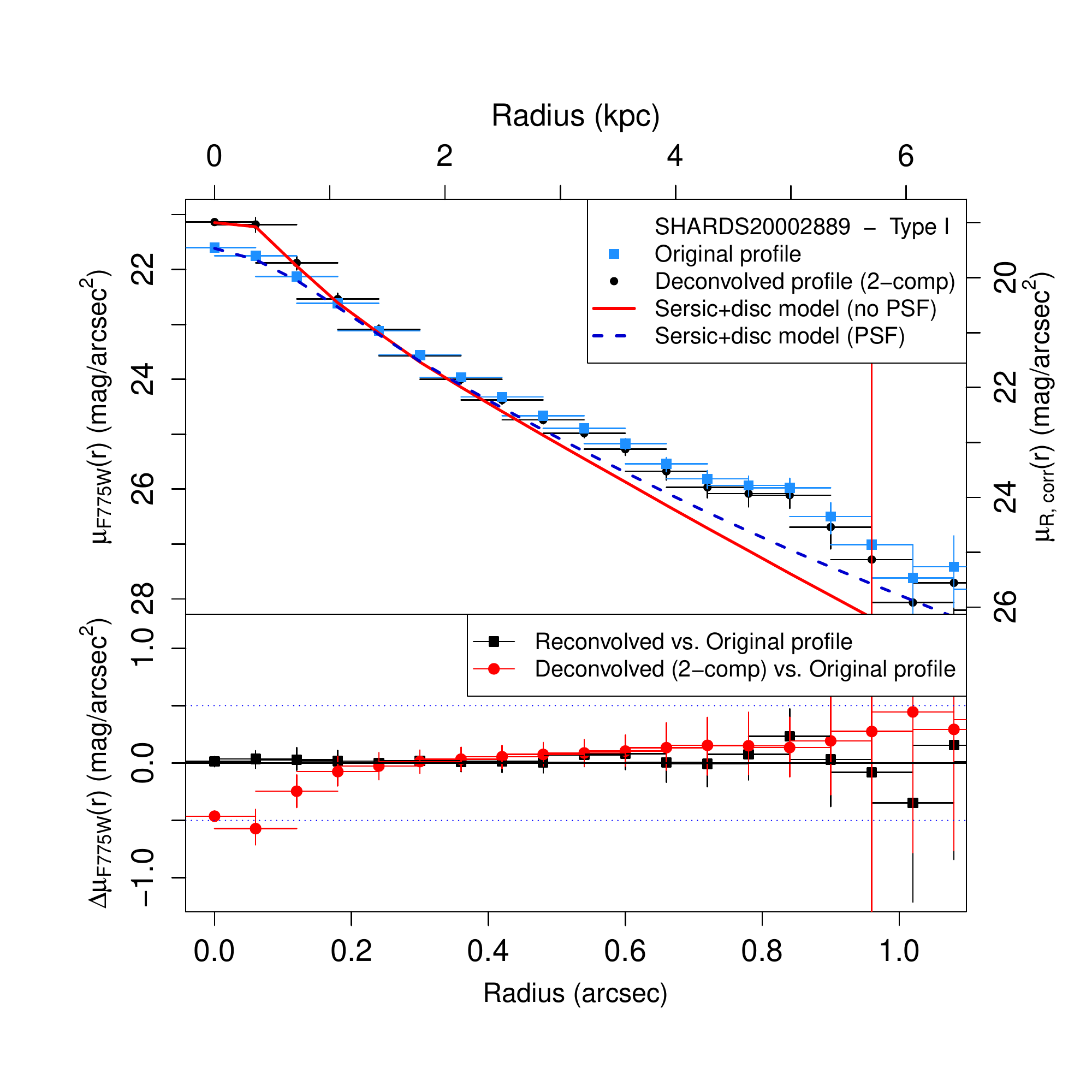}
\end{minipage}
\begin{minipage}{.49\textwidth}
\includegraphics[clip, trim=0.1cm 0.1cm 1cm 0.1cm, width=0.95\textwidth]{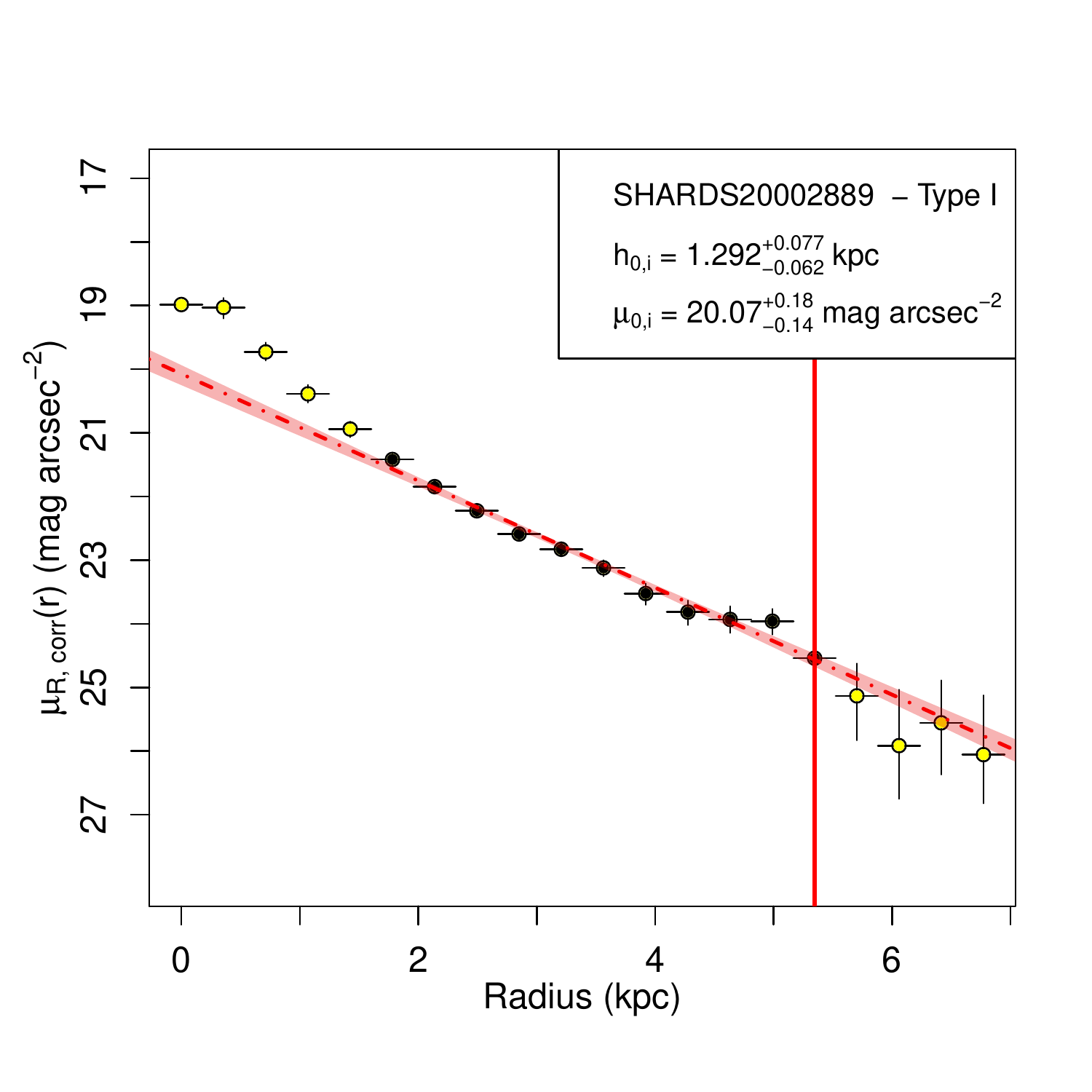}
\end{minipage}%

\vspace{-0.5cm}}
\caption[]{See caption of Fig.1. [\emph{Figure  available in the online edition}.]}         
\label{fig:img_final}
\end{figure}
\clearpage
\newpage

\textbf{SHARDS20002935:} S0 galaxy with a face-on orientation (see Table \ref{tab:fits_psforr}). We classified it as Type-I profile despite the noticeable bump found in the outskirts that cannot be explained by PSF dispersed light. Multiple and extensive masking was required due to the presence of multiple field objects. The profile shows a simple bulge + exponential disc structure. The small excess of light at the outermost part of the galaxy is not compatible with an exponential component. 

\begin{figure}[!h]
{\centering
\vspace{-0cm}

\begin{minipage}{.5\textwidth}
\hspace{1.2cm}
\begin{overpic}[width=0.8\textwidth]
{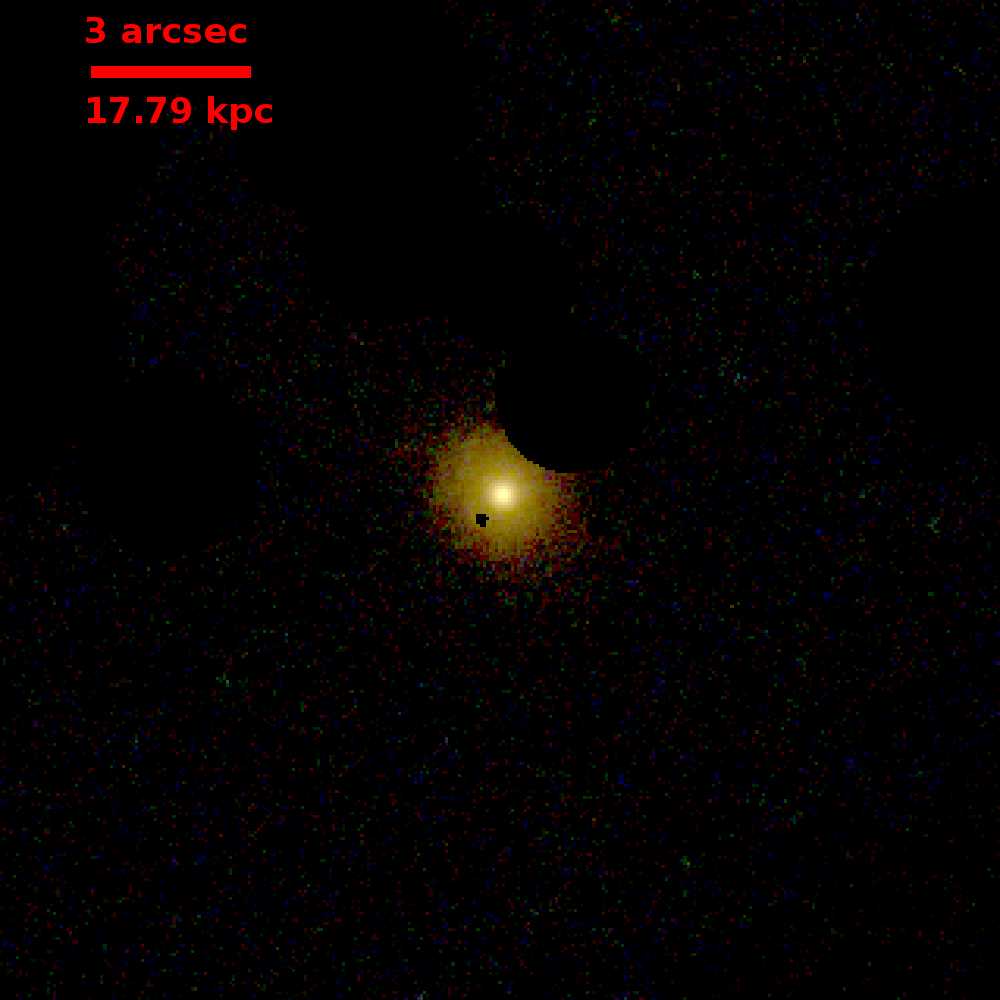}
\put(110,200){\color{yellow} \textbf{SHARDS20002935}}
\put(110,190){\color{yellow} \textbf{z=0.4745}}
\put(110,180){\color{yellow} \textbf{S0}}
\end{overpic}
\vspace{-1cm}
\end{minipage}%
\begin{minipage}{.5\textwidth}
\includegraphics[clip, trim=1cm 1cm 1.5cm 1.5cm, width=\textwidth]{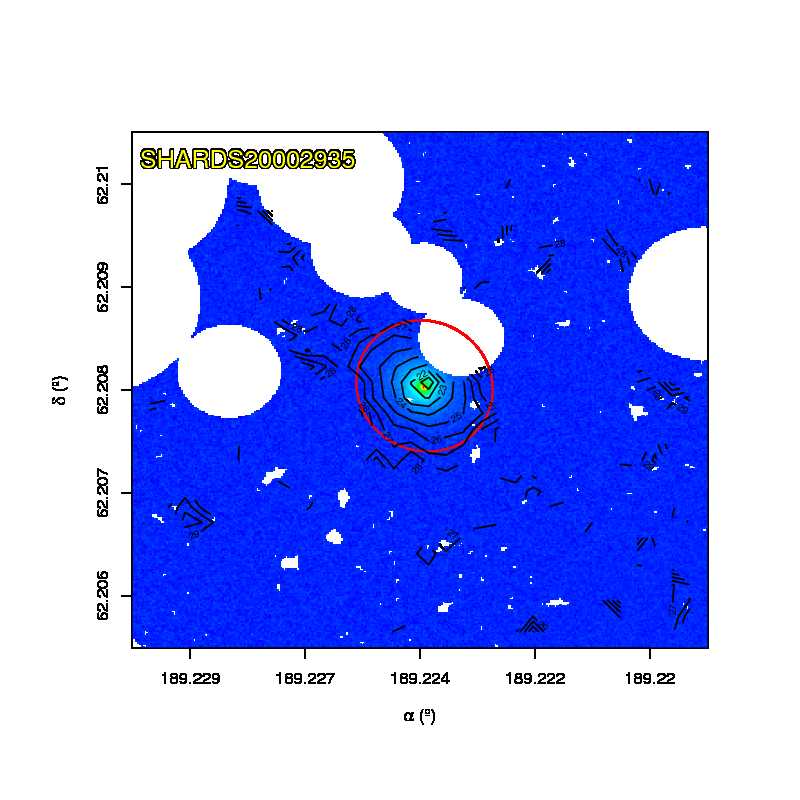}\vspace{-1cm}
\end{minipage}%

\begin{minipage}{.49\textwidth}
\includegraphics[clip, trim=0.1cm 0.1cm 0.1cm 0.1cm, width=\textwidth]{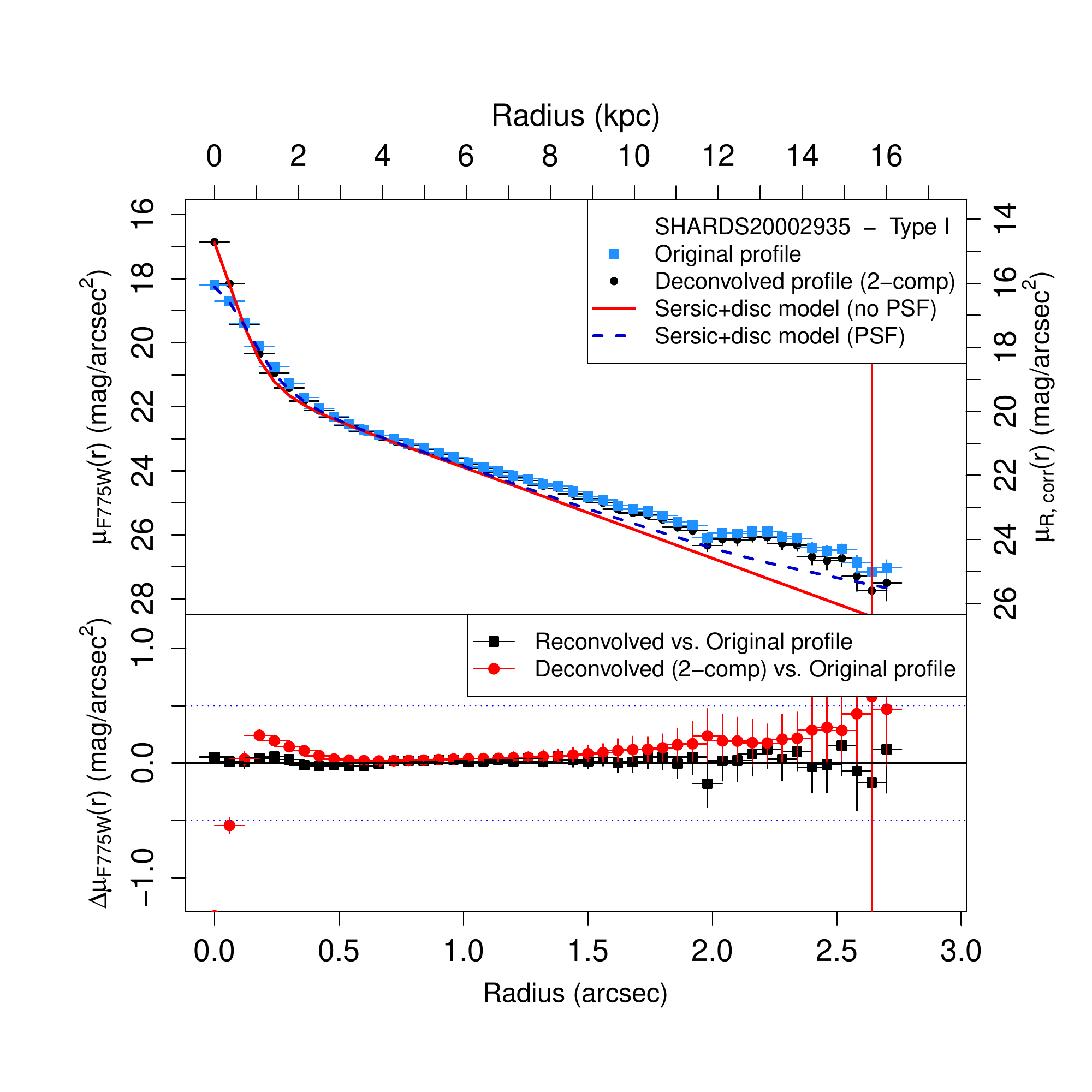}
\end{minipage}
\begin{minipage}{.49\textwidth}
\includegraphics[clip, trim=0.1cm 0.1cm 1cm 0.1cm, width=0.95\textwidth]{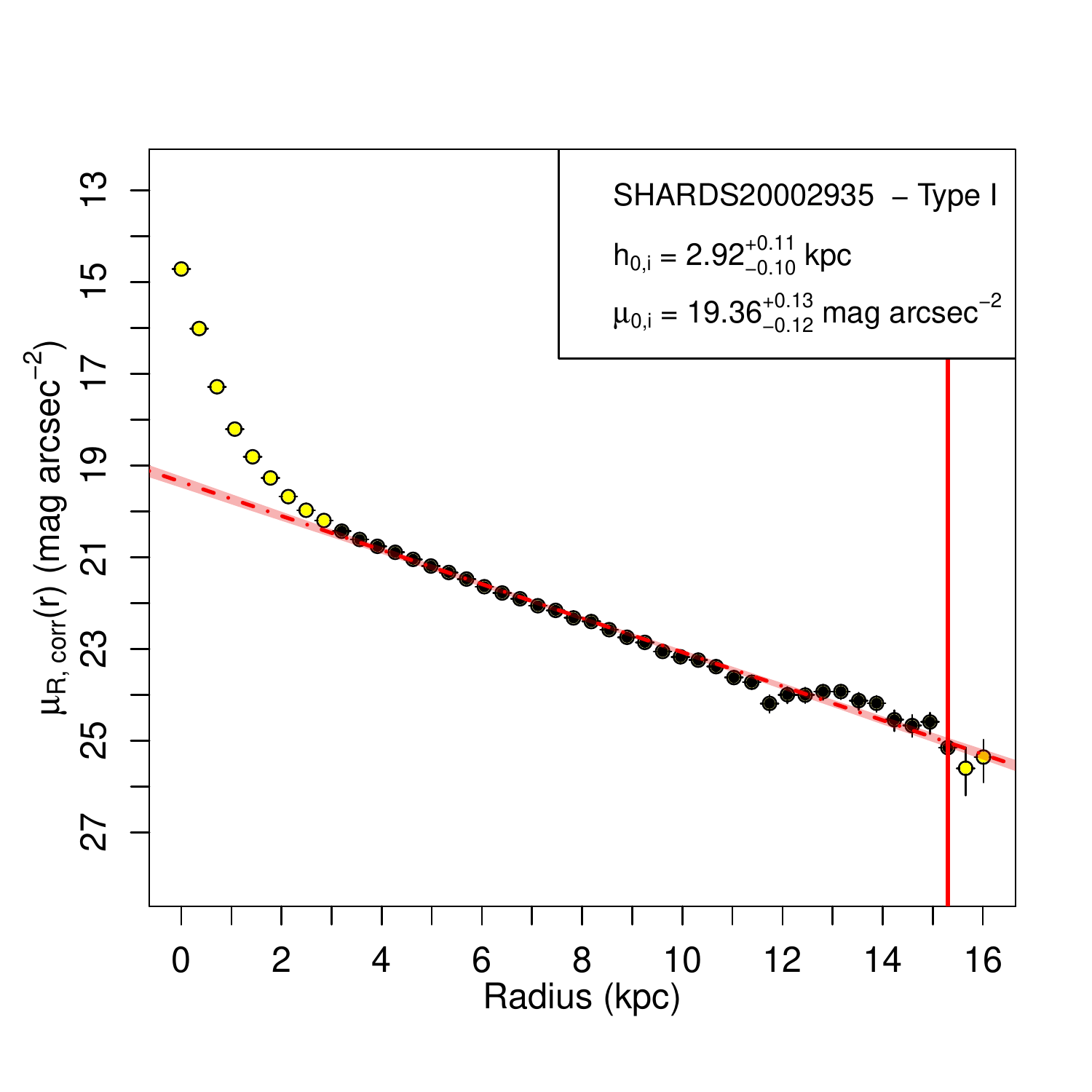}
\end{minipage}%

\vspace{-0.5cm}}
\caption[]{See caption of Fig.1. [\emph{Figure  available in the online edition}.]}         
\label{fig:img_final}
\end{figure}
\clearpage
\newpage

\textbf{SHARDS20002966:} S0 galaxy with a Type-I profile. The object presents medium inclination (see Table \ref{tab:fits_psforr}). The image did not require any masking inside the final fitting region. After inspection of the surface brightness profile, we found a typical bulge + exponential disc distribution. The disc appears to be featureless. The automatic break analysis does not reveal any significant breaks.

\begin{figure}[!h]
{\centering
\vspace{-0cm}

\begin{minipage}{.5\textwidth}
\hspace{1.2cm}
\begin{overpic}[width=0.8\textwidth]
{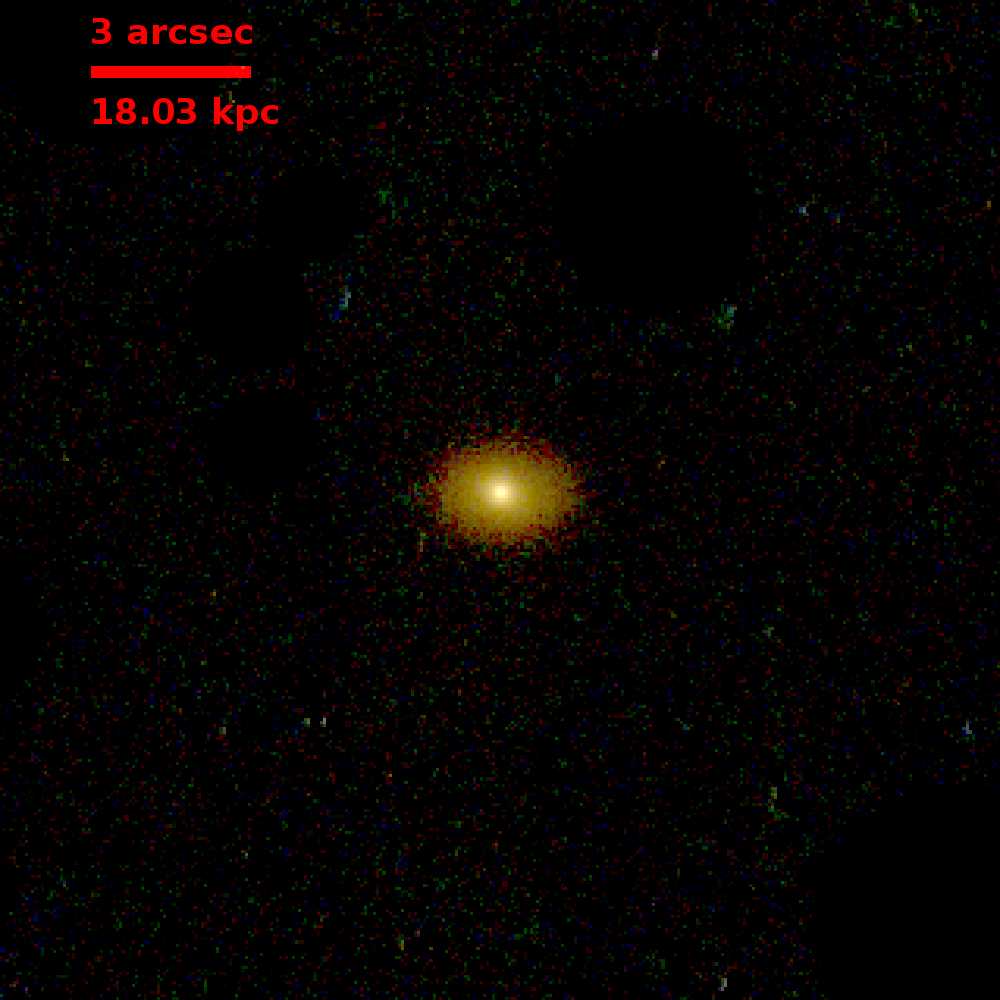}
\put(110,200){\color{yellow} \textbf{SHARDS20002966}}
\put(110,190){\color{yellow} \textbf{z=0.4854}}
\put(110,180){\color{yellow} \textbf{S0}}
\end{overpic}
\vspace{-1cm}
\end{minipage}%
\begin{minipage}{.5\textwidth}
\includegraphics[clip, trim=1cm 1cm 1.5cm 1.5cm, width=\textwidth]{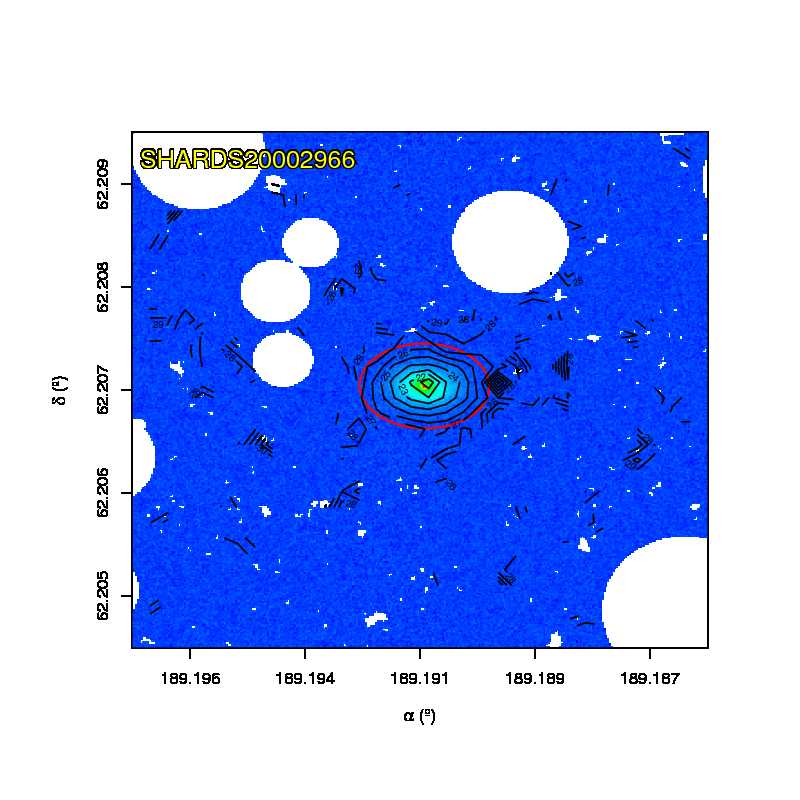}\vspace{-1cm}
\end{minipage}%

\begin{minipage}{.49\textwidth}
\includegraphics[clip, trim=0.1cm 0.1cm 0.1cm 0.1cm, width=\textwidth]{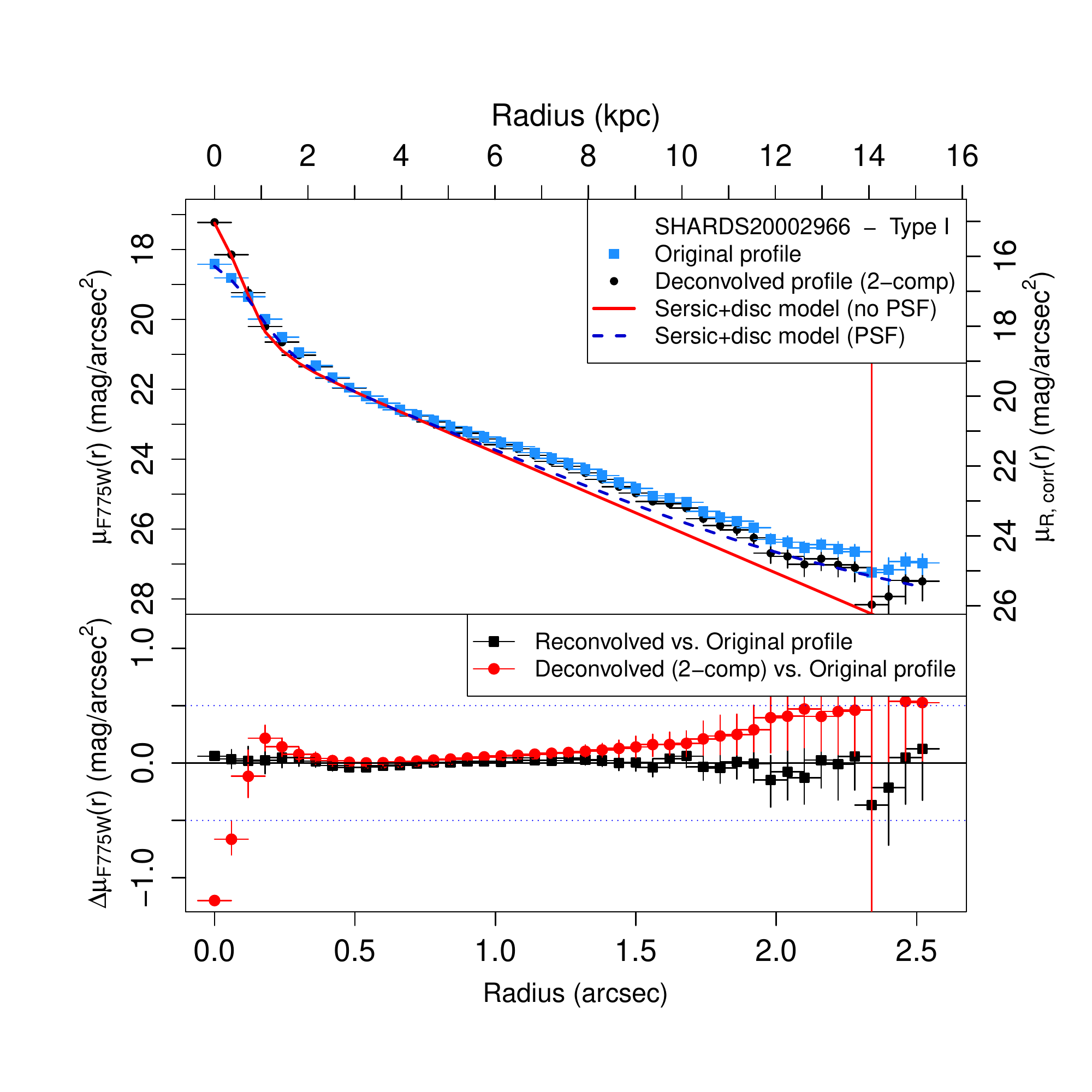}
\end{minipage}
\begin{minipage}{.49\textwidth}
\includegraphics[clip, trim=0.1cm 0.1cm 1cm 0.1cm, width=0.95\textwidth]{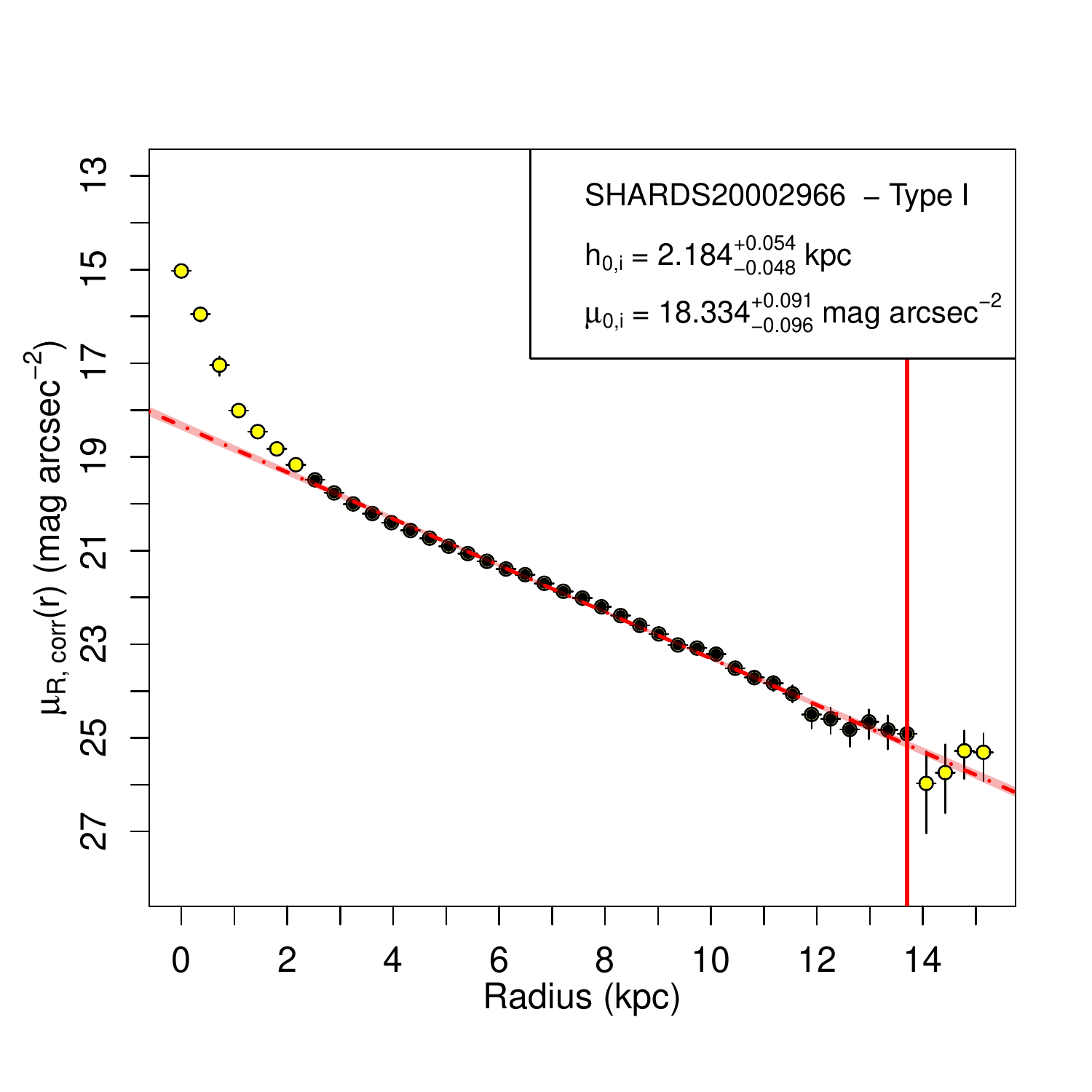}
\end{minipage}%

\vspace{-0.5cm}}
\caption[]{See caption of Fig.1. [\emph{Figure  available in the online edition}.]}         
\label{fig:img_final}
\end{figure}
\clearpage
\newpage

\textbf{SHARDS20002995:} S0 galaxy with a Type-II profile. Manual masking was required for three nearby sources, one of them an apparent satellite in accretion. The smooth surface brightness profile free from spikes ensure us that the masking was enough and successful. The inner region was removed for the fit to avoid the possible contamination from the bulge or an inner lens. The outer profile presents some irregularities at the outskirts ($\sim13 - 15$ kpc). The break presents smooth and statistically significant PDDs.  

\begin{figure}[!h]
{\centering
\vspace{-0cm}

\begin{minipage}{.5\textwidth}
\hspace{1.2cm}
\begin{overpic}[width=0.8\textwidth]
{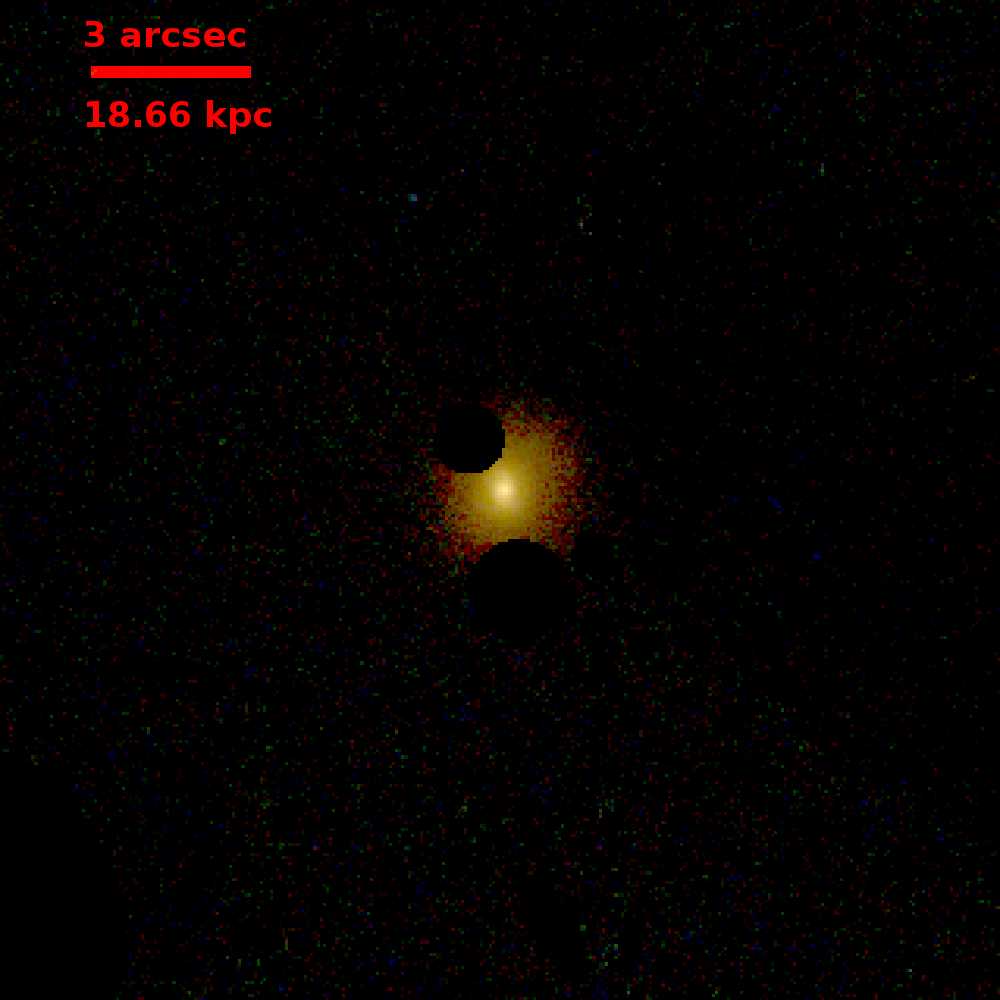}
\put(110,200){\color{yellow} \textbf{SHARDS20002995}}
\put(110,190){\color{yellow} \textbf{z=0.5185}}
\put(110,180){\color{yellow} \textbf{S0}}
\end{overpic}
\vspace{-1cm}
\end{minipage}%
\begin{minipage}{.5\textwidth}
\includegraphics[clip, trim=1cm 1cm 1.5cm 1.5cm, width=\textwidth]{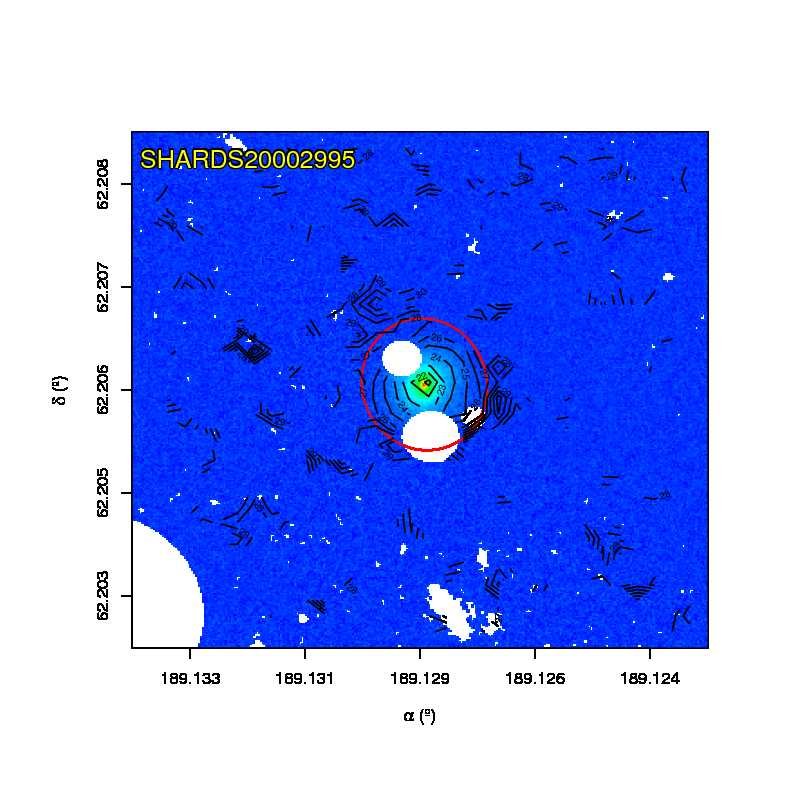}\vspace{-1cm}
\end{minipage}%

\begin{minipage}{.49\textwidth}
\includegraphics[clip, trim=0.1cm 0.1cm 0.1cm 0.1cm, width=\textwidth]{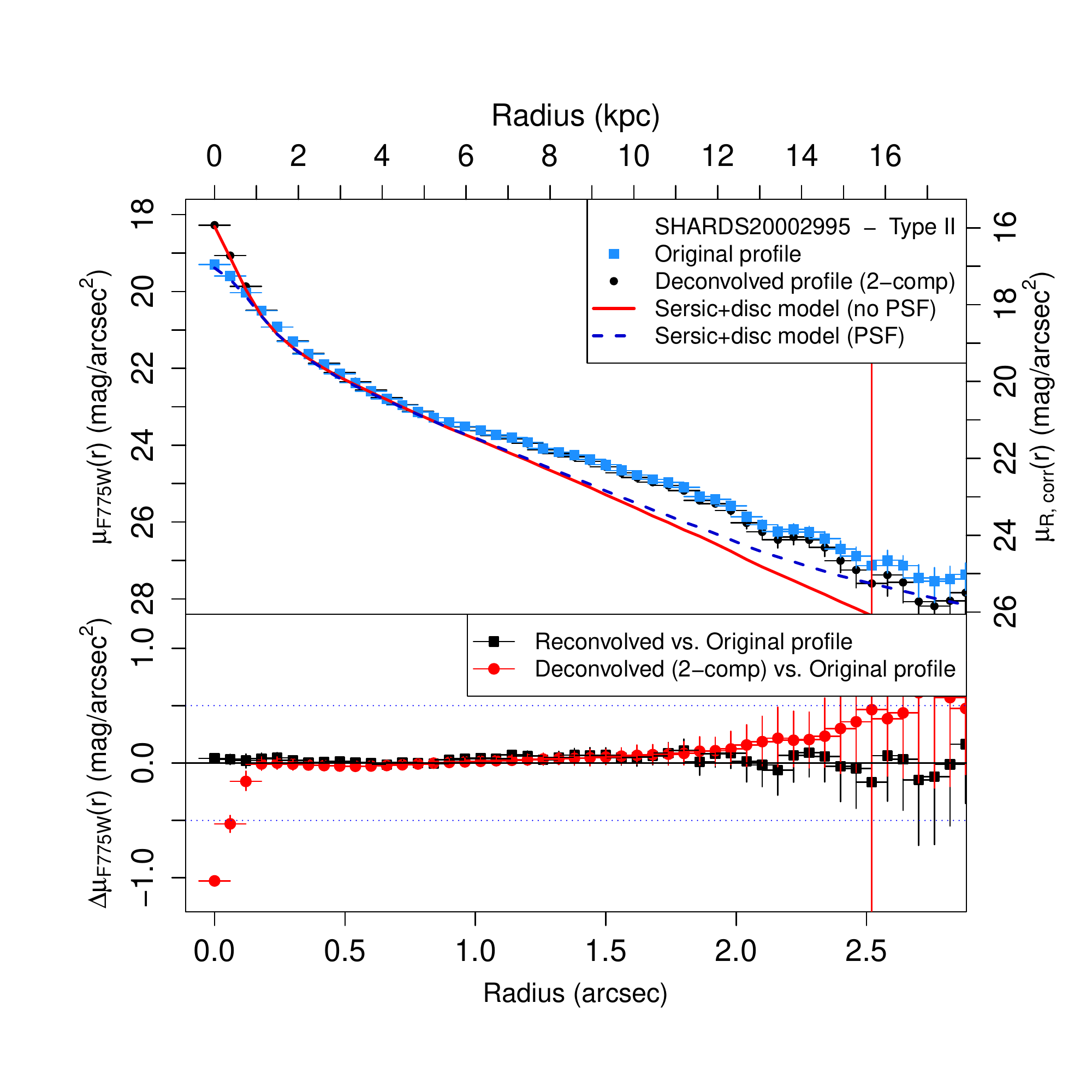}
\end{minipage}
\begin{minipage}{.49\textwidth}
\includegraphics[clip, trim=0.1cm 0.1cm 1cm 0.1cm, width=0.95\textwidth]{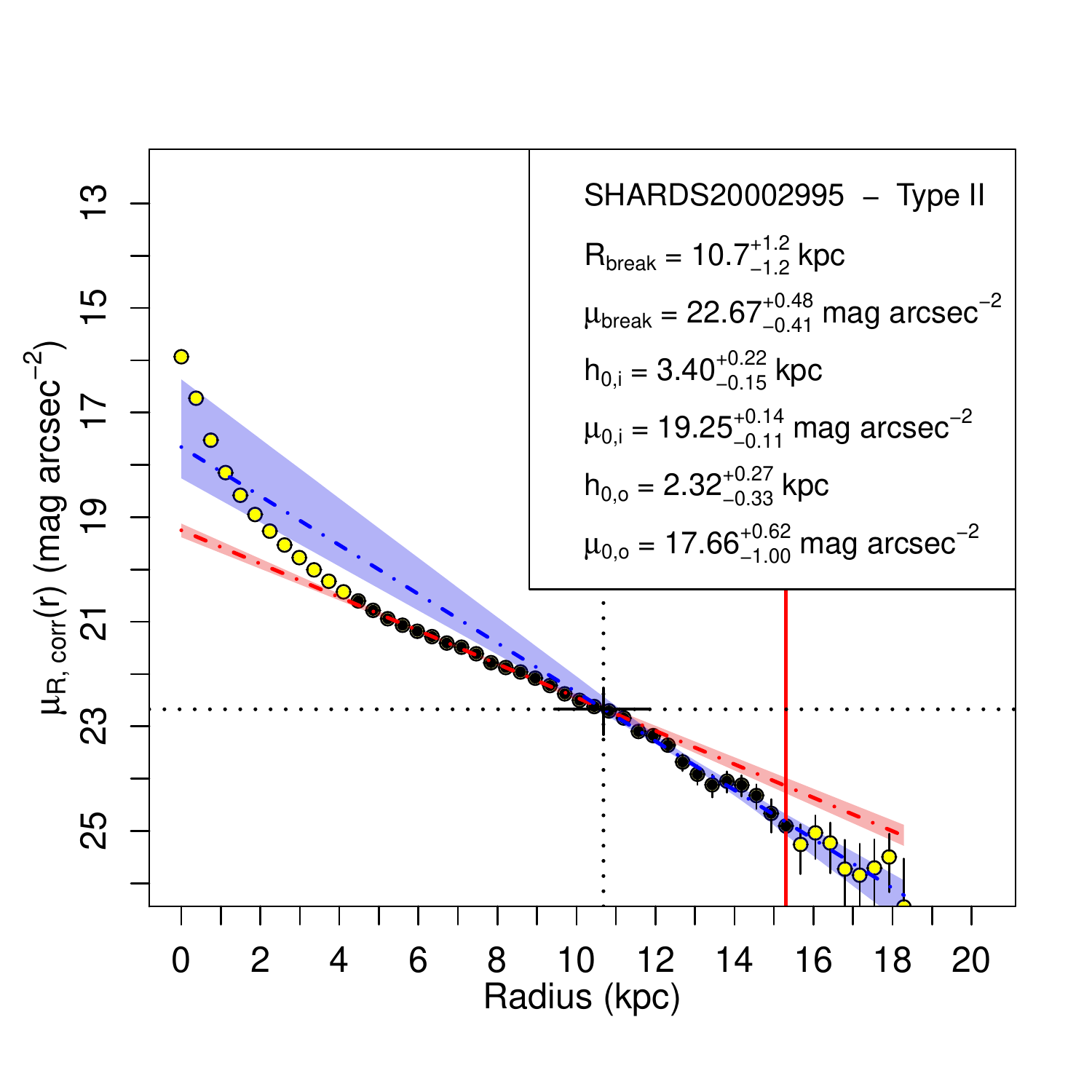}
\end{minipage}%

\vspace{-0.5cm}}
\caption[]{See caption of Fig.1. [\emph{Figure  available in the online edition}.]}         
\label{fig:img_final}
\end{figure}
\clearpage
\newpage

\textbf{SHARDS20003134:} S0 galaxy with a Type-II profile. Manual masking was required to avoid the contamination by compact bright object  located to the W. It was analysed by {\tt{ISOFIT}} instead of {\tt{ellipse}} due to its edge-on orientation (see Table \ref{tab:fits_psforr}). The PDDs for $h$ and $\mu_{0}$ show two clearly separated peaks corresponding to the two profiles. 

\begin{figure}[!h]
{\centering
\vspace{-0cm}

\begin{minipage}{.5\textwidth}
\hspace{1.2cm}
\begin{overpic}[width=0.8\textwidth]
{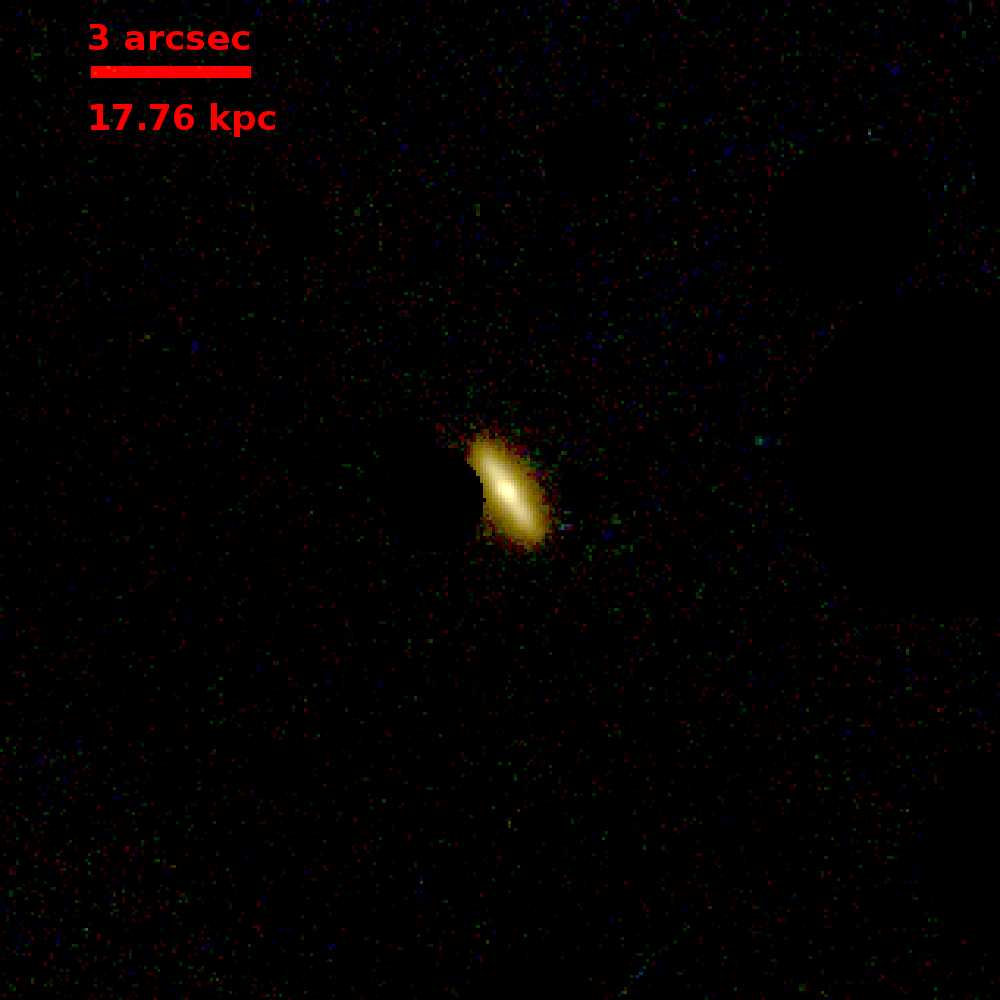}
\put(110,200){\color{yellow} \textbf{SHARDS20003134}}
\put(110,190){\color{yellow} \textbf{z=0.4725}}
\put(110,180){\color{yellow} \textbf{S0}}
\end{overpic}
\vspace{-1cm}
\end{minipage}%
\begin{minipage}{.5\textwidth}
\includegraphics[clip, trim=1cm 1cm 1.5cm 1.5cm, width=\textwidth]{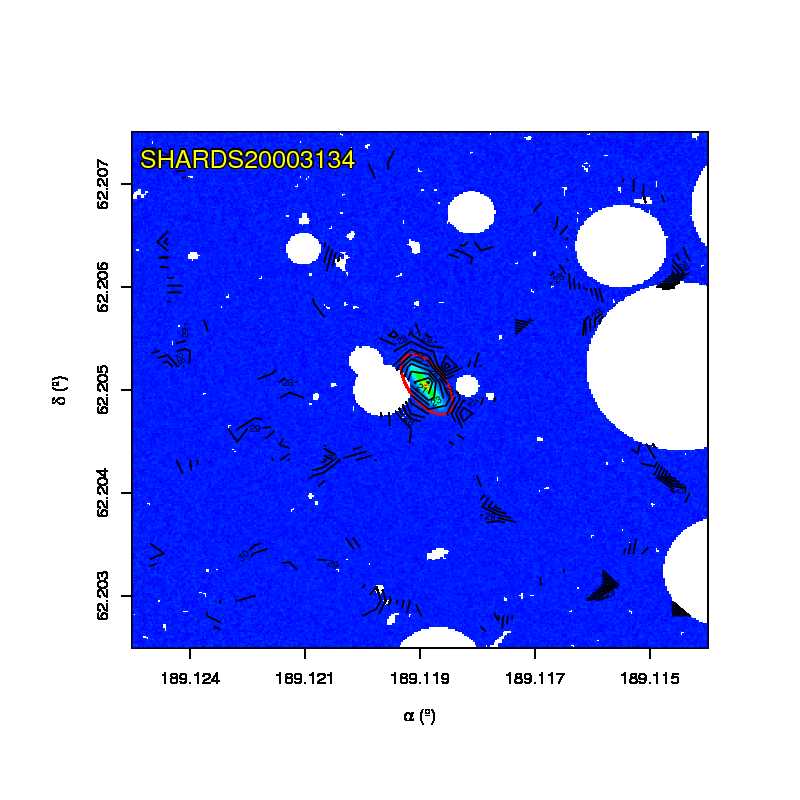}\vspace{-1cm}
\end{minipage}%

\begin{minipage}{.49\textwidth}
\includegraphics[clip, trim=0.1cm 0.1cm 0.1cm 0.1cm, width=\textwidth]{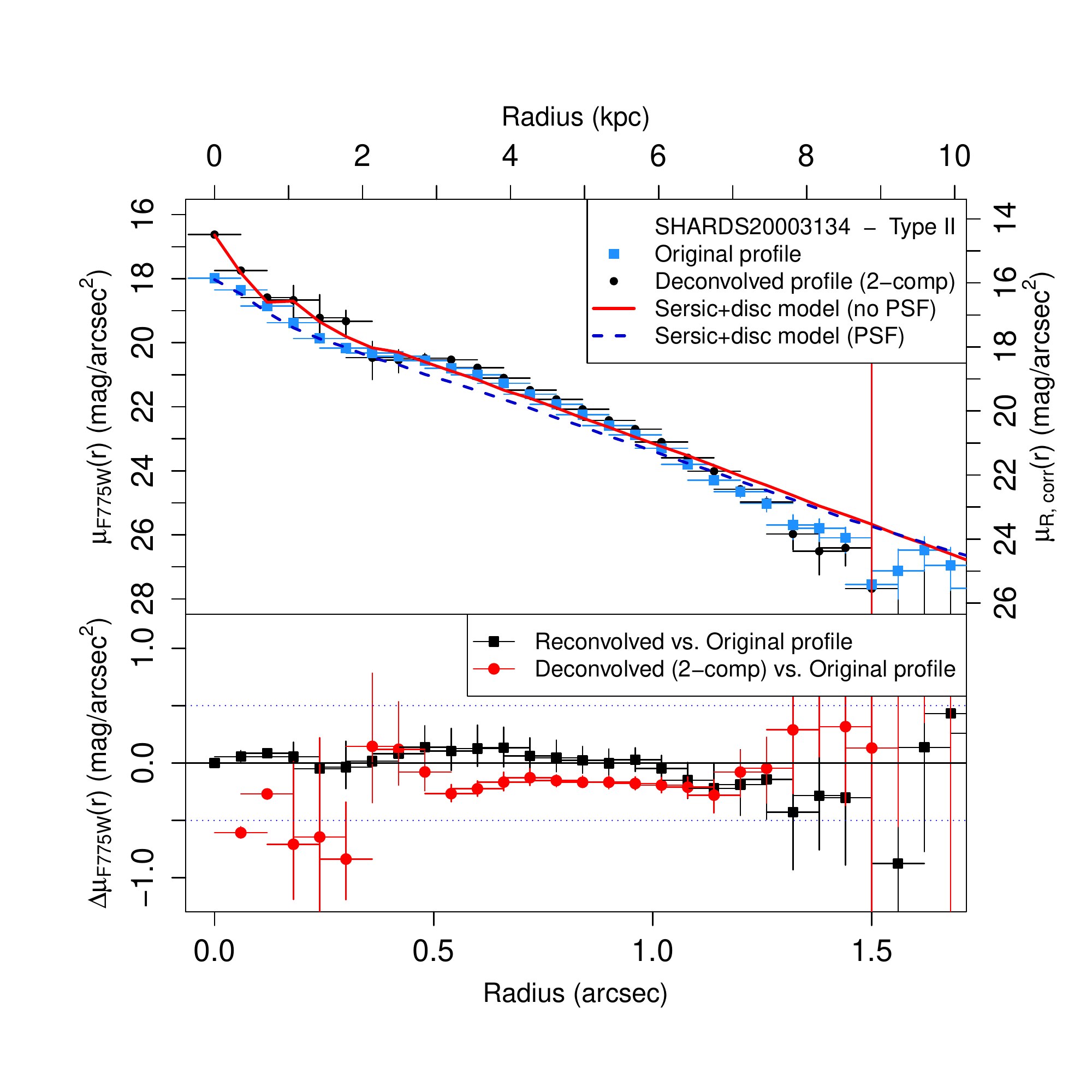}
\end{minipage}
\begin{minipage}{.49\textwidth}
\includegraphics[clip, trim=0.1cm 0.1cm 1cm 0.1cm, width=0.95\textwidth]{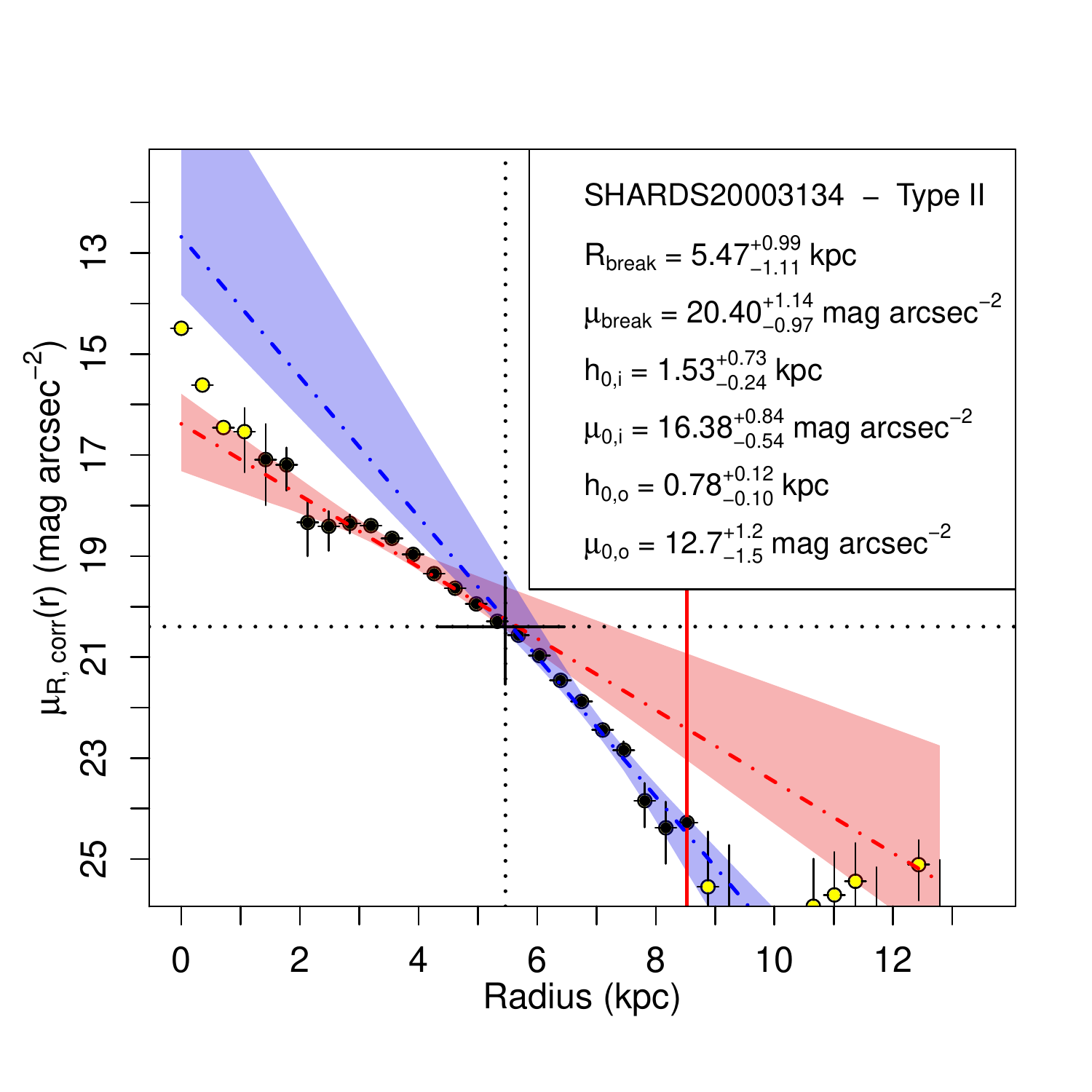}
\end{minipage}%

\vspace{-0.5cm}}
\caption[]{See caption of Fig.1. [\emph{Figure  available in the online edition}.]}         
\label{fig:img_final}
\end{figure}
\clearpage
\newpage

\textbf{SHARDS20003210:} S0 galaxy with a Type-III profile. The object presents a face-on orientation (see Table \ref{tab:fits_psforr}). The profile reveal a clear bulge + double exponential structure. The PDDs for $h$ and $\mu_{0}$ show two clearly separated peaks corresponding to the two profiles.

\begin{figure}[!h]
{\centering
\vspace{-0cm}

\begin{minipage}{.5\textwidth}
\hspace{1.2cm}
\begin{overpic}[width=0.8\textwidth]
{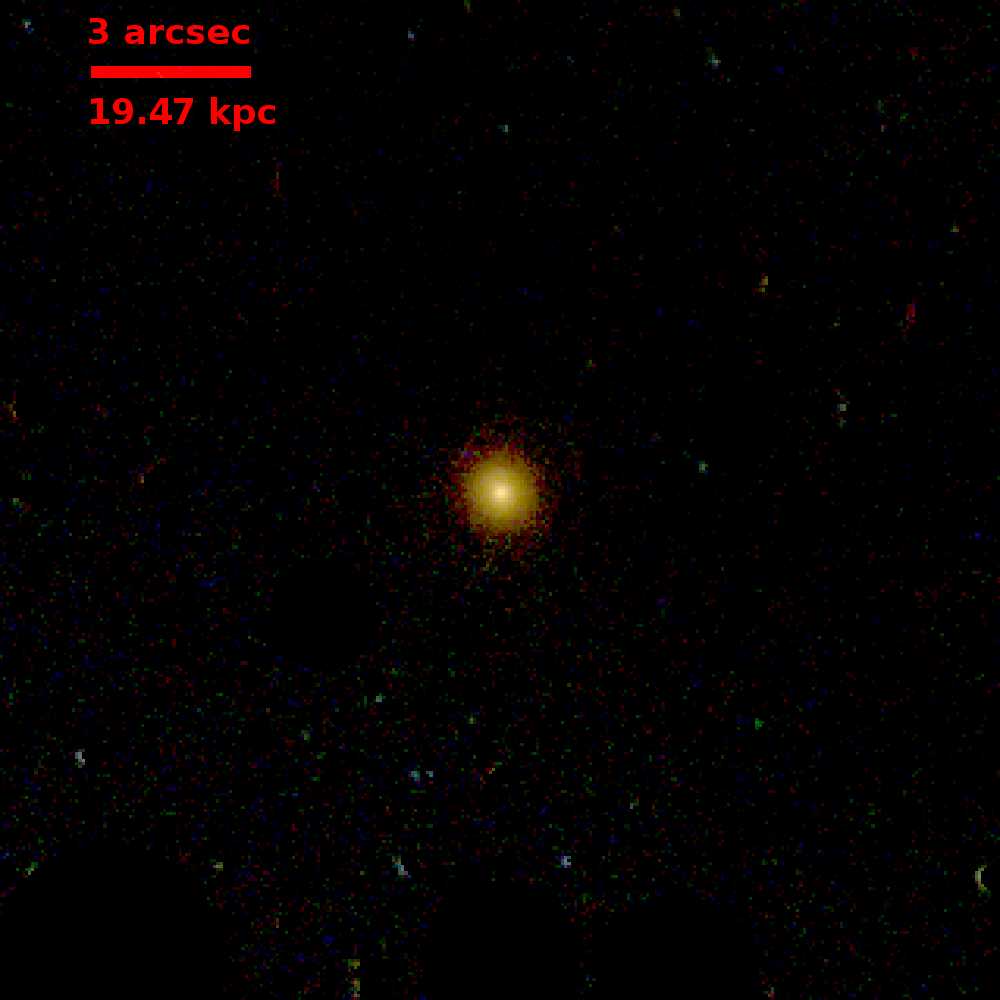}
\put(110,200){\color{yellow} \textbf{SHARDS20003210}}
\put(110,190){\color{yellow} \textbf{z=0.5631}}
\put(110,180){\color{yellow} \textbf{S0}}
\end{overpic}
\vspace{-1cm}
\end{minipage}%
\begin{minipage}{.5\textwidth}
\includegraphics[clip, trim=1cm 1cm 1.5cm 1.5cm, width=\textwidth]{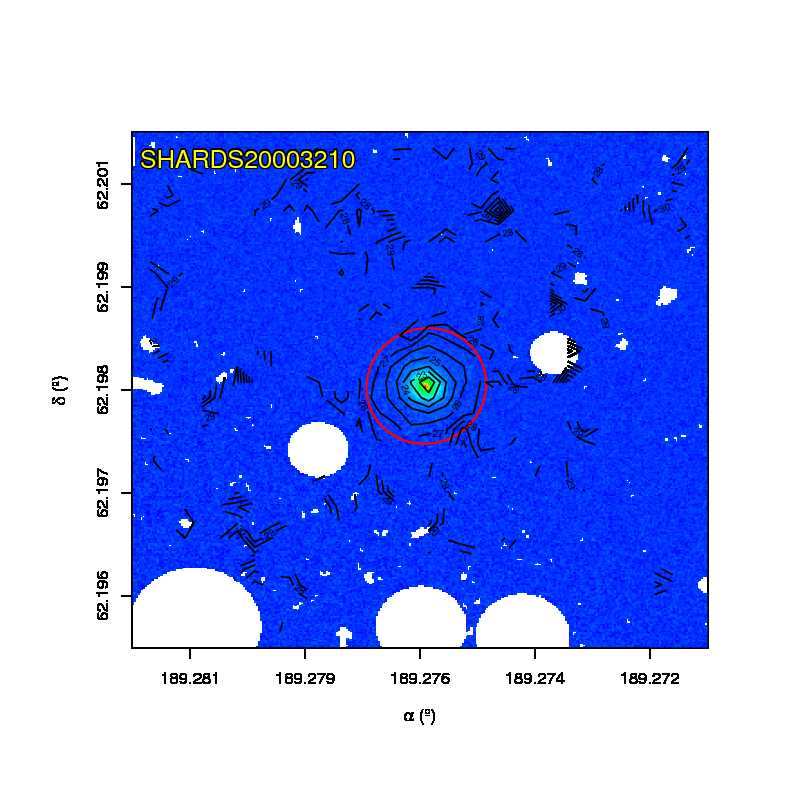}\vspace{-1cm}
\end{minipage}%

\begin{minipage}{.49\textwidth}
\includegraphics[clip, trim=0.1cm 0.1cm 0.1cm 0.1cm, width=\textwidth]{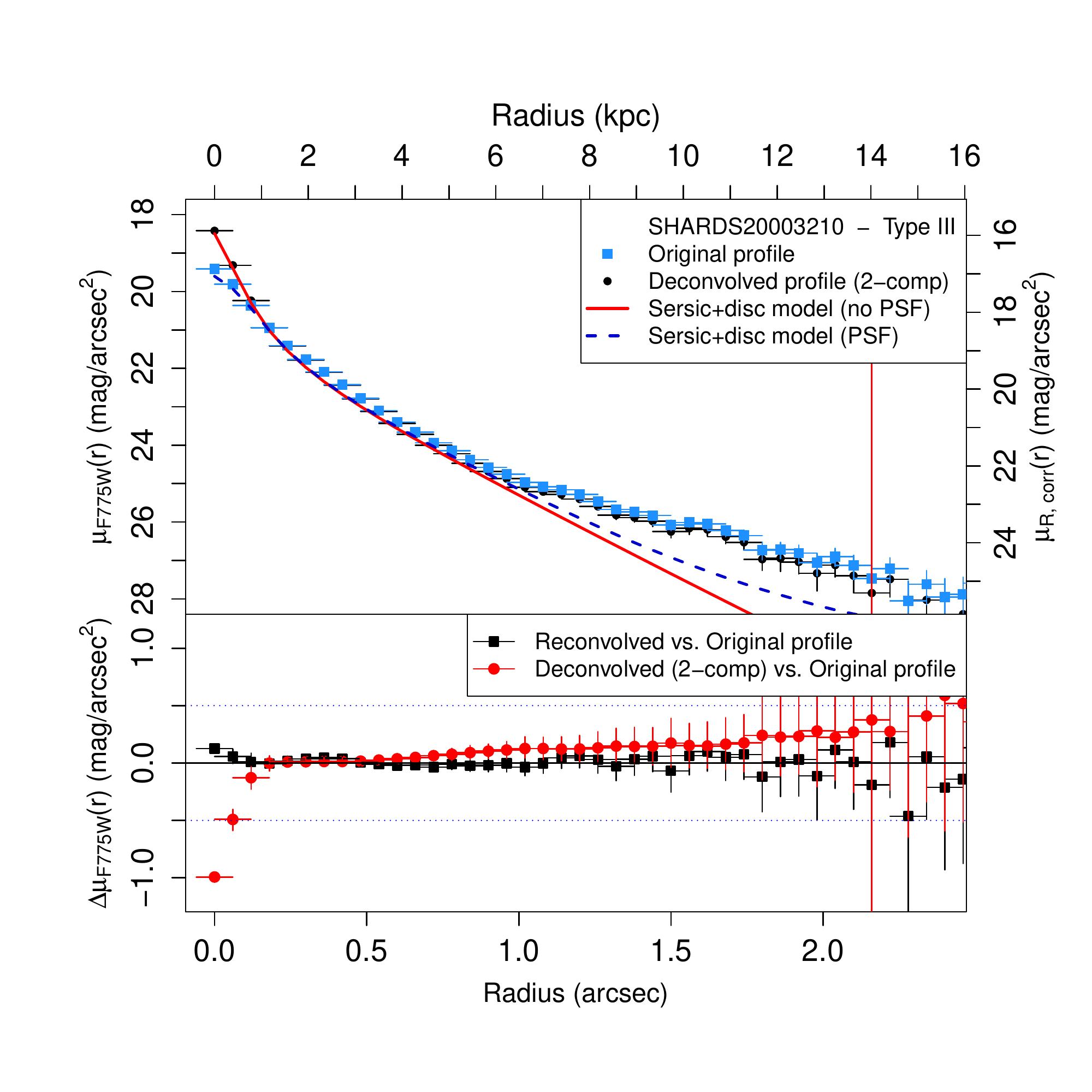}
\end{minipage}
\begin{minipage}{.49\textwidth}
\includegraphics[clip, trim=0.1cm 0.1cm 1cm 0.1cm, width=0.95\textwidth]{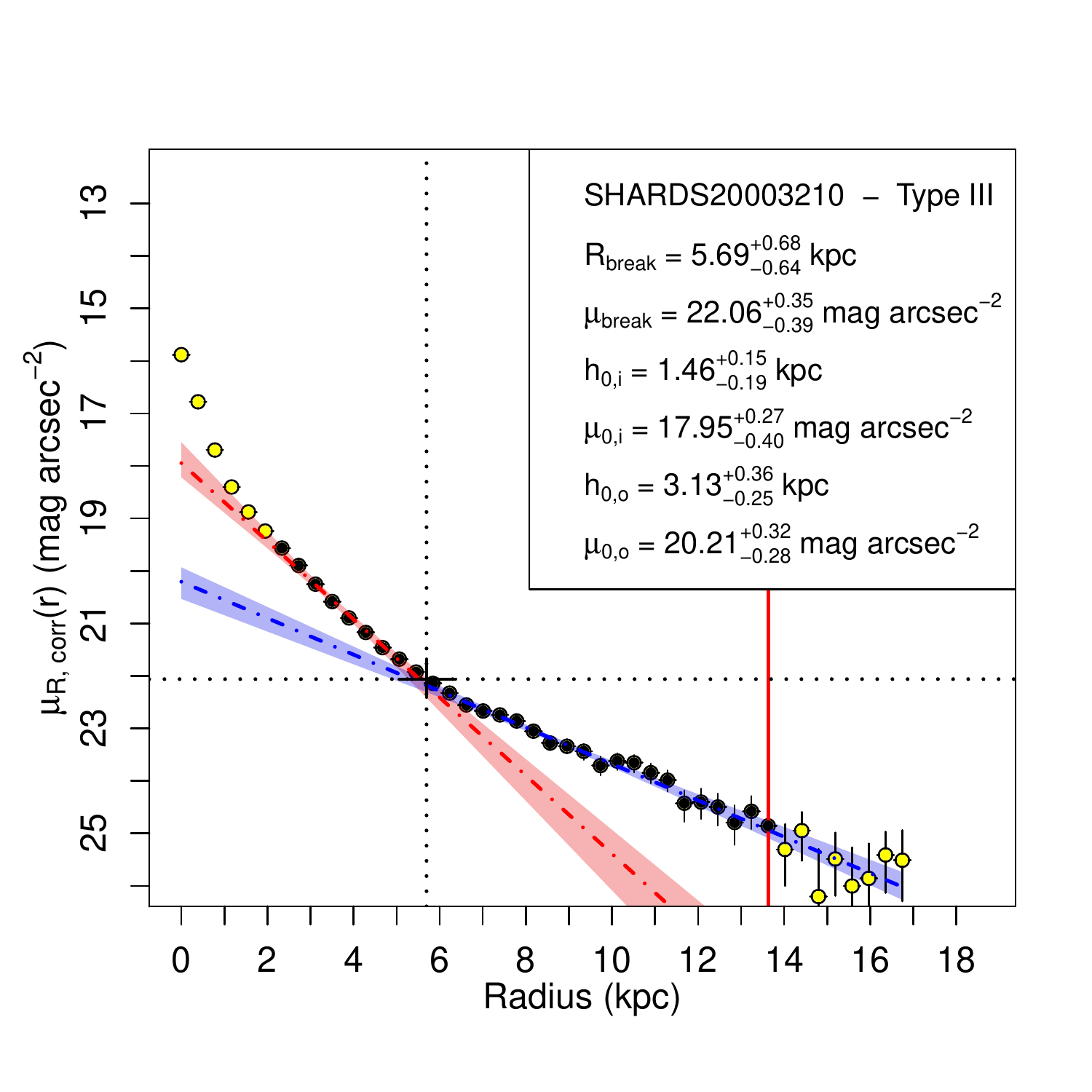}
\end{minipage}%

\vspace{-0.5cm}}
\caption[]{See caption of Fig.1. [\emph{Figure  available in the online edition}.]}         
\label{fig:img_final}
\end{figure}
\clearpage
\newpage

\textbf{SHARDS20003217:} S0 galaxy with a Type-I profile (see Table \ref{tab:fits_psforr}). The object presents a very low inclination, but shows a clear  bulge + exponential structure. A small bump at $\sim 3 - 4$ kpc in the surface brightness profile originates double peaked PDDs when performing automated break analysis, but the $p$-values associated to this bump were negligible.

\begin{figure}[!h]
{\centering
\vspace{-0cm}

\begin{minipage}{.5\textwidth}
\hspace{1.2cm}
\begin{overpic}[width=0.8\textwidth]
{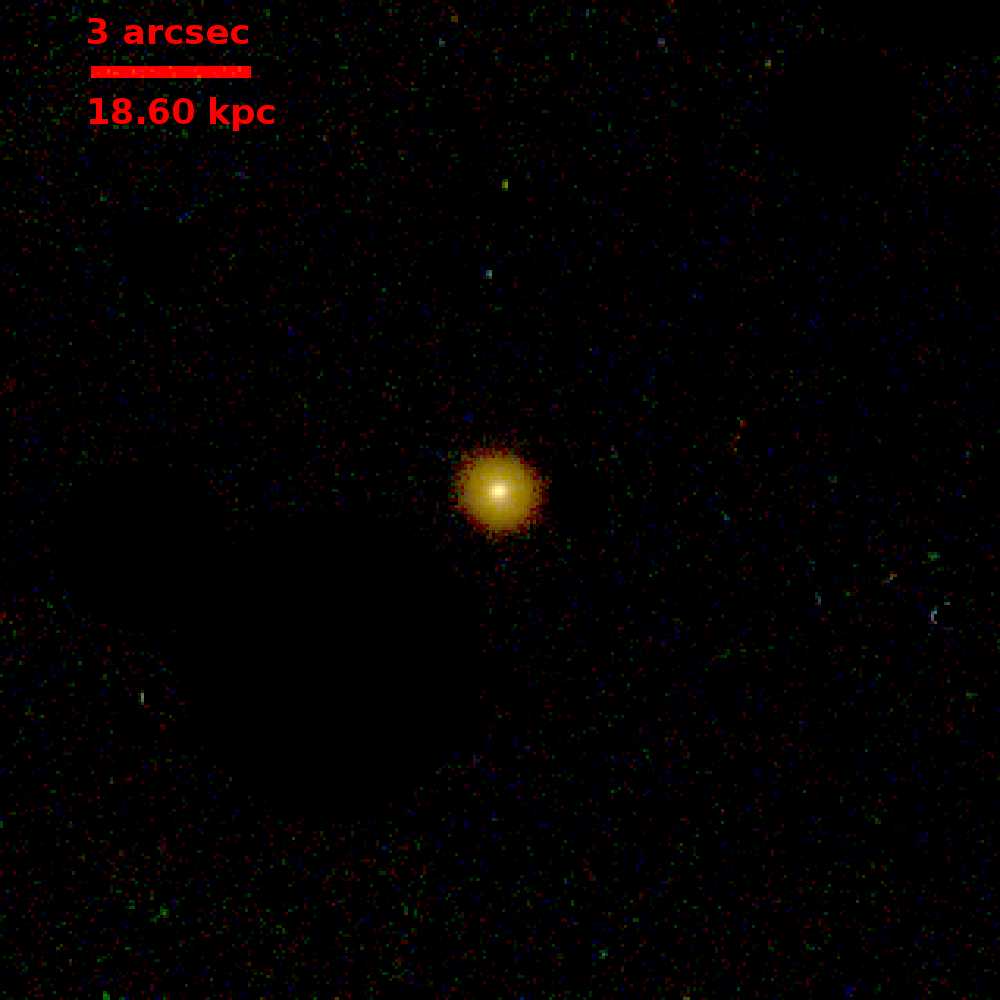}
\put(110,200){\color{yellow} \textbf{SHARDS20003217}}
\put(110,190){\color{yellow} \textbf{z=0.5153}}
\put(110,180){\color{yellow} \textbf{S0}}
\end{overpic}
\vspace{-1cm}
\end{minipage}%
\begin{minipage}{.5\textwidth}
\includegraphics[clip, trim=1cm 1cm 1.5cm 1.5cm, width=\textwidth]{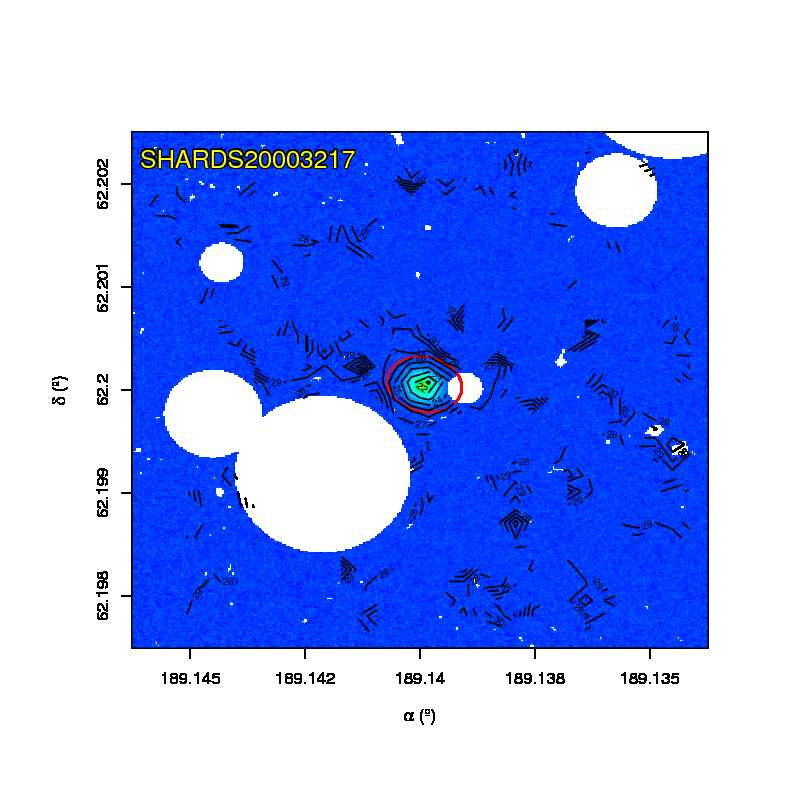}\vspace{-1cm}
\end{minipage}%

\begin{minipage}{.49\textwidth}
\includegraphics[clip, trim=0.1cm 0.1cm 0.1cm 0.1cm, width=\textwidth]{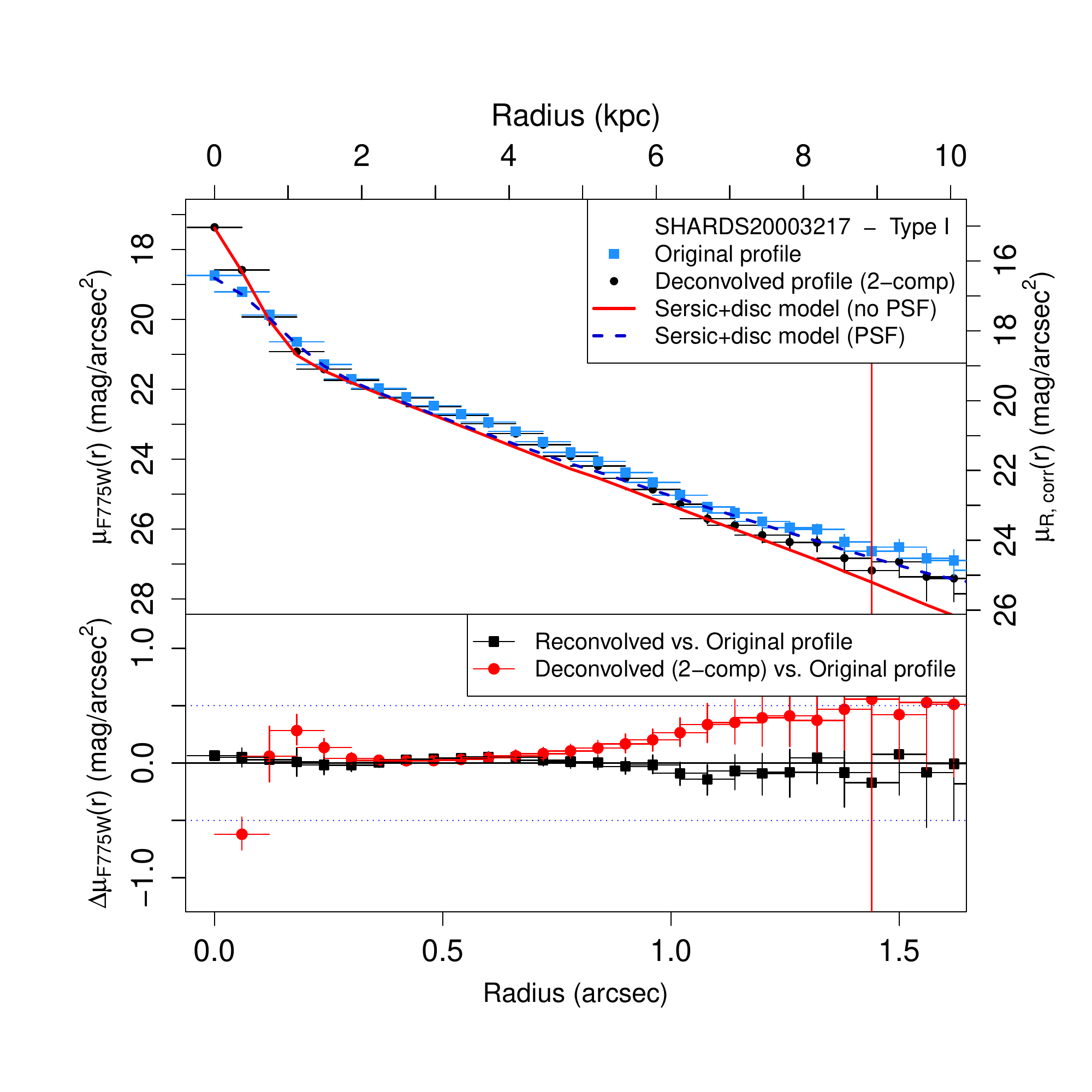}
\end{minipage}
\begin{minipage}{.49\textwidth}
\includegraphics[clip, trim=0.1cm 0.1cm 1cm 0.1cm, width=0.95\textwidth]{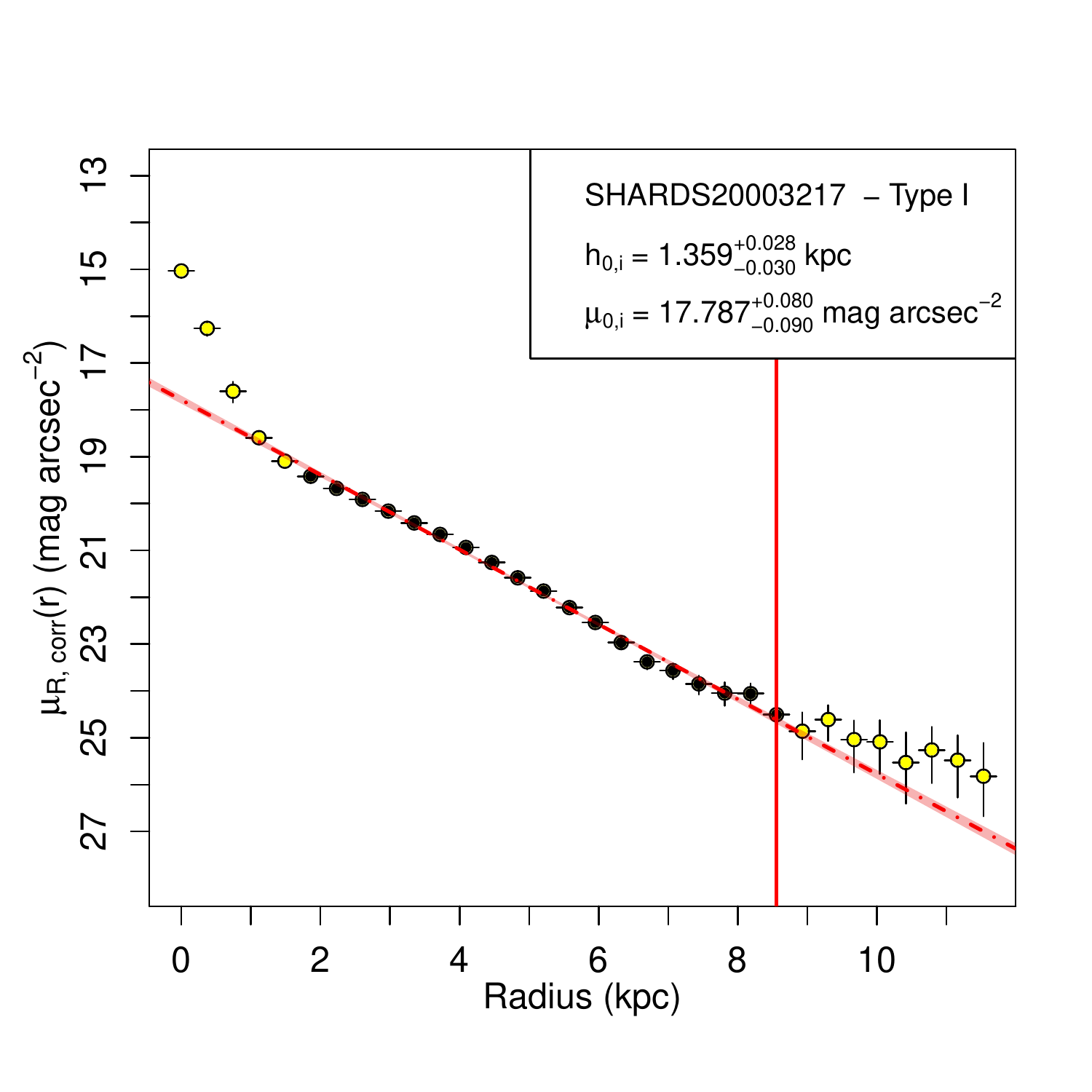}
\end{minipage}%

\vspace{-0.5cm}}
\caption[]{See caption of Fig.1. [\emph{Figure  available in the online edition}.]}         
\label{fig:img_final}
\end{figure}
\clearpage
\newpage

\textbf{SHARDS20003377:} S0 galaxy with a Type-I profile. It has a medium to high inclination (see Table \ref{tab:fits_psforr}). It was flagged as an AGN source (see Sect.\,\ref{Subsec:AGN}). The profile shows a simple bulge and exponential disc composition. The disc appears to be clear and featureless, with a small deviation from the exponential profile in the middle region ($\sim 4-6$ kpc). The automatic break analysis does not reveal any significant breaks. 

\begin{figure}[!h]
{\centering
\vspace{-0cm}

\begin{minipage}{.5\textwidth}
\hspace{1.2cm}
\begin{overpic}[width=0.8\textwidth]
{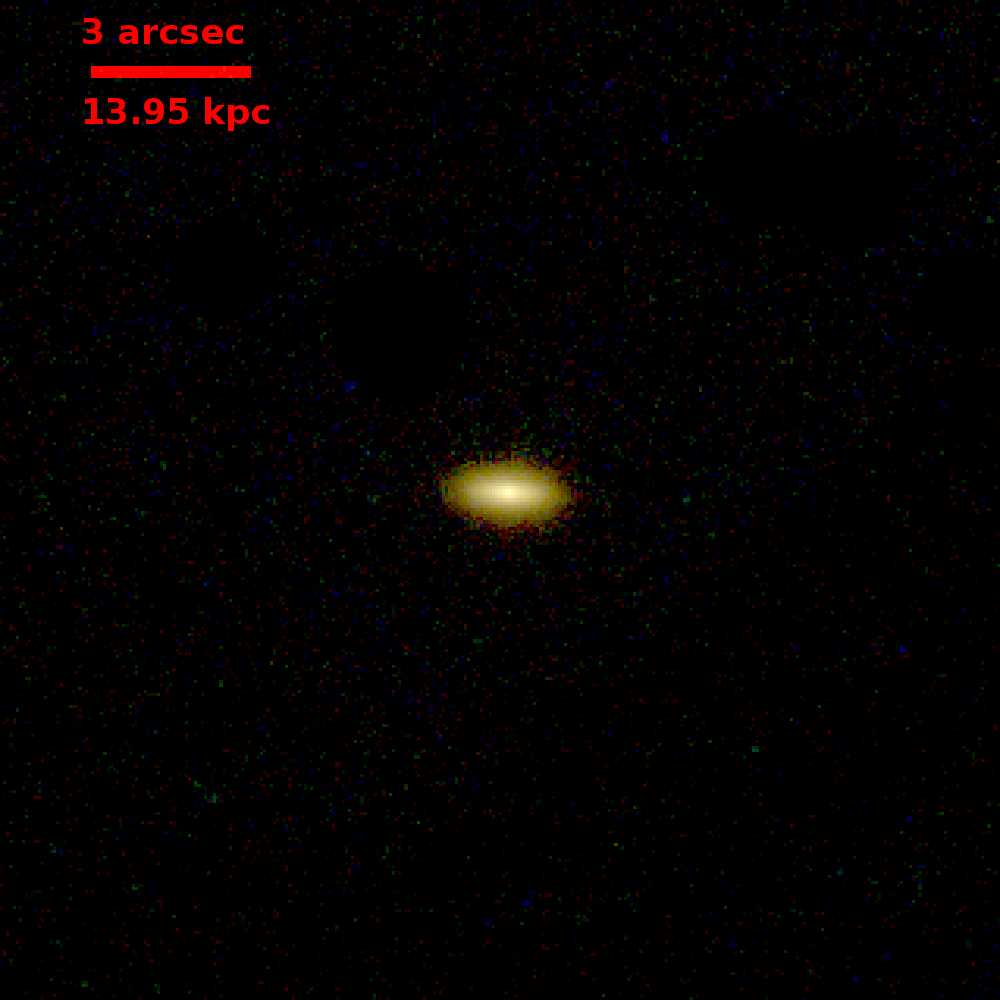}
\put(110,200){\color{yellow} \textbf{SHARDS20003377}}
\put(110,190){\color{yellow} \textbf{z=0.3198}}
\put(110,180){\color{yellow} \textbf{S0}}
\end{overpic}
\vspace{-1cm}
\end{minipage}%
\begin{minipage}{.5\textwidth}
\includegraphics[clip, trim=1cm 1cm 1.5cm 1.5cm, width=\textwidth]{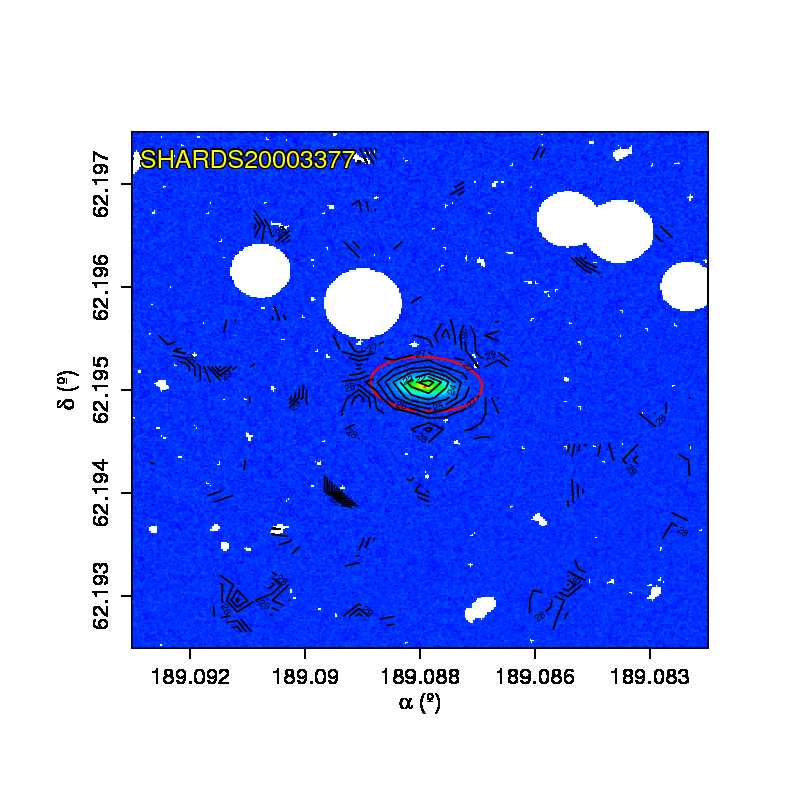}\vspace{-1cm}
\end{minipage}%

\begin{minipage}{.49\textwidth}
\includegraphics[clip, trim=0.1cm 0.1cm 0.1cm 0.1cm, width=\textwidth]{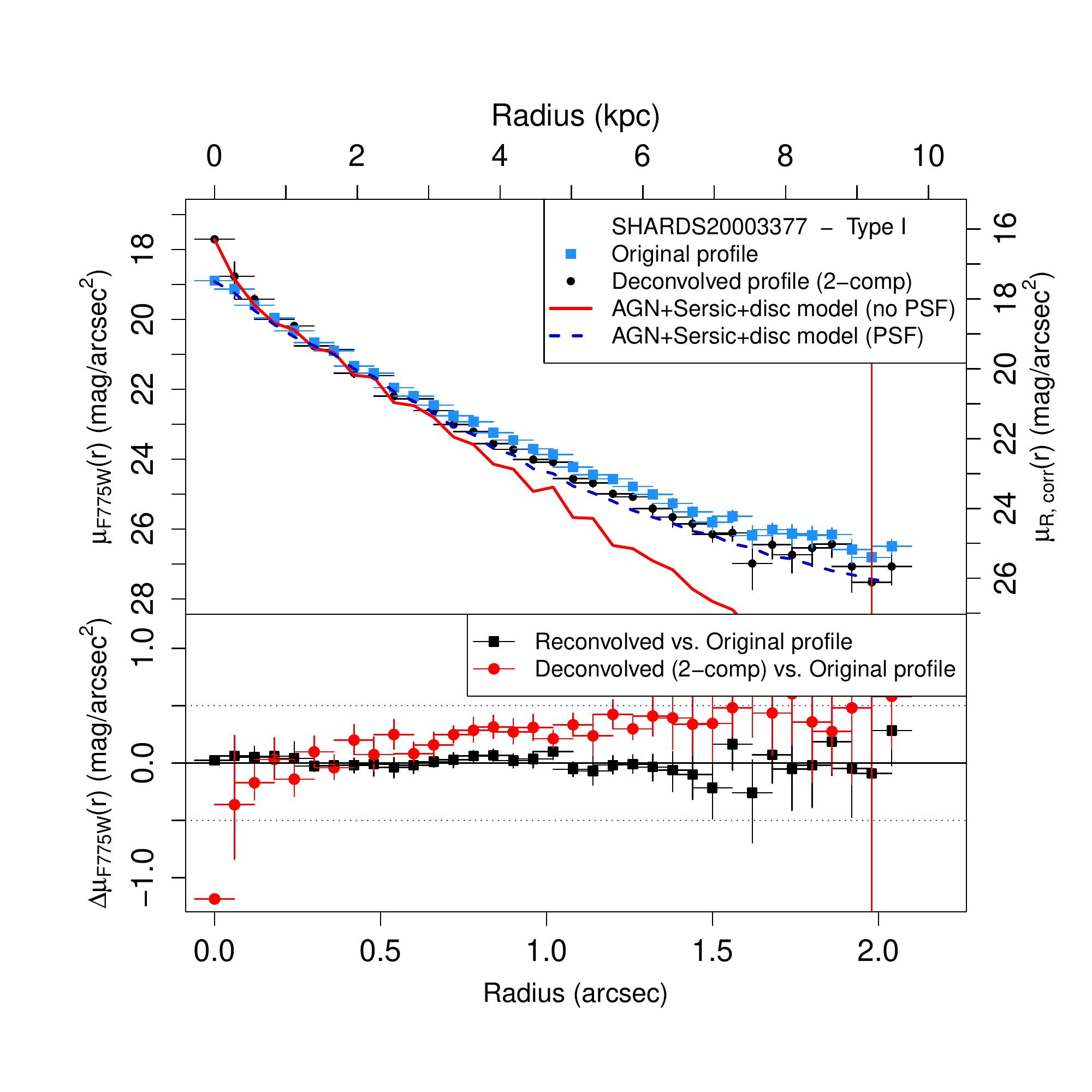}
\end{minipage}
\begin{minipage}{.49\textwidth}
\includegraphics[clip, trim=0.1cm 0.1cm 1cm 0.1cm, width=0.95\textwidth]{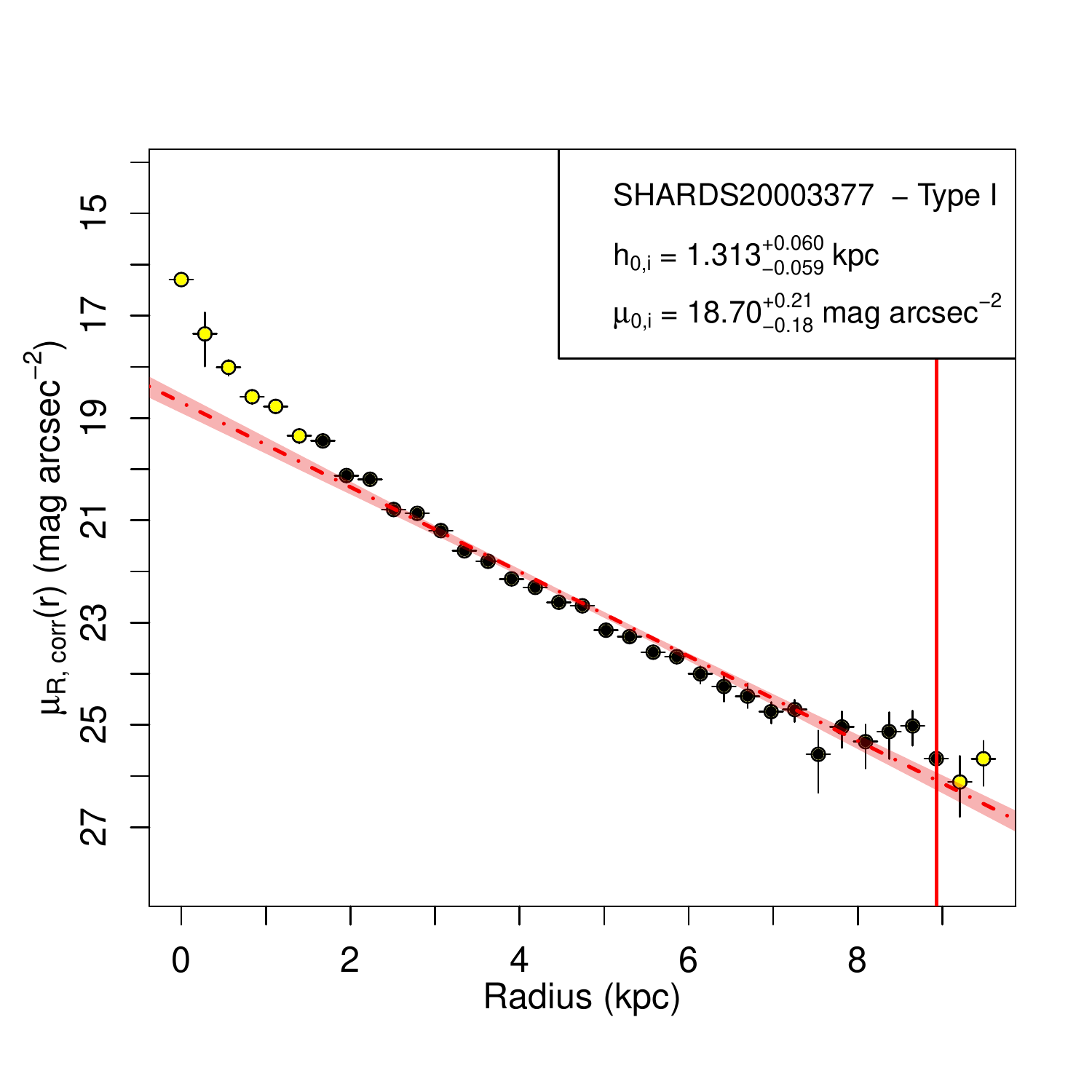}
\end{minipage}%

\vspace{-0.5cm}}
\caption[]{See caption of Fig.1. [\emph{Figure  available in the online edition}.]}         
\label{fig:img_final}
\end{figure}
\clearpage
\newpage

\textbf{SHARDS20003678:} Small E/S0 galaxy with with a Type-I disc and very low inclination (see Table \ref{tab:fits_psforr}). The profile shows a profile with a bulge + exponential disc structure. Due to the low resolution in this galaxy, the radial extension of the inner bulge is uncertain.

\begin{figure}[!h]
{\centering
\vspace{-0cm}

\begin{minipage}{.5\textwidth}
\hspace{1.2cm}
\begin{overpic}[width=0.8\textwidth]
{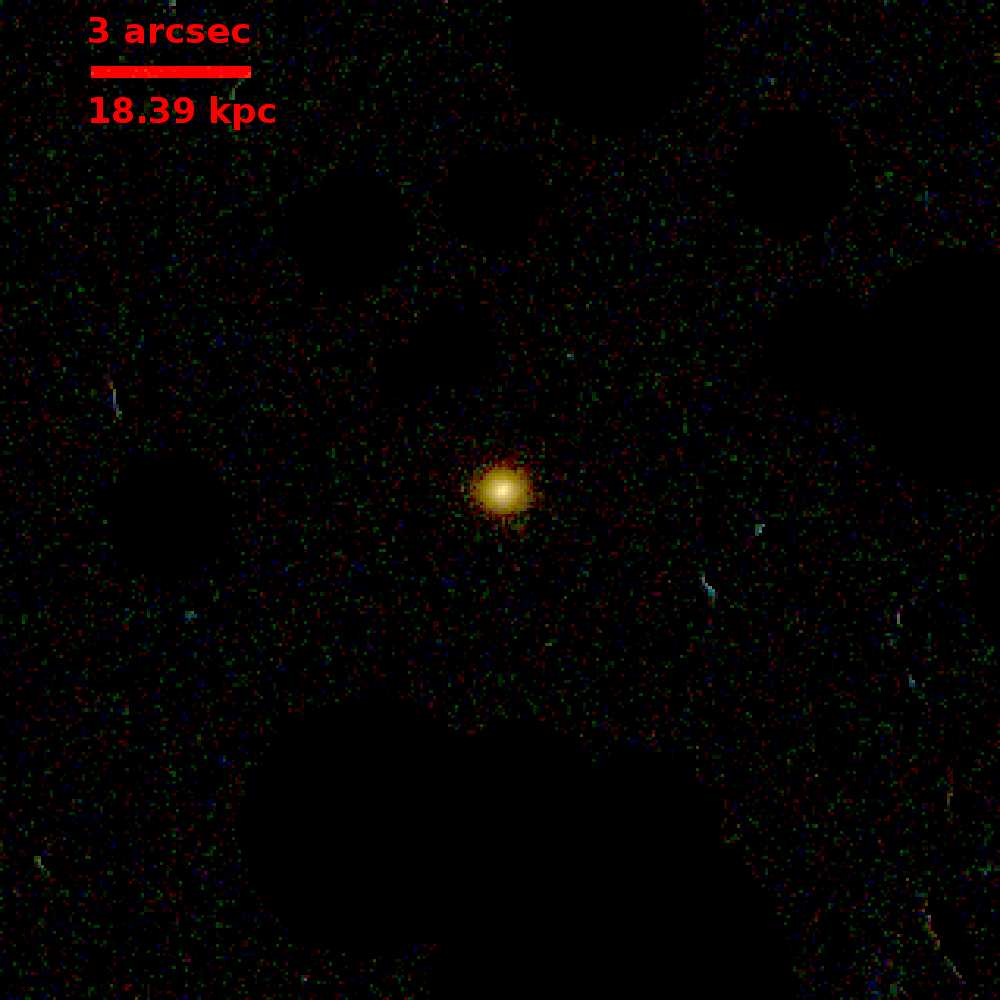}
\put(110,200){\color{yellow} \textbf{SHARDS20003678}}
\put(110,190){\color{yellow} \textbf{z=0.5039}}
\put(110,180){\color{yellow} \textbf{E/S0}}
\end{overpic}
\vspace{-1cm}
\end{minipage}%
\begin{minipage}{.5\textwidth}
\includegraphics[clip, trim=1cm 1cm 1.5cm 1.5cm, width=\textwidth]{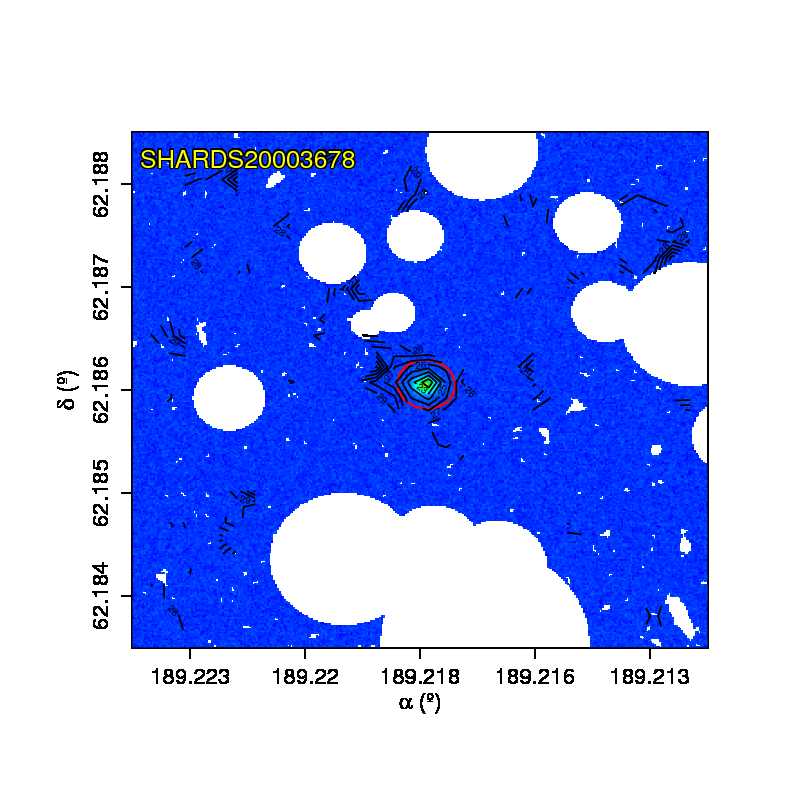}\vspace{-1cm}
\end{minipage}%

\begin{minipage}{.49\textwidth}
\includegraphics[clip, trim=0.1cm 0.1cm 0.1cm 0.1cm, width=\textwidth]{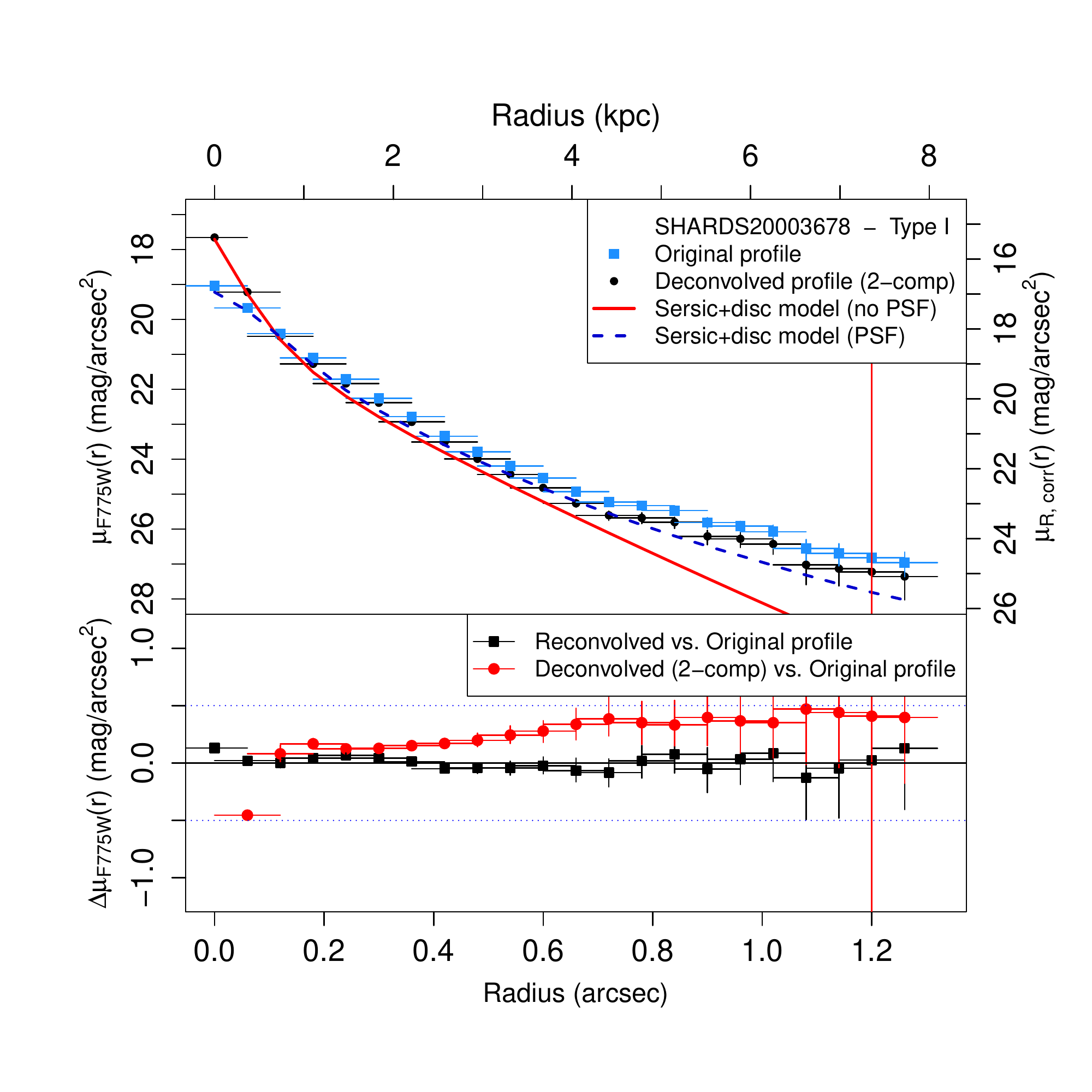}
\end{minipage}
\begin{minipage}{.49\textwidth}
\includegraphics[clip, trim=0.1cm 0.1cm 1cm 0.1cm, width=0.95\textwidth]{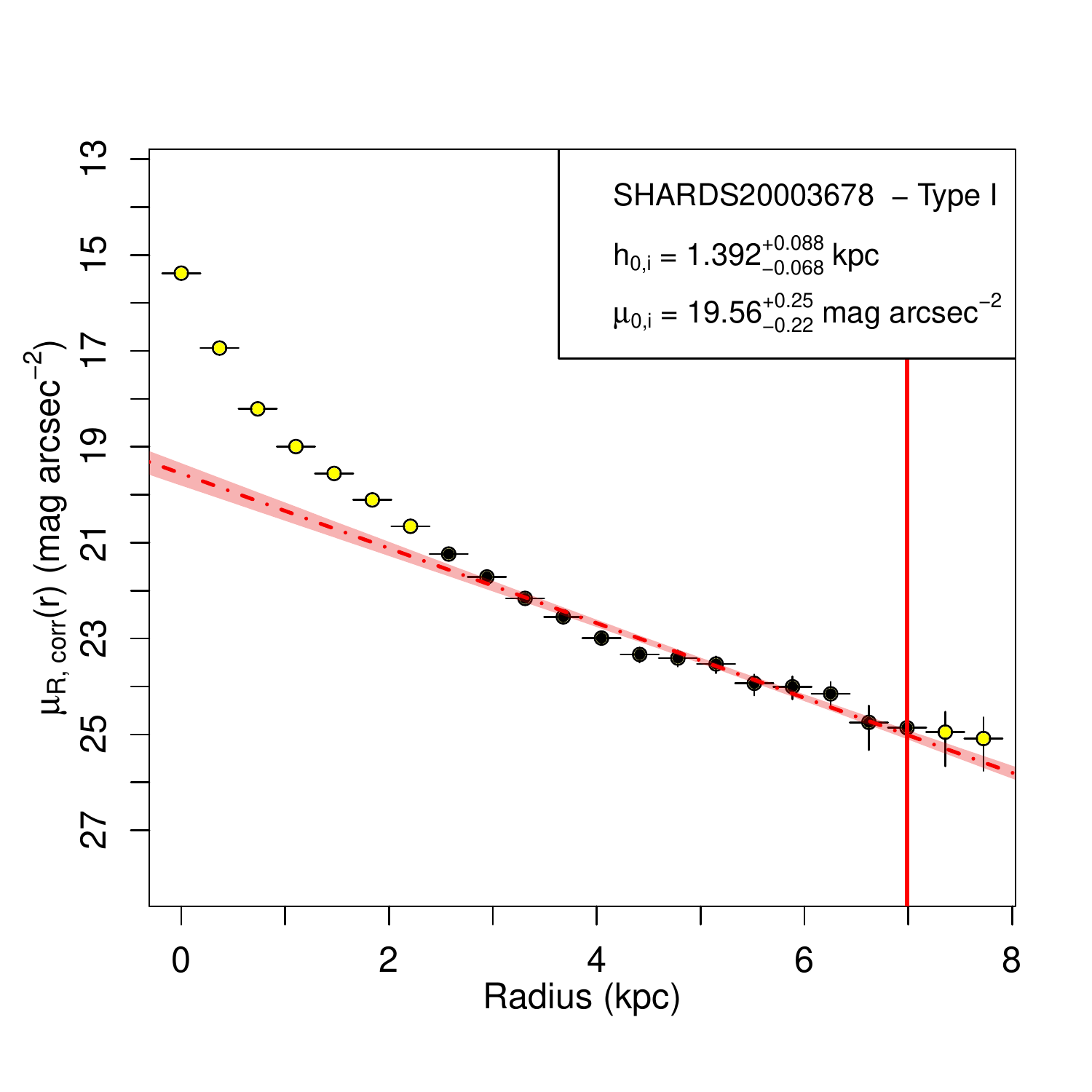}
\end{minipage}%

\vspace{-0.5cm}}
\caption[]{See caption of Fig.1. [\emph{Figure  available in the online edition}.]}         
\label{fig:img_final}
\end{figure}
\clearpage
\newpage

\textbf{SHARDS20004359} S0 galaxy with a Type-I profile. The object presents low inclination (see Table \ref{tab:fits_psforr}). Manual masking was applied to the image in many regions to avoid contamination from a tadpole irregular galaxy located to the North. The model greatly reduces residuals when adding an exponential profile to the free Sersic model, so we discarded the elliptical morphology for it. The profile generates too noisy PDDs to detect any statistically significant break. The visual analysis of the isophotes reveals a distortion between $\mu_{\mathrm{F775W}}\sim 25 - 26$ \magarc, that coincides with a step observed in the surface brightness profile at $\sim 14 - 16$ kpc. Finally the object was classified as a Type-I disc. 

\begin{figure}[!h]
{\centering
\vspace{-0cm}

\begin{minipage}{.5\textwidth}
\hspace{1.2cm}
\begin{overpic}[width=0.8\textwidth]
{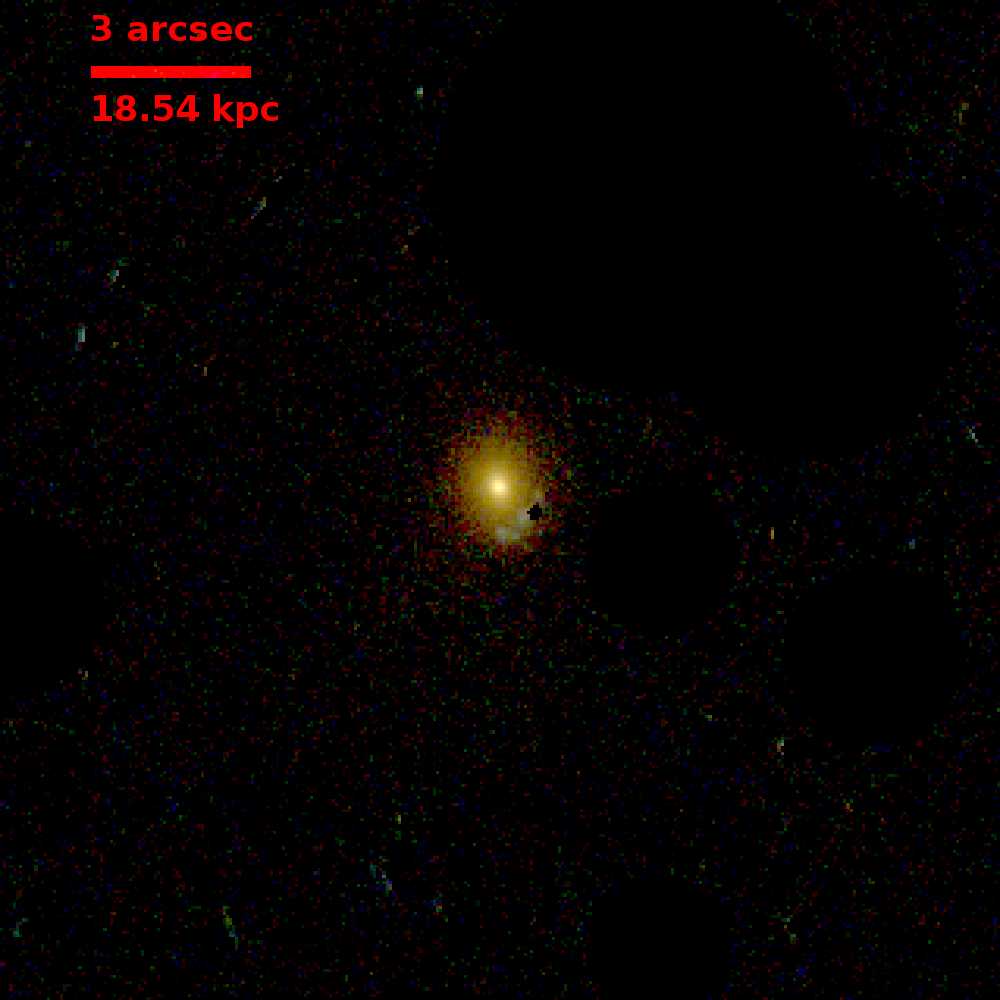}
\put(110,200){\color{yellow} \textbf{SHARDS20004359}}
\put(110,190){\color{yellow} \textbf{z=0.5120}}
\put(110,180){\color{yellow} \textbf{S0}}
\end{overpic}
\vspace{-1cm}
\end{minipage}%
\begin{minipage}{.5\textwidth}
\includegraphics[clip, trim=1cm 1cm 1.5cm 1.5cm, width=\textwidth]{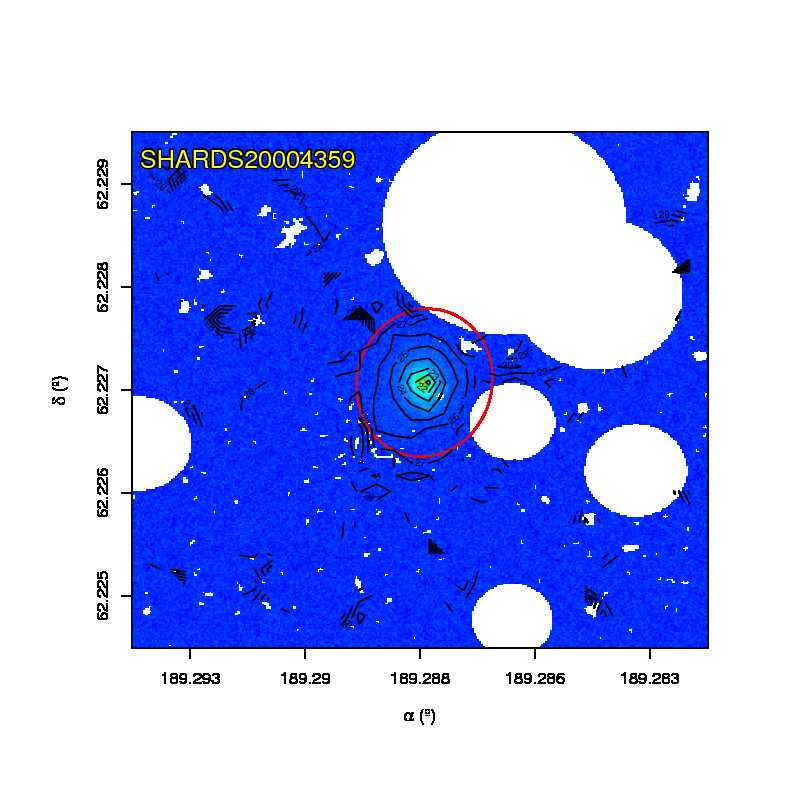}\vspace{-1cm}
\end{minipage}%

\begin{minipage}{.49\textwidth}
\includegraphics[clip, trim=0.1cm 0.1cm 0.1cm 0.1cm, width=\textwidth]{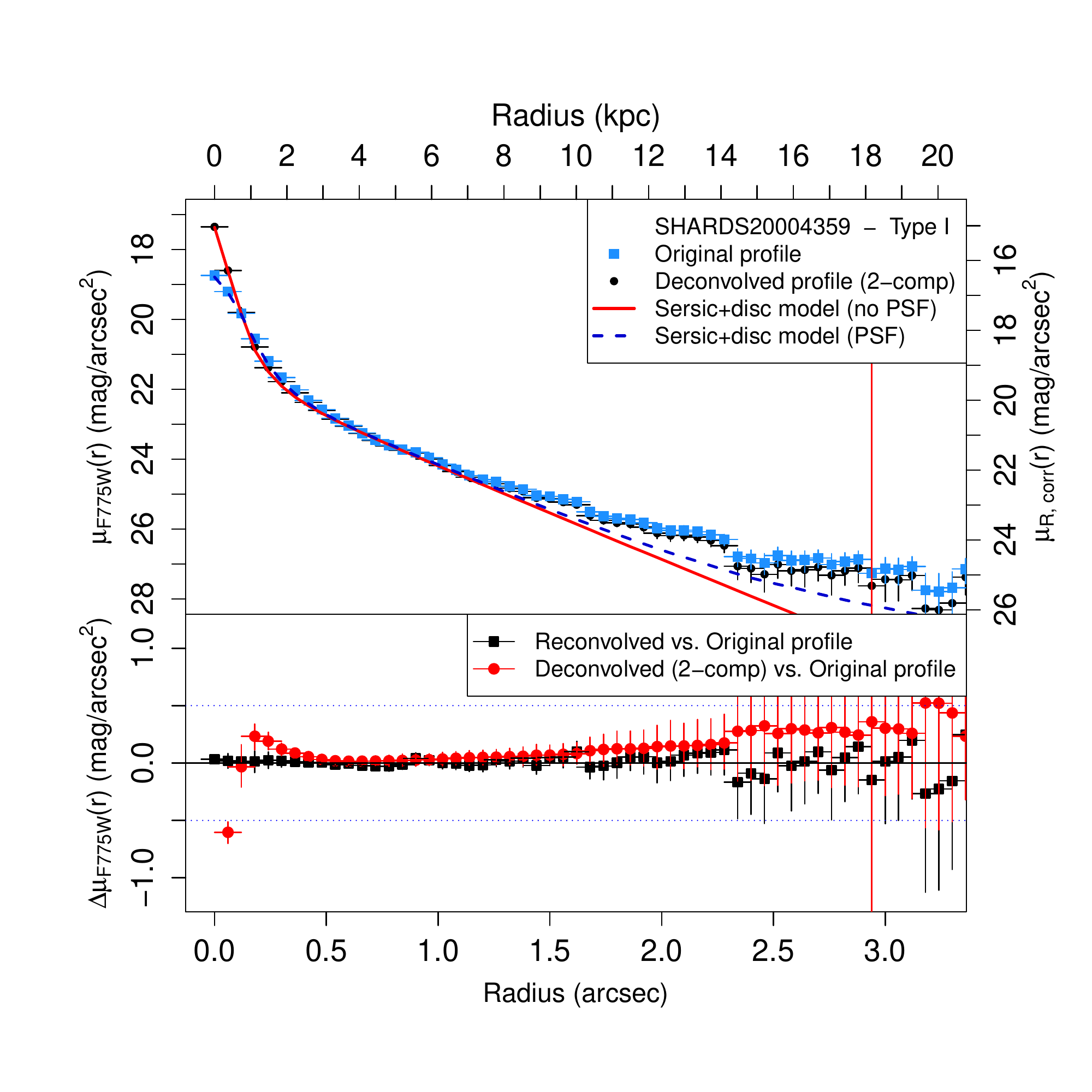}
\end{minipage}
\begin{minipage}{.49\textwidth}
\includegraphics[clip, trim=0.1cm 0.1cm 1cm 0.1cm, width=0.95\textwidth]{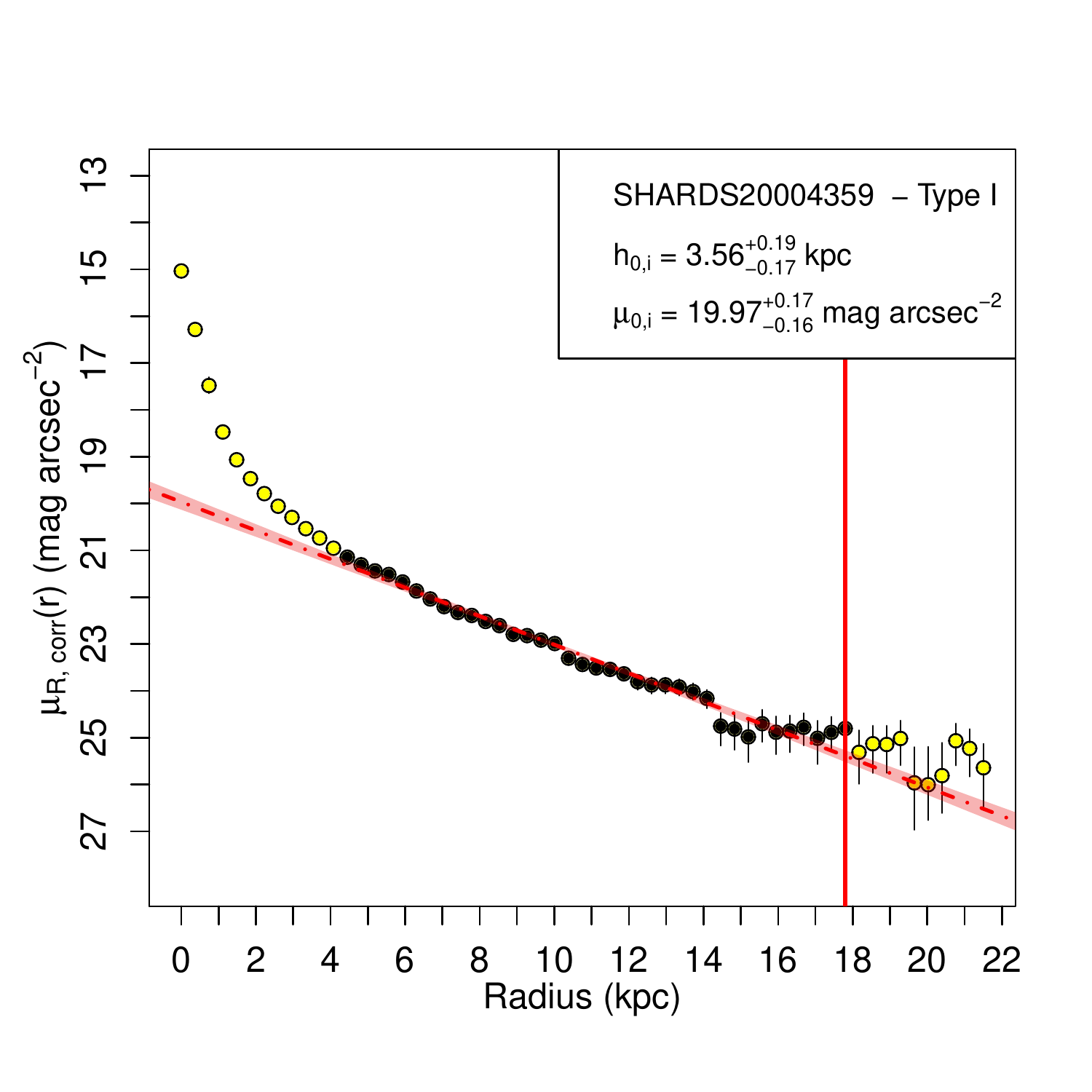}
\end{minipage}%

\vspace{-0.5cm}}
\caption[]{See caption of Fig.1. [\emph{Figure  available in the online edition}.]}         
\label{fig:img_final}
\end{figure}
\clearpage
\newpage

\textbf{SHARDS20004420:} E/S0 galaxy with a Type-I profile. The object presents low inclination (see Table \ref{tab:fits_psforr}). Manual masking was applied to several small field objects to avoid contamination. The innermost part of the galaxy present higher ellipticity that the outer parts when performing visual inspection of the image. The model greatly improves when adding an exponential disc to a free Sersic profile, so we classified it as an E/S0 instead of an elliptical. The PDDs of $h$ and $\mu_{0}$ present a two peaked distribution, although no significant no significant break is detected. Notice the small spike at $\sim 14$ kpc. 

\begin{figure}[!h]
{\centering
\vspace{-0cm}

\begin{minipage}{.5\textwidth}
\hspace{1.2cm}
\begin{overpic}[width=0.8\textwidth]
{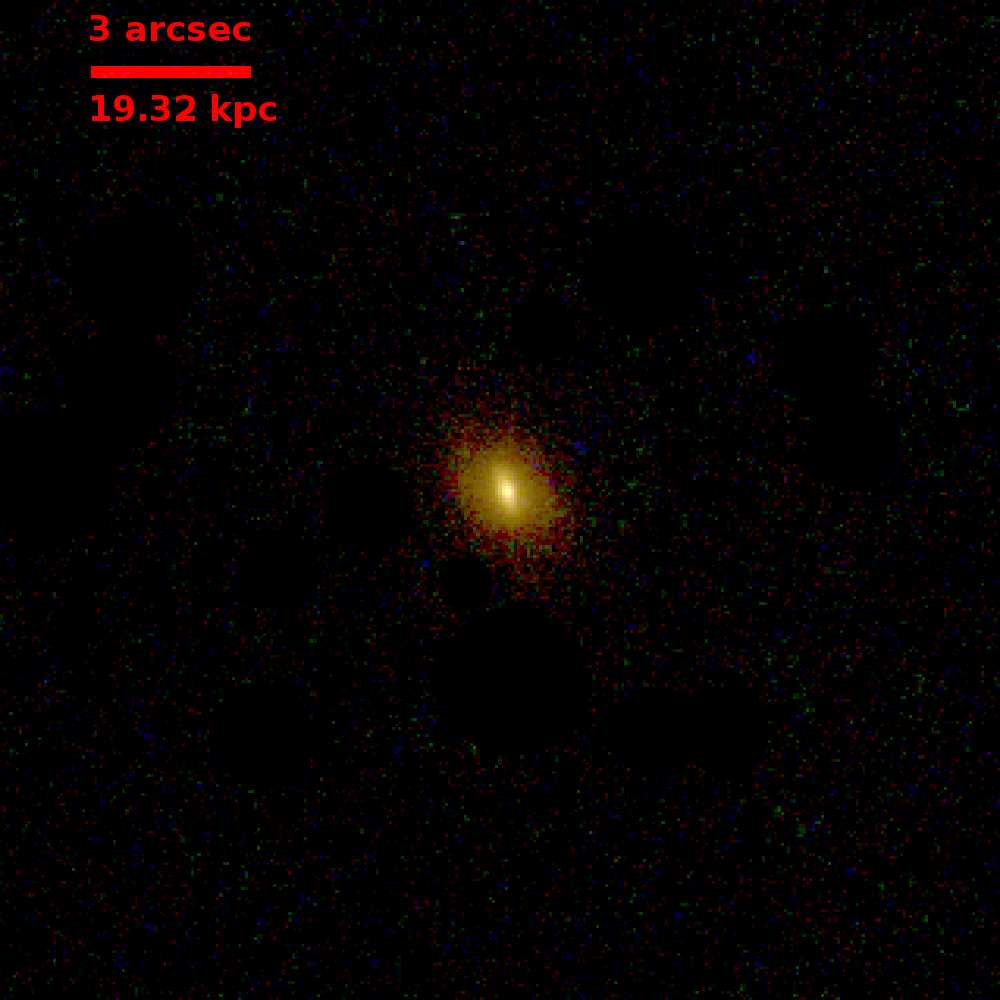}
\put(110,200){\color{yellow} \textbf{SHARDS20004420}}
\put(110,190){\color{yellow} \textbf{z=0.5556}}
\put(110,180){\color{yellow} \textbf{E/S0}}
\end{overpic}
\vspace{-1cm}
\end{minipage}%
\begin{minipage}{.5\textwidth}
\includegraphics[clip, trim=1cm 1cm 1.5cm 1.5cm, width=\textwidth]{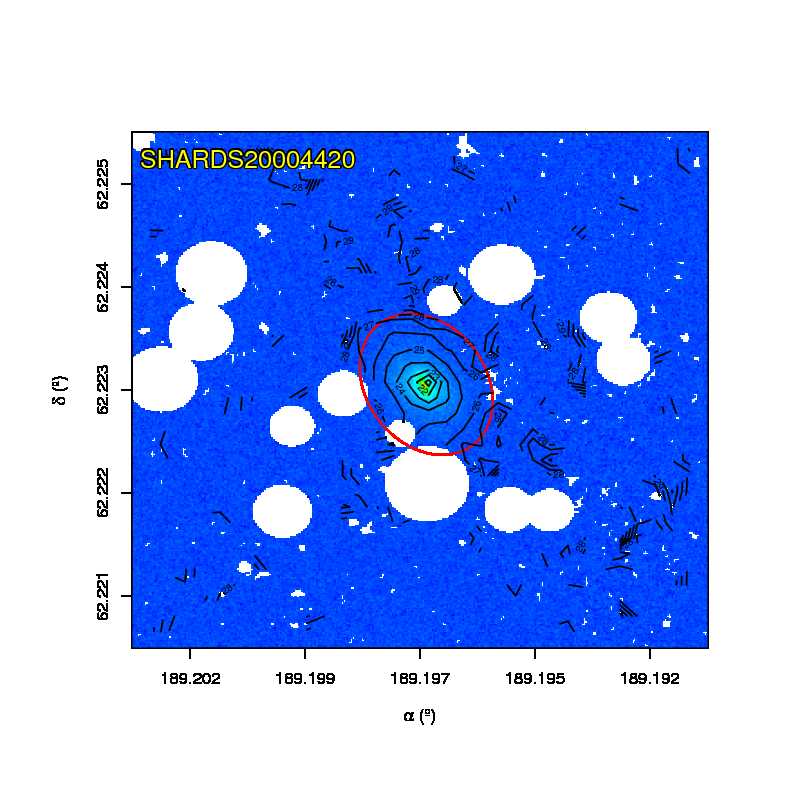}\vspace{-1cm}
\end{minipage}%

\begin{minipage}{.49\textwidth}
\includegraphics[clip, trim=0.1cm 0.1cm 0.1cm 0.1cm, width=\textwidth]{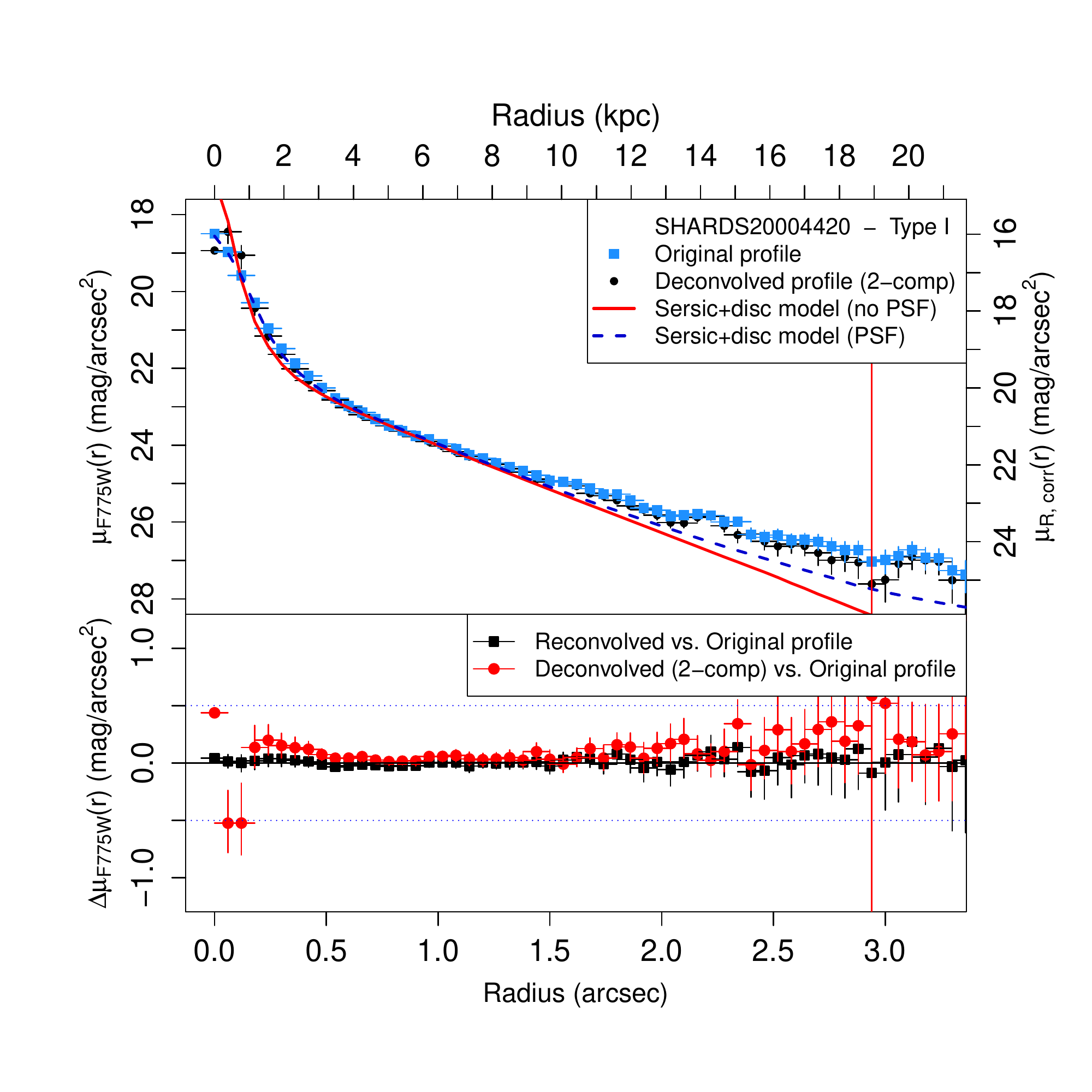}
\end{minipage}
\begin{minipage}{.49\textwidth}
\includegraphics[clip, trim=0.1cm 0.1cm 1cm 0.1cm, width=0.95\textwidth]{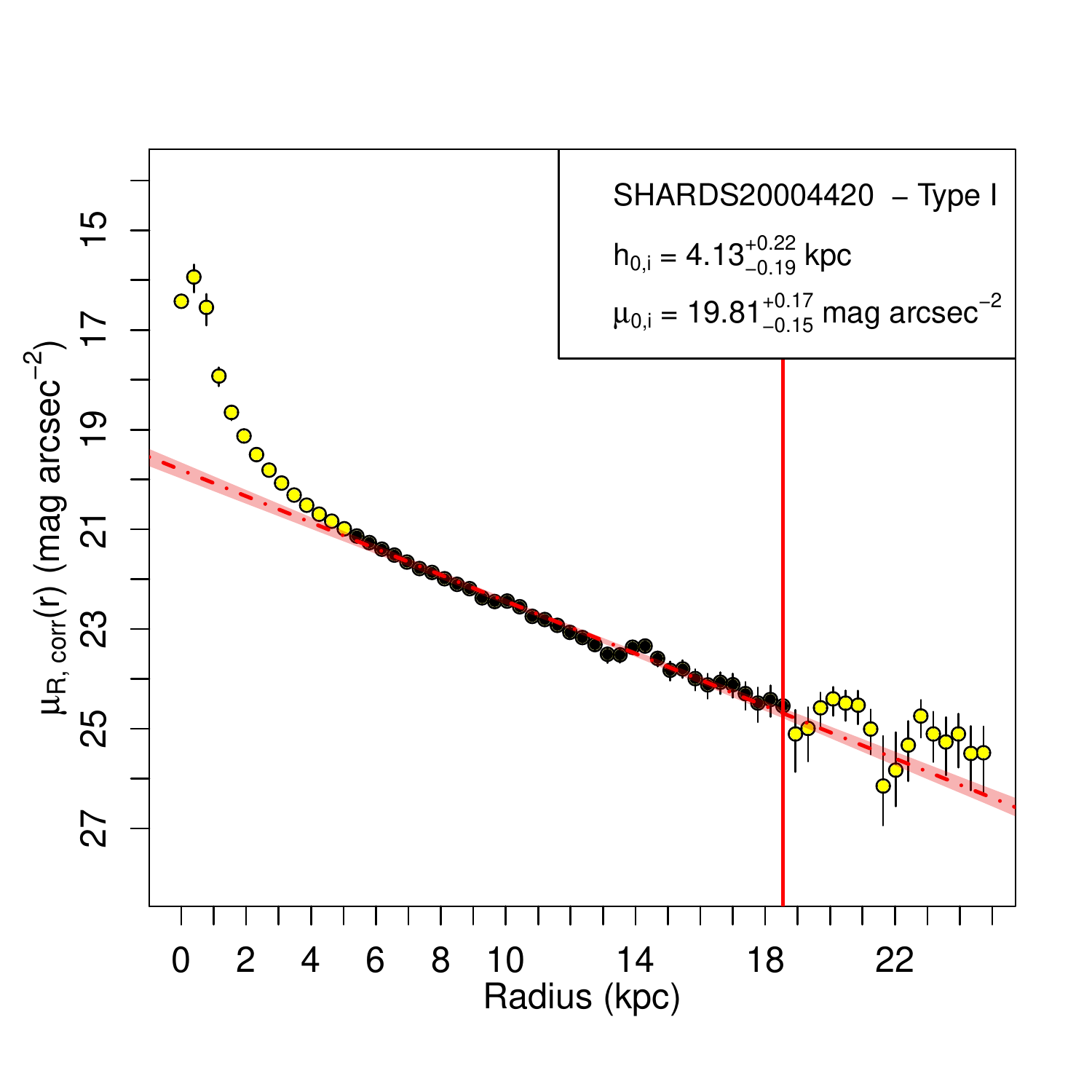}
\end{minipage}%

\vspace{-0.5cm}}
\caption[]{See caption of Fig.1. [\emph{Figure  available in the online edition}.]}         
\label{fig:img_final}
\end{figure}
\clearpage
\newpage

\textbf{SHARDS20004440:} E/S0 galaxy with a Type-I profile. It was flagged as an AGN source (see Sect.\,\ref{Subsec:AGN}). It has a low inclination (see Table \ref{tab:fits_psforr}). Multiple and extensive masking was applied to the outer parts of the galaxy, where tiny spot-like objects were detected via smoothing and re-masking. The excess of light detected before PSF correction appears to be caused by the the dispersed light from the centre.  The automatic break analysis does not reveal any significant breaks. 

\begin{figure}[!h]
{\centering
\vspace{-0cm}

\begin{minipage}{.5\textwidth}
\hspace{1.2cm}
\begin{overpic}[width=0.8\textwidth]
{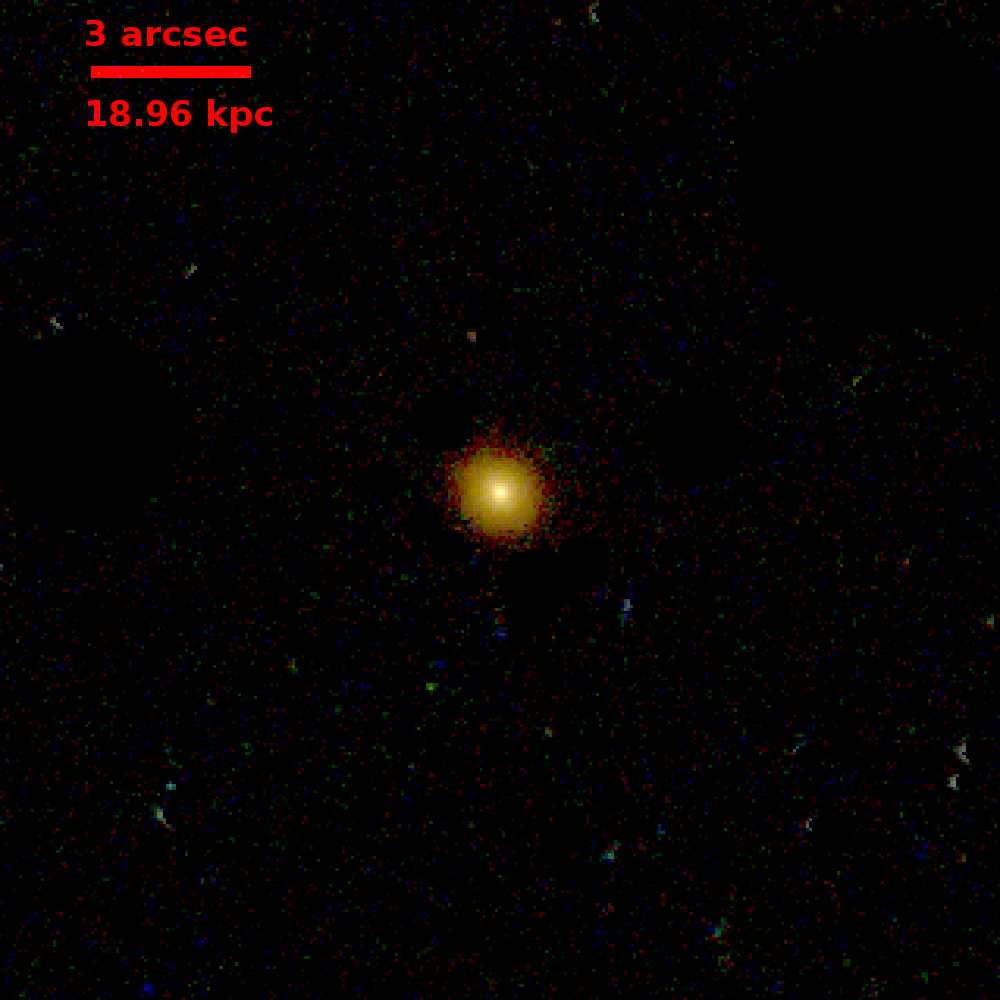}
\put(110,200){\color{yellow} \textbf{SHARDS20004440}}
\put(110,190){\color{yellow} \textbf{z=0.5337}}
\put(110,180){\color{yellow} \textbf{E/S0}}
\end{overpic}
\vspace{-1cm}
\end{minipage}%
\begin{minipage}{.5\textwidth}
\includegraphics[clip, trim=1cm 1cm 1.5cm 1.5cm, width=\textwidth]{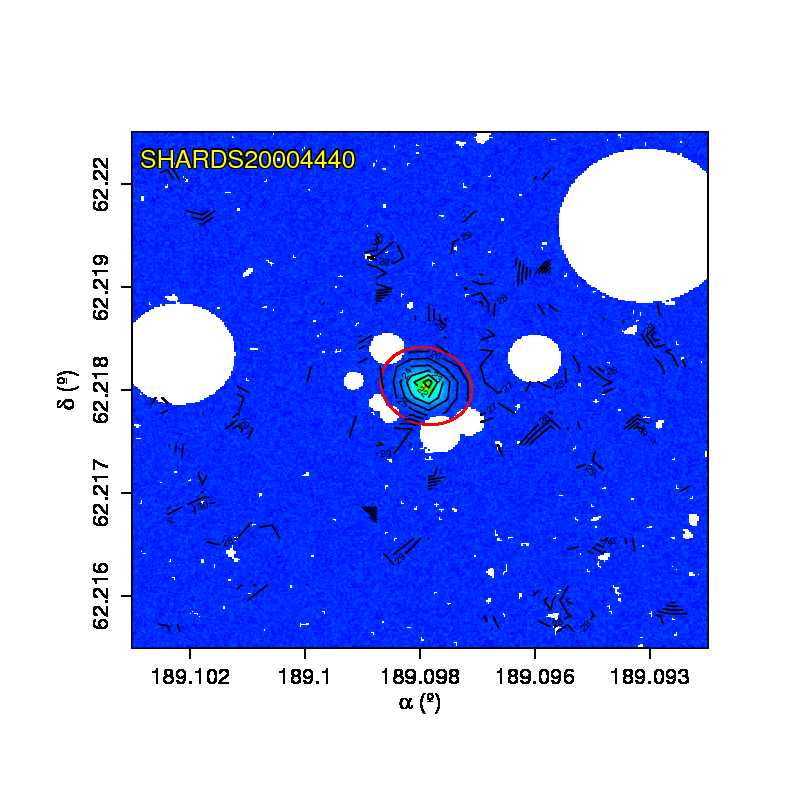}\vspace{-1cm}
\end{minipage}%

\begin{minipage}{.49\textwidth}
\includegraphics[clip, trim=0.1cm 0.1cm 0.1cm 0.1cm, width=\textwidth]{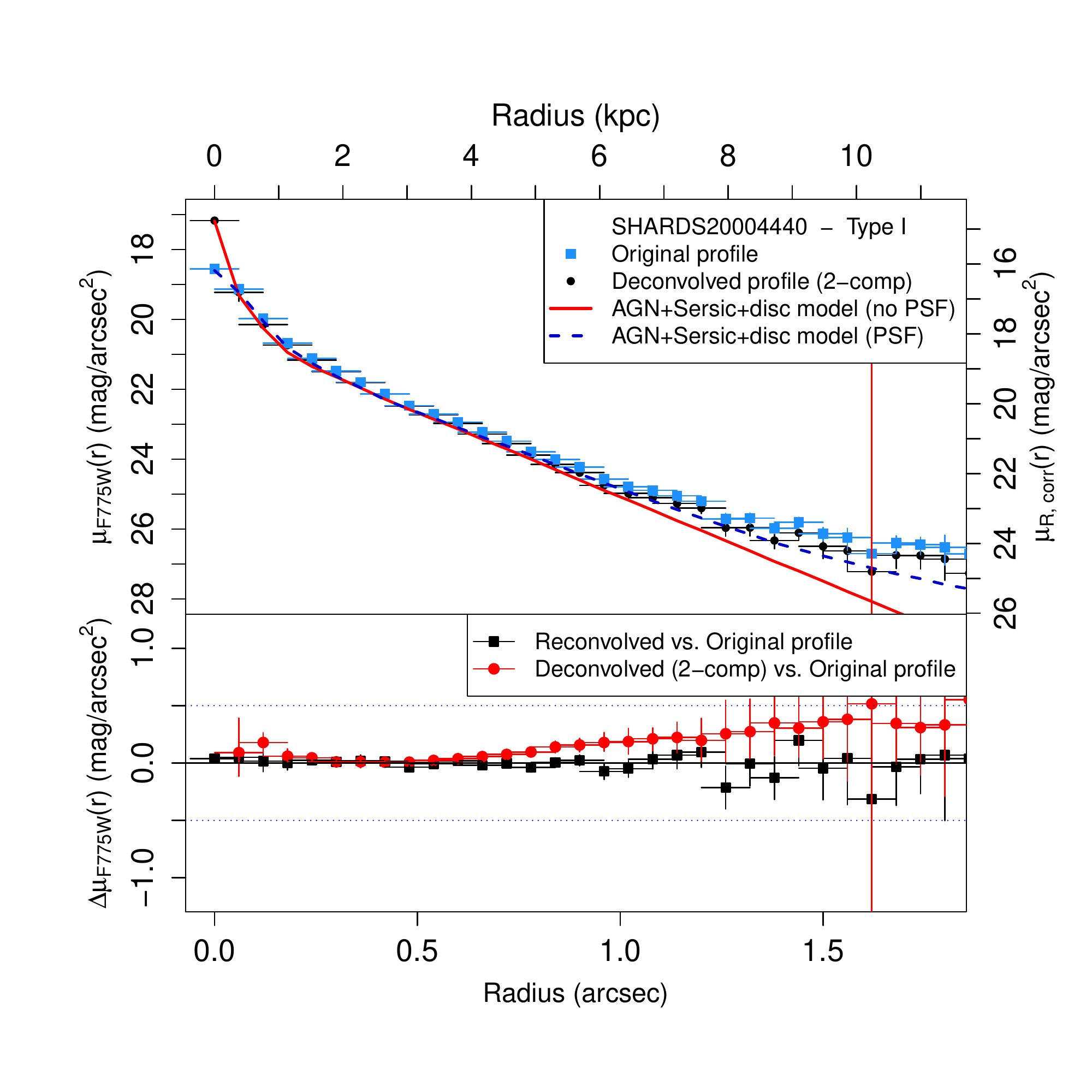}
\end{minipage}
\begin{minipage}{.49\textwidth}
\includegraphics[clip, trim=0.1cm 0.1cm 1cm 0.1cm, width=0.95\textwidth]{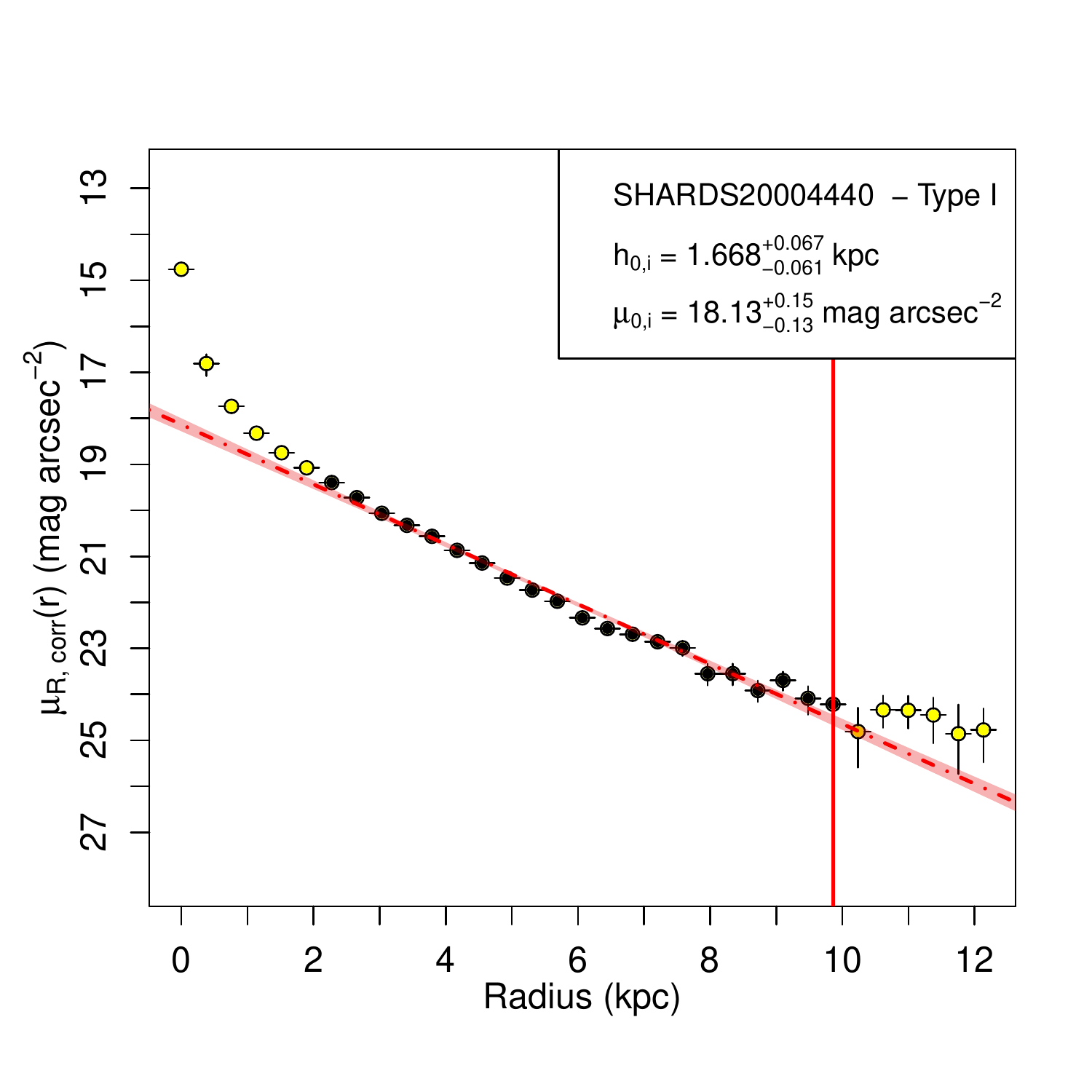}
\end{minipage}%

\vspace{-0.5cm}}
\caption[]{See caption of Fig.1. [\emph{Figure  available in the online edition}.]}         
\label{fig:img_final}
\end{figure}
\clearpage
\newpage

\textbf{SHARDS20011817:} Small S0 galaxy with a Type-I profile. The orientation is edge-on (see Table \ref{tab:fits_psforr}). No masking was needed into the fitting region. The surface brightness profile presents a small bulge + exponential disc distribution. The disc appears to be clear and featureless. {\tt{Elbow}} does not reveal any significant breaks. 

\begin{figure}[!h]
{\centering
\vspace{-0cm}

\begin{minipage}{.5\textwidth}
\hspace{1.2cm}
\begin{overpic}[width=0.8\textwidth]
{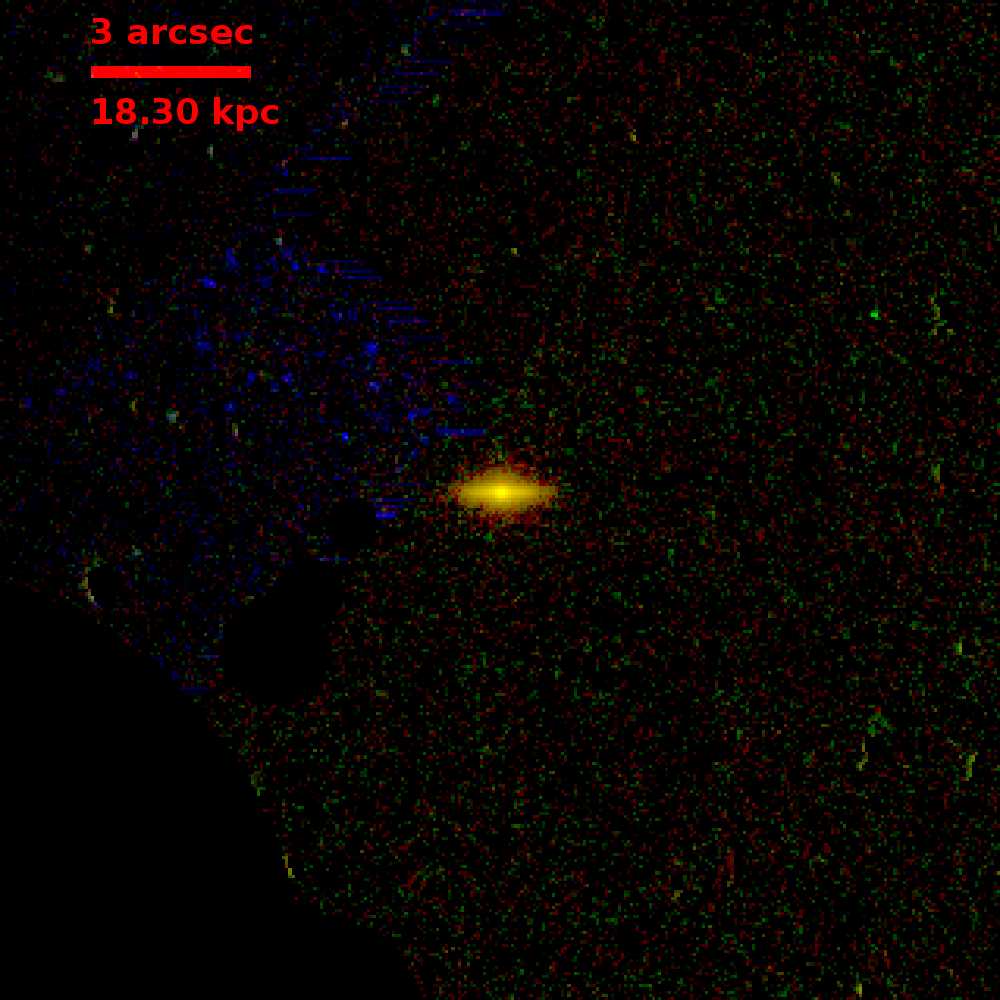}
\put(110,200){\color{yellow} \textbf{SHARDS20011817}}
\put(110,190){\color{yellow} \textbf{z=0.50}}
\put(110,180){\color{yellow} \textbf{S0}}
\end{overpic}
\vspace{-1cm}
\end{minipage}%
\begin{minipage}{.5\textwidth}
\includegraphics[clip, trim=1cm 1cm 1.5cm 1.5cm, width=\textwidth]{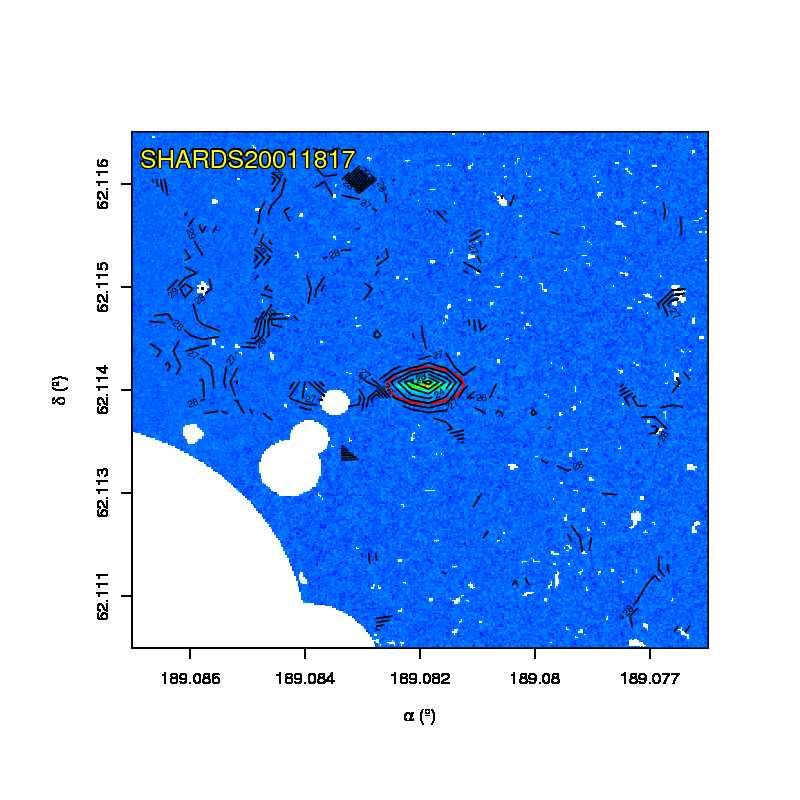}\vspace{-1cm}
\end{minipage}%

\begin{minipage}{.49\textwidth}
\includegraphics[clip, trim=0.1cm 0.1cm 0.1cm 0.1cm, width=\textwidth]{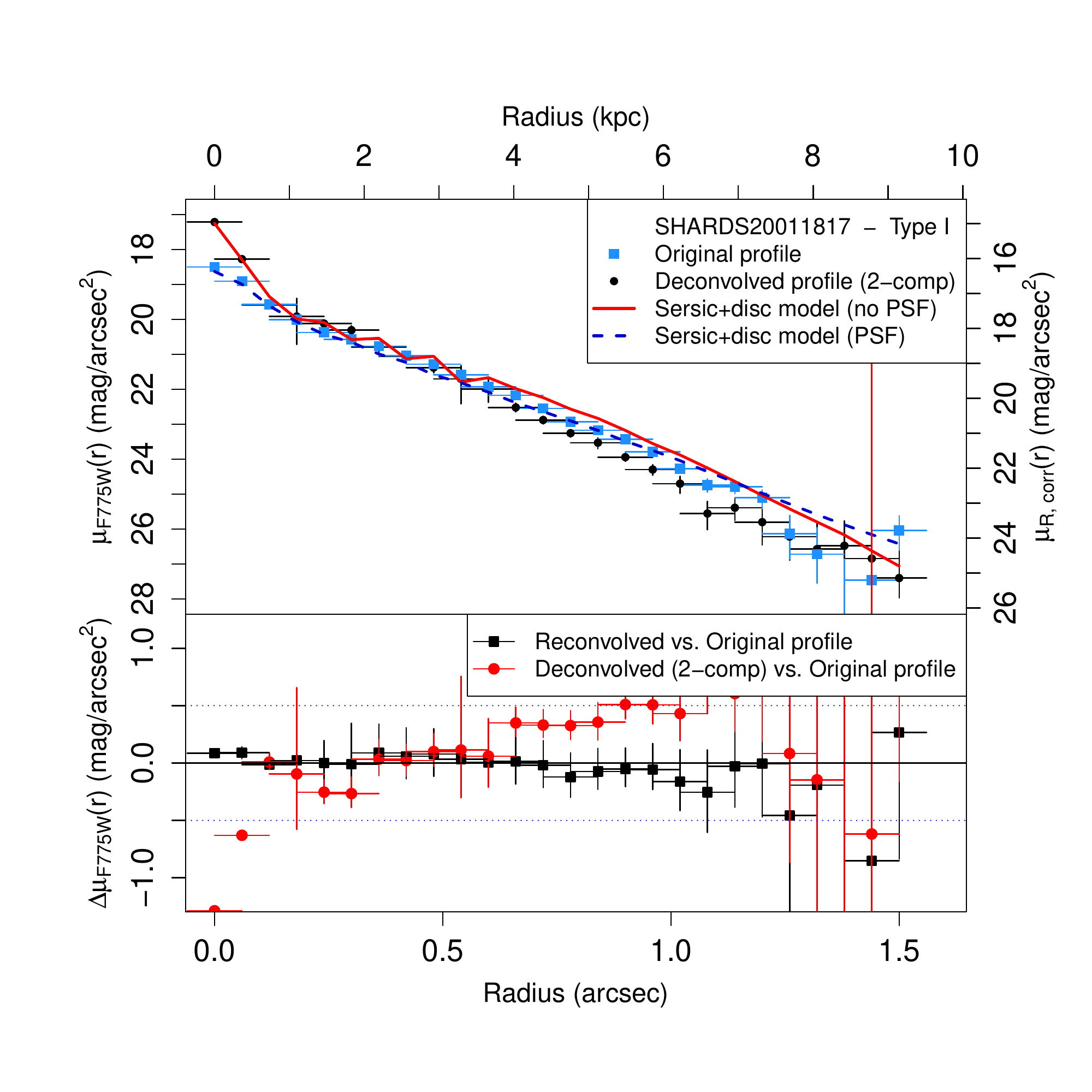}
\end{minipage}
\begin{minipage}{.49\textwidth}
\includegraphics[clip, trim=0.1cm 0.1cm 1cm 0.1cm, width=0.95\textwidth]{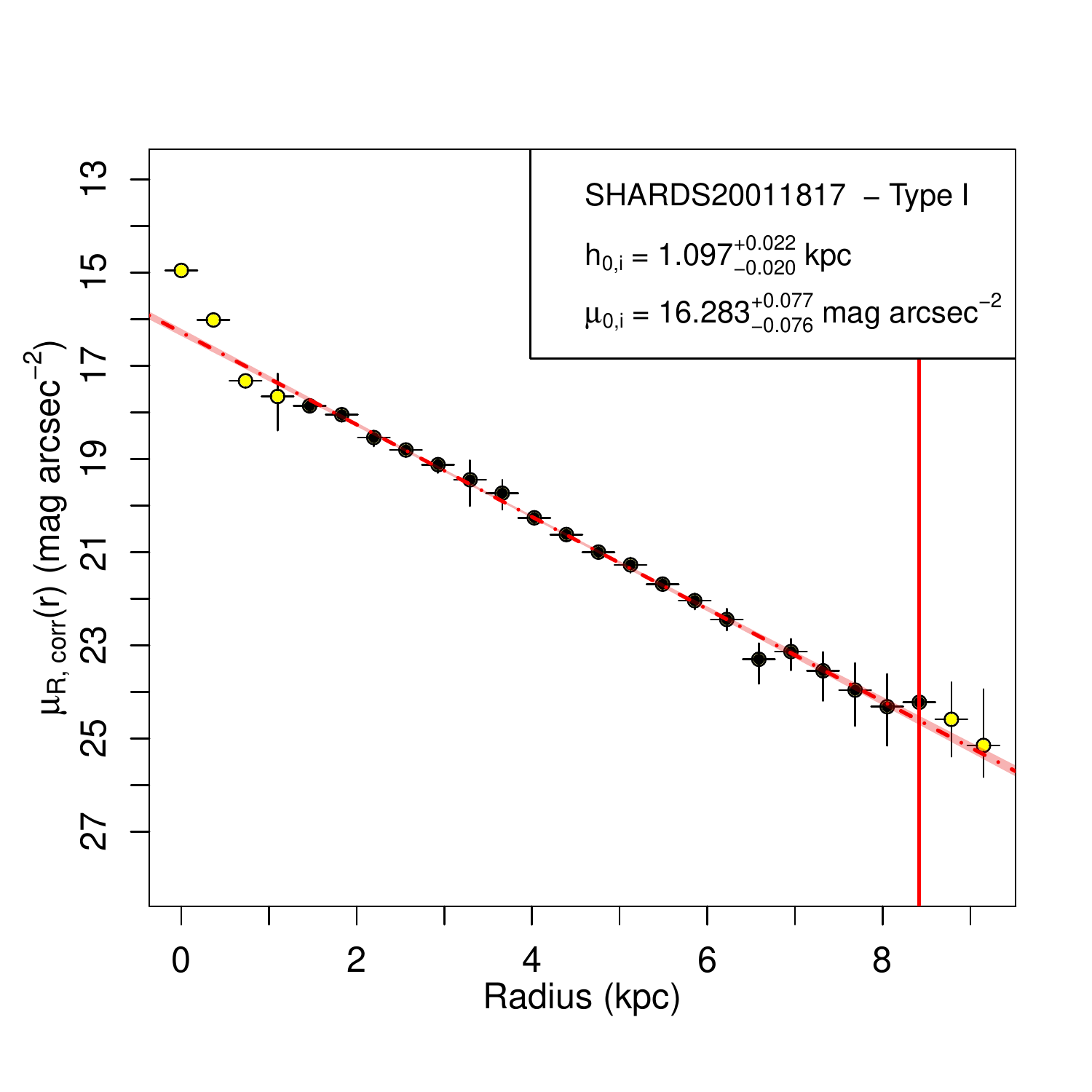}
\end{minipage}%

\vspace{-0.5cm}}
\caption[]{See caption of Fig.1. [\emph{Figure  available in the online edition}.]}         
\label{fig:img_final}
\end{figure}

\end{appendix}

\clearpage
\newpage
\twocolumn

\small